\documentstyle[prc,aps,epsf,eqsecnum,mbf]{revtex}
\def\eq#1{Eq.\ (\ref{#1})}

\def\bS{{\mbf S}}

\def\bL{{\mbf L}}
\def\bP{{\mbf P}}
\def\bR{{\mbf R}}
\def\bS{{\mbf S}}

\def\ba{{\mbf a}}
\def\bb{{\mbf b}}

\def\bk{{\mbf k}}
\def\bn{{\mbf n}}
\def\bp{{\mbf p}}
\def\bq{{\mbf q}}

\def\bfsigma{{\mbf \sigma}}
\def\bftau{{\mbf \tau}}

\def\CA{{\cal A}}

\def\CM{{\cal M}}
\def\CO{{\cal O}}
\def\CP{{\cal P}}
\def\CR{{\cal R}}

\def\CT{{\cal T}}

\def\CY{{\cal Y}}
\def\CZ{{\cal Z}}

\def\SM3{\Sigma N (3/2)}
\def\SN1{\Sigma N (1/2)}

\def\Sp{\Sigma^- p}
\def\Sn{\Sigma^0 n}
\def\Ln{\Lambda n}
\def\TS1{\hbox{}^3S_1}
\def\TD1{\hbox{}^3D_1}

\def\aagp{{\alpha \alpha^\prime}}

\def\lam#1{{#1 \over \lambda}}

\def\ttau{\bftau_1 \cdot \bftau_2}
\def\ttaup{(1+\ttau) P_F}
\def\ssigma{\bfsigma_1 \cdot \bfsigma_2}

\def\tint{{\rm int}}
\def\tlab{{\rm lab}}

\def\tcentral{{\rm central}}

\begin{document}
\draft
\preprint{KUNS-1703 / January, 2001}

\title   {
A Realistic Description of Nucleon-Nucleon and Hyperon-Nucleon
Interactions \\
in the $\mbf{SU_6}$ Quark Model
         }

\author  {
Y. Fujiwara$^1$\footnotemark[1]\footnotetext{$*$ E-mail address:
fujiwara@ruby.scphys.kyoto-u.ac.jp},
T. Fujita$^1$\footnotemark[2]
\footnotetext{$\hbox{}^\dagger$ Present address: Japan
Meteorological Agency, Chiyoda-ku Tokyo 100, Japan},
 M. Kohno$^2$, C. Nakamoto$^3$,
and Y. Suzuki$^4$
         }
\address {
$^1$Department of Physics, Kyoto University,
Kyoto 606-8502, Japan \\
$^2$Physics Division, Kyushu Dental College,
Kitakyushu 803-8580, Japan \\
$^3$Suzuka National College of Technology,
Suzuka 510-0294, Japan \\
$^4$Department of Physics,
Niigata University, Niigata 950-2181, Japan
         }

\maketitle

\bigskip
\bigskip

\begin{abstract}
We upgrade a $SU_6$ quark-model description for the nucleon-nucleon
and hyperon-nucleon interactions by improving the effective
meson-exchange potentials acting between quarks.
For the scalar- and vector-meson exchanges,
the momentum-dependent higher-order term is incorporated
to reduce the attractive effect of the central interaction
at higher energies. The single-particle potentials of the nucleon
and $\Lambda$, predicted by the $G$-matrix calculation,
now have proper repulsive behavior in the momentum
region $q_1=5~\hbox{-}~20~\hbox{fm}^{-1}$. A moderate contribution
of the spin-orbit interaction from the scalar-meson exchange
is also included. As to the vector mesons, a dominant contribution 
is the quadratic spin-orbit force generated from the $\rho$-meson
exchange. The nucleon-nucleon phase shifts at the non-relativistic energies
up to  $T_{\rm lab}=350~\hbox{MeV}$ are
greatly improved especially for the $\hbox{}^3E$ states.
The low-energy observables of the nucleon-nucleon and
the hyperon-nucleon interactions
are also reexamined. The isospin symmetry breaking and the
Coulomb effect are properly incorporated in the particle basis.
The essential feature of the $\Lambda N$-$\Sigma N$ coupling
is qualitatively similar to that obtained from the previous models.
The nuclear saturation properties and the single-particle
potentials of the nucleon, $\Lambda$, and $\Sigma$ are
reexamined through the $G$-matrix calculation. 
The single-particle potential of the $\Sigma$ hyperon is
weakly repulsive in symmetric nuclear matter.
The single-particle spin-orbit strength for the $\Lambda$ particle
is very small, in comparison with that of the nucleons,
due to the strong antisymmetric spin-orbit force generated
from the Fermi-Breit interaction.

\end{abstract}

\bigskip


\section{Introduction}

One of the most important purposes of studying
the nucleon-nucleon ($NN$) and hyperon-nucleon ($YN$) interactions
in the quark model is to understand comprehensively the fundamental
strong interaction in a natural picture,
in which the quark-gluon degree of freedom is relevant
to describe the short-range part of the interaction,
while the medium- and long-range parts of the interaction are dominated
by the meson-exchange processes.
We have recently achieved a simultaneous and realistic
description of the $NN$ and $YN$ interactions
in the resonating-group (RGM) formalism
of the spin-flavor $SU_6$ quark model \cite{RGMFa,RGMFb,PRL,FSS,SCAT}.
In this approach the effective quark-quark ($qq$) interaction
is built by combining a phenomenological quark-confining potential
and the colored version of the Fermi-Breit (FB) interaction
with minimum effective meson-exchange potentials (EMEP)
of scalar and pseudo-scalar meson nonets directly coupled to quarks.
The flavor symmetry breaking for the $YN$ system is explicitly
introduced through the quark-mass dependence of the Hamiltonian.
An advantage of introducing the EMEP at the quark level lies
in the stringent relationship of the flavor dependence
appearing in the various $NN$ and $YN$ interaction pieces.
In this way we can utilize our rich knowledge of the $NN$
interaction to minimize the ambiguity of model parameters,
which is crucial since the present experimental data
for the $YN$ interaction are still very scarce.

In this study we upgrade our model \cite{RGMFa,RGMFb,PRL,FSS,SCAT} by
incorporating such interaction pieces provided
by scalar and vector mesons as the spin-orbit ($LS$),
quadratic spin-orbit ($QLS$),
and the momentum-dependent Bryan-Scott terms.
Introduction of these pieces to the EMEP is primarily
motivated by the insufficient description
of the experimental data by previous models.
First, some discrepancy of the $NN$ phase shifts in previous
models requires the introduction of vector mesons. For example,  
$\hbox{}^3D_2$ phase shift in the model FSS \cite{FSS} is
more attractive than experiment by $10^\circ$ around $T_{\rm lab}\sim
300~\hbox{MeV}$. This implies that the one-pion tensor force
is too strong in our previous models.
In the standard one-boson exchange potentials (OBEP),
the strong one-pion tensor force
is partially weakened by the $\rho$ meson tensor force.
We use the $QLS$ force of vector mesons
from the reasons given below. Furthermore, some phase shifts
of other partial waves deviate from the empirical ones
by a couple of degrees.  
Another improvement is required as for the central
attraction. The $G$-matrix calculation using the quark-exchange
kernel explicitly \cite{GMAT} shows that energy-independent attraction,
dominated by $\epsilon$-meson exchange, is unrealistic,
since in our previous models the single particle (s.p.) potentials
in symmetric nuclear matter show a strongly attractive behavior
in the momentum region $q_1=5~\hbox{-}~20~\hbox{fm}^{-1}$.
We have shown in \cite{LSRGM} that this flaw can be removed
by introducing the momentum-dependent higher-order term
of scalar-meson exchange potentials, the importance of which was
first pointed out by Bryan and Scott \cite{BR67}.
In the higher energy region, the $LS$ term of the scalar mesons
also makes an appreciable contribution, in addition to this
momentum dependent term.

Another purpose of the present investigation is
to examine the charge symmetry breaking (CSB) and
the Coulomb effect from the viewpoint of the quark model.
It is well known that the $\hbox{}^1S_0$ phase shift
of the $pp$ interaction is slightly less attractive than
that of the $np$ interaction. This charge independence
breaking (CIB) is partially explained by the so-called
pion-Coulomb correction \cite{MA00},
which implies 1) the small mass difference
of the neutron and the proton, 2) the mass difference of the
charged pion and the neutral pion, and 3) the Coulomb effect.
Furthermore, it was claimed long ago that the $\Lambda p$ interaction
should be more attractive than the $\Lambda n$ interaction,
since the binding energy of the $0^+$ ground state
of $\hbox{}^4_\Lambda \hbox{He}$ is fairly larger
than that of $\hbox{}^4_\Lambda \hbox{H}$ \cite{JU73}.
The CSB energy of 350 keV in these isodoublet hypernuclei is much larger
than $\sim$ 100 keV CSB effect seen
in the $\hbox{}^3\hbox{H}$-$\hbox{}^3\hbox{He}$ binding energy difference
after the correction of the $pp$ Coulomb energy
in $\hbox{}^3\hbox{He}$ is made.
The early version of the Nijmegen potential \cite{NA73} already focused
on this CSB in the OBEP including the pion-Coulomb correction
and the correct threshold energies of the $\Lambda N$-$\Sigma N$ coupling
in the particle basis.
The RGM calculation using the particle basis is rather cumbersome,
since all the spin-flavor factors of the quark-exchange kernel
should be recalculated by properly incorporating the $z$-components
of the isospin quantum numbers.
Furthermore, there is a problem inherent in the
RGM formalism: the internal energies of the clusters
are usually not properly reproduced when a unique model Hamiltonian
is used. We have given in \cite{GRGM} a convenient
prescription to avoid this problem without spoiling the
exact effect of the Pauli principle.
For the Coulomb effect, we calculate the full exchange kernel without
any approximation.
The pion-Coulomb correction and the correct treatment
of the threshold energies in the particle basis are found to be
very important for the detailed description of the
low-energy observables in the $\Sigma N$ - $\Lambda N$ coupled-channel
problem.

With these renovations of EMEP and the framework, we have again
searched for model parameters in the isospin basis
to fit the most recent result of the $NN$ phase shifts,
the deuteron binding energy, the $\hbox{}^1S_0$ $NN$ scattering length,
and the low-energy $YN$ total cross section data.
This model is named fss2 since it is based on our previous
model FSS \cite{PRL,FSS,SCAT}.
The agreement of the phase-shift
parameters in the $NN$ sector is greatly improved.
The model fss2 shares the good reproduction of the $YN$ scattering
data and the essential features of the $\Lambda N$-$\Sigma N$ coupling
with our previous models \cite{RGMFa,RGMFb,PRL,FSS,SCAT}.
The single-particle (s.p.) potentials of $N$, $\Lambda$
and $\Sigma$ are predicted through the $G$-matrix calculation. \cite{GMAT}
The strength of the s.p. spin-orbit potential is also
examined by using these $G$-matrices \cite{SPLS}.

In the next section we first recapitulate the formulation
of the $(3q)$-$(3q)$ Lippmann-Schwinger RGM (LS-RGM) \cite{LSRGM} and
the $G$-matrix calculation \cite{GMAT} using the explicit
quark-exchange kernel.
Section II\,B introduces a new EMEP Hamiltonian for fss2
in the momentum representation.
This serves to clarify the difference
between the present model fss2 and the previous two models,
FSS and RGM-H \cite{PRL,FSS,SCAT}.
The spatial part of the quark-exchange kernel and the spin-flavor
factors in the EMEP sector are given in Appendices A and B, respectively.
The model parameters determined in the isospin basis
are discussed in Sec.\,II\,C.
Short comments are given in Sec.\,II\,D with respect to the
special treatment in the particle basis, including
the Coulomb force in the momentum representation.
Section III presents results and discussions.
We first discuss in Sec.\,III\,A the $NN$ phase shifts,
differential cross sections and the polarization for the
energies $T_{\rm lab} \leq 800~\hbox{MeV}$.
Special attention is paid to the effect of inelastic channels,
which is not taken into account in the present framework.
The five invariant amplitudes for the $pp$ scattering
are also examined at the highest energy $T_{\rm lab}=800~\hbox{MeV}$,
in order to clarify the behavior of the s.p. potentials
in the asymptotic momentum region and to find a clue to the
missing ingredients in the present framework.
The deuteron properties and the effective-range parameters of
the $NN$ system are discussed in Sec.\,III\,B.
A simple parameterization of the deuteron wave functions
is given in Appendix C.
Sections III\,C, D, and E discuss the phase-shift behavior
and the characteristic features of the $\Sigma^+ p$,
$\Lambda N$, and $\Sigma^- p$ systems, respectively.
In Sec.\,III\,E, the pion-Coulomb correction and
the $\Sigma^- p$ inelastic capture ratios
at rest and in flight are also discussed in the particle basis.
The $YN$ cross sections in the low- and
intermediate-energy regions are discussed in Sec.\,III\,F.
The $G$-matrix calculation using fss2 is presented in Sec.\,III\,G. 
This includes the discussion of the nuclear saturation curve,
the density dependence of the s.p. potentials and the Scheerbaum
factors of the s.p. spin-orbit strength in symmetric nuclear matter.
The final section is devoted to a summary.

\section{Formulation}

\subsection{The Lippmann-Schwinger formalism
for ${\protect\mbf (3q)}$ - ${\protect\mbf (3q)}$ RGM
and the ${\protect\mbf G}$-matrix equation}

A new version of our quark model employs the Hamiltonian
which includes the interactions generated from 
the scalar (S), pseudoscalar (PS) and vector (V)
meson exchange potentials acting between quarks:
\begin{eqnarray}
H=\sum^6_{i=1} \left( m_ic^2+{\bp^2_i \over 2m_i}-T_G \right)
+\sum^6_{i<j} \left( U^{\rm Cf}_{ij}+U^{\rm FB}_{ij}
+\sum_\beta U^{{\rm S}\beta}_{ij}
+\sum_\beta U^{{\rm PS}\beta}_{ij}
+\sum_\beta U^{{\rm V}\beta}_{ij} \right).
\label{fm1}
\end{eqnarray}
Here $U^{\rm Cf}_{ij}$ is a confinement
potential with a quadratic power law, and $U^{\rm FB}_{ij}$ is
the full FB interaction with explicit quark-mass dependence.
It is important to note that this confinement potential
gives a vanishing contribution to the baryon-baryon interaction,
since we assume $(0s)^3$ harmonic oscillator wave functions
with a common width parameter $b$ for the internal cluster
wave functions. Also, all the contributions from the FB
interaction are generated from the quark-exchange diagrams,
since we assume color-singlet cluster wave functions.
These features are all explained
in our previous publications \cite{FSS}.
When the calculations are made in the particle basis,
the Coulomb force is also introduced at the quark level.
The RGM equation for the parity-projected relative
wave function $\chi^\pi_\alpha(\bR)$ is derived
from the variational
principle $\langle\delta\Psi|E-H|\Psi\rangle=0$,
and it reads \cite{FSS}
\begin{equation}
\left[~\varepsilon_\alpha + {\hbar^2 \over 2\mu_\alpha}
\left({\partial \over \partial {\bR}} \right)^2~\right]
\chi^\pi_\alpha({\bR})=\sum_{\alpha^\prime} \int d {\bR}^\prime
~G_\aagp({\bR}, {\bR}^\prime; E)
~\chi^\pi_{\alpha^\prime}({\bR}^\prime)\ ,
\label{fm2}
\end{equation}
where $G_{\alpha \alpha^\prime}({\bR}, {\bR}^\prime; E)$ is
composed of various pieces of the interaction kernels
as well as the direct potentials of EMEP:
\begin{eqnarray}
G_\aagp({\bR}, {\bR}^\prime; E)
= \delta({\bR}-{\bR}^\prime)
\sum_\beta \sum_\Omega V^{\Omega\beta}
_{D\,\alpha \alpha^\prime}({\bR})
+ \sum_\Omega \CM^{\Omega}_\aagp ({\bR}, {\bR}^\prime)
-\varepsilon_\alpha~\CM^N_\aagp({\bR}, {\bR}^\prime)\ . \qquad
\label{fm3}
\end{eqnarray}
The subscript $\alpha$ stands for a set of quantum
numbers of the channel wave function; $\alpha=
\left[1/2(11)\,a_1, 1/2(11)a_2 \right]$ $SS_zYII_z;\CP$,
where $1/2(11) a$ is the spin and $SU_3$ quantum number
in the Elliott notation $(\lambda \mu)$, $a$ ($=YI$) is the flavor label
of the octet baryons ($N=1(1/2),~\Lambda=00,
~\Sigma=01$ and $\Xi=-1(1/2)$),
and $\CP$ is the flavor-exchange phase \cite{NA95}.
In the $NN$ system with $a_1a_2=NN$, $\CP$ becomes redundant,
since it is uniquely determined by the isospin as $\CP=(-1)^{1-I}$.
These are the channel specification scheme in the isospin basis.
In the particle basis, necessary modification should be made
for the flavor degree of freedom.
The relative energy $\varepsilon_\alpha$ in the
channel $\alpha$ is related to the total energy $E$ of the system
in the center-of-mass (c.m.) system
through $\varepsilon_\alpha=E-E^{\tint}_a$.
Here $E^{\tint}_a=E^{\tint}_{a_1}+E^{\tint}_{a_2}$ with $a=a_1a_2$.
In \eq{fm3} the sum over $\Omega$ for the direct term
implies various contributions of interaction types
for the meson-exchange potentials, while $\beta$ specifies
the meson species. On the other hand, $\Omega$ for the exchange
kernel $\CM_{\alpha\alpha'}^{\Omega}({\bR}, {\bR}^\prime)$ involves
not only the exchange kinetic-energy ($K$) term but also
various pieces of the FB interaction, as well as several
components of EMEP.
The RGM equation (\ref{fm2}) is solved in the Lippmann-Schwinger
formalism developed in \cite{LSRGM}
(which we call LS-RGM).
In this formalism, we first calculate
the basic Born kernel defined through
\begin{eqnarray}
M^{\rm B}_\aagp (\bq_f, \bq_i; E)
& = & \langle\,e^{i \bq_f \cdot {\bR}}\,\vert
\,G_\aagp ({\bR}, {\bR}^\prime; E)
\,\vert\,e^{i \bq_{\,i}\cdot {{\bR}}^\prime} \rangle \nonumber \\
& = & \sum_\beta \sum_\Omega M_{D\,\aagp}^{\Omega\beta}(\bq_f, \bq_i)
+\sum_\Omega M_\aagp^\Omega(\bq_f, \bq_i)\,\CO^\Omega(\bq_f, \bq_i)
-\varepsilon_\alpha~M_{\aagp}^N(\bq_f, \bq_i)\ \ ,
\label{fm4}
\end{eqnarray}
where $\varepsilon_\alpha$ is the relative energy
in the final channel (in the prior form).
Each component of the Born kernel \eq{fm4} is given
in terms of the transferred momentum $\bk=\bq_f-\bq_i$ and
the local momentum $\bq=(\bq_f+\bq_i)/2$.
In \eq{fm4} the space-spin
invariants $\CO^\Omega=\CO^\Omega(\bq_f, \bq_i)$ are given
by $\CO^{\tcentral}=1$ and
\begin{eqnarray}
& & \CO^{LS} = i \bn \cdot \bS\ ,\quad
\CO^{LS^{(-)}} = i \bn \cdot \bS^{(-)}\ ,\quad 
\CO^{LS^{(-)}\sigma} = i \bn \cdot \bS^{(-)}\,P_\sigma\ ,
\nonumber \\
& & \hbox{with} \quad \bn=[ \bq_i \times \bq_f ]\ ,\quad 
\bS={1 \over 2}(\bfsigma_1+\bfsigma_2)\ ,\quad
\bS^{(-)}={1 \over 2}(\bfsigma_1-\bfsigma_2)\ ,\quad
P_\sigma={1+\bfsigma_1
\cdot \bfsigma_2 \over 2}\ .
\label{fm5}
\end{eqnarray} 
For the tensor and $QLS$ parts, it would be convenient to take four
natural operators defined by
\begin{equation}
\CO^{T} = S_{12}(\bk, \bk)\ ,\quad
\CO^{T^\prime} = S_{12}(\bq, \bq)\ ,\quad 
\CO^{T^{\prime \prime}} = S_{12}(\bk, \bq)\ ,\quad
\CO^{QLS} = S_{12}(\bn, \bn)\ ,
\label{fm6}
\end{equation}
where $S_{12}(\ba, \bb)=(3/2) [\,({\bfsigma}_1 \cdot \ba)
(\bfsigma_2 \cdot \bb)+(\bfsigma_2 \cdot \ba)
(\bfsigma_1 \cdot \bb)\,]
-(\bfsigma_1 \cdot \bfsigma_2)(\ba \cdot \bb)$.
The direct Born kernel $M_{D\,\aagp}^{\Omega\beta}(\bq_f, \bq_i)$ in
\eq{fm4} is explicitly given in Sec.\,II\,B.
The exchange Born kernel $M_{\aagp}^{(\Omega)}(\bq_f, \bq_i)$
is given in Appendix B of \cite{LSRGM} for the FB interaction
and in Appendix A for the EMEP.
The LS-RGM equation is given by
\begin{eqnarray}
T_{\gamma \alpha}(\bp, \bq; E)
=V_{\gamma \alpha}(\bp, \bq; E)
+\sum_\beta {1 \over (2\pi)^3} \int d \bk
~V_{\gamma \beta}(\bp, \bk; E)
{2 \mu_\beta \over \hbar^2}
{1 \over k_\beta^2-k^2+i \varepsilon}
~T_{\beta \alpha}(\bk, \bq; E)\ ,
\label{fm7}
\end{eqnarray}
where the quasi-potential $V_{\gamma \alpha}(\bp, \bq; E)$ or
more generally $V_{\gamma \beta}(\bp, \bq; E)$ is calculated from
\begin{equation}
V_{\gamma \beta}(\bk, \bk^\prime; E)
={1 \over 2} \left[~M^{\rm B}_{\gamma \beta}(\bk, \bk^\prime; E)
+(-1)^{S_\beta}\CP_\beta
~M^{\rm B}_{\gamma \beta}(\bk, -\bk^\prime; E)~\right]\ .
\label{fm8}
\end{equation}
After the standard procedure of the partial-wave
decomposition,\footnote{We use the Gauss-Legendre
20-point quadrature formula to carry out the numerical integration
for the partial-wave decomposition of \eq{fm8}.}
the LS-RGM equation (\ref{fm7}) is solved
by the Noyes-Kowalski method \cite{NO65,KO65}.
The singularity at $k=k_\beta$ is avoided by separating
the momentum region into two pieces.
The intermediate $k$-integral over $0 \leq k \leq k_\beta$ is
carried out using the Gauss-Legendre 15-point quadrature formula
and the integral over $k_\beta \leq k < \infty$ using
the Gauss-Legendre 30-point quadrature formula
through the mapping $k=k_\beta+\tan (\pi(1+x)/4)$.

The LS-RGM equation (\ref{fm7}) is straightforwardly
extended to the $G$-matrix equation by a trivial replacement
of the free propergator with the ratio of the angle-averaged
Pauli operator and the energy denominator:
\begin{eqnarray}
G_{\gamma \alpha}(\bp, \bq; K, \omega)
=V_{\gamma \alpha}(\bp, \bq; E)
+\sum_\beta {1 \over (2\pi)^3} \int d~\bk
~V_{\gamma \beta}(\bp, \bk; E)
{Q_\beta(k, K) \over e_\beta(k, K; \omega)}
~G_{\beta \alpha}(\bk, \bq; K, \omega)\ \ .
\label{fm9}
\end{eqnarray}
Since a detailed description of this formalism
is already given in \cite{GMAT}, there is no need to
repeat other equations. The formula to calculate the
Scheerbaum factor for the s.p. spin-orbit potential
by using the $G$-matrix solution is also
given in \cite{SPLS}. 
We only repeat how we deal with the energy dependence of the
quasi-potential $V_{\gamma \alpha}(\bp, \bq; E)$ in
the $G$-matrix equation (\ref{fm9}). The total energy of the two
interacting particles in the nuclear medium is not conserved.
Since we only need the diagonal $G$-matrices for calculating
s.p. potentials and the nuclear-matter properties in the
lowest-order Brueckner theory, we simply use
\begin{eqnarray}
\varepsilon_\gamma=E_a^{\rm int} -E_c^{\rm int}
+{\hbar^2 \over 2\mu_\alpha}q^2\ \ ,
\label{fm10}
\end{eqnarray}
both in $V_{\gamma \alpha}(\bp, \bq; E)$ and $V_{\gamma \beta}
(\bp, \bk; E)$ in \eq{fm9}.
The meaning and the adequacy of this procedure are discussed
in \cite{GRGM} by using a simple model.

\subsection{Effective meson-exchange potentials for fss2}

The EMEP at the quark level is most easily formulated
in the momentum representation, by using the second-order
perturbation theory with respect to the quark-baryon vertices.
We employ the following $qq$ interaction,
which is obtained through the non-relativistic reduction
of the one-boson exchange amplitudes
in the parameter  $\gamma=(m/2m_{ud})$ (where $m$ is the
exchanged meson mass and $m_{ud}$ is the up-down quark mass):
\begin{eqnarray}
U^{\rm S}(\bq_f, \bq_i) & = & gg^\dagger {4\pi \over \bk^2+m^2}
\left\{~-1+{\bq^2 \over 2m_{ud}^2}
-{1 \over 2m_{ud}^2}i\bn \cdot \bS \right\}\ , \nonumber \\ [3mm]
U^{\rm PS}(\bq_f, \bq_i) & = & -ff^\dagger {1 \over m_{\pi^+}^2}
{4\pi \over \bk^2+m^2} \left[~(\bfsigma_1 \cdot \bk)
(\bfsigma_2 \cdot \bk)-(1-c_\delta) (m^2+ \bk^2)
{1 \over 3}(\bfsigma_1 \cdot \bfsigma_2)~\right]\ , \nonumber \\ [3mm]
U^{\rm V}(\bq_f, \bq_i) & = & {4\pi \over \bk^2+m^2} \left\{
f^e {f^e}^\dagger \left(~1+{3\bq^2 \over 2m_{ud}^2}\right)
-f^m {f^m}^\dagger {2 \over (m_{ud} m)^2}
\left[~(\bfsigma_1 \cdot \bn)(\bfsigma_2 \cdot \bn)-(1-c_{qss})
{1 \over 3}\bn^2 (\bfsigma_1 \cdot \bfsigma_2)~\right] \right.
\nonumber \\ [2mm]
& & \left. -\left(f^m {f^e}^\dagger + f^e {f^m}^\dagger \right)
{2 \over m_{ud} m} i\bn \cdot \bS~\right\}
\label{new1}
\end{eqnarray}
Here $\bk=\bq_f-\bq_i,~\bq=(1/2)(\bq_f+\bq_i)$, and the quark-meson
coupling constants are expressed in the operator form in the
flavor space \cite{FU92,FU97}. For example, the product of the
two different coupling-constant operators $g$ and $f$ are
expressed as
\begin{eqnarray}
gf^\dagger = \left\{
\begin{array}{c}
g_1 f_1 \\
g_8 f_8 \Sigma_a \lambda_a(i) \lambda_a(j) \\
\end{array}
\right.
~\hbox{for}
~\left\{
\begin{array}{c}
\hbox{singlet~mesons}\\
\hbox{octet~mesons}\\
\end{array}
\right. \ ,
\label{new2}
\end{eqnarray}
where $\lambda_a(i)$ represents the Gell-Mann matrix for
particle $i$.
For the realistic description, the meson mixing between
the flavor singlet and octet mesons is very important,
which implies
\begin{eqnarray}
f_{\eta^\prime}=f_1 \cos \theta+f_8 \sin \theta \lambda_8 \quad,
\qquad f_\eta=-f_1 \sin \theta+f_8 \cos \theta \lambda_8 \ ,
\label{new3}
\end{eqnarray}
instead of $f_1$ and $f_8 \lambda_8$ in \eq{new2} for the PS mesons.
Similar transformation is also applied to the V-meson
coupling constants.
The $SU_3$ parameters of the EMEP coupling constants
are therefore $f_1$, $f_8$, and $\theta$.
The S-meson exchange EMEP in \eq{new1} involves not only
the attractive leading term, but also the
momentum-dependent $\bq^2$ term and the $LS$ term.  
The PS-meson exchange operator is the same as before, but
the parameter $c_\delta$ is introduced only for the one-pion
exchange, in order to reduce the very strong effect
of the delta-function type contact term involved in the
spin-spin interaction. The case $c_\delta=1$ corresponds to 
the full expression, while $c_\delta=0$ corresponds to
the case with no spin-spin contact term.
The V-meson exchange potential is composed of
the electric-type term, the magnetic-type term
and the cross term. In the electric term, the central
force generated by the $\omega$-meson exchange potential
is usually most important, and it also includes the $\bq^2$-type
momentum-dependent term.
As to the introduction of the vector-meson EMEP to the quark model,
some discussion already addressed the problem
of double counting, especially with the strong short-range
repulsion from the $\omega$ meson \cite{YA90}.
We will not discuss this problem here, but take a standpoint
to avoid this  double counting problem for the short-range repulsion
and the $LS$ force, by simply choosing appropriate coupling
constants for vector mesons.
The magnetic term is usually important for the isovector $\rho$ meson,
and yields the spin-spin, tensor and $QLS$ terms in the standard OBEP.
The choice in \eq{new1} is to keep only the $QLS$ term with the
spin-spin term proportional to $\bL^2$, the reason for which is
discussed below. Finally, the cross term between the electric and magnetic
coupling constants leads to the $LS$ force for the $qq$ interaction.
The antisymmetric $LS$ ($LS^{(-)}$) force
with $\bS=(\bfsigma_1-\bfsigma_2)/2$ is not generated from EMEP,
because the flavor operator in \eq{new2} is the Gell-Mann matrix
and also because the mass difference between the up-down
and strange quark masses is ignored in \eq{new1}.

We should keep in mind that these EMEP are by no means
a theoretical consequence of the real meson-exchange
processes taking place between quarks. First of all,
the static approximation used to derive the meson-exchange
potentials between quarks is not permissible,
since the masses of S mesons and V mesons are more than
twice as heavy as the quark mass $m_{ud}\sim$ 300 - 400 MeV.
Since the parameter $\gamma$ is not small,
the non-relativistic reduction is not justified.
Also, the very strong S-meson central attraction is just a replacement
of the real processes of the $2\pi$ exchange,
the $\pi \rho$ exchange, the $\Delta$ excitations and so forth.
The V mesons are supposed to behave as composite particles
of the $(q{\bar q})$ pairs.
Furthermore, the choice of terms in \eq{new1} is quite ad hoc and
phenomenological. We should consider \eq{new1} as an effective
interaction to simulate the residual interaction
between quarks, which is not taken into account by the
FB interaction. 

The calculation of the basic Born kernel in \eq{fm4} for
each term of \eq{new1} becomes rather involved, if we use   
the standard technique of calculating the exchange
kernel via the generator-coordinate kernel (GCM kernel).
This becomes especially tedious, when the $qq$ interaction
involves the non-static $\bq^2$ dependence and
the second-order term of $\bq$ as in the $QLS$ force.
We have developed in \cite{LSRGM} a new technique
to calculate the Born kernel directly from the
two-body interaction in the momentum representation.
In this technique, there is no need to calculate
the GCM kernel. Since the final expression is
rather lengthy for the exchange kernel, it is relegated
to Appendix A. Here we only show the direct term,
which is particularly useful to see the
main characteristics of the EMEP introduced
in the present model:
\begin{eqnarray}
& & M^{\rm S}_D(\bq_f, \bq_i) = g^2 {4\pi \over \bk^2+m^2}e^{-{1 \over 3}
(b\bk)^2} \left\{~X^C_{0D_+}\left[~-1+{1 \over 2(3m_{ud})^2}
\left(\bq^2+{9 \over 2b^2} \right)
~\right] -{3 \over 2(3m_{ud})^2}X^{LS}_{0D_+}
i\bn \cdot \bS~\right\}\ , \nonumber \\
& & M^{\rm PS}_D(\bq_f, \bq_i) = -f^2 {1 \over m_{\pi^+}^2}
{4\pi \over \bk^2+m^2}e^{-{1 \over 3}(b\bk)^2} X^T_{0D_+}
\left[~(\bfsigma_1 \cdot \bk)
(\bfsigma_2 \cdot \bk) -(1-c_\delta)(m^2+\bk^2){1 \over 3}
(\bfsigma_1 \cdot \bfsigma_2)\right]\ , \nonumber \\
& & M^{\rm V}_D(\bq_f, \bq_i) = {4\pi \over \bk^2+m^2}e^{-{1 \over 3}
(b\bk)^2} \left\{~(f^e)^2~X^C_{0D_+}\left[~1+{3 \over 2(3m_{ud})^2}
\left(\bq^2+{9 \over 2b^2} \right)
~\right] \right. \nonumber \\
& & - (f^m)^2 {2 \over (3m_{ud} m)^2} X^T_{0D_+}
\left[~(\bfsigma_1 \cdot \bn)(\bfsigma_2 \cdot \bn)
-(1-c_{qss})\left({\bn^2 \over 3}+{\bk^2 \over b^2}\right)
(\bfsigma_1 \cdot \bfsigma_2)
+{3 \over 2b^2}[\bfsigma_1 \times \bk]
\cdot [\bfsigma_2 \times \bk]~\right] \nonumber \\ [0mm]
& & \left. - 2f^m f^e {2 \over 3m_{ud} m} X^{LS}_{0D_+}
i\bn \cdot \bS~\right\}\ .
\label{new4}
\end{eqnarray}
Here $X^\Omega_{0D_+}$ represents the spin-flavor factors
related to the spin-flavor operators in \eq{new1}.
The Gaussian factor $\exp \{-(b\bk)^2/3 \}$ appearing
in \eq{new4} represents the form factor effect of the $(0s)^3$ cluster
wave functions.
The finite size effect of the baryons also appears as the constant
zero-point oscillation terms accompanied with the $\bq^2$ terms,
appearing in the S- and V-meson contributions.
For the $QLS$ force, the same effect appears as the tensor force
having the form $[\bfsigma_1 \times \bk]\cdot
[\bfsigma_2 \times \bk]$. The magnitude of this term is about
one third, if we compare this with the strength from
the original tensor term appearing at the level of $qq$ interaction.
The advantage of using the $QLS$ force in \eq{new1},
instead of the tensor force, is that we can avoid
the $\pi$-$\rho$ cancellation of the tensor force
for the coupling term of the $S$ and $D$ waves.
The $\epsilon_1$ parameter of the $NN$ interaction is very
sensitive to this coupling strength.

\subsection{Determination of parameters}

We have four quark-model parameters; the harmonic-oscillator
width parameter $b$ for the $(3q)$ clusters, the up-down quark
mass $m_{ud}$, the strength of the quark-gluon
coupling constant $\alpha_S$, and the mass ratio
of the strange to up-down quarks $\lambda=(m_s/m_{ud})$.
A reasonable range of the values for these parameters
in the present framework is $b=0.5$ - 0.6 fm,
$m_{ud}=300$ - 400 MeV/$c^2$, $\alpha_S \sim 2$,
and $\lambda=1.2$ - 1.7.
Note that we are dealing with the constituent quark model
with explicit mesonic degree of freedom.
The size of the system determined from the $(3q)$ wave function
with $b$ (the rms radius of the $(3q)$ system
is equal to $b$) is related to the quark distribution,
which determines the range in which the effect
of the FB interaction plays an essential role
through the quark-exchange kernel.
The internal energies of the clusters
should be calculated from the same Hamiltonian as used in the
two-baryon system, and contain not only the quark contribution
but also various EMEP contributions.
The value of $\alpha_S$ is naturally correlated with $b$, $m_{ud}$,
and other EMEP parameters.
This implies that $\alpha_S$ in our framework
is merely a parameter, and has very little to do with
the real quark-gluon coupling constant of QCD.

For the EMEP part, we have three parameters $f_1$, $f_8$, and $\theta$ for
each of the S, PS, Ve (vector-electric) and Vm (vector-magnetic)
terms. It is convenient to use the coupling constants
at the baryon level, in order to compare our result with the predictions
by other OBEP models.
These are related to the coupling constants at the quark level
used in Eqs.\ (\ref{new1}) and (\ref{new4}) through
a simple relationship
\begin{eqnarray}
& & f^{\rm S}_1=3 g_1 \quad, \qquad f^{\rm S}_8=g_8 \quad,
\qquad f^{\rm PS}_1=f_1 \quad,
\qquad f^{\rm PS}_8={5 \over 3} f_8 \ ,\nonumber \\
& & f^{\rm Ve}_1=3 f^e_1 \quad, \qquad f^{\rm Ve}_8=f^e_8 \quad,
\qquad f^{\rm Vm}_1=f^m_1 \quad, \qquad f^{\rm Vm}_8={5 \over 3} f^m_8 \ .
\label{new5}
\end{eqnarray}
Through this replacement, the leading term for each meson
in \eq{new4} precisely coincides with
that of the OBEP with Gaussian form factors.
In the present framework,
the S-meson masses are also considered to be free parameters
within some appropriate ranges.
We further introduce three extra parameters, $c_\delta$ the
strength factor for the delta-function type spin-spin contact
term of the one-pion exchange potential (OPEP),
$c_{qss}$ the strength factor for the spin-spin term
of the $QLS$ force, and $c_{qT}$ the strength factor for the
tensor term of the FB interaction.
These parameters are introduced to improve the fit of
the $NN$ phase shifts to the empirical data, as is discussed below. 

We determine these parameters by fitting
the most recent result of the phase shift
analysis SP99 \cite{SAID} for the $np$ scattering
with the partial waves $J \leq 2$ and
the incident energies $T_{\rm lab} \leq 350~\hbox{MeV}$,
under the constraint of the deuteron binding energy
and the $\hbox{}^1S_0$ $NN$ scattering length,
as well as to reproduce the available data
for the low-energy $YN$ total cross sections.
The result is shown in Table I.
The parameters of the previous model FSS are also
shown for comparison. The $\chi^2$ value used
in the parameter search is defined through
\begin{eqnarray}
\sqrt{\chi^2}=\left\{{1 \over N}\sum_{i=1}^N
\left( \delta_i^{cal}-\delta_i^{exp}\right)^2\right\}^{1 \over 2}\ ,
\label{new6}
\end{eqnarray}
where no experimental error bars are employed because
the energy-dependent solution of the phase-shift
analysis does not give them.
In \eq{new6} the sum over $i=1$ - $N$ is with respect to
various angular momenta and energies,
and the mixing parameters, $\epsilon_1$ and $\epsilon_2$,
are also included in the unit of degrees.
The value $\sqrt{\chi^2}$ therefore gives some measure
for the averaged deviation of the calculated
phase shifts from the empirical values.
Using the parameter set in Table I,
we have obtained $\sqrt{\chi^2}=0.655^\circ$ for
the $np$ scattering. The best solution in our previous models
is $\sqrt{\chi^2} \sim 3^\circ$ in FSS.
Since the present model fss2 is a renovated version of FSS,
we summarize in the following only the changes
and new points of fss2, in comparison with the model FSS:
\begin{enumerate}
\item[1)] In the original expression of the meson-exchange
potentials between quarks, the momentum-dependent Bryan-Scott
term appears in the combination of $\bq^2-\bk^2/4$ for
the S meson and $3\bq^2-\bk^2/4$ for the V meson.
We find that these $\bk^2/4$ terms (usually replaced
by $\bk^2=-m^2$) play a rather characterless
role in making the whole interaction slightly repulsive.
With these terms, the energy-dependence of
the $\hbox{}^1S_0$ and $\hbox{}^3S_1$ phase shifts
becomes too strong to keep the value of $b$ in
the reasonable range.
(The value of $b$ turns out to be too small,
about $b \sim 0.4$ fm to
compensate the strong energy dependence.)
We therefore drop all these $\bk^2/4$ terms
in the present calculation.
\item[2)] We ignore the $QLS$ force from the S-mesons,
since it is very weak. The S-meson EMEP therefore consists
of the leading term with $-1$ in \eq{new4}, the momentum-dependent
Bryan-Scott term and the $LS$ term. This $LS$ term yields
an appreciable contribution at medium and higher energies,
which consequently reduces the value of $b$ from the previous
value $\geq 0.6~\hbox{fm}$ to a smaller value $\sim 0.56~\hbox{fm}$.
\item[3)] The reduction of the spin-spin contact term
for the PS mesons is introduced only for the pion
with the smallest mass.
For the other heavier PS mesons, we assume the full
strength factor $c_\delta=1$.
The reduction from 1 for the pion improves
the fit of the $NN$ $\hbox{}^1P_1$ phase shift
to a great extent. (Otherwise, the repulsion at
higher energies is insufficient for this partial wave.)
We introduce $c_\delta$ only for pion, since
the effect of the present $(3q)$-cluster folding corresponds
to a very low value of the
cut-off mass $\Lambda\sim 800$ - 900 MeV for
the pion form factor in OBEP.
It is well known that such a low value of $\Lambda$ converts
even the sign of the medium-range part of the OPEP
if the full strength of the contact term
is introduced.
The factor $c_\delta < 1$ also reduces the very strong
repulsion generated from the one pion spin-spin contact
term for the $S$-wave states of the $NN$ system.
In the present framework, this repulsion is almost 300 MeV
if $c_\delta=1$ is assumed.
Furthermore, the value of $c_\delta$ has a strong
influence on the internal energies of single baryons.
It reduces the very large contribution of the pion
to the $N$-$\Delta$ and $\Lambda$-$\Sigma$ mass
difference, the latter helping us
to keep $\lambda=(m_s/m_{ud})$ at the moderate value.
(Otherwise, we obtain $\lambda \sim 1$.)
If we do not introduce $c_\delta$ and the
parameters $c_{qss}$, $c_{qT}$ discussed below,
the $\sqrt{\chi^2}$ value cannot be improved
by more than $1.5^\circ$.
The contribution of $\eta$ and $\eta^\prime$ mesons
was necessary in the previous models in order to make
the $\hbox{}^3S$ central force relatively more repulsive than
the $\hbox{}^1S$ central force.
In the present framework, it turns out that the introduction of
these $\eta$ mesons is not convenient for the subtle balance
of the central and tensor forces.
We therefore take out all these $\eta$-meson contributions.
The well-known too strong repulsion
of the $NN$ $\hbox{}^1S$ central force from the
color-magnetic interaction of the
FB interaction \cite{ZH94} is remedied by assuming
two different masses for the isovector $\delta$ meson,
i.e., $m_\delta=720~\hbox{MeV}/c^2$ for the $NN$ system
and $m_\delta=846~\hbox{MeV}/c^2$ for the $YN$ system
(see comment 3) in Table I). 
\item[4)] As is discussed at the end of the preceding subsection,
the present model fss2 is the $QLS$ dominant model.
This implies that we use the $QLS$ force to reduce the too strong
OPEP tensor force, instead of the tensor force itself.
The main reason for this choice is that the $NN$ mixing
parameter $\epsilon_1$ is very difficult
to reproduce if the cancellation of the one pion tensor
force and the $\rho$-meson tensor force
is too strong for the $S$-wave and $D$-wave coupling.
Another question is how this $QLS$ force is incorporated
into the model. We find that the $QLS$ spin-spin
term $\bn^2 (\bfsigma_1 \cdot \bfsigma_2)$ in \eq{new1} plays
a favorable role in improving the fit of the $NN$ phase shifts.
This term corresponds
to the $(\bfsigma_1 \cdot \bfsigma_2)\bL^2$ term
in the Hamada-Johnstone potential \cite{HJ62}.
Since the full introduction of this term results in
too vigorous behavior, we introduce a reduction
factor $c_{qss}$, which turns out around $c_{qss} \sim 0.6$.
The two-pole formula for the $\rho$-meson exchange potential,
introduced in \cite{ST94}, is found to give a favorable result.
We further find that the short-range tensor force
is still too weak. We avoid this difficulty
simply by increasing the strength of the tensor term
of the FB interaction with the factor $c_{qT}$.
The value $c_{qT} \sim 3$ seems to be reasonable.
If we carry out the parameter search with $c_{qT}=1$,
the value of $\sqrt{\chi^2}$ cannot be improved
by more than $1.3^\circ \sim 1.0^\circ$, mainly due to
the disagreement of $\epsilon_1$.
We should note, however, that the introduction
of the V mesons is a rather minor change
from our previous models. With the exception
of $f^{\rm Vm}_8=2.577$, the V-meson coupling
constants in Table I are around one,
which is less than half of the coupling constants in the
standard OBEP. In particular, the isospin
dependent $LS$ force from the $\rho$ meson
is exactly zero, since $f^{\rm Ve}_8$ is fixed at zero.
The short-range repulsion in the
$NN$ interaction is still mainly described
by the color-magnetic term of the FB interaction.
The dominant effect of the V mesons is almost
solely the $\rho$ meson $QLS$ force, which is the
reason we call fss2 the $QLS$ dominant model.
\item[5)] The following five parameters in Table I are
directly related to the reproduction of the
low-energy $YN$ cross sections; $\lambda=(m_s/m_{ud})$,
$\theta^{\rm S}$, $\theta^{\rm S}_4$, $m_\delta$, and $m_\kappa$.
Among them, the angle of the singlet-octet meson
mixing $\theta^{\rm S}$ of the S mesons are used to control
the relative strength of the central attraction
of the $NN$ and $YN$ interactions.
It was found before \cite{FSS} that,
once the $\theta^{\rm S}$ is determined
to fit the low-energy $\Lambda p$ cross section data,
the attraction of the $\Sigma N(I=3/2)$ channel
is too strong and the $\Sigma^+ p$ total cross sections
are overestimated. We therefore use a larger
value for $\theta^{\rm S}$ (which is denoted
by $\theta^{\rm S}_4$) only for the $\Sigma N (I=3/2)$ channel
in order to reduce the attraction,
which is the same prescription employed in the
previous models \cite{PRL,FSS}.
\item[6)] The largest ambiguity for determining
the parameters related to the $YN$ interaction lies
in the strength of the central attraction
in the $\Sigma N (I=1/2)$ $\hbox{}^3S_1$ channel \cite{LSRGM}.
If the phase-shift rise of the $\hbox{}^3S_1$ state is less
than $30^\circ$, the low-energy $\Sigma^- p$ elastic
total cross section becomes too small.
If this attraction is too strong, as in RGM-F \cite{RGMFb},
the $\hbox{}^3S_1$ phase shift shows a sudden decrease
from $180^\circ$ to 60 - $90^\circ$, and the behavior
of the $\Lambda p$ total cross sections
at the $\Sigma N$ threshold becomes a round peak,
instead of the cusp structure \cite{MI99}.
Furthermore, the strength of the central attraction
plays a crucial role even for the odd-parity state.
The $\Sigma N (I=1/2)$ $\hbox{}^3P_1$ phase shift
is attractive due to the exchange kinetic-energy
kernel; i.e., the effect of the Pauli principle \cite{NA95}.
This attraction is reinforced by the $LS$ force
in the diagonal channel, and also by the $LS^{(-)}$ force
acting between this channel and the $\hbox{}^1P_1$ channel.
This channel coupling also takes place between
the $\Sigma N (I=1/2)$ channel and the $\Lambda N$ channel.
This channel coupling is mainly determined by the strength
of the $LS^{(-)}$ force, which is directly related to the
magnitude of $\alpha_S$, but also considerably
influenced by the strength of the central attraction
in the $\Sigma N (I=1/2)$ channel.
In \cite{LSRGM}, we have clarified that the
central attraction of the previous models RGM-F and FSS
is so strong that the $\Sigma N (I=1/2)$
$\hbox{}^3P_1$ resonance is moved to the $\Lambda N$
$\hbox{}^1P_1$ channel. The consequence of this behavior
is the strong enhancement of the $\Lambda p$ total
cross sections in the cusp region.
On the contrary, the $P$-wave coupling in the model RGM-H
is less strong, and the agreement of the $\Lambda p$ total
cross sections is much better. (See Fig.\,10(a) in \cite{FSS} and
Table II in \cite{LSRGM}.)
Here we assume that the resonance stays
in the original $\Sigma N (I=1/2)$ $\hbox{}^3P_1$ channel,
and try to find the parameter set which gives the maximum
strength of the $\Sigma N (I=1/2)$ central attraction.
In practice, we assume $\sqrt{2/\pi}\alpha_S x^3 m_{ud}c^2
=440$ MeV ($x=(\hbar/m_{ud}cb)$ is the ratio of the Compton
wave-length of the up-down quarks to $b$) as in RGM-F
and FSS,\footnote{This value corresponds
to assuming the $N$-$\Delta$ mass difference 293.3 MeV only
by the FB interaction, as seen from Table III.
If we use the $\alpha_S$ value about 1.3 times larger,
the transition of the $P$-wave resonance
to the $\Lambda N$ $\hbox{}^1P_1$ channel
takes place in the present model.}
and adjust the value of $m_\delta$
for the $YN$ interaction, independently of the value
in the case of $NN$ interaction.
If we use a smaller value for $m_\delta$, the $\Sigma N (I=3/2)$
$\hbox{}^1S_0$ state becomes more attractive
and the $\Sigma N (I=1/2)$ $\hbox{}^3S_1$ state becomes
less attractive.
\item[7)] Another important change from the previous models FSS
and RGM-H is the relative strength
of the $\hbox{}^1S_0$ and $\hbox{}^3S_1$ attraction
in the $\Lambda N$ interaction.
The maximum phase-shift values
of the $\hbox{}^1S_0$ and $\hbox{}^3S_1$ states
in these models are about $46^\circ$ and $16^\circ$,
respectively, around $p_\Lambda \sim 200
~\hbox{MeV}/c$. The big difference of almost $30^\circ$ is
known to be unfavorable for the description
of the $s$-shell $\Lambda$-hypernuclei.
Detailed few-body calculations for these hypernuclei 
have recently been carried out by several groups
\cite{SH83,YA94,HI97,NE00}
by using various effective $\Lambda N$ interactions.
In these effective $\Lambda N$ interactions,
the effect of the $\Sigma N$ channel coupling
is usually renormalized.
These calculations imply that the phase-shift
difference of a little less than $10^\circ$ seems 
to be most appropriate.
We follow this suggestion and adjust the strength
of the $\Lambda N$ attraction such that
the $\hbox{}^1S_0$ and $\hbox{}^3S_1$ phase-shift
difference is less than $10^\circ$ and
the low-energy $\Lambda p$ cross sections are
correctly reproduced.
We can use the $\kappa$-meson mass to adjust this
phase-shift difference. Namely, if $m_{\kappa}$ is
smaller, then the $\Lambda N$ $\hbox{}^1S_0$ phase
shift becomes more attractive
and the $\hbox{}^3S_1$ phase shift becomes less
attractive.
\end{enumerate} 
In order to give an outline of the framework,
we summarize the difference of FSS and fss2 in Table II,
with respect to the meson species and interaction
types of EMEP included in the models.
Table III shows the quark and EMEP contributions
to the baryon mass difference
between $N$ and $\Delta$ ($\Delta E_{N\hbox{-}\Delta}
=E^{\rm int}_\Delta-E^{\rm int}_N$),
and the mass difference between $\Lambda$ and $\Sigma$
($\Delta E_{\Lambda \hbox{-}\Sigma}
=E^{\rm int}_\Sigma-E^{\rm int}_\Lambda$),
calculated in the isospin basis.
We note that various meson contributions
largely cancel each other and the net contribution
is roughly given by the quark contribution
from the color-magnetic term of the FB interaction.

\subsection{Calculation in the particle basis}

In this subsection we discuss some new features
required in the calculation in the particle basis.
Three different types of calculations are carried out in this paper.
\begin{enumerate}
\setlength{\itemsep}{0mm}
\item[1)] calculation in the isospin basis
\item[2)] calculation in the particle basis without the Coulomb force
\item[3)] calculation in the particle basis with the Coulomb force
\end{enumerate}

For the $NN$ interaction, the calculation
in the particle basis is rather straightforward. We use the
empirical baryon masses listed in Table \ref{table4}
and evaluate spin-flavor factors
for the charged pion and the neutral pion
separately in the isospin representation. The other spin-flavor
factors for heavier mesons and the FB interaction
are generated in the simple isospin relations.
The Coulomb force is introduced at the quark level
by using the quark charges. The exchange Coulomb kernel has the
same structure as the color-Coulombic term of the FB interaction.

Only complexity arises when we solve the LS-RGM equation in the
momentum representation. The standard technique by Vincent and
Phatak \cite{VP74} is employed to solve the Lippmann-Schwinger
equation in the momentum representation,
including the Coulomb force.
This technique requires introducing a cut-off radius $R_C$ for
the Coulomb interaction. In the RGM formalism, we have to introduce
this cut-off at the quark level,
in order to avoid violating the Pauli principle.
The two-body Coulomb force assumed in the present
calculation is therefore written as
\begin{eqnarray}
U^{CL}_{ij}=Q_i Q_j e^2 {1 \over r_{ij}} \Theta(R_C-r_{ij}) \ ,
\label{pa1}
\end{eqnarray}
where $\Theta$ is the Heaviside step function and $Q_i,~Q_j
=2/3$ for the up quark and $-1/3$ for the down and strange quarks.
The Coulomb contribution to the internal energies becomes zero
for the proton and $\Sigma^+$.
More explicitly, this can be given by
\begin{eqnarray}
E^{CL}_{\rm int}=X^{CL}_{0E} \sqrt{{2 \over \pi}}\alpha x m_{ud}c^2
\left(1-e^{-{1 \over 2}\left({R_C \over b}\right)^2}\right) \ ,
\label{pa2}
\end{eqnarray}
where $\alpha=(e^2/\hbar c) \sim 1/137$ is the hyperfine
coupling constant and the direct spin-flavor factor is expressed
as $X^{CL}_{0E}=\sum_{i=1,2}\left[Z_i(Z_i-1/3)/2-1/3\right]$ in
terms of the total charge $Z_i$ of the $i$-th baryon.
The basic Born kernel for the direct Coulomb term reads
\begin{eqnarray}
M^{CL}_D(\bq_f, \bq_i) = Z_1 Z_2 e^2 2\pi {R_C}^2
\left({2 \over kR_C}\sin {kR_C \over 2}\right)^2
e^{-{1 \over 3}(bk)^2}\qquad \hbox{with}
\qquad k=|\bq_f-\bq_i|\ ,
\label{pa3}
\end{eqnarray}
which corresponds to the direct Coulomb potential
\begin{eqnarray}
V_D(r)=Z_1 Z_2 e^2 {1 \over r}\left\{
\hbox{erf}\left(\sqrt{\gamma}r\right)
-{1 \over 2}\left[\hbox{erf}\left(\sqrt{\gamma}(r+R_C)\right)
+\hbox{erf}\left(\sqrt{\gamma}(r-R_C)\right)\right]\right\} \ .
\label{pa4}
\end{eqnarray}
Here $\hbox{erf}\,(x)=(2/\sqrt{\pi})\int^x_0 e^{-t^2} dt$ stands
for the error function and $\gamma=\mu \nu=(3/4b^2)$.
The exchange Coulomb kernel is also slightly modified
from the exact Coulomb kernel. This is given in Appendix A,
together with other EMEP kernels.
The value $R_C$ should be sufficiently large
to be free from any nuclear effect beyond $R_C$.
Then the final $S$-matrix is calculated from the condition
that the wave function obtained by solving
the Lippmann-Schwinger equation with the modified Coulomb
force is smoothly connected to
the asymptotic Coulomb wave function.
We take $R_C=9~\hbox{fm}$, although a much smaller value
seems to be sufficient.
Note that, even in the $np$ and $nn$ systems, we have
small contributions from the Coulomb interaction through
the exchange Coulomb kernel. The difference between the
calculations 2) and 3) for the system of chargeless
particles implies this effect. 

For the $YN$ interaction, more consideration is required
for the treatment of the threshold energies. We note that
the mass difference of $\Sigma^-$ and $\Sigma^+$ is
about 8 MeV and is fairly large. Figure \ref{thres} shows the
comparison of threshold relations in the isospin and
particle bases, evaluated in the non-relativistic
kinematics. The $\Lambda p$ system has the total
charge $Q=1$ and the $\Sigma^- p$ system $Q=0$. 
The direct Coulomb term exists only in the $\Sigma^+ p$
channel and the $\Sigma^- p$ channel.
The EMEP contribution to the $\Lambda$ - $\Sigma$ mass
difference is given in Table V, both in the isospin basis
and in the particle basis. We find rather large cancellation
between the neutral and charged pion contributions
to the $\Sigma^\pm$ - $\Sigma^0$ mass difference.
The pion-Coulomb correction yields the calculated
threshold energies given in Table VI in the non-relativistic
kinematics. Apparently the empirical mass difference
is not precisely reproduced.\footnote{The up-down quark-mass
difference does not help, because the total hypercharge
is conserved in the two-baryon systems.}
When we use the relativistic kinematics,
the empirical threshold energies are defined
by $E^{\rm exp}_{\rm th}=(p^{\rm cm}_{\rm th})^2/2\mu_{\rm inc}$
($\mu_{\rm inc}$ is the non-relativistic reduced mass
of the incident channel),
which are used in the present non-relativistic model
in the c.m. system. Table VII shows these energies in
the column of relativistic $E^{\rm exp}_{\rm th}$.
The difference between these $E^{\rm exp}_{\rm th}$ and the
calculated threshold energies $E^{\rm cal}_{\rm th}$ is
given in the last column. 
For the $\Sigma^- p$ - $\Sigma^0 n$ - $\Lambda n$ system,
$E^{\rm exp}_{\rm th}$ are calculated from those
in the $\Lambda n$ - $\Sigma^0n$ - $\Sigma^- p$ system.
The disagreement of the threshold energies between the
calculation and the experiment is a common feature of
any microscopic model. Fortunately, we have
a nice method to remedy this flaw without violating
the Pauli principle. As discussed in \cite{GRGM} in
detail, we only need to add a small correction term $\Delta G$
to the exchange kernel, in order to use the
empirical threshold energies $E^{\rm exp}_{\rm th}$ listed
in Table VII. The same technique is also applied to the
reduced mass corrections.

\section{Results and discussions}

\subsection{The ${\protect\mbf NN}$ result}

Figures \ref{npphase}(a) - \ref{npphase}(i) compare the $np$ phase shifts
and the mixing angles $\epsilon_J$ predicted by fss2 with
the recent phase-shift analysis SP99
by Arndt {\em et al.} \cite{SAID}
The parameter search and the calculation of
phase-shift parameters in this subsection
are all carried out in the isospin basis.
For comparison, the previous results by FSS are also shown
with the dotted curves. 
Here we examine the partial waves up to $J=4$ in
the energy range $T_{\rm lab}=0$ - 800 MeV.
For energies higher than 300 MeV, the inelasticity parameters
of SP99 are given for a measure of possible deviations
of the phase-shift values in the single-channel calculation.
The $\hbox{}^3D_2$ phase shift is greatly improved
by the $QLS$ component. Even in the other partial waves,
the improvement of the phase-shift parameters is usually
achieved. This includes 1) $\hbox{}^3P_0$, $\hbox{}^3P_1$,
and $\hbox{}^3G_4$ phase shifts, 2) $\hbox{}^3S_1$,
$\hbox{}^1S_0$, $\hbox{}^1P_1$, $\hbox{}^1F_3$,
and $\hbox{}^3H_4$ phase shifts at higher
energies $T_{\rm lab}=400$ - 800 MeV,
and 3) some improvement in $\hbox{}^3F_2$ phase shift
and $\epsilon_2$ mixing parameter.
On the other hand, $\hbox{}^3P_2$ and $\hbox{}^3D_3$ phase
shifts turn out worse and $\hbox{}^3F_4$ phase shift
is not much improved.
The disagreement of the $\hbox{}^3D_3$ phase shift
and the deviation of the $\hbox{}^3D_1$ phase shift
at the higher energies imply that our description
of the central, tensor and $LS$ forces
in the $\hbox{}^3E$ states requires further improvement.
The insufficiency in the $\hbox{}^3O$ partial
waves is probably related to the imbalance of the central force
and the $LS$ force in the short-range region.
The decomposition of the $\hbox{}^3P_J$ phase shifts
to the central, $LS$ and tensor components, shown in
Fig. \ref{phcom}, implies that the $\hbox{}^3O$ central
force is too repulsive at higher energies $T_{\rm lab}
\geq 400$ - 500 MeV. 
It should be noted that whenever the discrepancy
of the phase-shift parameters between the calculation
and the experiment is large, the inelasticity parameters
are also very large. In particular, the inelasticity
parameters of the $\hbox{}^3P_2$, $\hbox{}^1D_2$,
and $\hbox{}^3F_3$ states rise very rapidly as the energy increases,
and reach more than $20^\circ$ at $T_{\rm lab}=800$ MeV.
The elastic phase shift for each of these states
shows a dispersion-like resonance behavior at the energy
range from 500 MeV to 800 MeV. These are the well-know
di-baryon resonances directly related
to the $\Delta N$ threshold in the isospin $I=1$ channel.
The present single-channel calculation is not
capable of describing these resonances.

Table \ref{phase} tabulates the values of phase-shift
parameters in the energy range $T_{\rm lab}=25$ - 300 MeV,
in comparison with those of SP99 \cite{SAID} and other 
meson-exchange models. The results of OBEP, Paris and Bonn
potentials are cited from Table 5.2 in \cite{MA89}.
The partial waves only up to $J=2$ are considered.
If we calculate the $\chi^2$ values using these numbers,
we obtain $\sqrt{\chi^2}$= 0.59, 1.10, 1.40 and 1.32 for fss2,
OBEP, Paris and Bonn, respectively.
The reason we get such results is as follows.
In the meson exchange models, the accuracy of the
low energy phase shifts is less than $0.2^\circ$,
and the agreement with the experiment is excellent.
However, in higher energies the deviation from the
experiment increases, and
in some particular partial waves like $\hbox{}^1S_0$ and
$\hbox{}^3P_0$ states, it becomes more than $2^\circ$.
In the Paris potential, the $\hbox{}^1S_0$ phase shift
is apparently too repulsive.
This is, however, because the parameters of the Paris
potential is determined by the fit to the $pp$ phase shifts,
and the correction due to the CSB is probably not
taken into account in the numbers given here.
Every model has its own weak points. For example,
the tensor force of the Bonn potential is usually
very weak, which is reflected in the $\epsilon_1$ parameter
and in the too attractive behavior
of the $\hbox{}^3P_0$ phase shift.
(However, the recent CD Bonn potential \cite{MA00} fits
the $NN$ phase-shift parameters in the non-relativistic
energies almost perfectly, with various possible
corrections taken into account.) 
The weak point of our model lies in the $\hbox{}^3P_2$ and
$\hbox{}^3D_3$ phase shifts at the intermediate and
higher energies $T_{\rm lab}=300$ - 800 MeV.
The empirical $\hbox{}^3P_2$ phase shift gradually decreases
if we ignore the weak dispersion-like behavior.
Our result, however, decreases too rapidly.
Our $\hbox{}^3D_3$ phase shift is too attractive
by $4^\circ$ - $6^\circ$.

Figures \ref{npdif} and \ref{nppol} illustrate
the fss2 predictions of the differential
cross sections ($d\sigma/d\Omega$ in mb/sr) and
the polarizations ($P(\theta)$) for
the elastic $np$ scattering,
in comparison with experiment \cite{SAID}.
The same observables for the elastic $pp$ scattering
are plotted in Figs. \ref{ppdif} and \ref{pppol}.
The final calculation of these observables in this paper 
is carried out in the particle basis with the full Coulomb
force incorporated.
The corresponding figures by our previous model FSS are
given in Figs.\ 1, 2 of \cite{SCAT} and Figs.\ 2, 3
of \cite{LSRGM}. (Note that the plot of the differential
cross sections at higher energies in \cite{LSRGM} is
given in the log scale.)
We find some improvement in the differential cross sections.
First, the previous overestimation
of the $np$ differential cross sections at the forward angle
at $T_{\rm lab}=320$ MeV is corrected. Secondly, the bump
structure of the $np$ differential cross sections
around $\theta_{\rm cm}=130^\circ$ at energies $T_{\rm lab}
=300$ - 800 MeV has disappeared. 
The overestimation of the $pp$ differential cross sections
at $\theta_{\rm cm}=10^\circ$ - $30^\circ$ at energies $T_{\rm lab}
=140$ - 400 MeV is improved.
However, the essential difficulties of FSS and RGM-H, namely the
oscillatory behavior of the $np$ polarization
around $\theta_{\rm cm} \sim 110^\circ$ and
that of the $pp$ polarization around the
symmetric angle $\theta_{\rm cm}=90^\circ$
for higher energies $T_{\rm lab}\geq 400$ MeV are
not resolved.
Furthermore, the $pp$ differential cross sections show
a deep dip at angles $\theta_{\rm cm}
\leq 30^\circ$ and $\geq 150^\circ$ for $T_{\rm lab}\geq 500$ MeV.
The low-energy $pp$ cross sections
at $\theta_{\rm cm}=90^\circ$ for $T_{\rm lab}\leq 100$ MeV are
still overestimated.

In order to find a possible reason for the unfavorable
oscillations of our polarizations, we show
in Fig. \ref{inv} the five independent $pp$ invariant
amplitudes at the highest energy $T_{\rm lab}=800$ MeV.
They are composed of the real and imaginary parts
of $g_0$ (spin-independent central), $h_0$ ($LS$),
$h_n$ ( $(\bfsigma_1\cdot \widehat{\bn})
(\bfsigma_2\cdot \widehat{\bn})$-type tensor),
$h_k$ ( $(\bfsigma_1\cdot \widehat{\bk})
(\bfsigma_2\cdot \widehat{\bk})$-type tensor),
and $h_P$ ( $(\bfsigma_1\cdot \widehat{\bP})
(\bfsigma_2\cdot \widehat{\bP})$-type tensor)
invariant amplitudes. In Fig. \ref{inv} the
Coulomb force is neglected in the predictions
by the Paris potential. The result by SP99 is calculated
using only the real parts of the empirical
phase-shift parameters.
If we recall that the polarization is given by
the cross term contribution of the central, $LS$, and tensor
invariant amplitudes (i.e., $P(\theta)=2\Im m
\left[(g_0+h_n)(h_0)^*\right]$,
see Eq. (2.32) of \cite{LSRGM})
we find that the disagreement
in $\Im m~h_n$ and $\Re e~h_0$ is most serious.
Since the oscillatory behavior of $\Im m~h_n$ in SP99
also appears in $\Im m~h_k$ and $\Im m~h_P$,
it is possible that this is an oscillation caused by
the $NN$ - $\Delta N$ channel coupling through
the one pion spin-spin and tensor forces.
Figure \ref{inv} also shows the reason for the underestimation
of the differential cross sections
at $\theta_{\rm cm}\leq 30^\circ$.
Namely, The imaginary part of $g_0$ is too small both for fss2 and
the Paris potential, and the real part of $g_0$ is strongly
reduced in fss2.

Another application of the invariant amplitudes is
the $t^{\rm eff}\rho$ prescription for calculating
the s.p. potentials of the nucleons and hyperons
in nuclear matter. It is discussed
in \cite{LSRGM} that the s.p. potentials predicted by
the model FSS in the $G$-matrix calculation
show fairly strong attractive behavior in the momentum
interval $q_1=5\,\hbox{--}\,20~\hbox{fm}^{-1}$ for all the baryons.
In particular, $U_N(q_1)$ in the continuous prescription
becomes almost $-80$ MeV at $q_1=10~\hbox{fm}^{-1}$.  
This momentum interval corresponds to the incident
energy range $T_{\tlab}=500~\hbox{MeV}\,\hbox{--}\,8~\hbox{GeV}$
in the $NN$ scattering.
The $t^{\rm eff}\rho$ prescription is a convenient way to
evaluate the s.p. potentials in the asymptotic momentum
region in terms of the spin-independent invariant amplitude
at the forward angle $g_0(\theta=0)$.
Since the present model fss2 incorporates the momentum-dependent
Bryan-Scott term, the asymptotic behavior of
the s.p. potentials in the large momentum region is improved.
We can see this in Fig. \ref{trho}, where the s.p. potentials 
of $N$, $\Lambda$, and $\Sigma$ calculated in the $G$-matrix
approach are shown in the momentum
range $q_1=0$ - 10 $\hbox{fm}^{-1}$.
Figures \ref{trho}(a) and (b) show the result in the $QTQ$ prescription,
and Figs.\,\ref{trho}(c) and (d) in the continuous choice for
intermediate spectra.\footnote{In Fig.\,\protect\ref{trho} we have
employed the approximate angular integration given in Eq.\,(A.8)
of \protect\cite{GMAT}, while in Fig.\,\protect\ref{spnuc} the
numerical integration over the angle $\theta_2$ in Eq.\,(A.6)
of \protect\cite{GMAT} is explicitly carried out.}
Figures \ref{trho}(a) and (c) show
the real part of $U_B(q_1)$, and Figs.\,\ref{trho}(b) and (d) the
imaginary part.
In Figs.\,\ref{trho}(c) and (d), the solid curves
for the nucleon s.p. potential are compared with
the results by the $t^{\rm eff}\rho$ prescription
with respect to the $T$-matrices of fss2,
the Paris potential \cite{PARI},
and the empirical phase shifts SP99 \cite{SAID}.
The partial waves up to $J\leq 8$ are included in fss2 and
the Paris potential, and $J\leq 7$ in SP99.
The momentum points calculated correspond
to the energies $T_{\rm lab}=100$, 200, 400, 800, and 1,600 MeV.
We find that the real part of $U_N(q_1)$ nicely
reproduces the result of the $G$-matrix calculation
even at such a low energy as $T_{\rm lab}=100$ MeV.
On the other hand, the imaginary part by
the $t^{\rm eff}\rho$ prescription usually overestimates the
exact result especially at the lower energies.

\subsection{Deuteron properties and effective range parameters}

The deuteron properties are calculated by solving the LS-RGM equation
with respect to the relative wave functions $f_0(k)$ and $f_2(k)$ in
the momentum representation (see Appendix C). 
The properly normalized wave functions in the Sch{\"o}dinger picture
are not $f_\ell(k)$ but $F_\ell=\sqrt{N}\,f_\ell$, where $N$ represents
the normalization kernel \cite{FSS}.
The $S$-wave and $D$-wave wave functions in the coordinate
representation, $u(R)$ and $w(R)$, are then
obtained from the inverse Fourier transform of $F_\ell(k)$.
This process is most easily carried out by
expanding $F_\ell(k)$ in a series of Yukawa
functions $\sqrt{2/\pi}k/(k^2+{\gamma_j}^2)$ in the
momentum representation (see Appendix D in \cite{MA00}).
We choose $\gamma_j=\gamma+(j-1)\gamma_0$ with
$\gamma_0=0.9~\hbox{fm}^{-2}$ and $j=1$ - 11.
The $\gamma$ is the $S$-matrix pole $q=-i\gamma$,
from which the deuteron energy $\epsilon_d$ is
most accurately calculated by using the relativistic relation
\begin{eqnarray}
M_n+M_p-\epsilon_d=\sqrt{{M_n}^2-\gamma^2}
+\sqrt{{M_p}^2-\gamma^2} \ .
\label{deu1}
\end{eqnarray}
Figure \ref{deutf} shows the deuteron wave functions of fss2
in the coordinate and momentum representations,
compared with those of the Bonn model-C
potential \cite{MA89} (dotted curves)\footnote{The results
of the Bonn model-C potential in Fig.\,\ref{deutf} and
in Table \ref{deutt} are based on the parameterized
deuteron wave functions given in Table C.4 of \cite{MA89}.}.
We find that the difference between the two models is very small.
Table \ref{deutt} compares various deuteron properties
calculated in three different schemes. They are also
compared with the empirical values and
the predictions by the Bonn model-C potential.
The final value of the deuteron binding energy
for fss2 is $\epsilon_d=2.2309~\hbox{MeV}$.
If we use the non-relativistic energy
expression,\footnote{In Table \protect\ref{deutt},
the value of $\epsilon_d$ in the isospin basis is calculated using this
non-relativistic formula.}
$\epsilon_d=(\gamma^2/M_N)$ for
$\gamma^2=0.05376157~\hbox{fm}^{-2}$ in the full
calculation, we obtain $\epsilon_d=2.2295$ MeV and
the difference is 1.4 keV.
The differences within the deuteron parameters
calculated in the three different schemes
are very small, except for the binding energy $\epsilon_d$.
In particular, the exchange Coulomb kernel
due to the exact antisymmetrization at the quark level
gives an attractive effect to the binding energy,
and increases $\epsilon_d$  by 4.8 keV.
This is even larger than the relativistic
correction included in \eq{deu1}.
The deuteron $D$-state probability is $P_D=5.49~\%$ in fss2,
which is slightly smaller than 5.88 $\%$ in FSS \cite{FSS}.
These values are rather close
to the value $P_D=5.60~\%$ obtained by the Bonn model-C
potential \cite{MA89}.
The asymptotic $D/S$ state ratio $\eta$ and
the rms radius are very well reproduced.
On the other hand, the quadrupole moment is
too small by about 5 - 6$\%$.
There are some calculations \cite{AL78,KO83} which claim that the effect
of the meson-exchange currents on the dueteron
quadrupole moment is as large as $\Delta Q_d=0.01~\hbox{fm}^2$.
It is noteworthy that the Bonn model-C almost reproduces
the correct quadrupole moment, in spite of the fact that
the $D$-state probability is very close to ours.
(On the other hand, the quadrupole moment
of CD-Bonn \cite{MA00} is $Q_d=0.270~\hbox{fm}^2$
with a smaller value $P_D=4.85 \%$.)
For the magnetic moment, precise comparison
with the experimental value requires
a careful estimation of various corrections
arising from the meson-exchange currents
and the relativistic effect, etc.

Table \ref{effect} lists the $S$- and $P$-wave effective range
parameters for the $NN$ system, calculated in the three schemes.
Since the pion-Coulomb correction is not sufficient to explain
the full CIB effect existing in the $np$ and $pp$ $\hbox{}^1S_0$ states,
a simple prescription to multiply the flavor-singlet $S$-meson coupling
constant $f^S_1$ by a factor 0.9949  is adopted to reduce
the too large attraction of the $pp$ central force.
(This prescription is applied only to the calculation
in the particle basis.)
The underlined values of $a$ in Table \ref{effect} indicate
that they are fitted to the experimental values. 
We find that the pion-Coulomb correction
in the $np$ $\hbox{}^1S_0$ state
has a rather large effect on the scattering length parameter $a$.
The value $a=-23.76~\hbox{fm}$ in the isospin basis changes
to $a=-27.38~\hbox{fm}$ due to the effect of the pion mass
correction and the explicit use of the neutron and proton masses.
It further changes to $a=-27.87~\hbox{fm}$ due to the small effect
of the exchange Coulomb kernel.
These changes however should be carefully reexamined by readjusting
the binding energy of the deuteron in Table \ref{deutt}.
We did not carry out this program, since the reduction
of $f^{\rm S}_1$ to fit these values to the empirical
value $a=-23.748 \pm 0.010~\hbox{fm}$ does not help
much to reproduce the CIB of the $pp$ channel anyway.
We have to say that the improvement of the $NN$ $S$-wave
effective range parameters in the particle basis calculation
is not excellent, in spite of the large effort 
expended in incorporating the pion-Coulomb correction
in the microscopic RGM formalism.
This shortcoming might be related to the insufficient
description of the low-energy $pp$ differential
cross sections around $\theta_{\rm cm} \sim 90^\circ$,
observed in  Fig.\,\ref{ppdif}.
It was also pointed out by the Nijmegen
group \cite{BE88} that
the Coulomb phase shift should be improved
by the effects of two-photon exchange, vacuum polarization
and magnetic moment interactions, in order to describe
the $\hbox{}^1S_0$ phase shift precisely at energies
less than 30 MeV. These effects are not incorporated
in the present calculation.
The $P$-wave effective range parameters are also
given in Table \ref{effect}, in order to compare with
a number of empirical predictions.
The parameters of $\hbox{}^3P_2$ state are not given, since
the effective range expansion of this partial wave requires
a correction term related to the accidental $p^5$ low-energy
behavior of OPEP \cite{NA75}.

\subsection{${\protect\mbf \Sigma^+ p}$ system}

Figure \ref{phsp} displays the $S$- and $P$-wave
phase shifts of the $\Sigma N(I=3/2)$ system,
calculated in the isospin basis.
The results given by FSS (dashed curves) and
RGM-H (dotted curves) are also shown for comparison.
The $\hbox{}^1E$ and $\hbox{}^3O$ states
of the $\Sigma N(I=3/2)$ system belong to 
the (22) component in the flavor $SU_3$ representation,
which is common with the $^1E$ and $^3O$ states
of the $NN$ system \cite{NA95}.
The phase-shift behavior of these partial waves
therefore resembles that of the $NN$ system,
as long as the effect of the $SU_3$ symmetry breaking
is not significant.
Figure \ref{phsp} shows that the attraction
in the $\hbox{}^1S_0$ state is much weaker than that
of the $NN$ system, and the phase-shift peak is
about $26^\circ$ around $p_{\Sigma}=200~\hbox{MeV}/c$.
The $^3P_J$ phase shifts show the characteristic
energy dependence observed in the $NN$ phase shifts,
which is caused by competition among the central,
tensor and $LS$ forces.
The appreciable difference in the three models fss2, FSS
and RGM-H appears only in the $^3P_2$ state.
It is discussed in \cite{FJ96} that the attractive
behavior of the $^3P_2$ phase shift is closely related to
the magnitude of the $\Sigma^+ p$ polarization $P(\theta)$
at the intermediate energies.
The more attractive, the larger $P(\theta=90^\circ)$.
Since the attraction of fss2 is just between
FSS and RGM-F (see Fig.\,\ref{phsp}(b)), the polarization
curve at $p_{\Sigma}=450~\hbox{MeV}/c$ falls between
the two curves given by FSS and RGM-H (see Fig. 4 of \cite{FJ96}).
This implies our quark-model prediction $P(\theta=90^\circ)=0.1$ - 0.2 at
$p_{\Sigma}=450~\hbox{MeV}/c$.

On the other hand, the $^3E$ and $^1O$ states
of the $\Sigma N(I=3/2)$ system have the (30) symmetry
of the flavor $SU_3$ \cite{NA95}.
We have no information on the properties of these states
from the $NN$ interaction.
In our present framework of the quark model,
there is very little ambiguity in the phase
shifts of these states, as can be seen from Fig. \ref{phsp}.
Since the configuration $(0s)^6$ in the $^3S_1$ state
is almost forbidden by the effect of the Pauli principle,
the interaction in the $^3S_1$ state is strongly repulsive.
This property is common in all of our models,
and the phase shifts predicted by fss2, FSS and
RGM-H are almost the same.
On the other hand, the $^1P_1$ phase shift
is weakly attractive, which is caused
by the exchange kinetic-energy kernel due to the
Pauli principle.
This property is also common to all three models.
The Nijmegen hard core models D and F give a resonance
in the phase shift of this channel,
which induces enhancement in the $\Sigma^+p$ differential
cross sections at the forward and backward
angles. (See Fig.\,6 in \cite{SCAT}.)
Such behavior is not found in any of our quark models.

The effective range parameters of the $YN$ scattering
in the single-channel analysis are given
in Table \ref{effect} with some empirical values.
For the $\Sigma^+ p$ system, the empirical values given
in \cite{NA73} should be compared with the results
in the particle-basis calculation including the
Coulomb force. We find a reasonable agreement
both in the $\hbox{}^1S_0$ and $\hbox{}^3S_1$ states.

\subsection{${\protect\mbf \Lambda N}$ system}

The total cross section for the $\Lambda p$ elastic scattering
predicted by fss2 in the isospin basis is displayed
in Fig.\,\ref{lato}(a), together with the previous result
given by FSS. The new model and FSS reproduce experimental data
equally well in the low momentum
region $p_{\Lambda} \le 300$ MeV/$c$. 
The total cross section has a cusp structure
at the $\Sigma N (I=1/2)$ threshold, which is due to
the strong $\Lambda N$ - $\Sigma N (I=1/2)$
$^3S_1$ - $^3D_1$ coupling caused by the OPEP tensor force.
To see the dominant role of the tensor force
of the $\Sigma N$ - $\Lambda N$ coupling in more detail,
we show the $^3S_1$ and $^1S_0$ phase shifts in Fig.\,\ref{phlam1},
which are calculated by fss2 (left) and FSS (right) in the isospin basis.
A cusp structure at the $\Sigma N$ threshold is apparent
in the $^3S_1$ channel, while very small in the $^1S_0$ channel.
In both models, attraction in the $^1S_0$ state is stronger than
that in the $^3S_1$ state.
However, the differences between the strengths of attraction
in the $^1S_0$ and $^3S_1$ states vary. The old model FSS
gives $\delta (\hbox{}^1S_0) - \delta (\hbox{}^3S_1)
\sim 30^\circ$ at $p_{\Sigma} \sim 200$ MeV/$c$,
while fss2 predicts $\delta (\hbox{}^1S_0) - \delta (\hbox{}^3S_1)
\sim 10^\circ$. The difference
of $\delta (\hbox{}^1S_0) - \delta (\hbox{}^3S_1) \sim 10^\circ$ is
required from the few-body calculations\cite {SH83,YA94,HI97,NE00}
of $s$-shell $\Lambda$-hypernuclei, as discussed in 7) of Sec.\,II\,C.
The parameter search of fss2 is carried out under this constraint.

As seen in Fig.\,\ref{lato}(a),
FSS predicts especially large enhancement
of the total cross sections around 
the cusp at the $\Sigma N$ threshold. This is due to a rapid
increase of the $\Lambda N$ $\hbox{}^3P_1$ - $\Lambda N$
$\hbox{}^1P_1$ transition around the threshold.
Figure \ref{phlam2} shows the $S$-matrix
$S_{ij}=\eta_{ij}e^{2i\delta_{ij}}$ 
for the $\Lambda N$ - $\Sigma N(I=1/2)$
$^1P_1$ - $^3P_1$ channel coupling for
fss2 (upper), FSS (middle), and RGM-H (lower).
The result by FSS shows that the transmission
coefficient $\eta_{21}$, which corresponds
to the $\Lambda N$ $\hbox{}^1P_1
\rightarrow \Lambda N$ $\hbox{}^3P_1$ transition,
increases very rapidly around the $\Sigma N$ threshold
as energy increases. This increase of
the $\eta_{21}$ (and the resultant decrease of the reflection
coefficient $\eta_{11}$) is a common feature of all our models.
The strength of the transition, however, has some model dependence. 
The transition is stronger in FSS than in fss2 and RGM-H.
In particular, the resemblance of $S$-matrix in fss2 and
RGM-H is very outstanding.

The behavior of the diagonal phase shifts is largely
affected by the strength of the $\hbox{}^1P_1$ - $\hbox{}^3P_1$ channel
coupling (which is directly reflected in $\eta_{21}$).
The new model fss2 and RGM-H yield a broad resonance
in the $\Sigma N$ $\hbox{}^3P_1$ channel.
On the other hand, FSS (and also RGM-F) with stronger channel
coupling provides no resonance in this channel.
Instead, a step-like resonance appears
in the $\Lambda N$ $\hbox{}^1P_1$ channel.
This situation is summarized in Table \ref{reson}.
The location of the resonance is
determined by the strength of the $LS^{(-)}$ force
and the strength of the attractive central force
in the $\Sigma N(I=1/2)$ channel.
In the quark model, the central attraction 
by the S mesons is enhanced by the exchange kinetic-energy kernel,
which is attractive in the $\Sigma N$ $\hbox{}^3P_1$ channel due to the 
effect of the Pauli principle. The $LS$ and $LS^{(-)}$ forces also
contribute to increase this attraction.
In FSS and RGM-F, a single channel calculation for
the $\Sigma N (I=1/2)$ system yields a resonance
in the  $\hbox{}^3P_1$ state. When the channel coupling
to the $\Lambda N$ system is introduced, this resonance
transfers to the $\Lambda N$ channel
due to the strong $LS^{(-)}$ force generated from the
FB interaction. On the other hand, the central force
of RGM-H in the $\Sigma N(I=1/2)$ channel is weaker than
in FSS (see $V^C_{\Sigma N (1/2)}(\hbox{}^3S)$ in
Table II of \cite{LSRGM}).
The $LS^{(-)}$ force of RGM-H is also
weak ($\sqrt{2/\pi}\alpha_Sx^3m_{ud}c^2 =$ 296 MeV). 
Accordingly, the resonance stays in the $\Sigma N$ $\hbox{}^3P_1$ channel,
even if the channel coupling is incorporated.
In the new model fss2, the $LS$ term of the S-meson exchange is included.
Since EMEP yields no $LS^{(-)}$ force in the present
framework (see Sec.\,II\,B), the role of the FB $LS^{(-)}$ force becomes
less significant, in comparison with the very strong effect in FSS. 
This is the reason the resonance remains in the $\Sigma N$
$\hbox{}^3P_1$ channel in fss2.

In spite of these quantitative differences in the coupling
strength, the essential mechanism
of the $\Lambda N$ - $\Sigma N(I=1/2)$
$^1P_1$ - $^3P_1$ channel coupling is the same
for all of our models. It is induced by the strong
FB $LS^{(-)}$ force, which directly connects
the two $SU_3$ configurations
with the $(11)_s$ and $(11)_a$ representations
in the $P$-wave $I=1/2$ channel.
In order to determine the detailed phase-shift behavior of each channel,
including the position of the $P$-wave resonance,
one has to know the strength of the $LS^{(-)}$ force
and the strength of the attractive central force
in the $\Sigma N(I=1/2)$ channel.
This is possible only by the careful analysis of rich
experimental data for the $\Lambda p$ and $\Sigma^- p$ 
scattering observables in the $\Sigma N$ threshold region.

Let us discuss another important feature
of the $\Lambda N$ phase shifts
induced by the $P$-wave coupling due to the $LS^{(-)}$ force.
The $\Lambda N$ $\hbox{}^1P_1$ and $\hbox{}^3P_1$ phase shifts
in Fig. \ref{phlam2} show weakly attractive behavior over the
energies below the $\Sigma N$ threshold.
This is due to the dispersion-like (or step-like
for the $\Lambda N$ $\hbox{}^1P_1$ state in FSS) resonance
behavior with a large width. 
This attraction in the $P$ state can be observed by the forward to
backward ratio ($F/B$) of the $\Lambda p$ differential cross sections,
as is seen in Fig.\,\ref{lato}(b). Our new model fss2 and FSS
give $F/B>1$ below the $\Sigma N$ threshold,
which implies that the $P$-state $\Lambda N$ interaction
is weakly attractive as suggested by Dalitz {\em et al.} \cite{dalitz}.
In our models, this attraction originates from
the strong  $\Lambda N$ - $\Sigma N (I=1/2)$
$\hbox{}^1P_1$ - $\hbox{}^3P_1$ coupling
due to the FB $LS^{(-)}$ force \cite{SCAT}.

Some comments are in order with respect to particle-basis calculations
of the $\Lambda p$ and $\Lambda n$ scatterings.
From the energy spectrum
of the $\hbox{}^3_\Lambda\hbox{H}$ - $\hbox{}^3_\Lambda\hbox{He}$ 
isodoublet $\Lambda$-hypernuclei, it is inferred that
the $\Lambda p$ interaction is more attractive than
the $\Lambda n$ interaction. This CSB may have its origin
in the different threshold energies of the $\Sigma N$ particle
channels as in Fig. \ref{thres} and the Coulomb attraction
in the $\Sigma^- p$ channel for the $\Lambda n$ system.
The former effect increases the cross sections of
the $\Lambda p$ interaction
and the latter the $\Lambda n$ interaction.
However, the full calculation including the pion-Coulomb
correction using the correct threshold energies
in Table \ref{table7} yields very small difference
in the $S$-wave phase shifts.
In the energy region up to $p_{\Lambda}=200~\hbox{MeV}/c$,
the $\Lambda p$ phase shift is more attractive
than the $\Lambda n$ phase shift only by less
than $0.2^\circ$ in the $\hbox{}^1S_0$ state, and
by less than $0.4^\circ$ in the $\hbox{}^3S_1$ state.
This can also be seen from
the $\Lambda p$ and $\Lambda n$ effective range
parameters tabulated in Table \ref{effect},
obtained in the three different calculational schemes.
On the other hand, a rather large effect of CSB is
found in the $\Sigma^- p$ channel as discussed in the
next subsection. This is because the CSB effect is enhanced
by the strong $\Lambda N$ - $\Sigma N (I=1/2)$ channel-coupling
effect in the $\hbox{}^3S_1$ - $\hbox{}^3D_1$ state. 
Figure \ref{cusp}(a) displays the enlarged picture of
total $\Lambda N$ cross sections in the $\Sigma N$ threshold
region, calculated in the full scheme 3) with the Coulomb force.
The effect of different threshold energies
in the particle basis is clearly observed.
Figure \ref{cusp}(b) demonstrates the
behavior of the subthreshold reaction cross sections,
which shows a very strong channel dependence
related to the small difference of the threshold energies.
In contrast to this, the Coulomb effect
in the $\Sigma^- p$ channel is found to be very small
as long as the reactions from the $\Lambda N$ incident channel
are concerned.

\subsection{The ${\protect\mbf \Sigma^- p}$ system}

Since the $\Sigma^- p$ system is expressed
as $|\Sigma^- p\rangle=-\sqrt{2/3}|\Sigma N (I=1/2)\rangle
+\sqrt{1/3}|\Sigma N (I=3/2)\rangle$ in the isospin basis,
it is important to know first the phase-shift behavior
of the $\Sigma N (I=1/2)$ system.
The $SU_3$ decomposition of the $\Sigma N (I=1/2)$ state
\begin{eqnarray}
& & \Sigma N (I=1/2)={1 \over \sqrt{10}}\left[ 3(11)_s-(22) \right]
\qquad \quad \hbox{for} \qquad ^1E
\quad \hbox{and} \quad ^3O \quad \hbox{states}\ , \nonumber \\
& & \Sigma N (I=1/2)={1 \over \sqrt{2}}\left[(11)_a+(03)\right]
\qquad \qquad \hbox{for} \qquad ^3E \quad \hbox{and} \quad
^1O \quad \hbox{states}\ ,
\label{sig1}
\end{eqnarray}
is very useful to know the quark-model prediction for
the phase-shift behavior in the isospin basis \cite{NA95}.
The most compact $(0s)^6$ configuration
in the $(11)_s$ $SU_3$ state is completely Pauli forbidden.
This implies that the $\Sigma N (I=1/2)$ $\hbox{}^1S_0$ phase shift
is very repulsive due to the exchange kinetic-energy kernel \cite{FSS}.
On the other hand, $\Sigma N (I=1/2)$ $\hbox{}^3S_1$ phase shift
is expected to have attraction similar
to the $\Lambda N$ $\hbox{}^3S_1$ phase shift, as long as
the effect of the flavor symmetry breaking is not important.
(Note that $\Lambda N=\left[-(11)_a+(03)\right]
/\sqrt{2}$ for $\hbox{}^3E$ and $\hbox{}^1O$ states.)
Unfortunately, the last condition is applicable
only to the central force, since the one pion tensor force
introduces quite a few complexities in the channel
dependence in the $\Sigma N (I=1/2)$ - $\Lambda N$
$\hbox{}^3S_1$ - $\hbox{}^3D_1$ channel-coupling problem.
The strength of the $\Sigma N (I=1/2)$ central attraction,
discussed in the preceding subsection, should therefore
be examined carefully after this $S$ - $D$ wave channel
coupling is properly treated. 

Figure \ref{phsgn} displays the $\hbox{}^1S_0$ phase shift
and the $S$-matrix of the $\Lambda N$-$\Sigma N (I=1/2)$
$\hbox{}^3S_1$ - $\hbox{}^3D_1$ coupled-channel state.
The upper figures are the predictions by fss2, while the lower
ones by FSS.
The $\hbox{}^1S_0$ phase shift shows very strong Pauli repulsion,
similar to the phase shift of the $\Sigma N (I=3/2)$
$\hbox{}^3S_1$ state.
The $\Sigma N (I=1/2)$ $\hbox{}^3S_1$ phase shift 
predicted by fss2 starts from $180^\circ$ at $p_{\Sigma}=0$,
decreases moderately down to $\sim 160^\circ$ over
the momentum region $p_{\Sigma}\le 100$ MeV/$c$. 
Then it suddenly decreases down
to $\sim 80^\circ$ at around $p_{\Sigma}\sim 120$ MeV/$c$.
Beyond this momentum, a moderate decrease follows again.
The situation in FSS is rather different.
The $\hbox{}^3S_1$ phase shift predicted by FSS starts
from $0^\circ$ at $p_{\Sigma}=0$, and shows
a clear resonance behavior over the momentum
range $100 < p_{\Sigma}< 120$ MeV/$c$.
The peak value of the phase shift is about $\sim 60^\circ$.
In spite of the very different behavior of the diagonal
phase shifts predicted by fss2 and FSS, the $S$-matrices
are found to be very similar to each other.
Figure \ref{sgsmat} illustrates the Argand diagram showing the
energy dependence of the $S$-matrix
element $S_{33}=\eta_{33}e^{2i\delta_{33}}$.
Here $i=3$ represents the $\Sigma N (I=1/2)$
$\hbox{}^3S_1$ channel.
The resemblance of the circles given by fss2 and FSS is apparent.
This implies that the strength of the central attraction
in the $\Sigma N(I=1/2)$ channel is almost the same in fss2 and FSS.
Figure \ref{sgsmat} also shows that the peak value
of the RGM-H phase shift is about $45^\circ$,
which indicates that the attraction of RGM-H is weaker than
that of fss2 and FSS.

The very strong reduction of $\eta_{33}=|S_{33}|$ in
Fig.\,\ref{phsgn} is related to the large enhancement
of the transmission coefficients $\eta_{13}$ and $\eta_{23}$ to
the $\Lambda N$ $\hbox{}^3S_1$ ($i=1$) and  
$\Lambda N$ $\hbox{}^3D_1$ ($i=2$) channels
around $p_\Sigma \sim 120~\hbox{MeV}$.
These transmission coefficients are directly connected
to the $\Sigma^- p\rightarrow \Lambda n$ reaction cross
sections (which we call process C),
and the driving force for this transition is the one-pion
tensor force.
Table \ref{tabcrs} shows contributions from each partial wave
to this reaction cross section at $p_{\Sigma}=$ 160 MeV/$c$.
The sum of the contributions from the transitions $\hbox{}^3S_1
\rightarrow \hbox{}^3S_1$ and $\hbox{}^3S_1\rightarrow
\hbox{}^3D_1$ amounts to about 120 mb both in fss2 and FSS,
which is much larger than the contribution
from the transition $\hbox{}^1S_0
\rightarrow \hbox{}^1S_0$ (3.5 mb).
This is in accordance with the analysis of the $\Lambda p$ system
in the preceding subsection.
Namely, there is a large cusp structure
in the $\hbox{}^3S_1$ phase shift at the $\Sigma N$ threshold,
while a very small cusp is seen in the $\hbox{}^1S_0$ channel.
There is, however, some quantitative difference
in the detailed feature of this tensor coupling between fss2 and FSS.
In fss2, $\eta_{13}$ and $\eta_{23}$ are almost equal
to each other over the momentum region
where the experimental data
exist ($110 \le p_{\Sigma}\le 160$ MeV/$c$).
This feature is very similar to RGM-F (see Fig.\,6 in \cite{RGMFb}).
On the other hand, $\eta_{13}< \eta_{23}$ holds in FSS,
which is common with RGM-H (Fig.\,5(b) in \cite{FSS}).
We can also read this difference 
from Table \ref{tabcrs}.
The details of the cross section of the process C indicate that 
$\sigma(\hbox{}^3S_1\rightarrow \hbox{}^3S_1)$ $\sim$
$\sigma(\hbox{}^3S_1\rightarrow \hbox{}^3D_1)$ in fss2,
while $\sigma(\hbox{}^3S_1\rightarrow \hbox{}^3S_1)$
$<$ $\sigma(\hbox{}^3S_1\rightarrow \hbox{}^3D_1)$ in FSS.

For more detailed evaluation of $\Sigma^- p$ cross sections, 
it is important to take into account the pion-Coulomb correction.
In \cite{FU98}, we incorporated the Coulomb force
of the $\Sigma^- p$ channel correctly in the particle basis,
but the threshold energies of the $\Sigma^0 n$ and $\Lambda n$ channels
were not treated properly. As is discussed in Sec.\,II\,D,
we can now deal with the empirical threshold energies and
the reduced masses in the RGM formalism, without spoiling
the correct effect of the Pauli principle.
Since the threshold energies and reduced masses are calculated
from the baryon masses, these constitute a part
of the pion-Coulomb correction in the $YN$ interaction.
Although the pion-Coulomb correction may
not be the whole story of the CSB, it is
certainly a first step to improve the accuracy of
the model predictions calculated in the isospin basis.
We can easily imagine that the small difference of threshold
energies becomes important more and more
for the low-energy $\Sigma^- p$ scattering. In particular,
the charge-exchange total cross section $\Sigma^- p \rightarrow
\Sigma^0 n$ (which we call process B) does not satisfy
the correct $1/v^2$ law in the zero-energy limit,
if the threshold energies of
the $\Sigma^- p$ and $\Sigma^0 n$ channels are assumed to be equal.
We therefore used the prescription\cite{SW62} to multiply
the factor $(k_f/k_i)$, in order
to get $\sigma (\hbox{B})$ from $\bar{\sigma}(\hbox{B})$ which
is calculated by ignoring the difference of the
threshold energies.
Here $k_i$  and $k_f$ are the relative momentum in the initial and final
states, respectively.
We will see below that this prescription is not accurate
and overestimate $\sigma (\hbox{B})$ and $\sigma (\hbox{C})$.

The largest effect of the pion-Coulomb correction appears
in the calculation of the $\Sigma^- p$ inelastic capture
ratio at rest $r_R$ \cite{FU98}.
This observable is defined by \cite{SW62}
\begin{eqnarray}
r_R={1\over 4}\,r_{S=0} + {3 \over 4}\,r_{S=1}
\qquad \hbox{with} \qquad
\left. r_{S=0,1}\equiv {\sigma_{(S=0,1)}({\rm B}) \over
\sigma_{(S=0,1)}({\rm B})+\sigma_{(S=0,1)}({\rm C})}
\right|_{p_{\Sigma^-}=0} \ ,
\label{sig2}
\end{eqnarray}
where B and C denote the scattering processes $\Sigma^-p\rightarrow
\Sigma^0 n$ and  $\Sigma^-p \rightarrow \Lambda n$, respectively.
This quantity is the ratio of the production rates
of the $\Sigma^0$ and $\Lambda$ particles 
when a $\Sigma^-$ particle is trapped in the atomic orbit
of the hydrogen and interacts with the proton nucleus. 
For the accurate evaluation of $r_R$, we first determine
the effective range parameters of the low-energy $S$-matrix
using the multi-channel effective range theory.
These are given in Tables \ref{effect2} and \ref{effect3} with
respect to the $\hbox{}^1S_0$ and $\hbox{}^3S_1+\hbox{}^3D_1$ states
of the $\Sigma^- p$ - $\Sigma^0 n$ - $\Lambda n$ system, respectively.
The calculations are performed by using the particle basis with
and without the Coulomb force.
Figures \ref{part1} and \ref{part2} show the calculated phase shifts 
in the particle basis, when the Coulomb force is included. 
Also shown are the predictions by the effective range
formula, using the parameters given
in Tables \ref{effect2} and \ref{effect3}.
For the $\hbox{}^3S_1+\hbox{}^3D_1$ state, the effective range expansion
breaks down around $p_\Sigma \sim$ 140 MeV/$c$ due to the singularity
of the matrix inversion.
On the other hand, the effective range approximation works
excellently for the $\hbox{}^1S_0$ state.
The scattering length matrices $A$ ($A^c$) and $\CA$ ($\CA^c$) are
employed to calculate $r_R$ given
in Table \ref{rest} without (with) the Coulomb force.
If we compare the result with the empirical
values $r_R=0.33  \pm 0.05$ \cite{RO58},
$0.474 \pm 0.016$ \cite{HE68},
and $0.465 \pm 0.011$ \cite{ST70},
we find that  $r_R=0.442$ predicted by fss2 is
slightly smaller than the recent values
between 0.45 and 0.49.
The contribution from each spin state is also listed in
Table \ref{rest}.
We find $r_{S=0}\sim 0.9$, which indicates
that the $\sigma_{\rm C}$ is very small
in comparison with the $\sigma_{\rm B}$ in the spin-singlet state.
On the other hand, $r_{S=1} \sim 0.29$ implies that
most of the $\Sigma N$-$\Lambda N$ channel
coupling takes place in the spin-triplet state. 
We again find that the one-pion tensor
force is very important
in the $\Sigma N (I=1/2)$ - $\Lambda N$
$\hbox{}^3S_1$ - $\hbox{}^3D_1$ channel-coupling problem.
We also find that the effect of the Coulomb force plays
a minor role for this ratio \cite{FU98}.

The $\Sigma^- p$ inelastic capture ratio in flight $r_F$,
predicted by fss2 in the full calculation,
is illustrated in Fig.\,\ref{flight}.
Unlike $r_R$, this quantity is defined by
using the total cross sections $\sigma_T(\hbox{B})
=(1/4)\sigma_0(\hbox{B})+(3/4)\sigma_1(\hbox{B})$ and 
$\sigma_T(\hbox{C})=(1/4)\sigma_0(\hbox{C})
+(3/4)\sigma_1(\hbox{C})$ directly:
\begin{eqnarray} 
r_F={\sigma_T(\hbox{B}) \over \sigma_T(\hbox{B})+\sigma_T(\hbox{C})}\ .
\label{sig3}
\end{eqnarray}
This is a rather sensitive quantity which depends on the relative
magnitudes of the $\Sigma^- p \rightarrow \Sigma^0 n$ and
$\Sigma^- p \rightarrow \Lambda n$ total cross sections.
In the momentum region of $p_\Sigma \geq 100~\hbox{MeV}/c$,
we find that $r_F$ in the particle basis gives rather similar values,
irrespective of whether the Coulomb force is incorporated or not.
The empirical value of $r_F$ averaged over the momentum
interval $p_\Sigma=110$ - 160 MeV/$c$ is $r_F=  0.47 \pm 0.03$ \cite{EN66},
and is plotted in Fig.\,\ref{flight} by a cross.
We find that the prediction of fss2, $r_F=0.419$, is too small,
which is the same feature as observed in FSS ($r_F=0.41$) \cite{FU98}.
The main reason for this disagreement is that
our $\Sigma^- p \rightarrow \Lambda n$ cross sections
are too large.

\subsection{${\protect\mbf YN}$ cross sections}

Figure \ref{sgto} displays the low-energy cross sections
predicted by fss2 for the $\Sigma^- p$ and $\Sigma^+ p$ scattering.
The results of three different calculations are shown;
the full calculation in the particle basis including the Coulomb
force (solid curves), the calculation in the particle basis
without the Coulomb force (dashed curves), and the calculation
in the isospin basis (dotted curves).
The effect of the correct threshold energies and the Coulomb
force is summarized as follows. In the $\Sigma^+ p$ scattering,
the effect of the repulsive Coulomb force reduces the total
cross sections by 11 - 6 mb in the momentum
range $p_\Sigma=140$ - 180 MeV/$c$, where the experimental data
exist. (See Table \ref{xdif}.) On the other hand, the attractive
Coulomb force in the incident $\Sigma^- p$ channel increases
all the cross sections. An important feature of the present
calculation is the effect of correct threshold energy
of the $\Sigma^0 n$ channel. It certainly increases
the $\Sigma^- p \rightarrow \Sigma^0 n$ charge-exchange
cross section, but the prescription to multiply
the factor $(k_f/k_i)$ overestimates this effect.
The real increase is about 1/2 - 2/3 of this estimation.
Furthermore, we find that this change is accompanied with
the fairly large decrease of the $\Sigma^- p$ elastic
and $\Sigma^- p \rightarrow \Lambda n$ reaction cross sections.
Apparently, this effect is due to the conservation of the total
flux. The net effect of the Coulomb and threshold energies
becomes almost zero for the $\Sigma^- p$ elastic scattering.
The charge-exchange reaction cross section largely increases,
and the $\Sigma^- p \rightarrow \Lambda n$ reaction cross
section gains the moderate decrease.
Table \ref{xdif} summarizes this change of total cross sections
in the momentum range $p_\Sigma=110$ - 200 MeV/$c$.
We have also examined the effect of the threshold energies 
and the Coulomb force in a simple potential model, which
fits the low-energy phase shifts of the model FSS.
(The crosses shown in the FSS phase-shift curves
in Figs.\,\ref{phsp}(a), \ref{phlam1}, and \ref{phsgn} indicate
the predictions of this potential model.)
The cross section difference in this model is also shown
in Table \ref{xdif} for comparison. We find that both calculations
give very similar results.
Figure \ref{sgto} also shows the comparison with the experimental
data. The final result given by fss2 reproduces the experimental data
reasonably well, although the $\Sigma^- p \rightarrow
\Lambda n$ total reaction cross sections are somewhat too large.

Figure \ref{yndif} shows the predicted differential cross
sections by fss2 in the full calculation,
compared with the experimental data. 
For $\Sigma^+ p$ and $\Sigma^- p$ elastic differential
cross sections, the recent experimental data taken
at KEK \cite{E251,E289} are also compared.
The agreement between the calculation
and the experiment is satisfactory. 

We show in Fig. \ref{tot} the total cross sections
of the $NN$ and $YN$ scatterings in the full energy range.
The solid curves denote the fss2 result in the full calculation,
while the dotted curves in the isospin basis.
The ``total'' cross sections in the charged channels,
$pp$, $\Sigma^+ p$, and $\Sigma^- p$, are calculated by
integrating the differential cross sections over the angles
from $\cos \theta_{\rm min}=0.5$ to $\cos \theta_{\rm max}=-0.5$.
This is the reason the Coulomb result of the $pp$ total
cross sections at the higher energies $T_{\rm lab}\geq 400$ MeV
is very small, in comparison with the result in the isospin basis.
Since the $pp$ differential cross sections at these higher energies
are very much V-shaped (see Fig. \ref{ppdif}),
the non-Coulomb calculation in the isospin basis is more reliable. 
The further difference between the dotted curve
and the experiment in th $pp$ scattering is due to the inelastic
cross sections, which are zero in our single-channel calculation.
The experimental analysis of the $pp$ total inelastic cross sections
shows that they are about 20 mb at $T_{\rm lab}=800$ MeV.
For the $np$ scattering, the inelastic contribution is
smaller and is about 10 mb at the same energy. This implies that
our fss2 predicts the total elastic $NN$ cross sections
almost correctly up to the energies $T_{\rm lab} \leq 800$ MeV.
For the $YN$ total cross sections, the pion-Coulomb correction
is important only for the charged channels and the
low-energy $\Sigma^- p \rightarrow \Sigma^0 n$ reaction
cross sections.
 
\subsection{${\protect\mbf G}$-matrix calculation}

Figure \ref{matter} shows saturation curves
calculated for ordinary nuclear matter
with the $QTQ$ prescription as well as the continuous prescription
for intermediate spectra.
The results produced by the Paris potential \cite{PARI}
and the Bonn B potential \cite{BM90} are also shown for comparison.
The $k$-dependence of the nucleon, $\Lambda$
and $\Sigma$ s.p. potentials $U_B(k)$ obtained
with the continuous choice is shown in Fig.\,\ref{spnuc} at
three densities $\rho = 0.5 \rho_0$, $0.7\rho_0$ and $\rho_0$,
with $\rho_0$ = 0.17 fm$^{-3}$ being the normal density.
(These densities correspond to $k_F =1.07,~1.2$ and 1.35 fm$^{-1}$,
respectively.)
For comparison, the results of the Nijmegen soft-core potential
NSC89 \cite{NSC89} calculated by
Schulze {\em et al.} \cite{SCHU} are also shown.
The corresponding figures of the s.p. potentials predicted by
our previous model FSS are given in Figs.\,2 - 5 of \cite{GMAT}.
We find that fss2 gives a nucleon s.p. potential $U_N(k)$ very
similar to that of FSS except for the higher momentum
region $q_1 \geq 3~\hbox{fm}^{-1}$.
As is discussed at the end of Sec.\,III\,A,
the too attractive behavior of FSS in this momentum region
is corrected in fss2, owing to the effect
of the momentum-dependent Bryan-Scott terms
involved in the S-meson and V-meson exchange EMEP.
The saturation curve in Fig. \ref{matter} shows that
this improvement of the s.p. potential in the high-momentum
region has the favorable feature of moving the saturation density
to the lower side, as long as the calculation is carried out
in the continuous prescription. 
On the other hand, the saturation curve with the $QTQ$ prescription
suffers a rather large change in the transition
from FSS to fss2. The prediction in fss2 with the $QTQ$ prescription
is very similar to the prediction in Bonn model-B potential.
It is interesting to note that our fss2 result is rather close
to Bonn model-C  for the deuteron properties (see Table \ref{deutt}),
while to model-B for the nuclear saturation properties.
The model-B has a weaker tensor force than model-C, which is
a favorable feature for the nuclear saturation properties.  

We should keep in mind that the short-range part of our quark model
is mainly described by the quark-exchange mechanism.
The non-local character of this part is entirely different from
the usual V-meson exchange picture in the standard
meson-exchange models.
In spite of this large difference the saturation point
of our quark model does not deviate much from the Coester band,
which indicates that our quark model has similar saturation properties
with other realistic meson-exchange potentials.

Figures\,\ref{spnuc}(b) and \ref{spnuc}(c) show the momentum dependence
of the $\Lambda$ and $\Sigma$ s.p. potentials in nuclear matter obtained
from the quark-model $G$-matrices of fss2.
We find that the $U_\Lambda(q_1)$ predicted by fss2
and FSS (shown in Fig.\,3 of \cite{GMAT}) are again
very similar for $q_1 \leq 2~\hbox{fm}^{-1}$.
On the other hand, the repulsion of the $U_\Sigma(q_1)$ predicted
by FSS in Fig.\,5 of \cite{GMAT} is somewhat reduced.
The partial wave contributions
of the s.p. potentials \(U_{\Lambda} (q_1=0)\) and
\(U_{\Sigma} (q_1=0)\) in symmetric nuclear matter
at $k_F = 1.35~\hbox{fm}^{-1}$, predicted by fss2, are tabulated
in Table \ref{spcom}, together with the result of NSC89 \cite{SCHU}.
The corresponding analysis in FSS is given in Table 1 of \cite{GMAT}.
For the $\Lambda$ s.p. potential, the characteristic feature
of fss2 appears in the less attractive $\hbox{}^1S_0$ state
and the more attractive $\hbox{}^3S_1$ state,
in comparison with FSS.
The partial wave contributions of fss2 now become very similar
to those of NSC89, except for the $\hbox{}^3S_1+\hbox{}^3D_1$
contribution. The extra attraction of fss2 to NSC89
in the $\Lambda$ s.p. potential mainly comes
from this channel (15 - 16 MeV).
This is probably because the tensor coupling
is stronger in fss2 than in NSC89.
A minor excess of the attraction comes from
the $\hbox{}^1P_1+\hbox{}^3P_1$ and $\hbox{}^3P_2
+\hbox{}^3F_2$ states (2 - 3 MeV).

For the $\Sigma$ s.p. potential, it should be noted that
it is repulsive in the quark model,
reflecting the characteristic repulsion
in the $\hbox{}^3 S_1+\hbox{}^3D_1$ channel
of the isospin $I=3/2$ state (the Pauli repulsion).
The repulsive feature of the $\Sigma$ s.p. potential is
supported by Dabrowski's analysis \cite{DA99} of
the recent $(K^- ,\pi^{\pm})$ experimental data
at BNL \cite{BNL98}.
Quantitatively, the strength of the repulsion (which
is 21 MeV in FSS for $U_\Sigma(q_1=0)$) is
reduced to 7 MeV in fss2.
This change of the s.p. potential is mainly brought about by
the 7 MeV reduction of the $I=3/2$ $\hbox{}^3 S_1
+\hbox{}^3D_1$ repulsion and the 4 MeV increase
of the $I=1/2$ $\hbox{}^3 S_1+\hbox{}^3D_1$ attraction.
The latter feature is again related to the strong tensor
coupling in fss2. On the other hand, the repulsive contribution
of the $I=3/2$ $\hbox{}^3 S_1+\hbox{}^3D_1$ state in NSC89
is very weak, since this channel has a broad resonance
around $p_\Sigma=500$ - $800~\hbox{MeV}/c$ (see Fig.\,1 in \cite{FJ96}).
It is interesting to note that the attractive contributions
to the $\Lambda$ and $\Sigma$ s.p. potentials
from the $I=1/2$ $\hbox{}^3 S_1+\hbox{}^3D_1$ state
is more than 10 MeV stronger in fss2, compared with
those of NSC89. Although NSC89 is considered to be a model with
a strong $\Lambda N$ - $\Sigma N$ coupling,
the $\Lambda N$ - $\Sigma N$ coupling of fss2 is even stronger.
This feature should have some consequence in the energy spectra
of the $s$-shell $\Lambda$-hypernuclei, if fss2 is used in
the few-body calculations of these hypernuclei.

The imaginary parts of the $\Lambda$ and $\Sigma$ s.p. potentials
are shown in Fig.\,\ref{spnuc}(d) with respect to fss2.
These results are rather similar to the predictions
of FSS (see Fig.\,4 in ref. \cite{GMAT}).
In particular, $\Im m~U_\Sigma (q_1=0)$ at $k_F = 1.35$ fm$^{-1}$ is
$-13.9~\hbox{MeV}$ in fss2 and $-18.5~\hbox{MeV}$ in FSS.
These results are in accord with the calculations
by Schulze {\it et al.} \cite{SCHU} for NSC89.

By using the $G$-matrix solution of fss2, we can calculate
the Sheerbaum factor $S_B$, which represents strength
of the s.p. spin-orbit potential defined through \cite{SPLS}
\begin{eqnarray}
U_B^{\ell s} (r)= -\frac{\pi}{2}~S_B~\frac{1}{r}
\frac{d\rho (r)}{dr}~\mbox{\boldmath $\ell \cdot \sigma$}\ .
\label{gmat1}
\end{eqnarray}
The explicit expression of $S_B(q_1)$ (which
actually contains the momentum dependence) in terms of
the $G$-matrix is given in Eq.\,(50) of \cite{SPLS}.
Here we only consider $S_B=S_B(q_1=0)$ as the measure of the
s.p. spin-orbit strength in the bound states.
The quark model description of the $YN$ interaction
contains the antisymmetric
spin-orbit ($LS^{(-)}$) component which originates from
the FB $LS$ interaction. The large cancellation
between the $LS$ and $LS^{(-)}$ contributions
in the $\Lambda N$ isospin $I=1/2$ channel
leads to a small s.p. spin-orbit
potential for the $\Lambda$-hypernuclei.
A very small ratio $S_{\Lambda}/S_N \leq 1/10$ was
reported for FSS \cite{SPLS}.
In fss2 this cancellation is less prominent, since
the present S-meson EMEP yields the ordinary $LS$ component
but no $LS^{(-)}$ component (see Sec.\,II\,B). 
Since the total strength of the $LS$ force is fixed in
the $NN$ scattering, the FB contribution
of the $LS$ force is somewhat reduced.
This can easily be seen from the simple formula given in Eq.\,(52)
of \cite{SPLS}, which shows that in the Born approximation
the $FB$ $LS$ contribution to the Scheerbaum factor is
determined only by a single strength
factor $\alpha_S x^3 m_{ud}c^2 b^5$. The value of this factor
is 29.35 $\hbox{MeV}\cdot \hbox{fm}^5$ for fss2,
which is 3/5 of the value of FSS, 48.91 $\hbox{MeV}\cdot \hbox{fm}^5$.
We show in Table \ref{sbfac} the Scheerbaum
factors $S_B$ (at $q_1=0$) predicted
by the $G$-matrix calculation of fss2 in the continuous prescription,
with respect to the nuclear-matter densities $\rho = 0.5 \rho_0$,
$0.7\rho_0$ and $\rho_0$.
At the normal density $\rho_0$ with $k_F=1.35~\hbox{fm}^{-1}$,
we obtain for fss2 $S_N=-42.4$, $S_\Lambda=-11.1$ and
$S_\Sigma=-23.3$ ($\hbox{MeV}\cdot \hbox{fm}^5$), which gives 
the ratios $S_{\Lambda}/S_N \sim 0.26$ and $S_{\Sigma}/S_N \sim 0.55$.
We find that $S_\Lambda/S_N$ and $S_\Sigma/S_N$ become slightly
smaller for lower densities.  
Each contribution from the $LS$ and the $LS^{(-)}$ components
in the even- and odd-parity states
as well as $I=1/2$ and $I=3/2$ channels is shown in Table \ref{lscom}
for $k_F=1.35~\hbox{fm}^{-1}$.
The parenthesized numbers are the predictions by FSS.
We find that a prominent difference between fss2 and FSS
appears only in the $\hbox{}^3O$ contribution of $S_\Lambda$.
Namely, the 5 MeV reduction
of the $\hbox{}^3O$ $LS^{(-)}$ contribution and
the 2 MeV enhancement of the $\hbox{}^3O$ $LS$ contribution,
which explains the increase of $S_\Lambda=-3.5$ in FSS
to $S_\Lambda=-11.1$ in fss2.
In the recent experiment at BNL,
very small spin-orbit splitting is reported
in the energy spectra of $\hbox{}^9_\Lambda \hbox{Be}$
and $\hbox{}^{13}_\Lambda \hbox{C}$ \cite{TA00}.
A theoretical calculation of these $\Lambda N$ spin-orbit
splittings using OBEP $\Lambda N$ interactions
is carried out by Hiyama et al. \cite{HI00}
The present result of fss2 is not entirely favorable for these
experimental data.
An accurate experimental determination
of the s.p. spin-orbit strengths
is very important to figure out the relative significance
of the FB and EMEP contributions, both of which
apparently constitute the bare two-body $LS$ forces
of the baryon-baryon interaction.

\section{Summary}

The purpose of this investigation is to construct
a realistic model of the nucleon-nucleon ($NN$) and
hyperon-nucleon ($YN$) interactions, which describes
not only the baryon-baryon scattering quantitatively
in the wide energy region, but also reproduces
rich phenomena observed in few-baryon systems
and various types of infinite nuclear matter. 
We believe that the present framework, incorporating
both the quark and mesonic degrees of freedom
into the model explicitly, is very versatile,
since it is based on the natural
picture that the quarks and gluons are the most economical  
ingredients in the short-range region,
while the meson-exchange processes are dominant
in the medium- and long-range part of the interaction.
Since our quark model describes the short-range
repulsion (which is observed in many channels
of the baryon-baryon interactions) in terms
of the non-locality of the quark exchange kernel,
the effect of the short-range correlation is rather moderate,
compared with the standard meson-exchange potentials.
This can be seen in the magnitude of the Born amplitudes
used in solving the Lippmann-Schwinger resonating-group
equations (LS-RGM) \cite{LSRGM} and the Bethe-Goldstone
equations \cite{GMAT},
and also in the fairly reasonable reproduction of
the single-particle (s.p.) spin-orbit strengths
calculated in the Born approximation \cite{SPLS}.
In \cite{LSRGM}, we have seen that the Born amplitudes
of the quark model have almost the same order of magnitude
as the empirical scattering amplitudes obtained by solving
the LS-RGM equation. The s.p. spin-orbit
strength $S_N$ predicted by the $G$-matrix solution
of our quark model is almost equal to that in the Born
approximation \cite{SPLS}, in contrast to the standard
potential models like the Reid soft-core potential
with the strong short-range repulsive core \cite{SC76}.
Since the Born amplitudes in the quark model reflect
rather faithfully the characteristic features
of the LS-RGM solution, it is easy to find missing
ingredients that impair the model.

In this study we upgrade our previous
model FSS \cite{PRL,FSS} in two respects.
The first one is the renovation of the effective meson-exchange
potentials (EMEP) acting between quarks. We extend our model
to include not only the leading terms of the scalar
and pseudo-scalar mesons but also
the vector mesons with all possible standard terms
usually used in the non-relativistic one-boson exchange
potentials (OBEP). 
The second point is the exact incorporation
of the pion-Coulomb correction in the particle basis.
This includes the exact treatment of the
threshold energies and the Coulomb exchange kernel,
as well as the separate evaluation of the spin-flavor
factors of the charged- and neutral-pion exchange EMEP.
This improvement is necessary in order to study
the effect of the charge symmetry breaking
in the $NN$ and $YN$ interactions.
These two renovations require various mathematical techniques
which are specifically developed
in refs. \cite{LSRGM} and \cite{GRGM} for these purposes.
Appendix A in \cite{LSRGM} discusses
a convenient transformation formula of the RGM kernel,
which directly gives the Born kernel
for the momentum-dependent EMEP at the quark level.
A procedure to avoid the problem of threshold
energies in the RGM formalism is given in \cite{GRGM}.
The new model fss2 with these features has acquired much freedom
to describe the $NN$ and $YN$ interactions more accurately than FSS.
Three different types of calculations are carried out using fss2.
The first one is the calculation in the isospin basis,
which is used for determining the model parameters
and also for the $G$-matrix calculation.
The second and third calculations are performed
in the particle basis with and without the Coulomb force. 
When the Coulomb force is included, the standard technique
by Vincent and Phatak \cite{VP74} is employed to solve
the Lippmann-Schwinger equation in the momentum representation.

In the $NN$ system, the incorporation of the
momentum-dependent Bryan-Scott term \cite{BR67} and
the vector-meson EMEP improves the quantitative
agreement to the experimental data to a large extent.
The momentum-dependent Bryan-Scott term, included
in the scalar- and vector-meson EMEP, is favorable
in extending our quark-model description
of the $NN$ scattering at the non-relativistic
energies to the higher energies up
to $T_{\rm lab}=800$ MeV, and also in describing
reasonable asymptotic behavior of the s.p. potentials
in the high-momentum region.
For vector mesons, we avoid the double-counting
problem \cite{YA90} with the
Fermi-Breit (FB) contribution by choosing
small coupling constants around 1 especially
for the flavor-singlet coupling
constants $f^{\rm Ve}_1$ and $f^{\rm Vm}_1$.
Since we have also chosen $f^{\rm Ve}_8=0$,
the $LS$ contribution from the vector mesons
is very small. For the $\rho$- and $K^*$-meson
contributions, the selected value $f^{\rm Vm}_8 \sim
2.6$ through the parameter search is a standard
size usually assumed in OBEP. Although
the $(f^{\rm Vm}_8)^2$ term usually gives the
isovector spin-spin, tensor and quadratic
spin-orbit ($QLS$) terms, we only retain the $QLS$ term
with the $\bL^2$-type spin-spin term.
This choice is rather ad hoc, but favorable
since we do not want to introduce too strong
cancellation between the one-pion tensor force
and the $\rho$-meson tensor force
in the $\hbox{}^3S_1$ - $\hbox{}^3D_1$ coupling
term of the $NN$ interaction.
Since the $(3q)$ cluster wave function yields
a large cut-off effect for the singular part
of the one-pion exchange potential (OPEP),
we also introduce a reduction factor $c_\delta$ for
the spin-spin contact term of the OPEP central force,
and multiply the short-range tensor term
of the FB interaction by about factor 3.
With these phenomenological ingredients, the accuracy of the
model in the $NN$ sector has now become almost comparable 
to that of the OBEP models.
For the energies above the pion threshold, our single-channel
calculation of the $NN$ scattering seems to have given
nearly satisfactory results, which are visible in the
good reproduction of the differential cross sections
up to $T_{\rm lab}=800$ MeV. The polarizations for
the $np$ and $pp$ scattering have some unfavorable
oscillations in the energy range $T_{\rm lab}=400$ - 800 MeV,
but the improvement is a future work which definitely
requires the explicit introduction of the inelastic channels
such as the $\Delta N$ channel. 

The existing low-energy data for the $YN$ scattering
is well reproduced. This includes; 1) $\Lambda p$ total
cross sections at $p_\Lambda \leq 300$ MeV,
2) $\Sigma^+ p$ and $\Sigma^- p$ total and
differential cross sections at $p_\Lambda \leq 200$ MeV.
The phase-shift difference of the $\hbox{}^1S_0$ and
$\hbox{}^3S_1$ states of the $\Lambda N$ system
at the maximum values is kept less than $10^\circ$ in
fss2, which seems to be necessary to describe the
energy spectra of the hypertriton,
$\hbox{}^4_\Lambda \hbox{H}$ and $\hbox{}^4_\Lambda \hbox{He}$ systems.
In the cusp region of the $\Lambda p$ total cross
sections, the enhancement of the cross sections
by the FB $LS^{(-)}$ force is found in all versions
of our quark model. The $\hbox{}^3P_1$ resonance
of the $\Sigma N (I=1/2)$ state still remains
in the original channel,
which is a common feature found in both fss2 and RGM-H.
The strong $LS^{(-)}$ force is one of the characteristics
of the quark model, which is related to the
spin-flavor $SU_6$ character
of the $\Lambda N$ - $\Sigma N (I=1/2)$ channel coupling
in the quark model. On the other hand,
the $\hbox{}^3S_1$ - $\hbox{}^3D_1$ coupling
of the $\Lambda N$ - $\Sigma N (I=1/2)$ system is
caused by the one-pion tensor force.
We find that the $S$-matrix is very similar in fss2 and FSS,
although the phase-shift behavior
of the $\Sigma N (I=1/2)$ $\hbox{}^3S_1$ diagonal channel looks
very different in these two models.
We can conclude that the essential mechanism
of the $(S+D)$-wave and $P$-wave $\Lambda N$ - $\Sigma N
(I=1/2)$ channel couplings is unchanged among
all versions of our quark model. 

We find a very small effect of the charge symmetry breaking
for the $\Lambda p$ and $\Lambda n$ scattering,
in contrast to the predictions by the Nijmegen group \cite{NA73}.
The energy region we are concerned with is $p_\Lambda \leq 300
~\hbox{MeV}/c$, which is too far away from the $\Sigma N$ threshold
to be affected by the small difference of the threshold
energies $\sim 3.5$ MeV between $\Sigma^0 n$ and $\Sigma^- p$,
$\sim 2.0$ MeV between $\Sigma^+ n$ and $\Sigma^0 p$,
and the Coulomb attraction between $\Sigma^-$ and $p$.
On the other hand, the low-energy $\Sigma^- p$ scattering
suffers a large effect of these threshold energies and the
Coulomb attraction. In particular, we find that the effect
of the correct threshold energies is very important
for the detailed description of the low-energy $\Sigma^- p$
total cross sections and the $\Sigma^- p$ inelastic
capture ratio at rest.
The prescription to multiply $(k_f/k_i)$ factor to reproduce
the low-energy behavior of the $\Sigma^- p\rightarrow
\Sigma^0 n$ charge-exchange total cross sections, used in our previous
calculation in \cite{FU98}, is not accurate enough
to yield reliable estimates for the $\Sigma^- p$ elastic,
$\Sigma^- p\rightarrow \Sigma^0 n$ charge-exchange
and $\Sigma^- p\rightarrow \Lambda n$ reaction cross sections.
In the final calculation in the particle basis
with correct threshold energies,
the increase of the $\Sigma^- p\rightarrow
\Sigma^0 n$ reaction cross section is almost half of
the $(k_f/k_i)$ prescription in the momentum
range $p_\Sigma \leq 200~\hbox{MeV}/c$.
Furthermore, the $\Sigma^- p$ elastic cross section
and $\Sigma^- p\rightarrow \Lambda n$ reaction cross section
decrease fairly largely due to
the effect of the flux conservation. 
The net effect of the pion-Coulomb correction
on the $\Sigma^- p$ elastic scattering is negligible.
The $\Sigma^- p\rightarrow \Lambda n$ reaction cross section seems
to be still too large even in the present model fss2.
This is reflected in the rather small values
of the $\Sigma^- p$ inelastic capture ratios
at rest ($r_R=0.442$) and in flight ($r_F=0.419$).   

The $G$-matrix calculation using fss2 shows that our previous results
given by FSS are qualitatively pertinent.
In particular, the nucleon s.p. potentials in symmetric nuclear matter
are very similar to the predictions of other realistic $NN$ potentials.
The nuclear saturation curve predicted by fss2 resembles
the curve given by the Bonn model-B potential. It is interesting
to note that the deuteron properties of fss2 are rather close
to those of model-C, which is known to have a larger $D$-state
probability than model-B.
Since fss2 reproduces the $NN$ phase shifts at non-relativistic
energies quite well, the difference of the off-shell effect
between our quark model and the other OBEP models does not
seem to appear so prominently, as far as the nuclear saturation
curve is concerned.
Some interesting features of our quark model appear in predictions
for hyperon properties in nuclear medium.
The $\Lambda$ s.p. potential has a depth
of about $48$ MeV in the case of the continuous
prescription for intermediate energy spectra,
which is almost the same as the FSS prediction $46$ MeV \cite{GMAT}.
This value is slightly more attractive
than the value expected from the
experimental data of $\Lambda$-hypernuclei \cite{BMZ}.
The $\Sigma$ s.p. potential is repulsive, with the strength
of about 7 MeV, which is smaller than 21 MeV by FSS.
The origin of this repulsion is the strong Pauli repulsion
in the $\Sigma N (I=3/2)$ $\hbox{}^3S_1$ state.
This result seems to be consistent
with the indication from the analysis by Dabrowski \cite{DA99} of
the recent $(K^{-}, \pi^{\pm})$ experiments \cite{BNL98} at BNL.
Future experiments will be expected to settle the problem
of the $\Sigma$ s.p. potential.
One of the characteristic features of fss2 is the $LS$ force
generated from the scalar-meson EMEP. If this contribution is large,
the cancellation of the $LS$ and $LS^{(-)}$ components from the
FB interaction becomes less prominent
in the Scheerbaum factor $S_\Lambda$.
The fss2 model predicts the relative ratio
to $N$ about $S_\Lambda/S_N \sim 1/4$, which is larger
than the FSS value $S_\Lambda/S_N \sim 1/12$.
The density dependence of the $S_\Lambda/S_N$ ratio is
rather weak in fss2.
We should however keep in mind that this ratio is
for the infinite $N=Z$ system with the normal
density $\rho_0=0.17~\hbox{fm}^{-3}$.
We did not discuss in this paper the most appropriate relative
strength of the $LS$ terms which come from the FB interaction
and the scalar-meson EMEP, since this cannot be determined
only from the $NN$ data.
We definitely need more experimental information
concerning each contribution of the $LS$ and $LS^{(-)}$ forces
in the $YN$ interaction. 

Finally we note that it is an important future subject
to consider few-body systems including hyperons
in the scope of the quark-model baryon-baryon interactions.
The hypertriton calculation can be performed 
in the Faddeev formalism and the stochastic variational approach
by using the quark-exchange kernel directly
for the $NN$ and $YN$ interactions.
The study of hyperonic nuclear matter is also interesting,
since the $G$-matrix calculation
of $\Lambda \Lambda$ and $\Xi N$ interactions \cite{NA97}
in the models FSS and fss2 is now in progress.
Since the $\Sigma$ s.p. potential is repulsive
in the quark-model description,
the admixture of the $\Sigma$ particle is suppressed,
and this should affect the behavior
of the $\Lambda$ particles in dense hyperonic nuclear matter.


\acknowledgments

This research is supported by Japan Grant-in-Aid for Scientific
Research from the Ministry of Education, Science, Sports and
Culture (12640265).

\appendix

\section{EMEP exchange kernel}

In this appendix we extend the derivation of the EMEP exchange
kernel developed in Appendices A and B in \cite{LSRGM},
to deal with various interaction pieces of the V mesons,
including the $LS$ and $QLS$ terms.
The Coulomb exchange kernel and internal-energy contribution
from EMEP are also discussed.

The systematic evaluation of the quark-exchange kernel
is carried out by assuming a two-body
interaction
\begin{eqnarray}
U_{ij}=\sum_\Omega \alpha^\Omega w^{\Omega^\prime}_{ij}
~u^{\Omega^{\prime \prime}}_{ij}\ ,
\label{a1}
\end{eqnarray}
where $w^{\Omega^\prime}_{ij}$ represents the spin-flavor
part (the color part is $w^C_{ij}=1$ for
EMEP) and $u^{\Omega^{\prime \prime}}_{ij}$ the spatial part.
Four different types of the spin-flavor
factors $\Omega=C,~SS,~T,~LS$ are required for the most
general EMEP up to the V mesons; $w^C=1$,
$w^{SS}=(\bfsigma_1\cdot \bfsigma_2)$,
$w^T=[\sigma_1 \sigma_2]^{(2)}_\mu$,
and $w^{LS}=(\bfsigma_1+\bfsigma_2)/2$.
For the flavor octet mesons, these spin operators
should be multiplied with $(\lambda_i
\lambda_j)$, where $\lambda$ represents the Gell-Mann matrix
in the flavor $SU_3$ space. 
The spin-flavor factors $X^\Omega_{x\CT}$ are defined
by Eq.\,(A.3) of \cite{LSRGM} for each $w^\Omega_{ij}$ with
the quark exchange number $x=0,~1$ and the five interaction
types $\CT=E,~S,~S^\prime,~D_+,~D_-$ \cite{KI94}.
The non-central factors are defined by the reduced matrix elements
for the tensor operators of rank 1 and 2. For example, the
tensor operator is expressed as
\begin{eqnarray}
S_{12}(\bk, \bk) & = & 3 \left(\bfsigma_1 \cdot \bk \right)
\left(\bfsigma_2 \cdot \bk \right) 
-\left(\ssigma\right) \bk^2 \nonumber \\
& = & 3\sqrt{10}\left[\,\left[ \sigma_1 \times \sigma_2\right]^{(2)}
\CY_2(\bk)\,\right]^{(0)} \ \ ,
\label{a2}
\end{eqnarray}
where $\CY_{2\mu}(\bk)=\sqrt{4\pi/15}\,\bk^2 Y_{2\mu}
\left(\widehat{\bk}\right)$.
The reduced matrix elements of the spin operators at the baryon level
are assumed to be 1.
For the spatial part, we also
need three extra types $\Omega=C(1),~SS(1),~QLS$ listed
in Table \ref{space}.
This table shows the polynomial
functions $\widetilde{u}(\bk, \bq)$ accompanied
with the Yukawa function in the momentum
representation through
\begin{eqnarray}
u(\bk, \bq)={4\pi \over \bk^2+m^2}\,\widetilde{u}(\bk, \bq)\ \ ,
\label{a3}
\end{eqnarray}
and the spatial part of the Born kernel $M^\Omega_{1\CT}
(\bq_f, \bq_i)$ defined in Eq.\,(A.4) of \cite{LSRGM}
explicitly. The formulae Eqs.\,(A.18) - (A.21)
given in \cite{LSRGM} greatly simplify the procedure
to obtain these results.
The spatial functions $f^\Omega_\CT(\theta)$ are explicitly
given below.
 
In \eq{a1} the coefficients $\alpha^\Omega$ and the 
correspondence among $\Omega$, $\Omega^\prime$
and $\Omega^{\prime \prime}$ are tabulated in Table \ref{corr}.
The EMEP contribution of the exchange Born kernel in \eq{fm4} 
is calculated through
\begin{eqnarray}
M^\Omega(\bq_f, \bq_i)~\CO^\Omega(\bq_f, \bq_i)
=\alpha^\Omega \sum_\CT X^{\Omega^\prime}_{1\CT}
~M^{\Omega^{\prime \prime}}_{1\CT}(\bq_f, \bq_i)\ \ .
\label{a4}
\end{eqnarray}
The final result is as follows.
For the central part, we have $\Omega=C,~C(1),~SS,~SS(1)$ types with
\begin{eqnarray}
& & M^{C\left({{\rm S} \atop {\rm V}}\right)}(\bq_f, \bq_i)
=\left(\begin{array}{c}
-g^2 \\ [3mm]
f^2_e \\
\end{array}\right)
\sum_{\CT} X^C_{1\CT}~f^C_\CT(\theta) \ ,\nonumber \\
& & M^{C(1)\left({{\rm S} \atop {\rm V}}\right)}(\bq_f, \bq_i)
=2\gamma^2 \left(\begin{array}{c}
g^2 \\ [3mm]
3 f^2_e \\
\end{array}\right)
\sum_{\CT} X^C_{1\CT}~f^{C(1)}_\CT(\theta) \ ,\nonumber \\
& & M^{SS\left({{\rm PS} \atop {\rm V}}\right)}(\bq_f, \bq_i)
=\left(\begin{array}{c}
f^2 {1 \over 3}\left({m \over m_{\pi^+}}\right)^2 \\ [3mm]
f^2_m {2 \over 3} \\
\end{array}\right)
\sum_{\CT} X^{SS}_{1\CT}~f^{CD}_\CT(\theta) \ ,\nonumber \\
& & M^{SS(1)\left({{\rm S} \atop {\rm V}}\right)}(\bq_f, \bq_i)
=\left(\begin{array}{c}
g^2 {1 \over 3}\gamma^4 \\ [3mm]
-f^2_m {8 \over 3}\gamma^2 \\
\end{array}\right)
\sum_{\CT} X^{SS}_{1\CT}~f^{SS(1)}_\CT(\theta) \ \ .
\label{a5}
\end{eqnarray}
Here $\gamma=(m/2m_{ud})$ and $\cos \theta=(\widehat{\bq}_f \cdot
\widehat{\bq}_i)$.
In these central terms, the spin-flavor
factors $X^{C,~SS}_{1E}$ should be replaced
with $-X^{C,~SS}_{1S^\prime}$,
because of the subtraction of the internal-energy
contribution in the prior form.
The tensor parts of the PS and V mesons
are given by 
\begin{eqnarray}
M^{T\left({{\rm PS} \atop {\rm V}}\right)}(\bq_f, \bq_i)
=\left(\begin{array}{c}
f^2 \left({m \over m_{\pi^+}}\right)^2 \\[3mm]
-f^2_m \\
\end{array}\right) {1 \over 3m^2}
\sum^{\qquad \prime}_{\CT \neq E} X^T_{1\CT}~f^{TD}_\CT(\theta) \ \ ,
\label{a6}
\end{eqnarray}
where the V-meson contribution is also given
for completeness although this term is not used in fss2.
The EMEP $QLS$ contribution reads
\begin{eqnarray}
M^{QLS\left({{\rm S} \atop {\rm V}}\right)}(\bq_f, \bq_i)
=\left(\begin{array}{c}
g^2 {1 \over 3}\gamma^4 \\ [3mm]
-f^2_m{8 \over 3}\gamma^2 \\
\end{array}\right)
\left[\,X^T_{1D_+}~f^{QLS}_{D_+}(\theta)
- X^T_{1D_-}~f^{QLS}_{D_-}(\theta)\,\right] \ \ ,
\label{a7}
\end{eqnarray}
but also contains the tensor contribution
\begin{eqnarray}
M^{QT\left({{\rm S} \atop {\rm V}}\right)}(\bq_f, \bq_i)
=\left(\begin{array}{c}
g^2 {1 \over 3}\gamma^4 \\ [3mm]
-f^2_m{8 \over 3}\gamma^2 \\
\end{array} \right)
{1 \over 4m^2}\sum^{\qquad \prime}_{\CT \neq E}
X^T_{1\CT}~f^{QT}_\CT(\theta) \ \ ,
\label{a8}
\end{eqnarray}
which we call $\Omega=QT$ term.
In Eqs.\,(\ref{a7}) and (\ref{a8}), the $QLS$ contribution from
the S meson is also shown for completeness, although this term
is negligibly small in fss2. 
The $LS$ term has the contribution both from the S meson and
the V meson:
\begin{eqnarray}
M^{LS\left({{\rm S} \atop {\rm V}}\right)}(\bq_f, \bq_i)
=-\left(\begin{array}{c}
g^2 (b\gamma)^2 \\ [3mm]
-f_m f_e 4b^2\gamma \\
\end{array} \right)
\left[\,X^{LS}_{1D_+}~f^{LS}_{D_+}(\theta)
- X^{LS}_{1D_-}~f^{LS}_{D_-}(\theta)\,\right] \ \ .
\label{a9}
\end{eqnarray}
For the tensor and $QLS$ tensor terms in Eqs.\,(\ref{a6})
and (\ref{a8}), each interaction term with $\CT=S,~S^\prime,
~D_+,~D_-$ types should be rearranged to $\Omega=T,~T^\prime,
~T^{\prime \prime}$ types in \eq{fm6}, according to the rules
given in Eq.\,(B.13) or (B.17) of \cite{LSRGM}.  

The EMEP spatial functions $f^\Omega_\CT(\theta)$ used here
are defined by extending $f^{CN}_\CT(\theta)$,
$f^{SN}_\CT(\theta)$, and $f^{TN}_\CT(\theta)$ given
in Eq.\,(B.18) of \cite{LSRGM}.
The following four basic functions are
used in Table \ref{space}. \footnote{Note
that $f^C_\CT(\theta)=-f^{CN}_\CT(\theta)$
and $f^{CD}_\CT(\theta)=3\,f^{SN}_\CT(\theta)$ except for the
difference of $c_\delta$, but $f^{TD}_\CT(\theta)$ here contains
different numerical factors from those of $f^{TN}_\CT(\theta)$.}  
\begin{eqnarray}
& & f^{C}_\CT(\theta)=4\pi
\left({3 \over 2} \right)^{{3 \over 2}} \hbar c b^2
\left\{ \begin{array}{c}
\exp \left\{-{1 \over 3}
b^2 (\bq^2+\bk^2)\right\}
\widetilde{\CY}_{\alpha_E}(0) \\ [3mm]
\left({8 \over 11}\right)^{1 \over 2} \exp \left\{-{2 \over 11}
b^2 \left[{4 \over 3}(\bq^2+\bk^2)-\bk \cdot \bq \right]\right\}
\widetilde{\CY}_{\alpha_S}\left(
{1 \over \sqrt{11}} b |\bq+\bk| \right) \\ [3mm]
\left({1 \over 2}\right)^{1 \over 2} \exp \left\{-{1 \over 3}
b^2 \left(\bq^2+{1 \over 4} \bk^2 \right) \right\}
\widetilde{\CY}_{\alpha_{D_+}}\left({1 \over 2} b |\bk|\right) \\ [3mm]
\left({2 \over 3}\right)^{1 \over 2} \exp \left\{-{1 \over 3}
b^2 \bk^2 \right\}
\widetilde{\CY}_{\alpha_{D_-}}\left({1 \over \sqrt{3}} b
|\bq|\right) \\ [3mm]
\end{array}
\right. \hbox{for} \quad \CT= \left\{ \begin{array}{l}
E \\ [3mm]
S \\ [3mm]
D_+ \\ [3mm]
D_- \\ [3mm]
\end{array} \right.
\ \ ,\nonumber \\
& & f^{CD}_\CT(\theta)=f^{C}_\CT(\theta) \quad \hbox{with}
\quad \widetilde{\CY}_{\alpha_\CT}(\rho) \rightarrow
\widetilde{\CY}_{\alpha_\CT}(\rho)-{1 \over 2\alpha_\CT}
\ \ ,\nonumber \\
& & f^{LS}_\CT(\theta)=f^{C}_\CT(\theta) \quad \hbox{with}
\quad \widetilde{\CY}_{\alpha_\CT}(\rho) \rightarrow
\widetilde{\CZ}^{(1)}_{\alpha_\CT}(\rho)
\ \ ,\nonumber \\ [3mm]
& & f^{TD}_\CT(\theta)=-4\pi \left({3 \over 2} \right)^{{3 \over 2}}
\hbar c b^2
\left\{ \begin{array}{c}
\left({8 \over 11}\right)^{5 \over 2}
\exp \left\{-{2 \over 11}
b^2 \left[{4 \over 3}(\bq^2+\bk^2)-\bk \cdot \bq \right]\right\}
~\widetilde{\CZ}^D_{\alpha_S}\left(
{1 \over \sqrt{11}} b |\bq+\bk| \right) \\ [3mm]
\left({1 \over 2}\right)^{5 \over 2}
\exp \left\{-{1 \over 3}
b^2 \left(\bq^2+{1 \over 4} \bk^2 \right) \right\}
~\widetilde{\CZ}^D_{\alpha_{D_+}}\left({1 \over 2} b |\bk|\right) \\ [3mm]
\left({2 \over 3}\right)^{5 \over 2}
\exp \left\{-{1 \over 3} b^2 \bk^2 \right\}
~\widetilde{\CZ}^D_{\alpha_{D_-}}\left({1 \over \sqrt{3}} b
|\bq|\right) \\
\end{array} \right.
\hbox{for} \quad \CT= \left\{ \begin{array}{l}
S \\ [3mm]
D_+ \\ [3mm]
D_- \\ [3mm]
\end{array} \right. \ \ .
\label{a10}
\end{eqnarray}
The $S^\prime$-type spatial function $f^\Omega_{S^\prime}(\theta)$ is
obtained from $f^\Omega_S(\theta)$ by taking $\bk \rightarrow -\bk$.
There is no $E$-type possible for the non-central terms.
The coefficients $\alpha_\CT$ are given
by $\alpha_S=\alpha_{S^\prime}=(11/8) \alpha_E$,
$\alpha_{D_+}=2\alpha_E$, and $\alpha_{D_-}=(3/2)\alpha_E$,
with $\alpha_E=(mb)^2/2=(1/2)(mcb/\hbar)^2$.
For the spin-spin part of the one-pion exchange EMEP,
$\widetilde{\CY}_{\alpha_\CT}(\rho)
-(1/2\alpha_\CT)$ should be modified
into $\widetilde{\CY}_{\alpha_\CT}(\rho)
-c_\delta~(1/2\alpha_\CT)$.
The modified Yukawa functions $\widetilde{{\cal Y}}_\alpha(\rho)$,
$\widetilde{{\cal Z}}^{(1)}_\alpha(\rho)$,
and $\widetilde{{\cal Z}}_\alpha(\rho)$ are essentially given
by the error function of the imaginary argument:
\begin{eqnarray}
& & \widetilde{{\cal Y}}_\alpha(\rho)=e^{\alpha-\rho^2}
\int_0^1 e^{-\alpha/t^2+\rho^2 t^2}~d t\ \ , \nonumber \\
& & \widetilde{{\cal Z}}^{(1)}_\alpha(\rho)
={1 \over 2\alpha} e^{\alpha-\rho^2}
\int_0^1 e^{-\alpha/t^2+\rho^2 t^2}~t^2~d t\ \ , \nonumber \\
& & \widetilde{{\cal Z}}_\alpha(\rho)=e^{\alpha-\rho^2}
\int_0^1 e^{-\alpha/t^2+\rho^2 t^2}~t^4~d t\ \ .
\label{a11}
\end{eqnarray}
The other spatial functions appearing
in Eqs.\,(\ref{a5}) - (\ref{a9}) are defined by using
the four spatial functions in \eq{a10}:
\begin{eqnarray}
f^{C(1)}_\CT(\theta) & = & \left[{3 \over 8\alpha_E}
\left\{ \begin{array}{c}
1 \\ [3mm]
{5 \over 8} \\ [3mm]
{1 \over 2} \\ [3mm]
0 \\
\end{array} \right\}
+\left({1 \over 2m}\right)^2
\left\{ \begin{array}{c}
0 \\ [3mm]
{1 \over 4}(\bk+\bq)^2 \\ [3mm]
\bq^2 \\ [3mm]
\bk^2 \\
\end{array} \right\}
\right] f^C_\CT(\theta) \nonumber \\
& & -\left({3 \over 16}\right)^2
\left\{ \begin{array}{c}
0 \\ [3mm]
1 \\ [3mm]
0 \\ [3mm]
0 \\
\end{array} \right\}
f^{CD}_\CT(\theta)
-{3 \over 16}b^2
\left\{ \begin{array}{c}
0 \\ [3mm]
{1 \over 4}(\bk+\bq)^2 \\ [3mm]
0 \\ [3mm]
0 \\
\end{array} \right\}
f^{LS}_\CT(\theta)
\quad \hbox{for} \quad
\CT=\left\{ \begin{array}{c}
E \\ [3mm]
S \\ [3mm]
D_+ \\ [3mm]
D_- \\
\end{array} \right. \ , \nonumber \\
f^{SS(1)}_\CT(\theta) & = & -{1 \over 4\alpha_E}
\left\{ \begin{array}{c}
1 \\ [3mm]
{5 \over 8} \\ [3mm]
{1 \over 2} \\ [3mm]
0 \\
\end{array} \right\} f^{CD}_\CT(\theta)
-\left({1 \over 2m^2}\right)^2
\left\{ \begin{array}{c}
0 \\ [3mm]
0 \\ [3mm]
1 \\ [3mm]
1 \\
\end{array} \right\} \bn^2 f^{TD}_\CT(\theta) \nonumber \\
& & +{1 \over 2m^2}
\left\{ \begin{array}{c}
0 \\ [3mm]
{1 \over 4}(\bk+\bq)^2 \\ [3mm]
\bq^2 \\ [3mm]
\bk^2 \\
\end{array} \right\}
f^{LS}_\CT(\theta)
\quad \hbox{for} \quad
\CT=\left\{ \begin{array}{c}
E \\ [3mm]
S \\ [3mm]
D_+ \\ [3mm]
D_- \\
\end{array} \right. \ , \nonumber \\
f^{QLS}_{D_\pm}(\theta) & = & -{1 \over 4}\left({1 \over m}\right)^4
f^{TD}_{D_\pm}(\theta) \ \ , \nonumber \\
f^{QT}_\CT(\theta) & = & \left\{ \begin{array}{c}
{1 \over 2\alpha_E}{5 \over 8}
f^{TD}_S(\theta)-f^{LS}_S(\theta) \\ [3mm]
{1 \over 2\alpha_E}{1 \over 2}
f^{TD}_{D_+}(\theta)-f^{LS}_{D_-}(\theta) \\ [3mm]
-f^{LS}_{D_+}(\theta) \\
\end{array} \right.
\quad \hbox{for} \quad
\CT=\left\{ \begin{array}{c}
S \\ [3mm]
D_+ \\ [3mm]
D_- \\
\end{array} \right. \ \ .
\label{a12}
\end{eqnarray}

For the numerical calculations, it is convenient to include
the direct term also in the above expressions.
This can be achieved in Eqs.\,(\ref{a5}) - (\ref{a9}), 
if we further add $X^{\Omega^\prime}_{0D_+}
f^{\Omega^{\prime \prime}}_D(\theta)$ term, in addition
to the $X^{\Omega^\prime}_{1D_+}
f^{\Omega^{\prime \prime}}_{D_+}(\theta)$ term.
The direct-type spatial functions $f^\Omega_D(\theta)$ are given by
\begin{eqnarray}
f^C_D(\theta)={4\pi \over \bk^2+m^2} e^{-{1 \over3}(b \bk)^2}\ \ ,
\label{a13}
\end{eqnarray}
and
\begin{eqnarray}
f^\Omega_D(\theta)=f^C_D(\theta)
\left\{ \begin{array}{c}
-{\bk^2 \over m^2} \\ [3mm]
-1 \\ [3mm]
{1 \over 3\alpha_E} \\ [3mm]
\left({1 \over m}\right)^2\left({1 \over 9}\bq^2
+{1 \over 2b^2}\right) \\ [3mm]
\left({1 \over m}\right)^4\left({1 \over 9}\bn^2
+{1 \over 3b^2} \bk^2 \right) \\ [3mm]
{1 \over 9}\left({1 \over m}\right)^4 \\ [3mm]
-{1 \over 3\alpha_E} \\
\end{array} \right.
\quad \hbox{for} \quad
\Omega=\left\{ \begin{array}{c}
CD \\ [3mm]
TD \\ [3mm]
LS \\ [3mm]
C(1) \\ [3mm]
SS(1) \\ [3mm]
QLS \\ [3mm]
QT \\
\end{array} \right. \ \ .
\label{a14}
\end{eqnarray}

The Coulomb exchange kernel is very similar to the color-Coulombic
term of the FB interaction, as is discussed in Sec.\,II\,D.
Only difference is 1) $\alpha_S \rightarrow \alpha=(e^2/\hbar c)$,
2) the definition of the Coulomb spin-flavor factor
\begin{eqnarray}
X^{CL}_{x\CT}=C_x \langle z_x~\xi\,|\,\sum_{i<j}^\CT Q_i Q_j
\,|\,\xi \rangle \ \ ,
\label{a15}
\end{eqnarray}
and 3) the modification of the spatial
function $\widetilde{h}_0(\rho)$ in Eq.\,(B.5) of \cite{LSRGM},
by the effect of the Coulomb cut-off at $R_C$.
The last modification is achieved by
\begin{eqnarray}
& & \widetilde{h}_0(\rho) \rightarrow
\widetilde{h}_0(\rho) - g(x,\rho)\ ,\nonumber \\
& & g(x,\rho)=e^{-(\rho^2+x^2)} \int^1_0 e^{\rho^2 t^2}
\cos (2\rho xt)\,dt \ \ ,
\label{a16}
\end{eqnarray}
with $x=(1/\sqrt{2})(R_C/b),~(2/\sqrt{11})(R_C/b),~(1/2)(R_C/b)$
and $(1/\sqrt{3})(R_C/b)$ for the $\CT=E,~S~\hbox{or}~S^\prime,
~D_+$ and $D_-$ types, respectively.
The function $g(x,\rho)$ is expressed as
\begin{eqnarray}
g(x,\rho) & = & {\sqrt{\pi} \over 2\rho} e^{-\rho^2}
~\Im m~\hbox{erf}\,(x+i\rho) \nonumber \\
& = & {\sqrt{\pi} \over 2\rho} e^{-x^2}
~[\,\sin (2\rho x)~\Re e~w(\rho+ix)+\cos (2\rho x)
~\Im m~w(\rho+ix)\,]\ \ ,
\label{a17}
\end{eqnarray}
where $w(z)=e^{-z^2}~\hbox{erfc}\,(-iz)$ with $\hbox{erfc}\,(z)
=1-\hbox{erf}\,(z)$.
We note the simple relationship
\begin{eqnarray}
g(0, \rho) & = & {\sqrt{\pi} \over 2\rho}~\Im m~w(\rho)
=\widetilde{h}_0(\rho) \ ,\nonumber \\
g(x, 0) &  = & e^{-x^2} \ \ . 
\label{a18}
\end{eqnarray}
For example, the $\CT=E$ type spatial function is given
by $f^{CL}_E(\theta)=\sqrt{2/\pi}\alpha x m_{ud}c^2(4/3)f(\theta)
\left(1-e^{-(R_C/b)^2/2}\right)$, since $\widetilde{h}_0(0)=1$
(Cf. \eq{pa2}).

The EMEP contribution to the internal energies of the octet baryons
originates only from the central force.
It reads
\begin{eqnarray}
E^{\rm S}_{\rm int} & = & mg^2~X^C_{0E}
~\left(-1+{3\gamma^2 \over 4 \alpha_E}\right)~Y_{\alpha_E}(0)
-mg^2~X^{SS}_{0E}~{\gamma^4 \over 12\alpha_E}~Y^D_{\alpha_E}(0)
\ ,\nonumber \\
E^{\rm PS}_{\rm int} & = & {m \over 3}f^2 
\left({m \over m_{\pi^+}}\right)^2 X^{SS}_{0E}
~Y^D_{\alpha_E}(0)
\ , \nonumber \\
E^{\rm V}_{\rm int} & = & m{f_e}^2~X^C_{0E}
~\left(1+{9\gamma^2 \over 4 \alpha_E}\right)~Y_{\alpha_E}(0)
+m {f_m}^2~X^{SS}_{0E}~{2 \over 3}\left(1+{\gamma^2 \over \alpha_E}
\right)~Y^D_{\alpha_E}(0) \ \ ,
\label{a19}
\end{eqnarray}
where the values of the modified Yukawa functions
at the origin are given
by $Y_\alpha(0)=1/\sqrt{\pi \alpha}-e^\alpha\,\hbox{efrc}\,(\sqrt{\alpha})$
and $Y^D_\alpha(0)=Y_\alpha(0)-1/(2\alpha \sqrt{\pi \alpha})$.
For the $qq$ interaction in \eq{new1}, the last spin-spin terms
of $E^{\rm S}_{\rm int}$ and $E^{\rm V}_{\rm int}$ should be modified
into 0 and $m {f_m}^2~X^{SS}_{0E}~(2\gamma^2/3\alpha_E)
~Y^D_{\alpha_E}(0)$, respectively.

\section{Spin-flavor factors in the EMEP sector}

All the spin-flavor-color factors in the quark sector
are already published in refs.\,\cite{FU87}, \cite{LSRGM},
and \cite{FU97}.
Some simple spin-flavor factors including the direct factors
in the EMEP sector are discussed in our previous
publication \cite{FSS}, together with the details of the
singlet-octet meson mixing.\footnote{A different
notation $\Omega=CN,~TN$ is used in Appendix of \cite{FSS},
to specify $\Omega=C,~T$ used here.}  
Since the V mesons and the non-central terms
of the S mesons are not introduced in \cite{FSS},
the $LS$ factors are not shown.
Here we show these $LS$ factors and a complete list
of the spin-flavor factors in the isospin basis
with respect to the exchange terms in the EMEP sector.
The Coulomb exchange factors are also shown
only for the $NN$ system.
The detailed derivation of these factors 
and the spin-flavor factors in the particle
basis ($\pi^\pm$ - $\pi^0$ separation),
as well as the exchange Coulomb factors for the $YN$ system,
will be published elsewhere.

The $LS$ factors for the flavor singlet and octet mesons
are expressed by the electric-type ($e^e_{(\lambda \lambda)}$) and
the magnetic-type ($e^m_{(\lambda \lambda)}$) $SU_6$ unit vectors
as follows (we use the same notation as in Appendix of \cite{FSS}):

\bigskip

\noindent
[ $(\lambda \lambda)=(00)$ ]
\begin{eqnarray}
& & X^{LS (00)}_{0E}=2\ \ ,\qquad  X^{LS (00)}_{0D_+}=3 \ \ ,\nonumber \\
& & X^{LS (00)}_{1E}=X_N(S=1)-X^{LS (00)}_{1D_-} \ \ ,\nonumber \\
& & X^{LS (00)}_{1S}=X_N(S=1)+X^{LS (00)}_{1D_-} \ \ ,\nonumber \\
& & X^{LS (00)}_{1D_+}=2X^{LS (00)}_{1E} \ \ ,\nonumber \\
& & X^{LS (00)}_{1D_-}=-{1 \over 12}\left(e^{e\dagger}e^m
+e^{m\dagger}e^e\right) \ \ ,
\label{b1}
\end{eqnarray}
where $X_N(S=1)=-(1/12)\left(e^{e\dagger}e^e+e^{m\dagger}e^m\right)$.
We have simple relationships $X^{LS (00)}_{0E}
+X^{LS (00)}_{0D_+}=5$ and $\sum_{\CT} X^{LS (00)}_{1\CT}
=5\,X_N(S=1)$.

\bigskip

\noindent
[ $(\lambda \lambda)=(11)$ ]
\begin{eqnarray}
& & X^{LS (11)}_{0E}={8 \over 3}\ \ ,\qquad
X^{LS (11)}_{0D_+}=-6X^{LS (00)}_{1D_-}-2 \ \ ,\nonumber \\
& & X^{LS (11)}_{1E}={4 \over 3}X^{LS (00)}_{1E} \ \ ,\nonumber \\
& & X^{LS (11)}_{1S}={4 \over 3}X^{LS (00)}_{1S} \ \ ,\nonumber \\
& & X^{LS (11)}_{1D_+}=X^{LS (00)}_{1D_+}2\left(P_F-{1 \over 3}\right)
\ \ ,\nonumber \\
& & X^{LS (11)}_{1D_-}=-{2 \over 3}X^{LS (00)}_{1D_-}-2 \ \ .
\label{b2}
\end{eqnarray}

The flavor octet factors $X^{\Omega (11)}_{x\CT}$ are further
separated into the sum of the isoscalar, isovector,
and $I=1/2$ (strange) factors in the isospin representation.
This separation is achieved in two steps.
First we calculate the reduced matrix elements
\begin{eqnarray}
X^\Omega_{1 \CT}(\lambda)=(-9)\,{1 \over 3}
\,\langle~P_{36}~\xi^{SF}\,||
\,\sum_{i<j}^\CT w_{ij}^{SF}{{m_{ud}}^2 \over m_i m_j}
\,||\,\xi^{SF}\,\rangle \ \ ,
\label{b3}
\end{eqnarray}
with respect to $w^{SF}_{ij}=1,~\left(\bfsigma_i\cdot\bfsigma_j\right),
~(\bfsigma_i+\bfsigma_j)/2~\langle 1 || \bS || 1 \rangle^{-1}$,
and $\left[\sigma_i \times \sigma_j
\right]^{(2)} \langle 1 ||\left[\sigma_{B_1} \times \sigma_{B_2}
\right]^{(2)} || 1 \rangle^{-1}$.
Here the $SU_3$ symmetry breaking is introduced
by the parameter $\lambda$ through
\begin{eqnarray}
{m_{ud} \over m_i}={1 \over 3}\left(2+{1 \over \lambda}\right)
+\left(1-{1 \over \lambda}\right)Y_i \ \ .
\label{b4}
\end{eqnarray}
The flavor-singlet factors are obtained by setting $\lambda=1$:
$X^\Omega_{x\CT}(1)=X^{\Omega (00)}_{x\CT}$.
On the other hand, $X^\Omega_{x\CT}(-1/2)$ is the contribution
from the $\lambda_8$ vertex.
For the product of two coupling constant
operators $(f_1\cos \theta+f_8\sin \theta \lambda_{8i})
\,(f_1\cos \theta+f_8\sin \theta \lambda_{8j})$,
the isoscalar spin-flavor factors including the (quark) coupling
constants are obtained through
\begin{eqnarray}
W^\Omega_{x\CT}(\theta)={1 \over 3}\left(\sqrt{3}f_1\cos \theta
+f_8\sin \theta\right)
\left[\,X^\Omega_{x\CT}(1)\sqrt{3}f_1\cos \theta
+X^\Omega_{x\CT}\left(-{1 \over 2}\right)f_8\sin \theta \right] \ \ ,
\label{b5}
\end{eqnarray}
for the $\CT\neq E$ types.
(We do not need $X^\Omega_{1E}$, since it is replaced
by $-X^\Omega_{1S^\prime}$ due to the subtraction of the
internal energy part.) The other isoscalar factors are
obtained from $W^\Omega_{x\CT}(\theta-\pi/2)$.
The non-isoscalar part of the spin-flavor factors are
obtained from
\begin{eqnarray}
X^{\Omega (I \neq 0)}_{x\CT}=X^{\Omega (11)}_{x\CT}
-{1 \over 3}X^{\Omega}_{x\CT}\left(-{1 \over 2}\right) \ \ .
\label{b6}
\end{eqnarray}
(Note the hypercharge operator is given by $Y=\lambda_8/\sqrt{3}$.)
These are further separated into the isovector and $I=1/2$ factors
by using the standard recoupling techniques of the $SU_3$ algebra. 
We will specify this isospin dependence
by using names of the S mesons $\delta$ and $\kappa$ for $\Omega
=C,~LS$, and those of PS mesons $\pi$ and $K$ for $\Omega=SS,~T$.
These factors, however, can also be used for the vector mesons. 

In the following, $X^\Omega_{1 \CT}(\lambda)$ in \eq{b3} is
expressed as $X^\Omega_{1 \CT}$. The spin and isospin operators,
$\bfsigma_i$ and $\bftau_i$, are with respect to
the two baryons $B_i$ with $i=1$ or 2 in the
initial state, and the flavor exchange operator $P_F$ is supposed
to operate on the ket state.
For example, $\left(\ttau\right)=I(I+1)-11/4$ for
the $\Sigma N$ channel.
We assume $B_2=B_4=N$, unless otherwise specified.
The spin-flavor-color factor for the exchange normalization kernel
is given by $X_N=X^C_{1D_-}(\lambda=1)$.
The factors of the interaction type $\CT=S^\prime$ are all equal
to the $\CT=S$ type factors for the diagonal channels,
since they are Hermitian conjugate to each other.
For the $\Lambda N$ - $\Sigma N$ coupling terms, the $S$ type
and the $S^\prime$ type are interchanged if the initial state
and the final state are interchanged.
(See $\S$ 5 of \cite{FU97} for
the symmetries of the spin-flavor factors.) 
For $\Omega=LS$, these factors with $\CT=S,~S^\prime$ are
not shown since they are not necessary, as in \eq{a9}.

\subsection{Spin-Flavor factors of \protect\eq{b3}}

\noindent
[ $B_3 B_1=NN$ ]
\begin{eqnarray}
& & X^C_{1D_-}=X_N=-{3 \over 4} \left[ 1+{1 \over 9}\left(\ttau\right)
+\left(\ssigma\right){1 \over 9}\left(1+{25 \over 9}\ttau\right)
\right] \ \ ,\nonumber \\
& & X^C_{1D_+}=X^C_{1S}=4X_N \ \ ,\nonumber \\
& & \qquad \nonumber \\
& & X^{SS}_{1D_-}=-{1 \over 4} \left[ 9+\left(\ttau\right)
-\left(\ssigma\right) {1 \over 3}\left(1+{25 \over 9}\ttau \right)
\right] \ \ ,\nonumber \\
& & X^{SS}_{1D_+}=-\left(1+{1 \over 3}\ssigma \right)
\left(1+{1 \over 9}\ttau \right) \ \ ,\nonumber \\
& & X^{SS}_{1S}=3\left[ 1-{1 \over 9}\ttau-\left(\ssigma\right)
{1 \over 9}\left(1-{5 \over 9}\ttau \right) \right]
\ \ ,\nonumber \\
& & \qquad \nonumber \\
& & X^{LS}_{1D_-}=-{1 \over 2} \left(1+{5 \over 9}\ttau \right)
\ \ ,\nonumber \\ 
& & X^{LS}_{1D_+}=-{2 \over 3} \left(1+{1 \over 9}\ttau \right)
\ \ ,\nonumber \\ 
& & \qquad \nonumber \\
& & X^T_{1D_-}=-{1 \over 6} \left(1+{25 \over 9}\ttau \right)
\ \ ,\nonumber \\ 
& & X^T_{1D_+}=-{1 \over 3} \left(1+{1 \over 9}\ttau \right)
\ \ ,\nonumber \\ 
& & X^T_{1S}=-{1 \over 3} \left(1-{5 \over 9}\ttau \right) \ \ .
\label{b7}
\end{eqnarray}

\bigskip

\noindent
[ $B_3 B_1=\Lambda \Lambda$ ]
\begin{eqnarray}
& & X^C_{1D_-}=-{1 \over 4} \left[ 2+\lam{1}\left(1+\ssigma \right)
P_F \right] \ \ ,\nonumber \\
& & X^C_{1D_+}=-\left[ 1+\lam{1}
+\left(1+\ssigma \right) P_F \right] \ \ ,\nonumber \\
& & X^C_{1S}=-{1 \over 2} \left[ 3+\lam{1}
+\left(1+\lam{1}\right)\left(1+\ssigma \right) P_F \right]
\ \ ,\nonumber \\
& & \qquad \nonumber \\
& & X^{SS}_{1D_-}=-{1 \over 4}
\left[ 6+\lam{3}P_F-\left(\ssigma\right) \lam{1} P_F \right]
\ \ ,\nonumber \\
& & X^{SS}_{1D_+}=-\left[ 1+\left(\ssigma \right){1 \over 3\lambda}
\right]\ \ ,\nonumber \\
& & X^{SS}_{1S}={1 \over 2} \left[ 5-\left(\ssigma \right)
{1 \over 3\lambda} \right]\ \ ,\nonumber \\
& & \qquad \nonumber \\
& & X^{LS}_{1D_-}=-{1 \over 2} \left({1 \over 3}
+\lam{1} P_F \right)
\ \ ,\nonumber \\ 
& & X^{LS}_{1D_+}=-{2 \over 3\lambda}
\ \ ,\nonumber \\ 
& & \qquad \nonumber \\
& & X^T_{1D_-}=-{1 \over 2\lambda} P_F \ \ ,\nonumber \\ 
& & X^T_{1D_+}=-{1 \over 3\lambda} \ \ ,\nonumber \\ 
& & X^T_{1S}=-{1 \over 6\lambda} \ \ .
\label{b8}
\end{eqnarray}

\bigskip

\noindent
[ $B_3 B_1=\Sigma \Sigma$ ]
\begin{eqnarray}
& & X^C_{1D_-}=-{1 \over 2} \left[ 1+{1 \over 3}\left(\ttau\right)
+{1 \over 6\lambda}\ttaup \right]
-\left(\ssigma \right){1 \over 9} \left[ 1+{5 \over 3}\left(\ttau\right)
+{1 \over 12\lambda}\ttaup \right]
\ \ ,\nonumber \\
& & X^C_{1D_+}=-\left(1+\lam{1}\right)
\left[ 1+{1 \over 3}\left(\ttau\right) +\left(\ssigma \right){2 \over 9}
\left(1+{5 \over 3}\ttau \right) \right]-{1 \over 3}
\left(1+{1 \over9}\ssigma\right)\ttaup\ \ ,\nonumber \\
& & \qquad  \nonumber \\
& & X^C_{1S}=-{1 \over 2}\left(3+\lam{1}\right)
\left[ 1+{1 \over 3}\left(\ttau\right)+\left(\ssigma \right){2 \over 9}
\left(1+{5 \over 3}\ttau \right) \right] \nonumber \\
& & -{1 \over 6}\left(1+\lam{1} \right)
\left(1+{1 \over9}\ssigma \right) \ttaup \ \ ,\nonumber \\
& & \qquad  \nonumber \\
& & X^{SS}_{1D_-}=-{1 \over 4} \left\{ 6+2\left(\ttau\right)
+\lam{1}\ttaup
-\left(\ssigma \right){1 \over 9}\left[ 4+{20 \over 3}\left(\ttau\right)
+{1 \over 3\lambda}\ttaup \right] \right\}
\ \ ,\nonumber \\
& & X^{SS}_{1D_+}=-{1 \over 9} \left\{
-3\left(1-\lam{2}\right) \left(1-{1 \over 3}\ttau \right)
+4\ttaup \right. \nonumber \\
& & \left. + \left(\ssigma \right)
\left[ \left(2-\lam{1}\right)\left(1-{1 \over 3}\ttau\right)
+{4 \over 3}\ttaup \right] \right\} \ \ ,\nonumber \\
& & X^{SS}_{1S}={1 \over 2} \left\{
\left(1+\lam{2}\right)-\left(1-{2 \over 3\lambda}\right)\left(\ttau\right)
+{2 \over 3}\left(1+\lam{1}\right)\ttaup \right. \nonumber \\
& & \left. -\left(\ssigma \right){1 \over 9}\left[ \left(6-\lam{1}\right)
+{1 \over 3}\left(6-\lam{5}\right)\left(\ttau\right)-{2 \over 3}\left(
1+\lam{1}\right)\ttaup \right] \right\} \ \ ,\nonumber \\
& & \qquad \nonumber \\
& & X^{LS}_{1D_-}=-{1 \over 2} \left[ 1+{7 \over 9}\left(\ttau\right)
-{1 \over 9\lambda}\ttaup \right] \ \ ,\nonumber \\ 
& & X^{LS}_{1D_+}=-{2 \over 9} \left[ \left(2-\lam{1}\right)
+{1 \over 6}\left(5-\lam{7}\right)\left(\ttau\right)
+{4 \over 3}\ttaup \right] \ \ ,\nonumber \\ 
& & \qquad \nonumber \\
& & X^T_{1D_-}=-{1 \over 18} \left[4+{20 \over 3}\left(\ttau\right)
+{1 \over 3\lambda}\ttaup \right] \ \ ,\nonumber \\ 
& & X^T_{1D_+}=-{1 \over 9} \left[ \left(2-\lam{1}\right)
\left(1-{1 \over 3}\ttau \right)+{4 \over3}\ttaup \right]
\ \ ,\nonumber \\
& & X^T_{1S}=-{1 \over 18} \left[ \left(6-\lam{1}\right)
+{1 \over 3}\left(6-\lam{5}\right)\left(\ttau\right)
-{2 \over 3}\left(1+\lam{1}\right)\ttaup \right] \ \ .
\label{b9}
\end{eqnarray}

\bigskip

\noindent
[ $B_3 B_1=\Xi \Xi$ ]
\begin{eqnarray}
& & X^C_{1D_-}=X_N=-{1 \over 4}\left[
1+{1 \over 3}\left(\ttau\right) -\left(\ssigma \right){1 \over 9}
\left(1+{5 \over 3}\ttau\right) \right] \ \ ,\nonumber \\
& & X^C_{1D_+}=\lam{4} X_N\ \ ,\nonumber \\
& & X^C_{1S}=2\left(1+\lam{1}\right) X_N \ \ ,\nonumber \\
& & \qquad  \nonumber \\
& & X^{SS}_{1D_-}=-{1 \over 4}\left[
3+\left(\ttau\right) +\left(\ssigma \right){1 \over 9}
\left(1+{5 \over 3}\ttau \right) \right]\ \ ,\nonumber \\
& & X^{SS}_{1D_+}=-{2 \over 3\lambda} \left( 1+{1 \over 3}
\ssigma \right) \left(1-{1 \over 3}\ttau \right) \ \ ,\nonumber \\
& & X^{SS}_{1S}={1 \over 2} \left\{
\left(1+\lam{2}\right)-{1 \over 3}\left(1-\lam{2}\right)\left(\ttau\right)
+\left(\ssigma \right){1 \over 9}\left[ \left(1-\lam{2}\right)
-{1 \over 3}\left(1+\lam{10}\right)\left(\ttau\right) \right] \right\}
\ \ ,\nonumber \\
& & \qquad \nonumber \\
& & X^{LS}_{1D_-}=-{1 \over 9}\left(\ttau\right) \ \ ,\nonumber \\ 
& & X^{LS}_{1D_+}=-{4 \over 9\lambda} \left(1-{1 \over 3}\ttau \right)
\ \ ,\nonumber \\ 
& & \qquad \nonumber \\
& & X^T_{1D_-}={1 \over 18} \left(1+{5 \over 3}\ttau \right)
\ \ ,\nonumber \\ 
& & X^T_{1D_+}=-{2 \over 9\lambda} \left(1-{1 \over 3}\ttau \right)
\ \ ,\nonumber \\ 
& & X^T_{1S}=-{1 \over 18} \left[ \left(\lam{2}-1\right)
+{1 \over 3}\left(1+\lam{10}\right)\left(\ttau\right) \right] \ \ .
\label{b10}
\end{eqnarray}

\bigskip

\noindent
[ $B_3 B_1=\Lambda \Sigma$ and $\Sigma \Lambda$ ]
\begin{eqnarray}
& & X^C_{1D_-}={1 \over 4} \left[\lam{1} P_F
+\left(\ssigma\right){1 \over 3}
\left({10 \over 3}-\lam{1} P_F \right) \right]\ \ ,\nonumber \\
& & X^C_{1D_+}={5 \over 9}\left(1+\lam{1}\right)
\left(\ssigma\right)+\left(1-{1 \over 3}\ssigma \right)
P_F\ \ ,\nonumber \\
& & X^C_{1S}={5 \over 18}\left(3+\lam{1}\right)
\left(\ssigma\right)+{1 \over 2}\left(1+\lam{1}\right)
\left(1-{1 \over 3}\ssigma \right) P_F\ \ ,\nonumber \\
& & \qquad  \nonumber \\
& & X^{SS}_{1D_-}=-{1 \over 4} \left[
-\lam{3} P_F+\left(\ssigma\right)
{1 \over 9}\left(10-\lam{3} P_F\right) \right] \ \ ,\nonumber \\
& & X^{SS}_{1D_+}={1 \over 3} \left(\lam{1}+{1 \over 3}\ssigma
\right) \ \ ,\nonumber \\
& & X^{SS}_{1S}=\left\{
\begin{array}{l}
{1 \over 2} \left[ \lam{1} \left(1-2P_F\right)
+\left(\ssigma\right){1 \over 9}\left(3-\lam{10}+\lam{6}P_F
\right) \right] \\ [3mm]
{1 \over 2} \left[ \lam{1}-2P_F
+\left(\ssigma\right){1 \over 9}\left(-17+\lam{10}+6 P_F
\right) \right] \\
\end{array} \right.
\quad \hbox{for} \quad \left\{
\begin{array}{c}
\Lambda \Sigma \\ [3mm]
\Sigma \Lambda \\
\end{array} \right. \ \ ,\nonumber \\
& & \qquad \nonumber \\
& & X^{LS}_{1D_-}={1 \over 6} \left( 1+\lam{1} P_F \right)
\ \ ,\nonumber \\ 
& & X^{LS}_{1D_+}=-{1 \over 9} \left( 1-\lam{3} \right)
\ \ ,\nonumber \\ 
& & \qquad \nonumber \\
& & X^T_{1D_-}={1 \over 18} \left(10-\lam{3}P_F\right) \ \ ,\nonumber \\ 
& & X^T_{1D_+}={1 \over 9} \ \ ,\nonumber \\ 
& & X^T_{1S}=\left\{
\begin{array}{l}
-{1 \over 18} \left[ \left(12-\lam{5}\right)-\lam{6}P_F \right] \\ [3mm]
-{1 \over 18} \left[ \left(2+\lam{5}\right)-6P_F \right] \\
\end{array} \right.
\quad \hbox{for} \quad \left\{
\begin{array}{c}
\Lambda \Sigma \\ [3mm]
\Sigma \Lambda \\
\end{array} \right. \ \ .
\label{b11}
\end{eqnarray}

\subsection{Non-isoscalar spin-flavor factors} 

\noindent
[ $B_3 B_1=NN$ ]
\begin{eqnarray}
& & X^{\delta C}_{1D_-}=-{1 \over 12} \left[ 27-\left(\ttau\right)
+\left(\ssigma\right)\left(3-{25 \over 9}\ttau\right)
\right] \ \ ,\nonumber \\
& & X^{\delta C}_{1D_+}=-\left(1+{1 \over 9}\ssigma \right)
\left(1+{1 \over 3}\ttau \right) \ \ ,\nonumber \\
& & X^{\delta C}_{1S}={1 \over 3} \left[ 9-\left(\ttau\right)
-\left(\ssigma\right)\left(1-{5 \over 9}\ttau\right)
\right] \ \ ,\nonumber \\
& & \qquad \nonumber \\
& & X^{\pi SS}_{1D_-}=-{1 \over 4} \left[ 27-\left(\ttau\right)
-\left(\ssigma\right)\left(1-{25 \over 27}\ttau \right)
\right] \ \ ,\nonumber \\
& & X^{\pi SS}_{1D_+}=-{1 \over 9} \left[ 75+\left(\ttau\right)
+\left(\ssigma\right)\left(1+{61 \over 3}\ttau \right)
\right] \ \ ,\nonumber \\
& & X^{\pi SS}_{1S}=-{1 \over 3} \left[ 45+\left(\ttau\right)
+\left(\ssigma\right)\left(1+{85 \over 9}\ttau \right)
\right] \ \ ,\nonumber \\
& & \qquad \nonumber \\
& & X^{\delta LS}_{1D_-}=-{3 \over 2} \left(1-{5 \over 27}\ttau \right)
\ \ ,\nonumber \\
& & X^{\delta LS}_{1D_+}=-{2 \over 9} \left(1+{7 \over 3}\ttau \right)
\ \ ,\nonumber \\ 
& & \qquad \nonumber \\
& & X^{\pi T}_{1D_-}=-{1 \over 2} \left(1-{25 \over 27}\ttau \right)
\ \ ,\nonumber \\ 
& & X^{\pi T}_{1D_+}=-{1 \over 9} \left(1+{7 \over 3}\ttau \right)
\ \ ,\nonumber \\ 
& & X^{\pi T}_{1S}=-{1 \over 3} \left(1-{5 \over 9}\ttau \right) \ \ .
\label{b12}
\end{eqnarray}

\bigskip

\noindent
[ $B_3 B_1=\Lambda \Lambda$ ]
\begin{eqnarray}
& & X^{\delta C}_{1D_-}=-{3 \over 2} \ \ ,\nonumber \\
& & X^{\delta C}_{1D_+}=-1 \ \ ,\nonumber \\
& & X^{\delta C}_{1S}={5 \over 2} \ \ ,\nonumber \\
& & \qquad \nonumber \\
& & X^{\kappa C}_{1D_-}=-\left(1+{1 \over 3}\ssigma \right)
\ \ ,\nonumber \\
& & X^{\kappa C}_{1D_+}=-\left(1+{1 \over 3}\ssigma \right) P_F
\ \ ,\nonumber \\
& & X^{\kappa C}_{1S}=-{1 \over 2}\left[ 1+{1 \over 3}\left(\ssigma\right)
+\left(1+\ssigma \right) P_F \right]
\ \ ,\nonumber \\
& & \qquad \nonumber \\
& & X^{\pi SS}_{1D_-}=-{9 \over 2} \ \ ,\nonumber \\
& & X^{\pi SS}_{1D_+}=-\left[ 5+\left(1+\ssigma \right) P_F \right]
\ \ ,\nonumber \\
& & X^{\pi SS}_{1S}=-{3 \over 2} \left[ {19 \over 3}
+\left(1+\ssigma \right) P_F \right]
\ \ ,\nonumber \\
& & X^{K SS}_{1D_-}=-\left(3-{1 \over 3}\ssigma\right)
\ \ ,\nonumber \\
& & X^{K SS}_{1D_+}=-\left(1+{5 \over 3}\ssigma\right) P_F
\ \ ,\nonumber \\
& & X^{K SS}_{1S}=-{3 \over 2} \left[ 1-{1 \over 9}\left(\ssigma \right)
+\left(1+\ssigma\right) P_F \right]\ \ ,\nonumber \\
& & \qquad \nonumber \\
& & X^{\delta LS}_{1D_-}=-{1 \over 2} \ \ ,\nonumber \\ 
& & X^{\delta LS}_{1D_+}=0 \ \ ,\nonumber \\ 
& & \qquad \nonumber \\
& & X^{\kappa LS}_{1D_-}=-{4 \over 3} \ \ ,\nonumber \\ 
& & X^{\kappa LS}_{1D_+}=-{4 \over 3} P_F \ \ ,\nonumber \\ 
& & \qquad \nonumber \\
& & X^{\pi T}_{1D_-}=X^{\pi T}_{1D_+}=X^{\pi T}_{1S}=0 \ \ ,\nonumber \\
& & X^{K T}_{1D_-}=-{2 \over 3} \ \ ,\nonumber \\ 
& & X^{K T}_{1D_+}=-{2 \over 3} P_F \ \ ,\nonumber \\ 
& & X^{K T}_{1S}=-{1 \over 3} \ \ .
\label{b13}
\end{eqnarray}

\bigskip

\noindent
[ $B_3 B_1=\Sigma \Sigma$ ]
\begin{eqnarray}
& & X^{\delta C}_{1D_-}=-{1 \over 3} \left[ {1 \over 2}
\left(9-\ttau\right)+\left(\ssigma \right)\left(1-{5 \over 9}\ttau\right)
\right] \ \ ,\nonumber \\
& & X^{\delta C}_{1D_+}={1 \over 3}
\left\{ 1-\left(\ttau\right)-\left(2+\ttau\right) P_F
-\left(\ssigma\right){1 \over 9}\left[
2 \left(1-\ttau\right)+\left(2+\ttau\right)P_F \right] \right\} 
\ \ ,\nonumber \\
& & X^{\delta C}_{1S}={1 \over 3}
\left\{ {3 \over 2}\left(1-\ttau\right)+\ttaup
-\left(\ssigma\right){1 \over 9}\left[
3 \left(3+\ttau\right)-\ttaup \right] \right\} 
\ \ ,\nonumber \\
& & \qquad  \nonumber \\
& & X^{\kappa C}_{1D_-}=-\left(1-{1 \over 9}\ssigma\right)
\ \ ,\nonumber \\
& & X^{\kappa C}_{1D_+}=-\left[ 1+{2 \over 3}\left(\ttau\right)
+\left(\ssigma\right){1 \over 9}\left(7-{4 \over 3}\ttau\right)
\right] P_F \ \ ,\nonumber \\
& & X^{\kappa C}_{1S}={1 \over 6}
\left\{ 3+\left(\ttau\right)+\ttaup
-\left(\ssigma\right){1 \over 9}\left[
3+5\left(\ttau\right)-\ttaup \right] \right\} 
\ \ ,\nonumber \\
& & \qquad  \nonumber \\
& & X^{\pi SS}_{1D_-}=-{1 \over 2}\left(9-\ttau\right)
+\left(\ssigma \right){1 \over 3}\left(1-{5 \over 9}\ttau\right)
\ \ ,\nonumber \\
& & X^{\pi SS}_{1D_+}=-{1 \over 9} \left\{
5+\left(\ttau\right)+\left(17+16\ttau\right)P_F
+\left(\ssigma \right){1 \over 3}
\left[ 2\left( 1+5\ttau\right)+\left(35-8\ttau\right)P_F
\right] \right\} \ \ ,\nonumber \\
& & X^{\pi SS}_{1S}=-{1 \over 2} \left[
11+\left(\ttau\right)+{7 \over 3}\ttaup \right]
-\left(\ssigma \right){1 \over 9}\left[ 3+13\left(\ttau\right)
+{7 \over 6}\ttaup \right] \ \ ,\nonumber \\
& & \qquad \nonumber \\
& & X^{K SS}_{1D_-}=-\left(3+{1 \over 9}\ssigma \right)
\ \ ,\nonumber \\
& & X^{K SS}_{1D_+}=-{1 \over 9} \left[
21+20\left(\ttau\right)+\left(\ssigma\right)
\left(7+{2 \over 3}\ttau\right)
\right] P_F \ \ ,\nonumber \\
& & X^{K SS}_{1S}=-{5 \over 6} \left\{
3+\left(\ttau\right)+\ttaup
+\left(\ssigma \right){1 \over 9}\left[ {21 \over 5}
+7\left(\ttau\right)+\ttaup \right] \right\} \ \ ,\nonumber \\
& & \qquad \nonumber \\
& & X^{\delta LS}_{1D_-}=-{3 \over 2} \left(1-{7 \over 27}\ttau\right)
\ \ ,\nonumber \\ 
& & X^{\delta LS}_{1D_+}=-{1 \over 27} \left[ 4+11 \left(\ttau\right)
+2\left(8+\ttau\right) P_F \right] \ \ ,\nonumber \\ 
& & \qquad \nonumber \\
& & X^{\kappa LS}_{1D_-}=0 \ \ ,\nonumber \\ 
& & X^{\kappa LS}_{1D_+}={4 \over 9} \left(1+{7 \over 6}\ttau\right) P_F
\ \ ,\nonumber \\ 
& & \qquad \nonumber \\
& & X^{\pi T}_{1D_-}=-{2 \over 3} \left(1-{5 \over 9}\ttau\right)
\ \ ,\nonumber \\ 
& & X^{\pi T}_{1D_+}=-{2 \over 27} \left[ 1+5\left(\ttau\right)
+4\left(1-\ttau\right)P_F \right]
\ \ ,\nonumber \\
& & X^{\pi T}_{1S}={1 \over 9} \left[ -\left(3+\ttau \right)
+{1 \over 3}\ttaup \right] \ \ ,\nonumber \\
& & \qquad \nonumber \\
& & X^{K T}_{1D_-}={2 \over 9} \ \ ,\nonumber \\ 
& & X^{K T}_{1D_+}={2 \over 9} \left(1-{1 \over 3}\ttau\right)P_F
\ \ ,\nonumber \\
& & X^{K T}_{1S}={1 \over 9} \left[ 1+{5 \over 3}\left(\ttau\right)
+{2 \over 3}\ttaup \right] \ \ .
\label{b14}
\end{eqnarray}

\bigskip

\noindent
[ $B_3 B_1=\Xi \Xi$ ]
\begin{eqnarray}
& & X^{\delta C}_{1D_-}=-{1 \over 12}\left[
9-\left(\ttau\right)-\left(\ssigma \right)
\left(1-{5 \over 9}\ttau\right)
\right] \ \ ,\nonumber \\
& & X^{\delta C}_{1D_+}=0 \ \ ,\nonumber \\
& & X^{\delta C}_{1S}={1 \over 2} \left[1-{1 \over 3}\left(\ttau\right)
+\left(\ssigma \right){1 \over 9}\left(1-{1 \over 3}\ttau\right)
\right] \ \ ,\nonumber \\
& & \qquad  \nonumber \\
& & X^{\kappa C}_{1D_-}=-\left(2+{4 \over 9}\ssigma\right)
\ \ ,\nonumber \\
& & X^{\kappa C}_{1D_+}=-\left[1-{1 \over 3}\left(\ttau\right)
+\left(\ssigma \right){1 \over 9}\left(7+{5 \over 3}\ttau\right)
\right] P_F \ \ ,\nonumber \\
& & X^{\kappa C}_{1S}={1 \over 2} \left[1+{1 \over 3}\left(\ttau\right)
-\left(\ssigma \right){1 \over 9}\left(1+{5 \over 3}\ttau\right)
\right] \ \ ,\nonumber \\
& & \qquad  \nonumber \\
& & X^{\pi SS}_{1D_-}=-{1 \over 4}\left[
9-\left(\ttau\right) +\left(\ssigma \right){1 \over 3}
\left(1-{5 \over 9}\ttau \right) \right]\ \ ,\nonumber \\
& & X^{\pi SS}_{1D_+}=0 \ \ ,\nonumber \\
& & X^{\pi SS}_{1S}=-{1 \over 2} \left[
5+{1 \over 3}\left(\ttau\right)-\left(\ssigma \right){1 \over 9}
\left(1+{17 \over 3}\ttau \right) \right]
\ \ ,\nonumber \\
& & \qquad \nonumber \\
& & X^{K SS}_{1D_-}=-\left(6-{4 \over 9}\ssigma\right)
\ \ ,\nonumber \\
& & X^{K SS}_{1D_+}=-{1 \over 9} \left[
3-\left(\ttau\right)+\left(\ssigma \right)
\left(13+{23 \over 3}\ttau \right) \right] P_F
\ \ ,\nonumber \\
& & X^{K SS}_{1S}=-{5 \over 2} \left[
1+{1 \over 3}\left(\ttau\right)-\left(\ssigma \right){1 \over 9}
\left(1+{5 \over 3}\ttau \right) \right]
\ \ ,\nonumber \\
& & \qquad \nonumber \\
& & X^{\delta LS}_{1D_-}={1 \over 9}\left(\ttau\right) \ \ ,\nonumber \\ 
& & X^{\delta LS}_{1D_+}=0 \ \ ,\nonumber \\ 
& & \qquad \nonumber \\
& & X^{\kappa LS}_{1D_-}=-2 \ \ ,\nonumber \\ 
& & X^{\kappa LS}_{1D_+}=-{8 \over 9}\left(1-{1 \over 3}\ttau\right) P_F
\ \ ,\nonumber \\ 
& & \qquad \nonumber \\
& & X^{\pi T}_{1D_-}={1 \over 6} \left(1-{5 \over 9}\ttau \right)
\ \ ,\nonumber \\ 
& & X^{\pi T}_{1D_+}=0 \ \ ,\nonumber \\ 
& & X^{\pi T}_{1S}={1 \over 18} \left(1-{1 \over 3}\ttau\right)
\ \ ,\nonumber \\
& & X^{K T}_{1D_-}=-{8 \over 9} \ \ ,\nonumber \\ 
& & X^{K T}_{1D_+}=-{4 \over 9} \left(1-{1 \over 3}\ttau \right)
P_F \ \ ,\nonumber \\ 
& & X^{K T}_{1S}=-{2 \over 9} \left(1+{5 \over 3}\ttau\right) \ \ .
\label{b15}
\end{eqnarray}

\bigskip

\noindent
[ $B_3 B_1=\Lambda \Sigma$ and $\Sigma \Lambda$ ]
\begin{eqnarray}
& & X^{\delta C}_{1D_-}=-{5 \over 18}\left(\ssigma\right)
\ \ ,\nonumber \\
& & X^{\delta C}_{1D_+}={1 \over 9}\left(\ssigma\right)
\ \ ,\nonumber \\
& & X^{\delta C}_{1S}=\left\{
\begin{array}{l}
{1 \over 6}\left(\ssigma\right) \\ [3mm]
-\left[ {17 \over 18}\left(\ssigma\right)
+\left(1-{1 \over 3}\ssigma\right)P_F \right] \\
\end{array} \right.
\quad \hbox{for} \quad \left\{
\begin{array}{c}
\Lambda \Sigma \\ [3mm]
\Sigma \Lambda \\
\end{array} \right. \ \ ,\nonumber \\
& & \qquad  \nonumber \\
& & X^{\kappa C}_{1D_-}=0 \ \ ,\nonumber \\
& & X^{\kappa C}_{1D_+}=\left(1-{1 \over 9}\ssigma\right)P_F
\ \ ,\nonumber \\
& & X^{\kappa C}_{1S}=\left\{
\begin{array}{l}
{1 \over 2}\left[
1-{5 \over 9}\left(\ssigma\right)
-\left(1-{1 \over 3}\ssigma\right)P_F\right] \\ [3mm]
{1 \over 2}\left[ 1+{5 \over 3}\left(\ssigma\right)
+\left(1-{1 \over 3}\ssigma\right)P_F\right] \\
\end{array} \right.
\quad \hbox{for} \quad \left\{
\begin{array}{c}
\Lambda \Sigma \\ [3mm]
\Sigma \Lambda \\
\end{array} \right. \ \ ,\nonumber \\
& & \qquad  \nonumber \\
& & X^{\pi SS}_{1D_-}={5 \over 18}\left(\ssigma\right) \ \ ,\nonumber \\
& & X^{\pi SS}_{1D_+}={17 \over 9}\left(\ssigma\right)
+{7 \over 3}\left(1-{1 \over 3}\ssigma\right)P_F \ \ ,\nonumber \\
& & X^{\pi SS}_{1S}=\left\{
\begin{array}{l}
{13 \over 6}\left(\ssigma\right)
+{3 \over 2}\left(1-{1 \over 3}\ssigma\right)P_F \\ [3mm]
{79 \over 18}\left(\ssigma\right)+{7 \over 2}\left(1-{1 \over 3}
\ssigma\right)P_F \\
\end{array} \right.
\quad \hbox{for} \quad \left\{
\begin{array}{c}
\Lambda \Sigma \\ [3mm]
\Sigma \Lambda \\
\end{array} \right. \ \ ,\nonumber \\
& & \qquad \nonumber \\
& & X^{K SS}_{1D_-}=0 \ \ ,\nonumber \\
& & X^{K SS}_{1D_+}=\left({7 \over 3}-\ssigma\right)P_F
\ \ ,\nonumber \\
& & X^{K SS}_{1S}=\left\{
\begin{array}{l}
-{1 \over 2}\left(1-{25 \over 9}\ssigma\right)
+{5 \over 2}\left(1-{1 \over 3}\ssigma\right)P_F \\ [3mm]
-{1 \over 2}\left(1-{5 \over 9}\ssigma\right)
+{3 \over 2}\left(1-{1 \over 3}\ssigma\right)P_F \\
\end{array} \right.
\quad \hbox{for} \quad \left\{
\begin{array}{c}
\Lambda \Sigma \\ [3mm]
\Sigma \Lambda \\
\end{array} \right. \ \ ,\nonumber \\
& & \qquad \nonumber \\
& & X^{\delta LS}_{1D_-}=-{1 \over 6}
\ \ ,\nonumber \\ 
& & X^{\delta LS}_{1D_+}={1 \over 9} \left( 1- 2 P_F \right)
\ \ ,\nonumber \\ 
& & \qquad \nonumber \\
& & X^{\kappa LS}_{1D_-}=0 \ \ ,\nonumber \\ 
& & X^{\kappa LS}_{1D_+}={2 \over 3} P_F \ \ ,\nonumber \\ 
& & \qquad \nonumber \\
& & X^{\pi T}_{1D_-}=-{5 \over 9} \ \ ,\nonumber \\ 
& & X^{\pi T}_{1D_+}=-{1 \over 9}\left(1-2P_F\right) \ \ ,\nonumber \\ 
& & X^{\pi T}_{1S}=\left\{
\begin{array}{l}
-{2 \over 3} \\ [3mm]
-{1 \over 9}\left(1-3P_F\right) \\
\end{array} \right.
\quad \hbox{for} \quad \left\{
\begin{array}{c}
\Lambda \Sigma \\ [3mm]
\Sigma \Lambda \\
\end{array} \right. \ \ ,\nonumber \\ 
& & \qquad \nonumber \\
& & X^{K T}_{1D_-}= X^{K T}_{1D_+}=0 \ \ ,\nonumber \\ 
& & X^{K T}_{1S}=\left\{
\begin{array}{l}
{1 \over 9}\left(5+6P_F\right) \\ [3mm]
-{5 \over 9} \\
\end{array} \right.
\quad \hbox{for} \quad \left\{
\begin{array}{c}
\Lambda \Sigma \\ [3mm]
\Sigma \Lambda \\
\end{array} \right. \ \ .
\label{b16}
\end{eqnarray}

\subsection{The Coulomb exchange factors for the $NN$ system} 

\noindent
[ $B_3 B_4$ - $B_1 B_2$ = $pp$ - $pp$ ] \footnote{$X^{CL}_{1S}$ and
$X^{CL}_{1D_+}$ factors in Eq.\,(A2) of \cite{FSS} contain
misprints in the $\left(\protect\ssigma\right)$ terms.}
\qquad $X^{CL}_{0E}=0$\ \ ,\qquad $X^{CL}_{0D_+}=1$
\begin{eqnarray}
& & X^{CL}_{1D_-}=-{1 \over 54} \left[ 17+{65 \over 9}
\left(\ssigma\right) \right] \ \ ,\nonumber \\
& & X^{CL}_{1D_+}=-{2 \over 27} \left[ 5+{8 \over 9}
\left(\ssigma\right) \right] \ \ ,\nonumber \\
& & X^{CL}_{1S}=-{4 \over 27} \left[ 1+{7 \over 9}
\left(\ssigma\right) \right] \ \ ,\nonumber \\
& & X^{CL}_{1S}+X^{CL}_{1D_+}+X^{CL}_{1D_-}
=-{1 \over 6} \left[ 5+{17 \over 9}
\left(\ssigma\right) \right]
=1\cdot 1\cdot X_N(I=1) \ \ .
\label{b17}
\end{eqnarray}

\bigskip

\noindent
[ $B_3 B_4$ - $B_1 B_2$ = $np$ - $np$ ]
\qquad $X^{CL}_{0E}=-1/3$\ \ ,\qquad $X^{CL}_{0D_+}=0$
\begin{eqnarray}
& & X^{CL}_{1D_-}=-{5 \over 27} \left[ 1-{2 \over 9}
\left(\ssigma\right) \right]
+{1 \over 27} \left[ 1+{25 \over 9}
\left(\ssigma\right) \right] P_F \ \ ,\nonumber \\
& & X^{CL}_{1D_+}=-{2 \over 27} \left[ 1-{2 \over 9}
\left(\ssigma\right) \right]
-{1 \over 54} \left[ 1+{25 \over 9}
\left(\ssigma\right) \right] P_F \ \ ,\nonumber \\
& & X^{CL}_{1S}={7 \over 27} \left[ 1-{2 \over 9}
\left(\ssigma\right) \right]
-{1 \over 54} \left[ 1+{25 \over 9}
\left(\ssigma\right) \right] P_F \ \ ,
\label{b18}
\end{eqnarray}
where $P_F=P_\tau=(1+\ttau)/2$.

\bigskip

\noindent
[ $B_3 B_4$ - $B_1 B_2$ = $nn$ - $nn$ ]
\qquad $X^{CL}_{0E}=-2/3$\ \ ,\qquad $X^{CL}_{0D_+}=0$
\begin{eqnarray}
& & X^{CL}_{1D_-}=X^{CL}_{1D_+}=-{1 \over 2}X^{CL}_{1S}
\ \ ,\nonumber \\
& & X^{CL}_{1S}={4 \over 27} \left[ 2+{5 \over 9}
\left(\ssigma\right) \right] \ \ .
\label{b19}
\end{eqnarray}

\section{Deuteron wave functions}

The relative wave functions for the deuteron in the momentum
representation, $f_{\ell}(q) \sim 1/(\gamma^2+q^2)
T_{\ell,\hbox{-}}(q, -i\gamma, -\epsilon_d)$,
satisfy the homogeneous equation
\begin{eqnarray}
\left(\gamma^2+p^2\right)f_{\ell}(p)=-{2\mu \over \hbar^2}
{4\pi \over (2\pi)^3}\sum_{\ell^\prime}\int^\infty_0 q^2 dq
~V_{\ell \ell^\prime}(p, q, -\epsilon_d)~f_\ell(q) \ \ ,
\label{c1}
\end{eqnarray}
where $V_{\ell \ell^\prime}(p, q, -\epsilon_d)$ is the partial-wave
components of \eq{fm8}. Since $f_\ell(q)$ are the relative
wave functions of the RGM equation, one needs to renormalize them
through the square root of the normalization kernel \cite{FSS}.
This can be achieved by calculating
\begin{eqnarray}
F_{\ell}(q)=q f_{\ell}(q)+q \sum_N R_{N \ell}
\left(q, b^2/3\right)
{\gamma_N \over \sqrt{1+\gamma_N}+1}~J_{N\ell}\ \ ,
\label{c2}
\end{eqnarray}
where $R_{N\ell}(r, \nu)$ represents the radial part of
the harmonic-oscillator wave function with the width parameter $\nu$, 
and $\gamma_N=(1/3)^{N+2}$ with $N=0,~2,~4,\cdots$ are
the eigen-values of the exchange normalization kernel
for the $\hbox{}^3E$ states of the $NN$ system.
The harmonic-oscillator components $J_{N\ell}$ of $f_{\ell}(q)$ are
calculated from
\begin{eqnarray}
J_{N\ell}=\int^\infty_0 q^2 dq~R_{N \ell}
\left(q, b^2/3\right)~f_{\ell}(q)\ \ .
\label{c3}
\end{eqnarray}
The deuteron wave functions $u_{\ell}(r)$ in the coordinate
representation (customarily written as $u(r)=u_0(r)$ and $w(r)=u_2(r)$
for the $S$-wave and $D$-wave states, respectively) are obtained
from the Fourier transformation
\begin{eqnarray}
u_{\ell}(r)=i^\ell \sqrt{{2 \over \pi}}\int^\infty_0 dq
~(qr)j_\ell(qr)~F_{\ell}(q) \ \ .
\label{c4}
\end{eqnarray}
In particular, $f_{\ell}(q)$ are normalized such that
\begin{eqnarray}
\sum_\ell \int^\infty_0 dr~\left(u_{\ell}(r)\right)^2
=\sum_\ell \int^\infty_0 dq~\left(F_{\ell}(q)\right)^2
=1 \ \ .
\label{c5}
\end{eqnarray}

We follow the standard ansatz \cite{LA81,MA89,MA00}
for the simple parameterization of the deuteron wave functions:
\begin{eqnarray}
& & F_{\ell}(q)=\sum^n_{j=1} \left\{ \begin{array}{c}
C_j \\ [3mm]
D_j \\
\end{array}\right\}\sqrt{{2 \over \pi}}{q \over q^2+{\gamma_j}^2}
\quad \hbox{for} \quad \left\{ \begin{array}{c}
\ell=0 \\ [3mm]
\ell=2 \\
\end{array}\right. \ \ ,\nonumber \\ [3mm]
& & u_{\ell}(r)=\left\{ \begin{array}{l}
\sum^n_{j=1} C_j e^{-\gamma_j r} \\ [3mm]
\sum^n_{j=1} D_j e^{-\gamma_j r} \left(1+\frac{3}{\gamma_j r}
+\frac{3}{(\gamma_j r)^2}\right) \\
\end{array}\right. \quad \hbox{for} \quad \left\{ \begin{array}{c}
\ell=0 \\ [3mm]
\ell=2 \\
\end{array}\right. \ \ .
\label{c6}
\end{eqnarray}
The range parameters $\gamma_j$ are chosen as $\gamma_j=\gamma
+(j-1)\gamma_0$ with $\gamma_0=0.9~\hbox{fm}^{-1}$ and $n=11$.
The coefficients $C_j$ ($j=1$ - 10) and $D_j$ ($j=1$ - 8)
with $\gamma=0.23186542~\hbox{fm}^{-1}$ are
given in Table \ref{deutw} for the deuteron wave functions
in the full calculation. The other coefficients, namely,
the last $C_j$ and the last three $D_j$ should be
calculated from Eqs.\,(C.7) and (C.8) of \cite{MA89}.

\clearpage


\begin{table}[h]
\caption{
Quark-model parameters, $SU_3$ parameters of the EMEP, S-meson
masses, and some reduction factors $c_\delta$ etc. for the models
fss2 and FSS.
The $\rho$ meson in fss2 is treated in the
two-pole approximation, for which $m_1$ ($\beta_1$)
and $m_2$ ($\beta_2$) are shown below the table.
}
\label{table1}
\begin{center}
\renewcommand{\arraystretch}{1.4}
\setlength{\tabcolsep}{4mm}
\begin{tabular}{ccccc}
 &  $b$ (fm) & $m_{ud}$ (MeV/$c^2$)
 & $\alpha_S$ & $\lambda=m_s/m_{ud}$ \\
\hline
fss2 & 0.5562 & 400 & 1.9759 & 1.5512 \\
FSS    & 0.616 & 360 & 2.1742 & 1.526 \\
\hline
 & $f^{\rm S}_1$ & $f^{\rm S}_8$ & $\theta^{\rm S}$ &
$\theta^{\rm S}_4$ $^{\,1)}$ \\
\hline
fss2 & 3.48002 & 0.94459 & $33.3295^\circ$ & $55.826^\circ$ \\
FSS    & 2.89138 & 1.07509 & $27.78^\circ$ & $65^\circ$ \\
\hline
 & $f^{\rm PS}_1$ & $f^{\rm PS}_8$ & $\theta^{\rm PS}$ & \\
\hline
fss2 & $-$    & 0.26748 & $-$ & (no $\eta, \eta^\prime$) \\
FSS    & 0.21426 & 0.26994 & $-23^\circ$ &  \\
\hline
 & $f^{\rm Ve}_1$ & $f^{\rm Ve}_8$ & $f^{\rm Vm}_1$
 & $f^{\rm Vm}_8$ $^{\,2)}$ \\
\hline
fss2 & 1.050 & 0 & 1.000 & 2.577 \\
\hline
(MeV/$c^2$) & $m_\epsilon$ & $m_{S^*}$ & $m_\delta$
 & $m_\kappa$ \\
\hline
fss2 & 800 & 1250 & $846^{\,3)}$ & 936 \\
FSS    & 800 & 1250 & 970 & 1145 \\
\hline
 & $c_\delta$ & $c_{qss}$ & ${c_{qT}}^{\ 5)}$ &  \\
\hline
fss2 &  $0.4756^{\,4)}$ & 0.6352 & 3.139 & \\
FSS & 0.381 & $-$ & $-$ & \\
\end{tabular}
\end{center}

\begin{enumerate}
\setlength{\itemsep}{0mm}
\item[1)] $\theta^{\rm S}_4$ is used only for $\Sigma N (I=3/2)$.
\item[2)] $\theta^{\rm V}=35.264^\circ$ (ideal mixing) and
two-pole $\rho$ meson with $m_1$ ($\beta_1$) = 664.56 $\hbox{MeV}/c^2$
(0.34687) and $m_2$ ($\beta_2$) = 912.772 $\hbox{MeV}/c^2$ (0.48747)
\protect\cite{ST94} are used.
\item[3)] For the $NN$ system, $m_\delta=720~\hbox{MeV}/c^2$ is used.
\item[4)] Only for $\pi$, otherwise 1.
\item[5)] The enhancement factor for the Fermi-Breit tensor term.
\end{enumerate}
\end{table}

\begin{table}[h]
\caption{The interaction types and the meson species
introduced in the EMEP of the models fss2 and FSS.
$C$ represents the central force,
$SS$ the spin-spin force, $T$ the tensor force,
and $QLS$ the quadratic spin-orbit force. $C(BS)$ implies
that the momentum-dependent Bryan-Scott term is also
included for the central force.
The tensor term of the vector mesons is switched off.
}
\label{table2}
\begin{center}
\renewcommand{\arraystretch}{1.4}
\newcommand{\lw}[1]{\smash{\lower1.ex\hbox{#1}}}
\setlength{\tabcolsep}{4mm}
\begin{tabular}{cccc}
model & meson type & interaction type & mesons \\
\hline
        &  S & $C(BS)+LS$
        &  $\epsilon$, $S^*$, $\delta$, $\kappa$ \\ 
fss2    & PS & $SS+T$     &  $\pi$, $K$ \\
        & V  & $C(BS)+LS+QLS$ &  $\omega$, $\phi$, $\rho$, $K^*$ \\
\hline
FSS &  S & $C$ &  $\epsilon$, $S^*$, $\delta$, $\kappa$ \\
    & PS & $SS+T$ & $\eta^\prime$, $\eta$, $\pi$, $K$ \\
\end{tabular}
\end{center}
\end{table}

%
\begin{table}[h]
\caption{Quark and EMEP contributions
to the $N$ - $\Delta$ mass difference ($\Delta
E_{N \hbox{-} \Delta}$) and the $\Lambda$ - $\Sigma$ mass
difference ( $\Delta E_{\Lambda \hbox{-} \Sigma}$ ) in MeV,
calculated in the isospin basis. The model is fss2.
The mass ratio of strange to up-down quarks,
$\lambda=(m_s/m_{ud})=1.5512$, is employed to calculate
the quark contribution in $\Delta E_{\Lambda \hbox{-} \Sigma}$.
The details of the EMEP contribution
to $\Delta E_{\Lambda \hbox{-} \Sigma}$ are given
in Table \protect\ref{table5}.
See Table \protect\ref{table1} for the two-pole $\rho$-meson
parameters and the other EMEP parameters.}
\label{table3}
\begin{center}
\renewcommand{\arraystretch}{1.4}
\setlength{\tabcolsep}{6mm}
\begin{tabular}{cccc}
 & $\beta$ & $m_{\beta}$ (MeV/$c^2$) & $E$ (MeV) \\
\hline
 & Quark &  & 293.33 \\
 & $\delta$ & 720 & $-164.70$ \\
$\Delta E_{N\hbox{-}\Delta}$ & $\pi$ & 138.039 & 71.56 \\
 & $\omega$ & 781.940 & $-34.36$ \\
 & $\phi$ & 1019.413 & $-0.19$ \\
 & $\rho$ & two-pole & 80.59 \\
\hline
exp & 293.3 & total & 246.23 \\
\hline
$\Delta E_{\Lambda\hbox{-}\Sigma}$
 & Quark & ($\lambda=1.5512$) & 69.49 \\
 & EMEP  & $-$ & 7.98 \\
\hline
exp & 77.44 & total & 77.47 \\
\end{tabular}
\end{center}
\end{table}

\begin{table}[h]
\caption{The baryon masses employed in the particle-basis
calculation \protect\cite{PA98}.}
\label{table4}
\begin{center}
\renewcommand{\arraystretch}{1.4}
\setlength{\tabcolsep}{0mm}
\begin{tabular}{cc}
$B$ & $M_B$ (MeV/$c^2$) \\
\hline
$p$ & 938.2723 \\
$n$ & 939.565 \\
$\Lambda$ & 1115.683 \\
$\Sigma^+$ & 1189.37 \\
$\Sigma^0$ & 1192.642 \\
$\Sigma^-$ & 1197.449 \\
\end{tabular}
\end{center}
\end{table}

\begin{table}
\caption{
Details of the EMEP contributions to the baryon mass difference (in MeV)
in the isospin and particle bases.
In the particle basis, only the mass difference of the charged
and neutral pions is introduced.
}
\label{table5}
\begin{center}
\renewcommand{\arraystretch}{1.4}
\setlength{\tabcolsep}{6mm}
\begin{tabular}{ccccc}
$\beta$ & $m_{\beta}$ & isospin basis & \multicolumn{2}{c}{particle
basis} \\
\cline{3-3}
\cline{4-5} 
 & (MeV/$c^2$) & $\Delta E_{\Lambda \hbox{-} \Sigma}$ 
 & $\Delta E_{\Lambda \hbox{-} \Sigma^0}$ 
 & $\Delta E_{\Sigma^\pm \hbox{-} \Sigma^0}$ \\
\hline
$\delta$ & 846 & $-87.345$ & $-87.345$ &  0  \\
$\kappa$ & 936 &  75.072   &  75.072   &  0  \\
$\pi$ & 138.039 & 47.704   &   $-$     & $-$ \\
$\pi^\pm$ & 139.570 & $-$  &  23.747   & $-11.873$ \\
$\pi^0$   & 134.976 & $-$  &  24.061   & 12.031 \\
$K$ & 495.675       & $-32.716$ & $-32.716$ & 0 \\
$\omega$ & 781.940  & $-26.587$ & $-26.587$ & 0 \\
$\phi$ & 1019.413   & $-1.799$  & $-1.799$  & 0 \\
$\rho$ & two-pole   & 53.723    & 53.723    & 0 \\
$K^*$  & 893.880    & $-20.072$ & $-20.072$ & 0 \\
\hline
 total &   & 7.980  & 8.084  & 0.158  \\
\end{tabular}
\end{center}
\end{table}

\begin{table}
\caption{
Calculated threshold energies (in MeV) compared with the
empirical values in the non-relativistic kinematics.
Note that the effect of the charged and neutral pion mass
difference on the $n$ and $p$ internal
energies is zero.
}
\label{table6}
\begin{center}
\renewcommand{\arraystretch}{1.4}
\setlength{\tabcolsep}{4mm}
\begin{tabular}{cccccc}
channel & $E^Q$ & $E^M$ & $E^{CL}$ & $E^{\rm cal}_{\rm th}$
& $E^{\rm exp}_{\rm th}$ \\
\hline
$\Lambda N$ & 0 & 0 & 0 & 0 & 0 \\
$\Sigma N$  & 69.488 & 7.980 & 0 & 77.468 & 77.471 \\
\hline
$\Lambda p$   & 0 & 0 & 0 & 0 & 0 \\
$\Sigma^+ n$  & 69.488 & 7.926 & 0 & 77.415 & 74.980 \\
$\Sigma^0 p$  & 69.488 & 8.084 & 0 & 77.572 & 76.959 \\
\hline
$\Lambda n$   & 0 & 0 & 0 & 0 & 0 \\
$\Sigma^0 n$  & 69.488 & 8.084 &     0 & 77.572 & 76.959 \\
$\Sigma^- p$  & 69.488 & 7.926 & 2.066 & 79.480 & 80.473 \\
\hline
$\Sigma^- p$  & 0 & 0 & 0 & 0 & 0 \\
$\Sigma^0 n$  & 0 & 0.158 & $-2.066$ & $-1.908$ & $-3.514$ \\
$\Lambda n$   & $-69.488$   & $-7.926$ & $-2.066$ & $-79.480$ & $-80.473$ \\
\end{tabular}
\end{center}
\end{table}

\begin{table}
\caption{
The threshold energies in the relativistic kinematics,
used in the particle-basis calculation.
The momentum is measured in MeV/$c$ and the energy in MeV. 
The difference, $\Delta E^{\rm int}=
E^{\rm exp}_{\rm th}({\rm relativistic})
-E^{\rm cal}_{\rm th}$, is given in the last column.
Here $E^{\rm exp}_{\rm th}({\rm relativistic})=
(p^{\rm cm}_{\rm th})^2/2\mu_{\rm inc}$ with $\mu_{\rm inc}$ being
the non-relativistic reduced mass of the incident channel.
}
\label{table7}
\begin{center}
\renewcommand{\arraystretch}{1.4}
\setlength{\tabcolsep}{4mm}
\begin{tabular}{ccccccc}
channel & \multicolumn{3}{c}{relativistic}
& \multicolumn{2}{c}{non-relativistic} & dif \\
\cline{2-4}
\cline{5-6}
& $p^{\rm cm}_{\rm th}$
& $p^{\rm lab}_{\rm th}$ & $E^{\rm exp}_{\rm th}$
& $E^{\rm exp}_{\rm th}$ & $E^{\rm cal}_{\rm th}$
& $\Delta E^{\rm int}$ \\
\hline
$\Lambda p$   & 0  & 0 & 0 & 0 & 0 & $-$ \\
$\Sigma^+ n$  & 279.04 & 633.14 & 76.389 & 74.980 & 77.415 & $-1.026$ \\
$\Sigma^0 p$  & 282.77 & 642.20 & 78.443 & 76.959 & 77.572 &   0.871  \\
\hline
$\Lambda n$   & 0  & 0 & 0 & 0 & 0 & $-$ \\
$\Sigma^0 n$  & 282.87 & 641.93 & 78.441 & 76.959 & 77.572 & 0.869 \\
$\Sigma^- p$  & 289.38 & 657.79 & 82.093 & 80.473 & 79.480 & 2.613 \\
\hline
$\Sigma^- p$  & 0 & 0 & 0 & 0 & 0 & $-$ \\
$\Sigma^0 n$  &  $-$  & $-$ & $-3.652$  & $-3.514$ & $-1.908$
& $-1.744$ \\
$\Lambda n$   &  $-$  & $-$ & $-82.093$ & $-80.473$ & $-79.480$
& $-2.613$ \\
\end{tabular}
\end{center}
\end{table}

\begin{table}
\caption{The $np$ phase-shift parameters
calculated in the isospin basis (in degrees).
The results given by OBEP, Paris and Bonn
potentials are cited from Table 5.2 in \protect\cite{MA89}.
}
\label{phase}
\begin{center}
\renewcommand{\arraystretch}{1.2}
\setlength{\tabcolsep}{2mm}
\begin{tabular}{cccccccc}
State & Model & \multicolumn{6}{c}{$T_{\rm lab}~(\hbox{MeV})$} \\
\cline{3-8}
      &        & 25 & 50 & 100 & 150 & 200 & 300 \\
\hline
 & fss2 & 80.98 & 63.03 & 43.21 & 30.51
& 21.00 & 7.02 \\
 & OBEP & 80.32 & 62.16 & 41.99 & 28.96 & 19.04 & 4.07 \\
$\hbox{}^3S_1$ & Paris  & 80.35 & 62.28 & 42.26
 &  29.24 & 19.25 & 3.91 \\
 & Bonn & 80.30 & 62.19 & 42.27 & 29.64 & 20.31 & 7.06 \\
 & SP99 & 80.26 & 62.10 & 42.22 & 29.69 & 20.51 & 7.07 \\
\hline
 & fss2 & $-2.82$ & $-6.52$ & $-12.43$ & $-16.59$ & $-19.49$ & $-22.58$ \\
 & OBEP & $-2.99$ & $-6.86$ & $-12.98$ & $-17.28$ & $-20.28$ & $-23.72$ \\
$\hbox{}^3D_1$   &  Paris   &  $-2.95$ & $-6.77$ & $-12.85$
 & $-17.22$ & $-20.42$ & $-24.52$ \\
 & Bonn & $-3.03$ & $-6.98$ & $-13.25$ & $-17.64$ & $-20.62$ & $-23.43$ \\
 & SP99 & $-2.72$ & $-6.84$ & $-13.09$ & $-16.69$ & $-19.08$ & $-23.04$ \\
\hline
 & fss2 &  1.68 &  1.91 &   2.21 &   2.68 &   3.33 &   4.97 \\
 & OBEP &  1.76 &  2.00 &   2.24 &   2.58 &   3.03 &   4.03 \\
$\epsilon_1$ & Paris    &   1.69 &   1.89 &   2.14 &   2.59
 & 3.21 &  4.76 \\
 & Bonn &  1.82 &  2.08 &   2.29 &   2.54 &   2.82 &   3.19 \\
 & SP99 &  1.69 &  2.14 &   2.91 &   3.55 &   4.08 &   5.06 \\
\hline
 & fss2 & $-6.70$ & $-10.26$ & $-14.82$ & $-18.38$
 & $-21.57$ & $-27.32$ \\
 & OBEP & $-7.21$ & $-11.15$ & $-16.31$ & $-20.21$
 & $-23.47$  & $-28.70$ \\
$\hbox{}^1P_1$    &  Paris   &  $-7.11$ & $-10.95$  & $-15.72$
 & $-19.08$ & $-21.73$  & $-25.92$ \\
 &   Bonn &  $-6.90$  & $-10.48$ & $-15.11$ & $-18.88$
 & $-22.41$ & $-29.17$ \\
 &  SP99  &  $-6.71$  &  $-9.98$ & $-14.47$ & $-18.29$
 & $-21.56$ & $-26.62$ \\
\hline
 &  fss2  &   3.67  &   8.82 &  17.09 &  22.26 &  25.06 &  26.38 \\
 &  OBEP  &   3.88  &   9.29 &  17.67 &  22.57 &  24.94 &  25.36 \\
$\hbox{}^3D_2$  & Paris      &  3.96  &   9.60
 &  18.64 &  24.19  &  27.15 &  28.54 \\
 &   Bonn &   3.88  &   9.27 &  17.41 &  21.68 &  23.09 &  20.84 \\
 &  SP99  &   3.87  &   9.37 &  17.89 &  22.73 &  25.03 &  25.47 \\
\hline
 &  fss2  & 52.26   &  41.94 &  27.51 &  16.91 &   8.41 &  $-4.86$ \\
 &  OBEP  & 50.72   &  39.98 &  25.19 &  14.38 &   5.66 &  $-8.18$ \\
$\hbox{}^1S_0$  & Paris      &  48.38 &  38.12 &  23.88
 & 13.38  &  4.84   &  $-9.01$ \\
 &  Bonn  & 50.03   &  39.15 &  24.36 &  13.72 &   5.30 &  $-7.62$ \\
 &  SP99  & 51.30   &  41.88 &  28.24 &  16.95 &   7.74 &  $-5.49$ \\
\hline
 &  fss2  &  8.55   &  11.25 &   9.04 &   4.02 &  $-1.49$ & $-12.10$ \\
 &  OBEP  &  9.34   &  12.24 &   9.80 &   4.57 &  $-1.02$ & $-11.48$ \\
$\hbox{}^3P_0$   & Paris     &   9.21 &  11.93 &   9.83
 &  5.32  &  0.48   &  $-8.52$ \\
 &  Bonn  &  9.57   &  12.79 &  10.88 &   6.02 &   0.66 & $-9.66$ \\
 &  SP99  &  8.24   &  10.75 &   8.18 &   3.15 & $-1.95$ & $-11.63$ \\
\hline
 &  fss2  & $-5.23$   &  $-8.68$ & $-13.45$ & $-17.27$
 & $-20.77$ & $-27.26$ \\
 &  OBEP  & $-5.33$   &  $-8.77$ & $-13.47$ & $-17.18$
 & $-20.49$ & $-26.38$ \\
$\hbox{}^3P_1$ & Paris  &  $-5.27$ &  $-8.64$ & $-13.44$
 & $-17.35$ & $-20.91$  & $-27.30$ \\
 &  Bonn    &  $-5.17$  &  $-8.53$ & $-13.38$ & $-17.62$
 & $-21.73$ & $-29.87$ \\
 &  SP99  & $-4.75$  &  $-8.15$ & $-13.52$ & $-17.92$
 & $-21.64$ & $-28.06$ \\
\hline
 &  fss2  &  0.64   &   1.47 &   3.29 &   5.30 &   7.27 &  10.28 \\
 &  OBEP  &  0.68   &   1.58 &   3.34 &   4.94 &   6.21 &   7.49 \\
$\hbox{}^1D_2$ &  Paris      &   0.78 &   1.85 &   4.00
 &  5.90  &  7.47   &   9.19 \\
 &  Bonn &  0.72   &   1.72 &   3.76 &   5.62 &   7.04 &   8.32 \\
 &  SP99  &  0.64   &   1.59 &   3.60 &   5.60 &   7.33 &   9.75 \\
\end{tabular}
\end{center}
\end{table}

\addtocounter{table}{-1}

\begin{table}
\caption{-continued}
\begin{center}
\renewcommand{\arraystretch}{1.2}
\setlength{\tabcolsep}{2mm}
\begin{tabular}{cccccccc}
State & Model & \multicolumn{6}{c}{$T_{\rm lab}~(\hbox{MeV})$} \\
\cline{3-8}
      &        & 25 & 50 & 100 & 150 & 200 & 300 \\
\hline
 &  fss2  &  2.58   &   6.26 &  12.43 &  15.92 &  17.36 &  16.97 \\
 &  OBEP  &  2.62   &   6.14 &  11.73 &  14.99 &  16.65 &  17.40 \\
$\hbox{}^3P_2$ &  Paris      &   2.61 &   5.97 &  11.34
 &  14.68 &  16.39  &  16.74 \\
 &   Bonn &   2.54  &  5.89  &  11.14 &  14.24 &  15.93 &  17.22 \\
 &  SP99  &   2.70  &  5.93  &  10.92 &  14.11 &  16.05 &  17.83 \\
\hline
 &  fss2  &   0.10  &  0.32  &  0.72  &  0.98  &  1.08  &  0.75 \\
 &  OBEP  &   0.11  &  0.34  &  0.77  &  1.04  &  1.10  &  0.52 \\
$\hbox{}^3F_2$ &  Paris      &  0.11  &  0.36  &  0.79
 &  1.05  &   1.06  &  0.49 \\
 &   Bonn &   0.11  &  0.35  &  0.81  &  1.14  &  1.28  &  0.87 \\
 &  SP99  &   0.09  &  0.33  &  0.85  &  1.19  &  1.31  &  0.90 \\
\hline
 &  fss2  &  $-0.82$  & $-1.77$  & $-2.85$  & $-3.22$
 & $-3.24$  & $-2.94$ \\
 &  OBEP  &  $-0.86$  & $-1.82$  & $-2.84$  & $-3.05$
 & $-2.85$  & $-2.02$ \\
$\epsilon_2$     & Paris & $-0.87$  & $-1.80$  & $-2.73$  & $-2.90$
 &  $-2.74$ &  $-2.14$ \\
 &   Bonn   &  $-0.85$  & $-1.77$  & $-2.74$  & $-2.97$
 & $-2.84$  & $-2.23$ \\
 &  SP99  &  $-0.70$  & $-1.48$  & $-2.37$  & $-2.71$
 & $-2.74$  & $-2.29$ \\
\end{tabular}
\end{center}
\end{table}

\begin{table}[h]
\caption{Deuteron properties by fss2 in three different
calculational schemes, compared with th predictions of
the Bonn model-C potential \protect\cite{MA89} and the experiment.}
\label{deutt}
\begin{center}
\renewcommand{\arraystretch}{1.5}
\setlength{\tabcolsep}{3mm}
\begin{tabular}{ccccccc}
& isospin basis & \multicolumn{2}{c}{particle basis} & Bonn C
& Expt. & Ref. \\
\cline{3-4}
& & Coulomb off & Coulomb on & & & \\
\hline
$\epsilon_d$ (MeV) & 2.2250 & 2.2261  & 2.2309 & fitted
& 2.224644 $\pm$ 0.000046 & \protect\cite{DU83} \\
$P_D$ ($\%$)       & 5.490   & 5.490   & 5.494  & 5.60 & & \\
$\eta=A_D/A_S$     & 0.02527 & 0.02527 & 0.02531 & 0.0266 &
0.0256 $\pm$ 0.0004 & \protect\cite{RO90} \\
rms (fm) & 1.9598   & 1.9599 & 1.9582  & 1.968 &
1.9635 $\pm$ 0.0046 & \protect\cite{DU83} \\
$Q_d$ (fm$\hbox{}^2$) & 0.2696 & 0.2696 & 0.2694 & 0.2814 &
0.2860 $\pm$ 0.0015 & \protect\cite{BI79} \\
$\mu_d$ ($\mu_N$) & 0.8485  & 0.8485 & 0.8485 & 0.8479 & 0.85742 & \\
\end{tabular}
\end{center}
\end{table}

\begin{table}[h]
\caption{Effective range parameters of fss2
for the $NN$ and $YN$ interactions
in the single-channel formula.
For the $pp$ and $nn$ systems, the calculation
in the particle basis uses $f_1^S \times 0.9949$,
in order to incorporate the effect of the charge independence
breaking. Unit of length is in $\hbox{fm}^{2\ell+1}$ in $a$,
$\hbox{fm}^{-2\ell+1}$ in $r$ and $\hbox{fm}^{-2\ell}$ in $P$
for the partial wave $\ell$.
The experimental values are taken from \protect\cite{DU83},
\protect\cite{MI90}, \protect\cite{HO98}, \protect\cite{GO99},
\protect\cite{BE88}, \protect\cite{MA00} for $NN$,
\protect\cite{NA73} for $\Sigma^+ p$,
and \protect\cite{alex68}, \protect\cite{sechi68} for $\Lambda p$.
}
\label{effect}
\begin{center}
\renewcommand{\arraystretch}{1.4}
\setlength{\tabcolsep}{4mm}
\begin{tabular}{cccccc}
 & & isospin basis & \multicolumn{2}{c}{particle basis} & Expt. \\
\cline{4-5}
 & &              & Coulomb off & Coulomb on & \\
\hline
   &  $a$  & $-23.76$ & $-17.80$ & \underline{$-7.810$}
& $-7.8063 \pm 0.0026$ \\
$pp$ $\hbox{}^1S_0$ & $r$ & 2.584 & 2.675 & 2.574
& $2.794 \pm 0.0014$ \\
   &  $P$  & 0.0393  & 0.0423  & 0.0334 &  \\
\hline
   &  $a$  & $-2.740$ & $-2.876$ & $-3.004$ &
$-4.82 \pm 1.11$, $-2.71 \pm 0.34$ \\
$pp$ $\hbox{}^3P_0$ & $r$ & 3.867 & 3.831 & 3.312 &
$7.14 \pm 0.93$, $3.8 \pm 1.1$ \\
   &  $P$  & $-0.014$  & $-0.0130$  & $-0.0125$ &  \\
\hline
   &  $a$  & 1.740 & 1.821 & 2.112 &
$1.78 \pm 0.10$, $1.97 \pm 0.09$ \\
$pp$ $\hbox{}^3P_1$ & $r$ & $-8.196$ & $-8.159$ & $-8.269$ &
$-7.85 \pm 0.52$, $-8.27 \pm 0.37$ \\
   &  $P$  & 0.0009 & 0.0010  & $-0.0063$ &  \\
\hline
\hline
   &  $a$  & $-23.76$ & $-18.04$ & $-18.05$ & $-18.5 \pm 0.3$,
$-18.9 \pm 0.4$ \\
$nn$ $\hbox{}^1S_0$ & $r$ & 2.584 & 2.672 & 2.672 & $2.75 \pm 0.11$ \\
   &  $P$  & 0.0393  & 0.0423  & 0.0423 &  \\
\hline
   &  $a$  & $-2.740$ & $-2.881$ & $-2.881$ &  \\
$nn$ $\hbox{}^3P_0$ & $r$ & 3.867 & 3.823 & 3.822 &  \\
   &  $P$  & $-0.0140$  & $-0.0131$  & $-0.0131$ &  \\
\hline
   &  $a$  & 1.740 & 1.823 & 1.823&  \\
$nn$ $\hbox{}^3P_1$ & $r$ & $-8.196$ & $-8.151$ & $-8.152$ &  \\
   &  $P$  & 0.0009 & 0.0010  & 0.0010 &  \\
\hline
\hline
   &  $a$  & \underline{$-23.76$} & $-27.38$ & $-27.87$
& $-23.748 \pm 0.010$ \\
$np$ $\hbox{}^1S_0$ & $r$ & 2.584 & 2.528 & 2.525 & $2.75 \pm 0.05$ \\
   &  $P$  & 0.0393  & 0.0324  & 0.0324 &  \\
\hline
   &  $a$  & $-2.740$ & $-2.466$ & $-2.466$ &  \\
$np$ $\hbox{}^3P_0$ & $r$ & 3.867 & 3.929 & 3.929 &  \\
   &  $P$  & $-0.0140$  & $-0.0186$  & $-0.0186$ &  \\
\hline
   &  $a$  & 5.399 & 5.400 & 5.395 & $5.424 \pm 0.004$ \\
$np$ $\hbox{}^3S_1$ & $r$ & 1.730  & 1.730 & 1.730 & $1.759 \pm 0.005$ \\
   &  $P$  & $-0.010$ & $-0.0096$  & $-0.0097$ &  \\
\hline
   &  $a$  & 2.824 & 2.826 & 2.826 &  \\
$np$ $\hbox{}^1P_1$ & $r$ & $-6.294$ & $-6.299$ & $-6.299$ &  \\
   &  $P$  & $-0.0058$  & $-0.0058$  & $-0.0058$ &  \\
\hline
   &  $a$  & 1.740 & 1.582 & 1.582 &  \\
$np$ $\hbox{}^3P_1$ & $r$ & $-8.196$ & $-8.185$ & $-8.185$ &  \\
   &  $P$  & 0.0009 & 0.0004  & 0.0004 &  \\
\end{tabular}
\end{center}
\end{table}

\addtocounter{table}{-1}

\begin{table}[h]
\caption{-continued}
\label{table8-3}
\begin{center}
\renewcommand{\arraystretch}{1.4}
\setlength{\tabcolsep}{4mm}
\begin{tabular}{cccccc}
 & & isospin basis & \multicolumn{2}{c}{particle basis} & Expt. \\
\cline{4-5}
 & & & Coulomb off & Coulomb on & \\
\hline
 &  $a_s$  & $-2.51$ & $-2.48$ & $-2.27$ & $-2.42 \pm 0.30$ \\
$\Sigma^+ p$ & $r_s$ & 4.91 & 5.03 & 4.56 & $3.41 \pm 0.30$ \\
 &  $a_t$  & 0.727 & 0.727 & 0.834 & $0.709 \pm 0.001$ \\
 &  $r_t$  & $-1.20$ & $-1.29$ & $-2.82$ & $-0.783 \pm 0.003$ \\
\hline
 &  $a_s$  & $-2.58$ & $-2.58$ & $-2.59$ & $-1.8$, $-2.0$ \\
$\Lambda p$ & $r_s$  &    2.83 &    2.83 & 2.83 & $2.8$, $5.0$ \\
 &  $a_t$  & $-1.60$ & $-1.60$ & $-1.60$ & $-1.6$, $-2.2$ \\
 &  $r_t$  & 3.00    &    3.00 &    3.00 & $3.3$, $3.5$ \\
\hline
 &  $a_s$  & $-2.58$ & $-2.55$ & $-2.56$ & \\
$\Lambda n$ & $r_s$  &    2.83 &    2.84 & 2.86 & \\
 &  $a_t$  & $-1.60$ & $-1.57$ & $-1.57$ & \\
 &  $r_t$  & 3.00    &    3.04 &    3.03 & \\
\hline
 &  $a_s$  & 1.41 & 1.73 & 1.20 & \\
$\Sigma^- p$ & $r_s$  & $-11.0$ & $-26.6$ & $-15.0$ & \\
 &  $a_t$  &    1.33 &   0.802 & 0.914 & \\
 &  $r_t$  & $-39.0$ & $-4.90$ & $-22.7$ & \\
\end{tabular}
\end{center}
\end{table}

\begin{table}
\caption{A summary of the phase-shift behavior
in each channel related to the $\Lambda N$ - $\Sigma N(I=1/2)$
$\hbox{}^1P_1$ - $\hbox{}^3P_1$ channel coupling. 
``disp'' in the table denotes  
a dispersion-like resonance, while ``step''
denotes a step-like resonance.}
\label{reson}
\begin{center}
\renewcommand{\arraystretch}{1.4}
\begin{tabular}{ccccc}
       & RGM-F & FSS & RGM-H & fss2 \\
\hline 
$\Lambda N$ $\hbox{}^1P_1$ & step & step & disp & disp \\
$\Lambda N$ $\hbox{}^3P_1$ & disp & disp & disp & disp \\
$\Sigma N$ $\hbox{}^1P_1 $ & $\delta<0$  & $\delta<0$ &
$\delta \sim 0 \rightarrow 60^\circ$ & $\delta <0$ \\
$\Sigma N$ $\hbox{}^3P_1 $ & $\delta<0$  & $\delta<0$ &
 $\delta \sim 40^\circ$ & $\delta\sim 40 ^\circ$ \\   
%
\end{tabular}
\end{center}
\end{table}

\begin{table}[h]
\caption{Contributions to the $\Sigma^-p$ scattering total cross sections 
from each partial wave at $p_{\Sigma}=160$  MeV/$c$,
calculated in the isospin basis.
The results given by fss2 and FSS are listed in the unit 
of mb. A, B and C denote the scattering
processes $\Sigma^-p\rightarrow \Sigma ^-p$,
$\Sigma^-p\rightarrow \Sigma ^0n$
and $\Sigma^-p\rightarrow \Lambda n$, respectively.
In the process B, the factor $(k_f/k_i)$ is not included.
}
\label{tabcrs}
\newcommand{\lw}[1]{\smash{\lower2.ex\hbox{#1}}}
\begin{center}
\renewcommand{\arraystretch}{1.4}
\begin{tabular}{ccccccc}
                           & \multicolumn{2}{c}{A} 
                           & \multicolumn{2}{c}{B} 
                           & \multicolumn{2}{c}{C} \\
\cline{2-3}
\cline{4-5}
\cline{6-7}
                           & fss2 & FSS & fss2 & FSS & fss2 & FSS \\
\hline
 $^1S_0$ $\rightarrow$ $^1S_0$ & 7.3   & 6.8   & 28.3 & 26.5 & 3.5  & 3.4 \\ 
 $^3S_1$ $\rightarrow$ $^3S_1$ & 118.2 & 117.2 & 54.6 & 55.8 & 60.2 & 41.6\\
 $^3S_1$ $\rightarrow$ $^3D_1$ & 0.1   & 0.2   & 0.3  & 0.4  & 60.3 & 78.0\\
 $^3D_1$ $\rightarrow$ $^3S_1$ & 0.1   & 0.2   & 0.3  & 0.4  & 0.1  & 0.1 \\
 $^3P_0$ $\rightarrow$ $^3P_0$ & 0.5   & 0.6   & 1.3  & 1.4  & 0.2  & 0.4 \\
 $^1P_1$ $\rightarrow$ $^1P_1$ & 0.1   & 0.3   & 0.5  & 0.8  & 2.0  & 2.7 \\ 
 $^1P_1$ $\rightarrow$ $^3P_1$ & 0.1   & 0.3   & 0.1  & 0.2  & 0.5  & 2.4 \\ 
 $^3P_1$ $\rightarrow$ $^1P_1$ & 0.1   & 0.3   & 0.1  & 0.2  & 8.2  & 21.7\\
 $^3P_1$ $\rightarrow$ $^3P_1$ & 4.3   & 2.4   & 4.6  & 2.2  & 12.2 & 15.0\\
 $^3P_2$ $\rightarrow$ $^3P_2$ & 0.1   & 0.1   & 0.0  & 0.0  & 0.1  & 0.1 \\
 $^3P_2$ $\rightarrow$ $^3F_2$ & 0.0   & 0.0   & 0.0  & 0.0  & 3.3  & 4.0 \\ 
\hline
 total ($J\le 3$) & 131.2 & 128.5 & 90.2 & 87.9 & 151.1 & 169.9\\
\end{tabular}
\end{center}
\end{table}
%

\begin{table}
\caption{
Scattering length ($A$) and effective ranges ($R$) matrices of fss2
for the low-energy $\Sigma^- p$ scattering in $\hbox{}^1S_0$ state.
Unit of length is in fm. $A^c$ and $R^c$ imply the Coulomb case
in the particle basis.
}
\label{effect2}
\begin{center}
\renewcommand{\arraystretch}{1.4}
\setlength{\tabcolsep}{5mm}
\begin{tabular}{ccccc}
& $A$ & $R$ & $A^c$ & $R^c$ \\
\hline
$\Sp \rightarrow \Sp$ & $0.2644$  & 2.197  & $0.2238$  & 2.349    \\
$\Sp \rightarrow \Sn$ & $-1.287$  & 1.997  & $-1.227$  & 1.845    \\
$\Sp \rightarrow \Ln$ & $-0.1692$ & 7.231  & $-0.1650$ & 6.683    \\
$\Sn \rightarrow \Sn$ & $-0.9617$ & 2.916  & $-1.066$  & 3.416    \\   
$\Sn \rightarrow \Ln$ &  0.0887   &$-6.165$& $ 0.0755$ & $-5.016$ \\
$\Ln \rightarrow \Ln$ & $-0.0850$ & 11.39  & $-0.0867$ & 13.50    \\ 
\end{tabular}
\end{center}
\end{table}

\begin{table}
\caption{
Scattering length ($\CA$) and effective
ranges ($\CR$) matrices of the reduced expansion
for the low-energy $\Sigma^- p$ scattering
in the $\hbox{}^3S_1+ \hbox{}^3D_1$ state,
calculated by fss2 using the particle basis.
The $\hbox{}^3D_1$ channel of the $\Sigma^- p$ state
is eliminated. $\CA^c$ and $\CR^c$ imply
the Coulomb case in the particle basis.
Unit of length is in $\hbox{fm}^{\ell+\ell^\prime+1}$ in $\CA$,
$\CA^c$, and in $\hbox{fm}^{1-\ell-\ell^\prime}$ in $\CR$, $\CR^c$,
where $\ell$ and $\ell^\prime$ are the orbital angular momenta
of the initial and final channels, respectively.
}
\label{effect3}
\begin{center}
\renewcommand{\arraystretch}{1.4}
\setlength{\tabcolsep}{5mm}
\begin{tabular}{ccccc}
& $\CA$ & $\CR$ & $\CA^c$ & $\CR^c$ \\
\hline
$\Sp\,\TS1 \rightarrow \Sp\,\TS1$    &
$-14.91$    &  2.196    &   143.7    &  2.527   \\
$\Sp\,\TS1 \rightarrow \Sn\,\TS1$    &
11.25       & $-2.594$  &  $-112.7$  & $-3.274$ \\
$\Sp\,\TS1 \rightarrow \Sn\,\TD1$    &
$-6.500$    &  9.071    &  59.83     &  9.204   \\
$\Sp\,\TS1 \rightarrow \Ln\,\TS1$    &
5.589       &  22.30    &  $-56.37$  &  29.97   \\
$\Sp\,\TS1 \rightarrow \Ln\,\TD1$    &
$-2.742$    &  31.81    &  27.36     &  44.52   \\
$\Sn\,\TS1 \rightarrow \Sn\,\TS1$    &
$-7.395$    &  13.81    & 89.46      &  14.58   \\
$\Sn\,\TS1 \rightarrow \Sn\,\TD1$    &
3.953       &  13.92    &  $-47.82$  &  15.06   \\   
$\Sn\,\TS1 \rightarrow \Ln\,\TS1$    &
$-4.032$    &  $-153.4$ &  44.39     & $-164.7$ \\
$\Sn\,\TS1 \rightarrow \Ln\,\TD1$    &
1.979       & $-264.2$  & $-21.55$   & $-283.8$ \\
$\Sn\,\TD1 \rightarrow \Sn\,\TD1$    &
$-2.166$    &  28.12    &  25.50     &  31.76   \\   
$\Sn\,\TD1 \rightarrow \Ln\,\TS1$    &
1.990       & $-88.32$  &  $-23.89$  & $-102.8$ \\
$\Sn\,\TD1 \rightarrow \Ln\,\TD1$    &
$-0.9492$   & $-208.6$  &  11.63     & $-236.2$ \\
$\Ln\,\TS1 \rightarrow \Ln\,\TS1$    &
$-1.981$    &   1664    &  22.23     &   1821   \\ 
$\Ln\,\TS1 \rightarrow \Ln\,\TD1$    &
0.9603      &   2783    & $-10.80$   &   3055   \\
$\Ln\,\TD1 \rightarrow \Ln\,\TD1$    &
$-0.4598$   &   4681    &  5.253     &   5151   \\  
\end{tabular}
\end{center}
\end{table}

\begin{table}[h]
\caption{
$\Sp$ zero-energy total cross sections
$k_i \sigma_S$ ($\hbox{mb}\cdot \hbox{fm}^{-1}$)
for each spin-state ($S=0$ and 1) and $k_i \sigma_T$ for reactions
B: $\Sigma^- p \rightarrow \Sigma^0 n$
and C: $\Sigma^- p \rightarrow \Lambda n$.
The divergent factor $C_0$ is taken out in the Coulomb case.
The inelastic capture ratio $r_R$ at rest and $r_F$ in flight
are also given. Two versions of the quark model, FSS and fss2,
are used.}
\label{rest}
\begin{center}
\renewcommand{\arraystretch}{1.5}
\setlength{\tabcolsep}{3mm}
\begin{tabular}{ccccccc}
%
& \multicolumn{3}{c}{FSS without Coulomb}
& \multicolumn{3}{c}{fss2 without Coulomb} \\
\cline{2-4}
\cline{5-7}
& $\begin{array}{c} \hbox{B} \\ \end{array}$ &
$\begin{array}{c} \hbox{C} \\ \end{array}$ &
$r_S$, $r_F$ \quad $r_R$ &
$\begin{array}{c} \hbox{B} \\ \end{array}$ &
$\begin{array}{c} \hbox{C} \\ \end{array}$ &
$\begin{array}{c} r_S,~r_F \\ \end{array}$ \quad $r_R$ \\
\hline
$\begin{array}{c}
k_i \sigma_0 \\
k_i \sigma_1 \\
\end{array}$ &
$\begin{array}{c}
104  \\ 143 \\ \end{array}$ &
$\begin{array}{c}
4.4 \\ 340 \\ \end{array}$ &
$\left. \begin{array}{c}
0.959 \\ 0.296 \\ \end{array}\right\} 0.462$ &
$\begin{array}{c}
59.8 \\ 66.7 \\ \end{array}$ &
$\begin{array}{c}
5.98 \\ 165 \\ \end{array}$ &
$\left. \begin{array}{c}
0.909 \\ 0.289 \\ \end{array} \right\}0.444$ \\
\hline
$k_i \sigma_T$ &
$\begin{array}{c} 133 \\ \end{array}$ &
$\begin{array}{c} 256 \\ \end{array}$ &
$\begin{array}{c} 0.342 \\ \end{array}$ &
$\begin{array}{c} 65.0 \\ \end{array}$ &
$\begin{array}{c} 125 \\ \end{array}$ &
$\begin{array}{c} 0.342 \\ \end{array}$ \\
\hline
& \multicolumn{3}{c}{FSS with Coulomb}
& \multicolumn{3}{c}{fss2 with Coulomb} \\
\cline{2-4}
\cline{5-7}
 &
$\begin{array}{c} \hbox{B} \\ \end{array}$ &
$\begin{array}{c} \hbox{C} \\ \end{array}$ &
$r_S$, $r_F$ \quad $r_R$ &
$\begin{array}{c} \hbox{B} \\ \end{array}$ &
$\begin{array}{c} \hbox{C} \\ \end{array}$ &
$\begin{array}{c} r_S,~r_F \\ \end{array}$ \quad $r_R$ \\
\hline
$\begin{array}{c}
(C_0)^{-1} k_i \sigma_0 \\
(C_0)^{-1} k_i \sigma_1 \\
\end{array}$ &
$\begin{array}{c}
95 \\ 71 \\ \end{array}$ &
$\begin{array}{c}
4.4 \\ 168 \\ \end{array}$ &
$\left. \begin{array}{r}
0.955 \\ 0.296 \\ \end{array} \right\} 0.461$ &
$\begin{array}{c}
47.7 \\ 38.3 \\ \end{array}$ &
$\begin{array}{c}
4.98 \\ 94.6 \\ \end{array}$ &
$\left. \begin{array}{c}
0.905 \\ 0.288 \\ \end{array} \right\} 0.442$ \\
\hline
$(C_0)^{-1} k_i \sigma_T$ &
$\begin{array}{c}  77 \\ \end{array}$ &
$\begin{array}{c} 127 \\ \end{array}$ &
$\begin{array}{c} 0.377 \\ \end{array}$ &
$\begin{array}{c} 40.6 \\ \end{array}$ &
$\begin{array}{c} 72.2 \\ \end{array}$ &
$\begin{array}{c} 0.360 \\ \end{array}$
\end{tabular}
\end{center}
\end{table}

\begin{table}[h]
\caption{Corrections to the total cross sections in mb;
1) pot: The Coulomb and threshold corrections used for
the parameter search, 2) fss2: The pion-Coulomb correction
(with the correct threshold energies) calculated with $\Delta G$ term
included in the particle basis.}
\label{xdif}
\begin{center}
\renewcommand{\arraystretch}{1.4}
\setlength{\tabcolsep}{5mm}
\begin{tabular}{ccccccccc}
$ p_{\Sigma}$ & \multicolumn{2}{c}{$\Sigma^-p$~elastic}
& \multicolumn{2}{c}{$\Sigma^-p\rightarrow \Sigma ^0n$}
& \multicolumn{2}{c}{$\Sigma^-p\rightarrow \Lambda n$}
& \multicolumn{2}{c}{$\Sigma^+p$~elastic} \\
\hline
(MeV/$c$) & pot & fss2 & pot & fss2 & pot & fss2 & pot & fss2 \\
\hline
 110 &  1.0 & $-2$ & 47.9 & 48 & $-$26.8 & $-26$ & $-$8.2 & $-16$ \\
 120 &  0.5 & $-2$ & 37.8 & 38 & $-$21.2 & $-19$ & $-$8.0 & $-15$ \\
 130 &  0.3 & $-2$ & 30.3 & 31 & $-$17.1 & $-14$ & $-$7.3 & $-12$ \\
 140 &  0.2 & $-2$ & 24.8 & 26 & $-$14.0 & $-11$  & $-$6.6 & $-11$ \\
 150 &  0.2 & $-1$ & 20.5 & 22 & $-$11.7 & $-8$  & $-$5.9 & $-9$  \\
 160 &  0.2 & $-2$ & 17.2 & 19 & $-$9.9  & $-7$  & $-$5.4 & $-8$  \\
 170 &  0.2 & $-2$ & 14.5 & 16 & $-$8.5  & $-5$  & $-$5.0 & $-7$  \\
 180 &  0.1 & $-1$ & 12.4 & 13 & $-$7.4  & $-4$  & $-$4.7 & $-6$  \\
 190 &  0.1 & $-2$ & 10.8 & 12 & $-$6.5  & $-3$  & $-$4.4 & $-5$  \\
 200 &  0.0 & $-3$ &  9.4 & 10 & $-$5.7  & $-3$  & $-$4.3 & $-5$  \\
\end{tabular}
\end{center}
\end{table}

\begin{table}[h]
\caption{$\Lambda$ and $\Sigma$ s.p. potentials
in nuclear matter with $k_F = 1.35~\hbox{fm}^{-1}$,
calculated from our quark-model (fss2) $G$-matrices
in the continuous prescription for intermediate spectra.
Predictions of the Nijmegen soft-core potential
(NSC89) \protect\cite{NSC89} is also shown
for comparison \protect\cite{SCHU}.
}
\label{spcom}
\renewcommand{\arraystretch}{1.4}
\setlength{\tabcolsep}{3mm}
\begin{center}
\begin{tabular}{cccccccc}
 & \multicolumn{2}{c}{$U_\Lambda(0)$ \hspace{1em}[MeV]} &
 & \multicolumn{4}{c}{$U_\Sigma(0)$ \hspace{1em}[MeV]} \\
\cline{2-3} \cline{5-8}
 & fss2 & NSC89 & & \multicolumn{2}{c}{fss2}
 & \multicolumn{2}{c}{NSC89} \\ \hline
$I$ & 1/2 & 1/2 &  & 1/2 & 3/2 & 1/2 & 3/2 \\ \hline
$\hbox{}^1S_0$ & $-14.7$ & $-15.3$ &  & $6.7$
& $-9.2$ & $6.7$ & $-12.0$ \\
$\hbox{}^3S_1+\hbox{}^3D_1$ & $-28.4$ & $-13.0$ &  & $-24.0$ & $41.2$
 & $-14.9$ & $6.7$\\
$\hbox{}^1P_1+\hbox{}^3P_1$ & $2.2$ & $3.6$ &  & $-6.4$ & $3.3$
 & $-3.5$ & $3.9$ \\
$\hbox{}^3P_0$ & $-0.4$ & $0.2$ &  & $2.9$ & $-2.2$ & $2.6$
& $-2.0$ \\
$\hbox{}^3P_2+\hbox{}^3F_2$ & $-5.7$ & $-4.0$  &  & $-1.6$ & $-2.5$
& $-0.5$ & $-1.9$\\ \hline
 subtotal  &   &  & & $-23.7$ & $31.3$ & $-9.8$ & $-5.5$ \\
total & $-47.9$ & $-29.8$ & & \multicolumn{2}{c}{$+7.6$}
 & \multicolumn{2}{c}{$-15.3$} \\
\end{tabular}
\end{center}
\end{table}
\begin{table}[h]
\caption{The nuclear-matter density dependence of the 
Scheerbaum factors $S_B$ for $N$,
$\Lambda$ and $\Sigma$, predicted by the $G$-matrices
of fss2 in the continuous prescription.
The unit is $\hbox{MeV}\cdot \hbox{fm}^5$.
}
\label{sbfac}
\bigskip
\renewcommand{\arraystretch}{1.4}
\setlength{\tabcolsep}{4mm}
\begin{center}
\begin{tabular}{@{}cc@{\extracolsep{\fill}}cccc}
$k_F$ (fm$^{-1}$) & & $1.07$ & $1.20$ & $1.35$ & ratio \\
\hline
      $N$ &  & $-44.5$ & $-43.6$ & $-42.4$ & ( 1 ) \\
$\Lambda$ &  & $-11.2$  & $-11.2$  & $-11.1$  & ($\sim {1 \over 4}$) \\
$\Sigma$  &  & $-22.6$ & $-22.9$ & $-23.3$ & ($\sim {1 \over 2}$) \\
\end{tabular}
\end{center}
\end{table}
\begin{table}[h]
\caption{Decomposition
of $S_{\Lambda} = -11.1~\hbox{MeV} \cdot \hbox{fm}^5$ and $S_{\Sigma}
= -23.3~\hbox{MeV} \cdot \hbox{fm}^5$ at $k_F = 1.35~\hbox{fm}^{-1}$
into various contributions. The model is fss2.
Contributions in FSS are also shown in parentheses
for comparison.
The unit is $\hbox{MeV} \cdot \hbox{fm}^5$.}
\label{lscom}
\begin{center}
\setlength{\tabcolsep}{4mm}
\renewcommand{\arraystretch}{1.4}
\begin{tabular}{@{}c@{\extracolsep{\fill}}cccccc}
 &  & \multicolumn{2}{c}{$I=1/2$} & &
  \multicolumn{2}{c}{$I=3/2$}\\
 \cline{3-4} \cline{6-7}
 &  & odd & even &
 & odd & even \\ \hline
 $S_{\Lambda}$ & $LS$    & $-19.1~(-17.1)$ & $-0.2~(0.6)$ & & --- & --- \\
 & $LS^{(-)}$  & $7.9~(12.7)$  & $0.3~(0.3)$   & & --- & --- \\ \hline
 $S_{\Sigma}$  & $LS$    & $1.3~(2.7)$  & $-0.3~(0.1)$ &
 & $-12.4~(-11.9)$ & $-1.5~(-1.5)$ \\
 & $LS^{(-)}$  & $-9.6~(-10.5)$ & $-0.4~(-0.6)$  &
 & $-0.4~(0.1)$ & $-0.1~(-0.0)$ \\
\end{tabular}
\end{center}
\end{table}

\begin{table}[h]
\caption{
The spatial part of the exchange Born amplitudes defined by
Eq.\,(A.4) of \protect\cite{LSRGM}.
The polynomial part $\widetilde{u}({\protect\mbf k},
{\protect\mbf q})$ of
the two-body force in \protect\eq{a3} is also shown.
The coefficients $\alpha$, $\epsilon$, $\Delta=2pq/(1-\tau^2)$,
and the vectors ${\protect\mbf \rho}_{\CT}
=({\protect\mbf V}/\protect\sqrt{2\mu}b)$,
${\protect\mbf \sigma}_{\CT}=({\protect\mbf A}/\protect\sqrt{2\mu}b)$ are
calculated from Eq.\,(A.14) of \protect\cite{LSRGM}
by setting $x=1$ and $\mu=3/2$ for each interaction
type $\protect\CT$.
The factor $\Delta$ is non-zero only
for the $\protect\CT=D_\pm$ types,
and $\epsilon \neq 0$ only for the $\protect\CT=S,~S^\prime$ types.  
The basic spatial functions $f^\Omega_{\protect\CT}
(\theta)$ with $\protect\CT=C,~CD,~LS,~TD$ are
defined by \protect\eq{a10}.
}
\label{space}
\begin{center}
\setlength{\tabcolsep}{4mm}
\renewcommand{\arraystretch}{1.6}
\begin{tabular}{ccc}
$\Omega$ & $\widetilde{u}({\mbf k}, {\mbf q})$ &
$M^{\Omega}_{1\CT}({\mbf q}_f, {\mbf q}_i)$ \\
\hline
$C$ & 1 & $f^C_\CT(\theta)$ \\
$SS$ & ${\mbf k}^2$ & $-m^2\,f^{CD}_\CT(\theta)$ \\
$C(1)$ & ${\mbf q}^2$ & ${3 \over 4b^2}\left(1-{\alpha \over 2\mu}
+{1 \over 3}b^2{{\mbf \sigma}_\CT}^2 \right)\,f^{C}_\CT(\theta)
-m^2\left({\epsilon \over 4\mu}\right)^2\,f^{CD}_\CT(\theta)
-m^2{\epsilon \over 4\mu}b^2({\mbf \rho}_\CT \cdot {\mbf \sigma}_\CT)
\,f^{LS}_\CT(\theta)$ \\
$SS(1)$ & ${\mbf n^2}$ & $-{m^2 \over 2b^2}\left(1-{\alpha \over 2\mu}\right)
\,f^{CD}_\CT(\theta)+{m^2 \over 2}{{\mbf \sigma}_\CT}^2
\,f^{LS}_\CT(\theta)
-\left({\Delta \over 2\mu^2}\right)^2{\mbf n^2}\,f^{TD}_\CT(\theta)$ \\
$T$ & $\CY_{2\mu}({\mbf k})$ & $-f^{TD}_\CT(\theta)
~\CY_{2\mu}({\mbf \rho}_\CT)$ \\
$QLS$ & $\CY_{2\mu}({\mbf n})$ & $-{m^2 \over 4}\,f^{LS}_\CT(\theta)
~\CY_{2\mu}({\mbf \sigma}_\CT)
+{1 \over 4b^2}\left(1-{\alpha \over 2\mu}\right)\,f^{TD}_\CT(\theta)
~\CY_{2\mu}({\mbf \rho}_\CT)
-\left({\Delta \over 2\mu^2}\right)^2\,f^{TD}_\CT(\theta)
~\CY_{2\mu}({\mbf n})$ \\
$LS$ & $i{\mbf n}$ & $\left({mb \over \mu}\right)^2{\Delta \over 2}
\,f^{LS}_\CT(\theta)~i{\mbf n}$ \\
\end{tabular}
\end{center}
\end{table}
\begin{table}[h]
\caption{
The coefficients $\alpha^\Omega$ and the correspondence
among $\Omega$, $\Omega^\prime$, $\Omega^{\prime \prime}$ in the
two-bodey force \protect\eq{a1}. The column $\beta$ implies
the meson types and $\gamma=(m/2m_{ud})$.
}
\label{corr}
\begin{center}
\setlength{\tabcolsep}{4mm}
\renewcommand{\arraystretch}{1.6}
\begin{tabular}{ccccc}
$\beta$ & $\Omega$ & $\alpha^\Omega$ & $w^{\Omega^\prime}$
& $w^{\Omega^{\prime \prime}}$ \\
\hline
   & $C$     & $-g^2$ & $w^C$ & $u^C$ \\
   & $C(1)$  & $g^2{2\gamma^2 \over m^2}$ & $w^C$ & $u^{C(1)}$ \\
S  & $SS(1)$ & $g^2{\gamma^4 \over 3m^4}$ & $w^{SS}$ & $u^{SS(1)}$ \\
   & $QLS$   & $g^2{\gamma^4 \over 3m^4}$ & $w^{T}$ & $u^{QLS}$ \\
   & $LS$    & $-g^2{2\gamma^2 \over m^2}$ & $w^{LS}$ & $u^{LS}$ \\
\hline
PS & $SS$   & $-f^2{1 \over 3m^2_{\pi^+}}$ & $w^{SS}$ & $u^{SS}$ \\
   & $T$    & $-f^2{1 \over 3m^2_{\pi^+}}$ & $w^{T}$ & $u^{T}$ \\
\hline
   & $C$     & ${f_e}^2$ & $w^C$ & $u^C$ \\
   & $C(1)$  & ${f_e}^2{6\gamma^2 \over m^2}$   & $w^C$    & $u^{C(1)}$ \\
   & $SS$    & $-{f_m}^2{2 \over 3m^2}$         & $w^{SS}$ & $u^{SS}$ \\
V  & $SS(1)$ & $-{f_m}^2{8\gamma^2 \over 3m^4}$ & $w^{SS}$ & $u^{SS(1)}$ \\
   & $T$     & ${f_m}^2{1 \over 3m^2}$          & $w^{T}$  & $u^{T}$ \\
   & $QLS$   & $-{f_m}^2{8\gamma^2 \over 3m^4}$ & $w^{T}$  & $u^{QLS}$ \\
   & $LS$    & $-f_m f_e{8\gamma \over m^2}$    & $w^{LS}$ & $u^{LS}$ \\
\end{tabular}
\end{center}
\end{table}
\begin{table}[h]
\caption{
The coefficients $C_j$ and $D_j$ in \protect\eq{c6} for
the parameterized deuteron wave functions.
The model is fss2, calculated in the particle basis
with the full Coulomb exchange kernel. The number
of parameters is $n=11$, but the last $C_j$ and
the last three $D_j$ (the parenthesized values) should be
calculated from Eqs.\,(C.7) and (C.8) of \protect\cite{MA89}.
}
\label{deutw}
\begin{center}
\setlength{\tabcolsep}{4mm}
\renewcommand{\arraystretch}{1.4}
\begin{tabular}{ccrr}
$j$ & $\gamma_j$ ($\hbox{fm}^{-1}$) & $C_j$ ($\hbox{fm}^{-1/2}$)
& $D_j$ ($\hbox{fm}^{-1/2}$) \\
\hline
 1 &   0.23186542   &   $0.88177292969E+00$  &   $0.22317366018E-01$  \\
 2 &   1.13186542   &  $-0.22759285797E+00$  &  $-0.47989721024E+00$  \\
 3 &   2.03186542   &  $-0.87378082999E-01$  &   $0.70358390560E+00$  \\
 4 &   2.93186542   &  $-0.19214145234E+02$  &  $-0.19602848976E+02$  \\
 5 &   3.83186542   &   $0.19019661123E+03$  &   $0.16245688580E+03$  \\
 6 &   4.73186542   &  $-0.10079545619E+04$  &  $-0.75342203360E+03$  \\
 7 &   5.63186542   &   $0.28344069046E+04$  &   $0.19989675989E+04$  \\
 8 &   6.53186542   &  $-0.44819643416E+04$  &  $-0.30666624647E+04$  \\
 9 &   7.43186542   &   $0.40462956321E+04$  &  ($0.27047041824E+04$) \\
10 &   8.33186542   &  $-0.19571100406E+04$  & ($-0.12779605335E+04$) \\
11 &   9.23186542   &  ($0.39477713946E+03$) &  ($0.25127320956E+03$) \\
\end{tabular}
\end{center}
\end{table}

\bigskip


\begin{figure}
\begin{center}
\setlength{\unitlength}{0.8mm}
\begin{picture}(140,140)(0,0)
\thicklines
\put(0,40){\line(1,0){40}}
\put(50,40){\line(1,0){40}}
\put(100,40){\line(1,0){40}}
\put(0,114.98){\line(1,0){40}}
\put(0,116.959){\line(1,0){40}}
\put(50,117.442){\line(1,0){40}}
\put(100,116.959){\line(1,0){40}}
\put(100,120.472){\line(1,0){40}}
\put(0,42){\shortstack{$\Lambda p$}}
\put(35,42){\shortstack{0}}
\put(50,42){\shortstack{$\Lambda N$}}
\put(85,42){\shortstack{0}}
\put(100,42){\shortstack{$\Lambda n$}}
\put(135,42){\shortstack{0}}
\put(100,32){\shortstack{$(-80.473)$}}
\put(10,20){\shortstack{$Q$=1}}
\put(50,20){\shortstack{isospin basis}}
\put(110,20){\shortstack{$Q$=0}}
\put(0,105){\shortstack{$\Sigma^+ n$~~74.980}} 
\put(0,120){\shortstack{$\Sigma^0 p$~~~76.959}} 
\put(50,120){\shortstack{$\Sigma N$~~77.471}} 
\put(100,100){\shortstack{$(-3.514)$}}
\put(100,107){\shortstack{$\Sigma^0 n$~~76.959}} 
\put(100,123){\shortstack{$\Sigma^- p$~~80.473}} 
\end{picture}
\end{center}
\bigskip
\caption{
Threshold relations of the $\Lambda N$ - $\Sigma N$ coupled-channel
system in the non-relativistic kinematics.}
\label{thres}
\end{figure}

\clearpage

\ \ 
\vspace{-28mm}

\begin{figure}
\begin{minipage}{0.35\textwidth}
\epsfxsize=\textwidth
\epsffile{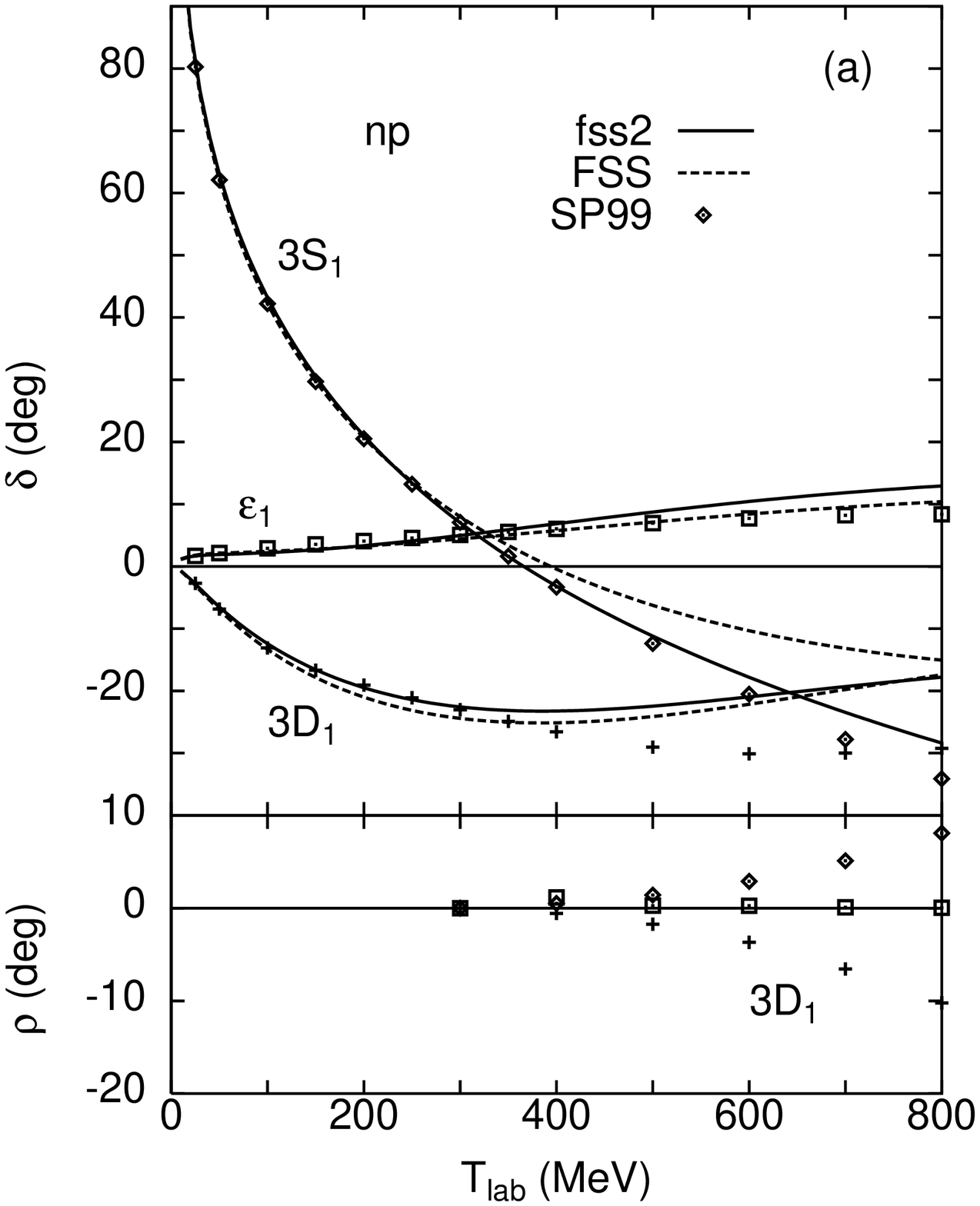}
\end{minipage}~%
\hspace{-20mm}
\hfill~%
\begin{minipage}{0.35\textwidth}
\epsfxsize=\textwidth
\epsffile{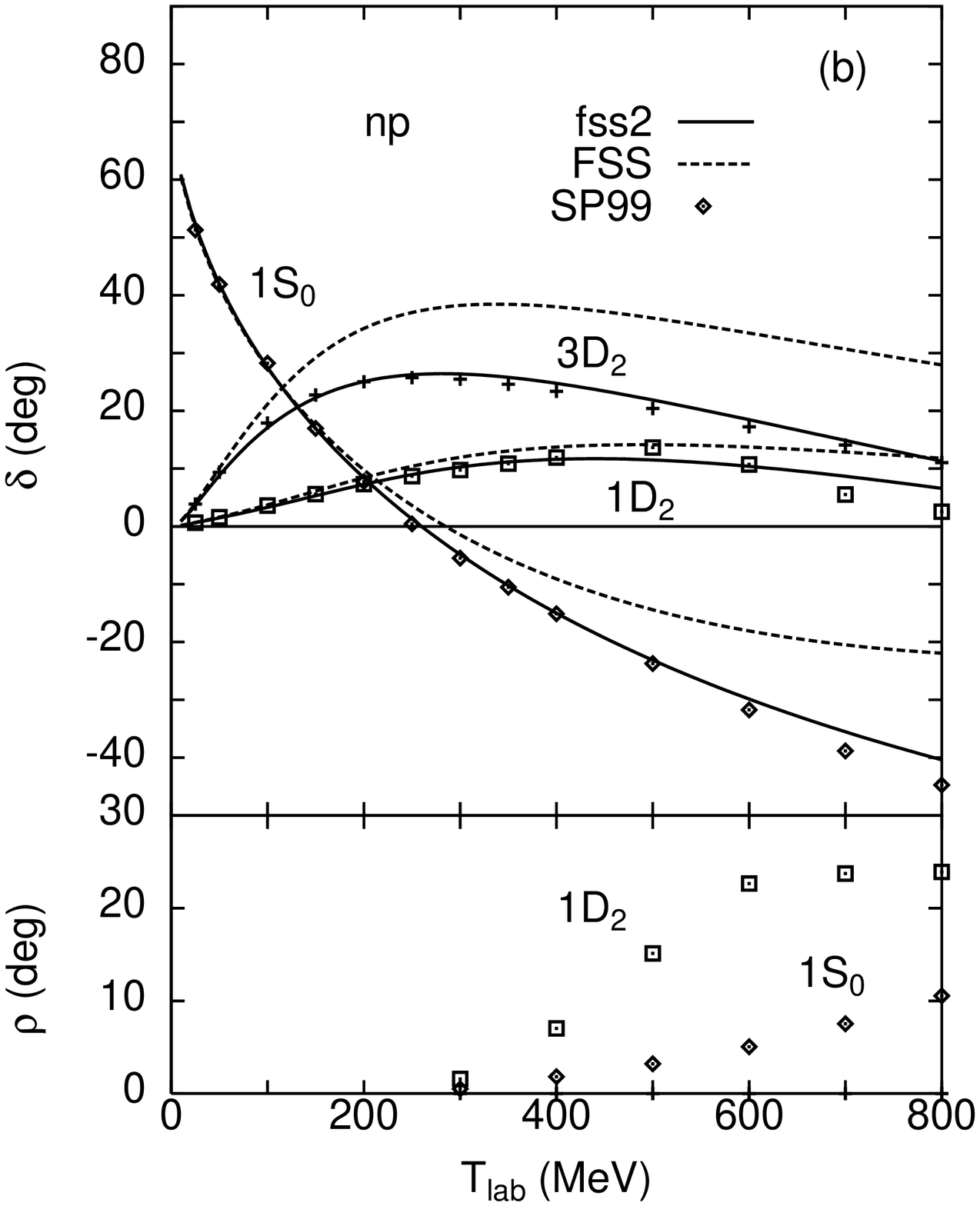}
\end{minipage}
\hspace{-20mm}
\hfill~%
\begin{minipage}{0.35\textwidth}
\epsfxsize=\textwidth
\epsffile{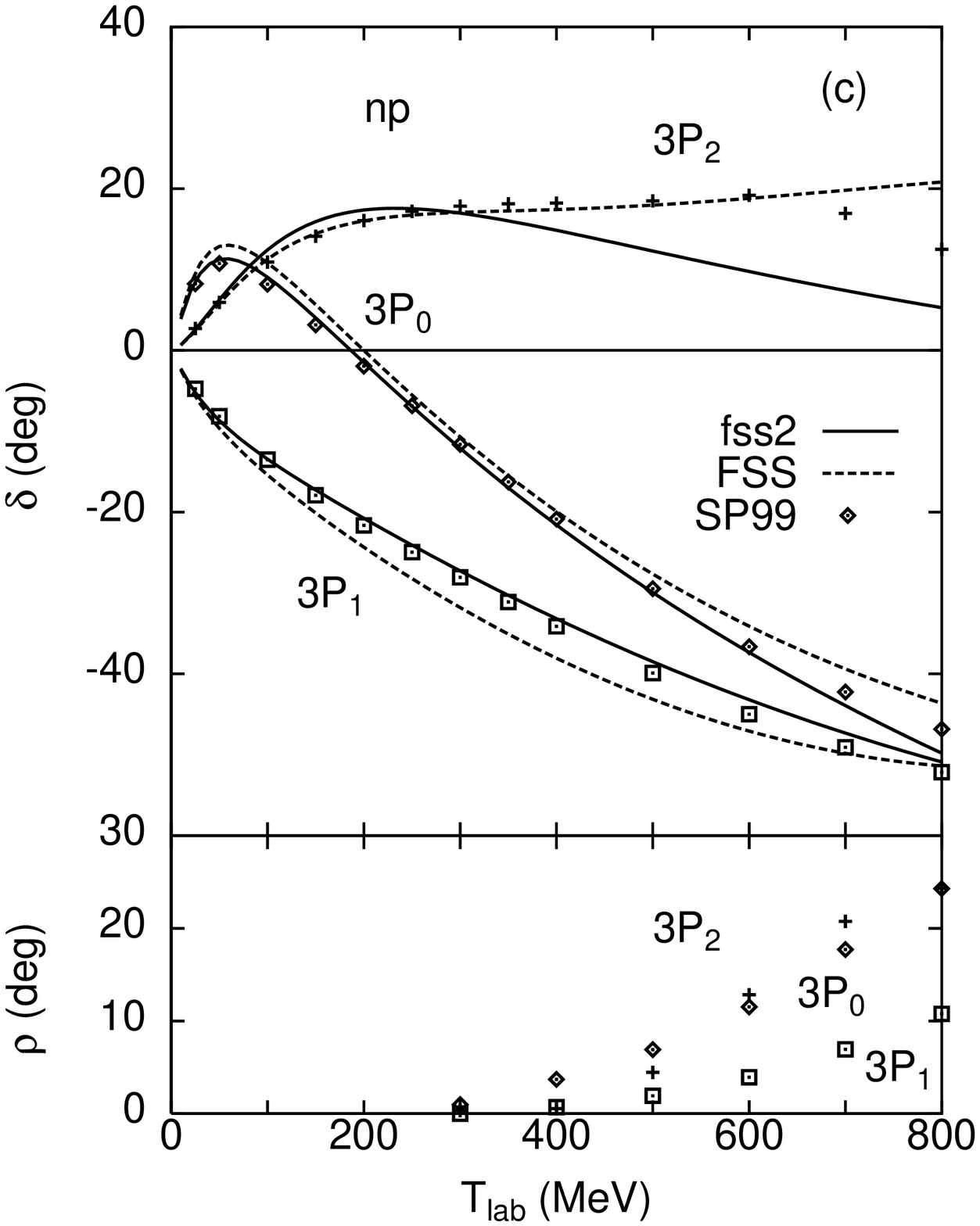}
\end{minipage}~%
\end{figure}

\vspace{-28mm}

\begin{figure}[h]
\begin{minipage}{0.35\textwidth}
\epsfxsize=\textwidth
\epsffile{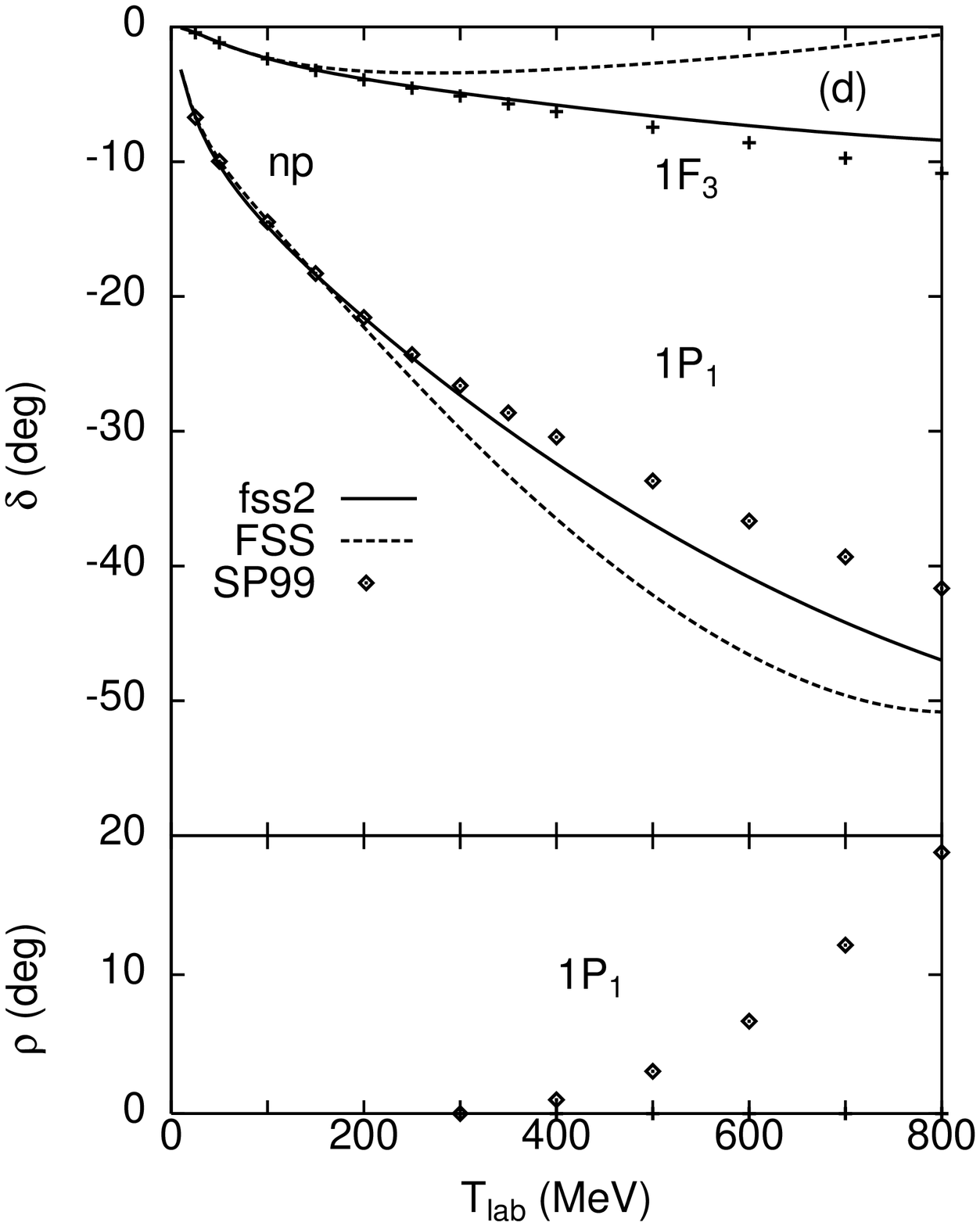}
\end{minipage}
\hspace{-20mm}
\hfill~%
\begin{minipage}{0.35\textwidth}
\epsfxsize=\textwidth
\epsffile{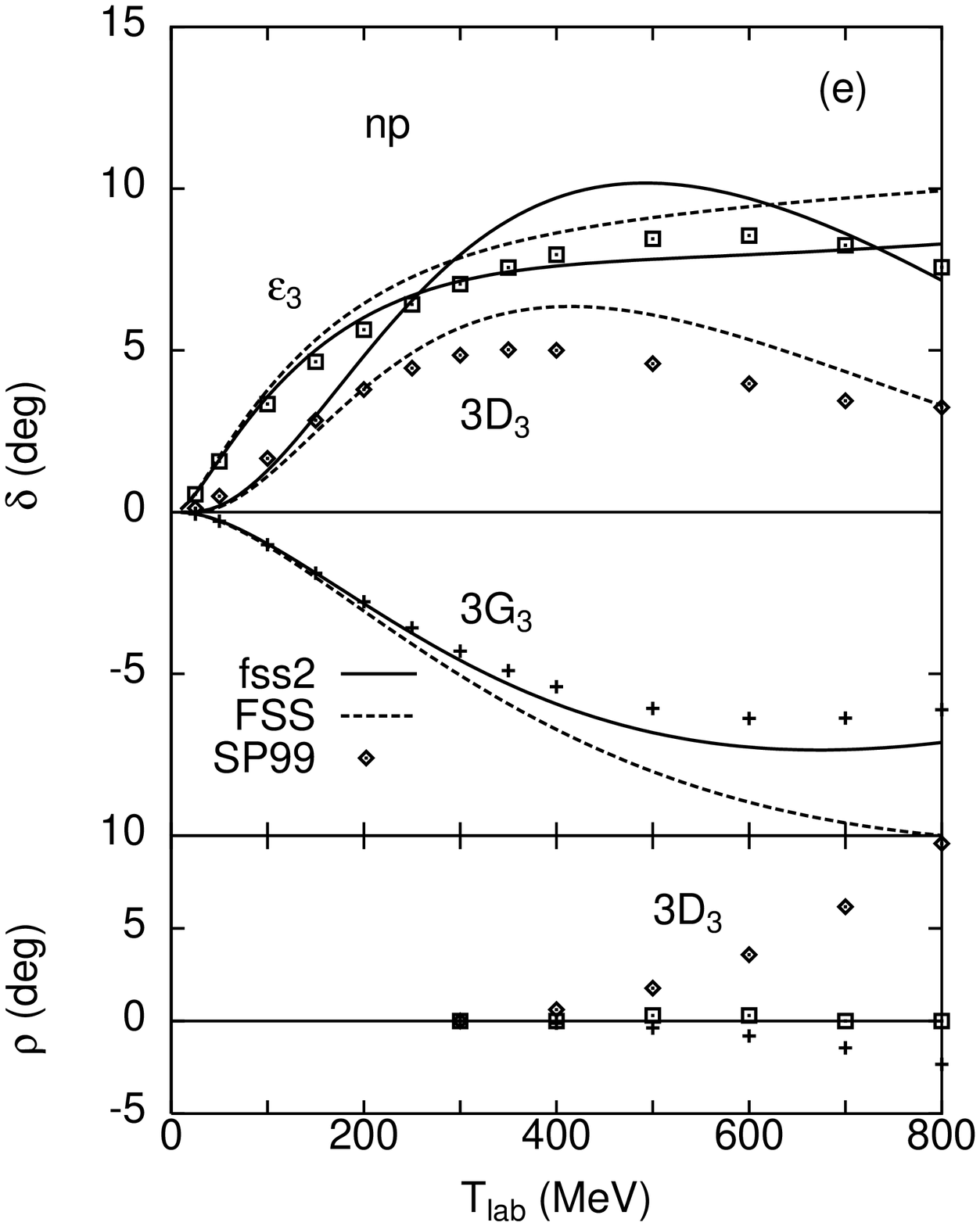}
\end{minipage}~%
\hspace{-20mm}
\hfill~%
\begin{minipage}{0.35\textwidth}
\epsfxsize=\textwidth
\epsffile{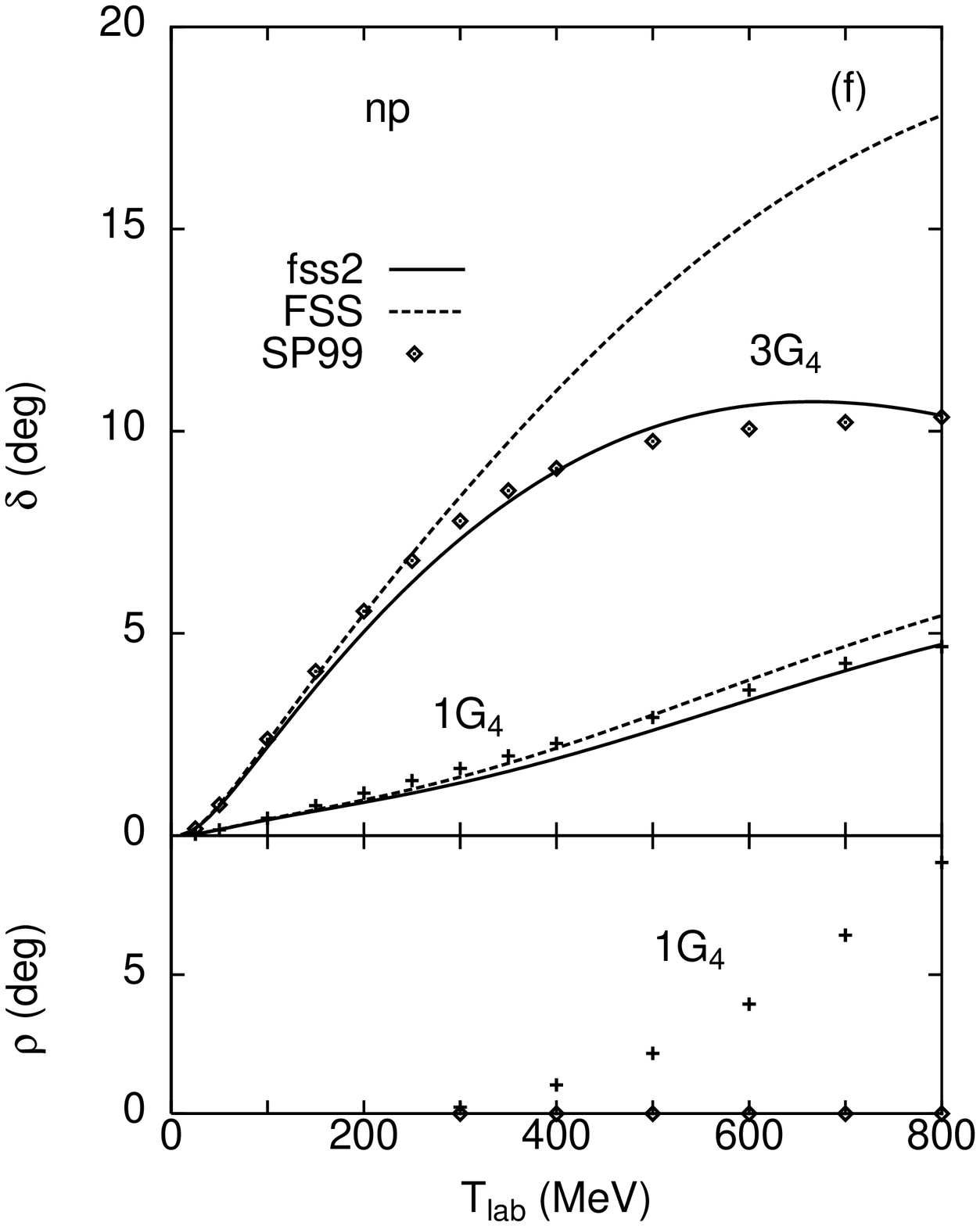}
\end{minipage}
\end{figure}

\vspace{-28mm}

\begin{figure}[h]
\begin{minipage}{0.35\textwidth}
\epsfxsize=\textwidth
\epsffile{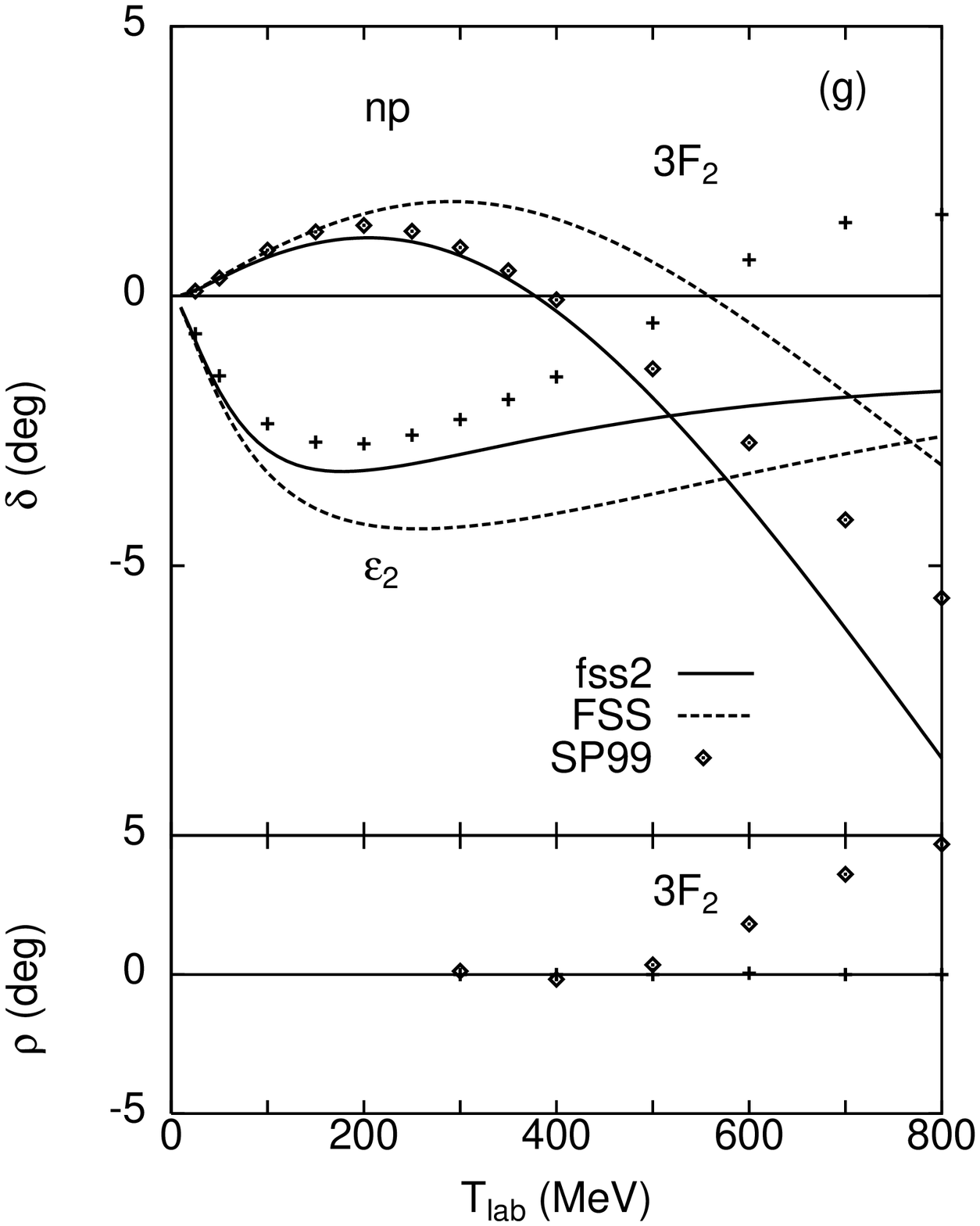}
\end{minipage}~%
\hspace{-20mm}
\hfill~%
\begin{minipage}{0.35\textwidth}
\epsfxsize=\textwidth
\epsffile{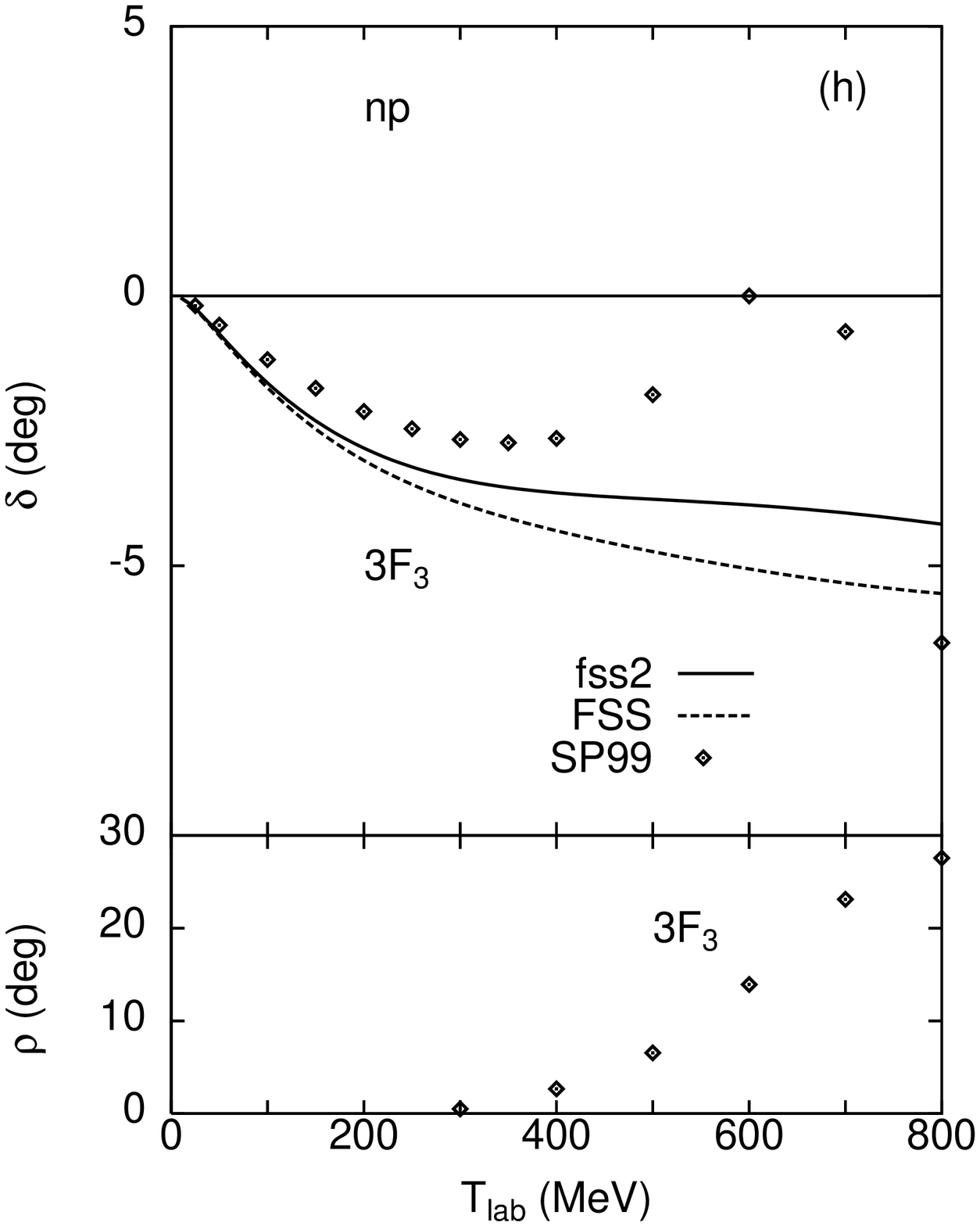}
\end{minipage}
\hspace{-20mm}
\hfill~%
\begin{minipage}{0.35\textwidth}
\epsfxsize=\textwidth
\epsffile{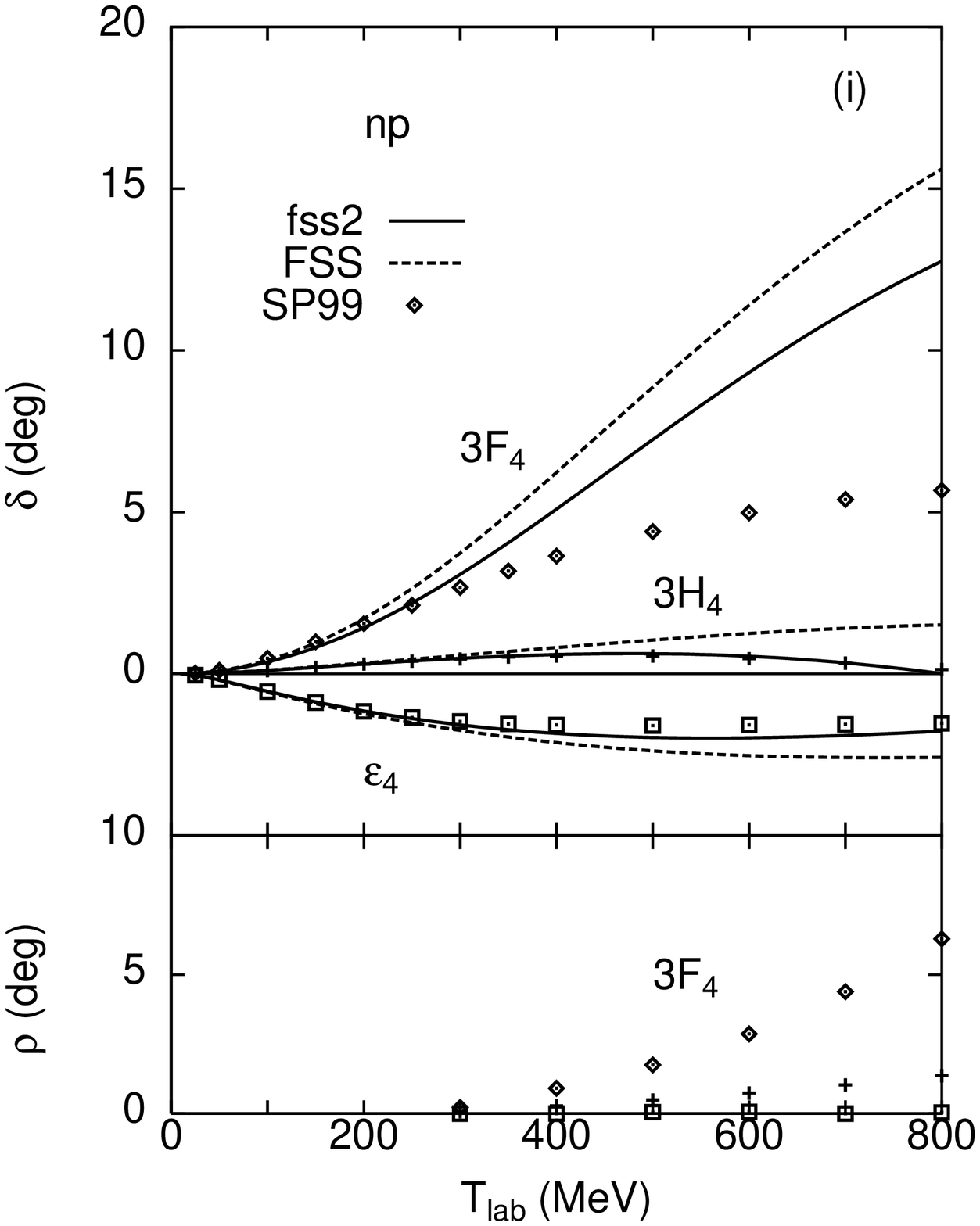}
\end{minipage}~%
\caption{Calculated $np$ phase shifts by fss2
in the isospin basis, compared with the
phase-shift analysis SP99 by Arndt {\em et al.} \protect\cite{SAID}
The dotted curves indicate the result given by FSS.
Some empirical inelasticity parameters $\rho$ of SP99
are also shown for $T_{\rm lab} \geq 300$ MeV, in order to
give a measure of possible deviations
of the phase-shift values in the single-channel calculation.
}
\label{npphase}
\end{figure}

\begin{figure}
\begin{minipage}{0.47\textwidth}
\epsfxsize=\textwidth
\epsffile{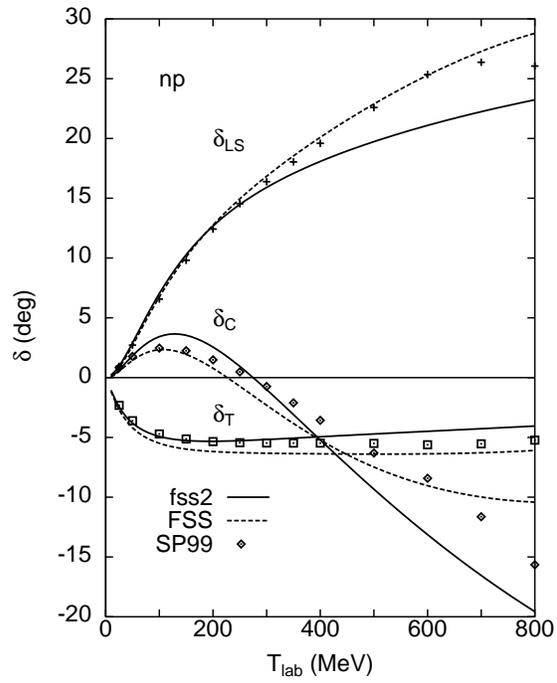}
\end{minipage}~%
\bigskip
\caption{Decomposition of the $\hbox{}^3P_J$ phase shifts
for the $np$ scattering
to the central ($\delta_C$), $LS$ ($\delta_{LS}$)
and tensor ($\delta_T$) components.
The results given by fss2 (solid curves) and FSS (dashed curves)
are compared with the decomposition of the empirical
phase shifts SP99 \protect\cite{SAID}.
}
\label{phcom}
\end{figure}

\clearpage

\begin{figure}[h]
\begin{minipage}{0.47\textwidth}
\epsfxsize=\textwidth
\epsffile{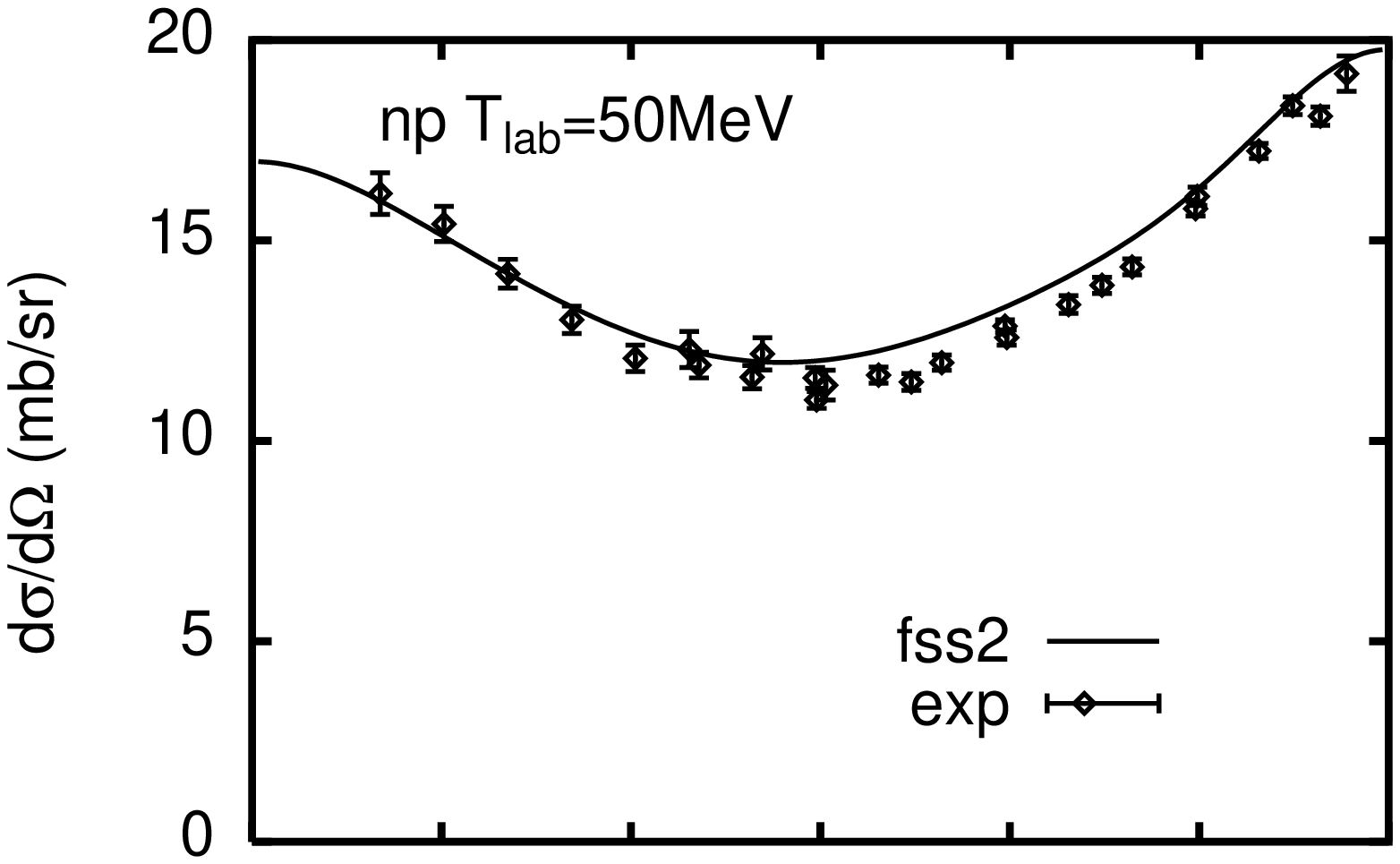}
\end{minipage}~%
\hspace{-29.49mm}
\begin{minipage}{0.47\textwidth}
\epsfxsize=\textwidth
\epsffile{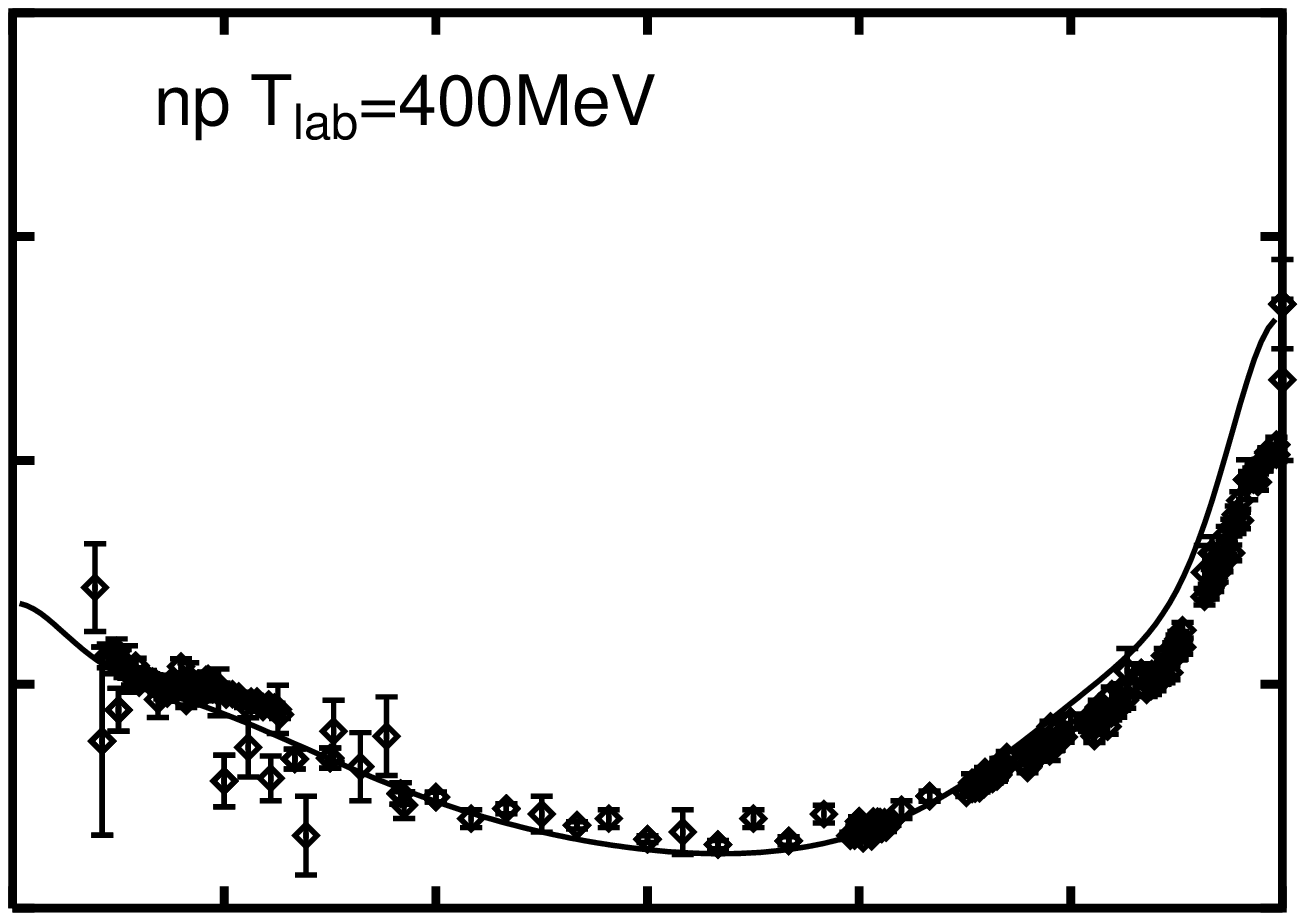}
\end{minipage}
\end{figure}

\vspace{-27.52mm}

\begin{figure}[h]
\begin{minipage}{0.47\textwidth}
\epsfxsize=\textwidth
\epsffile{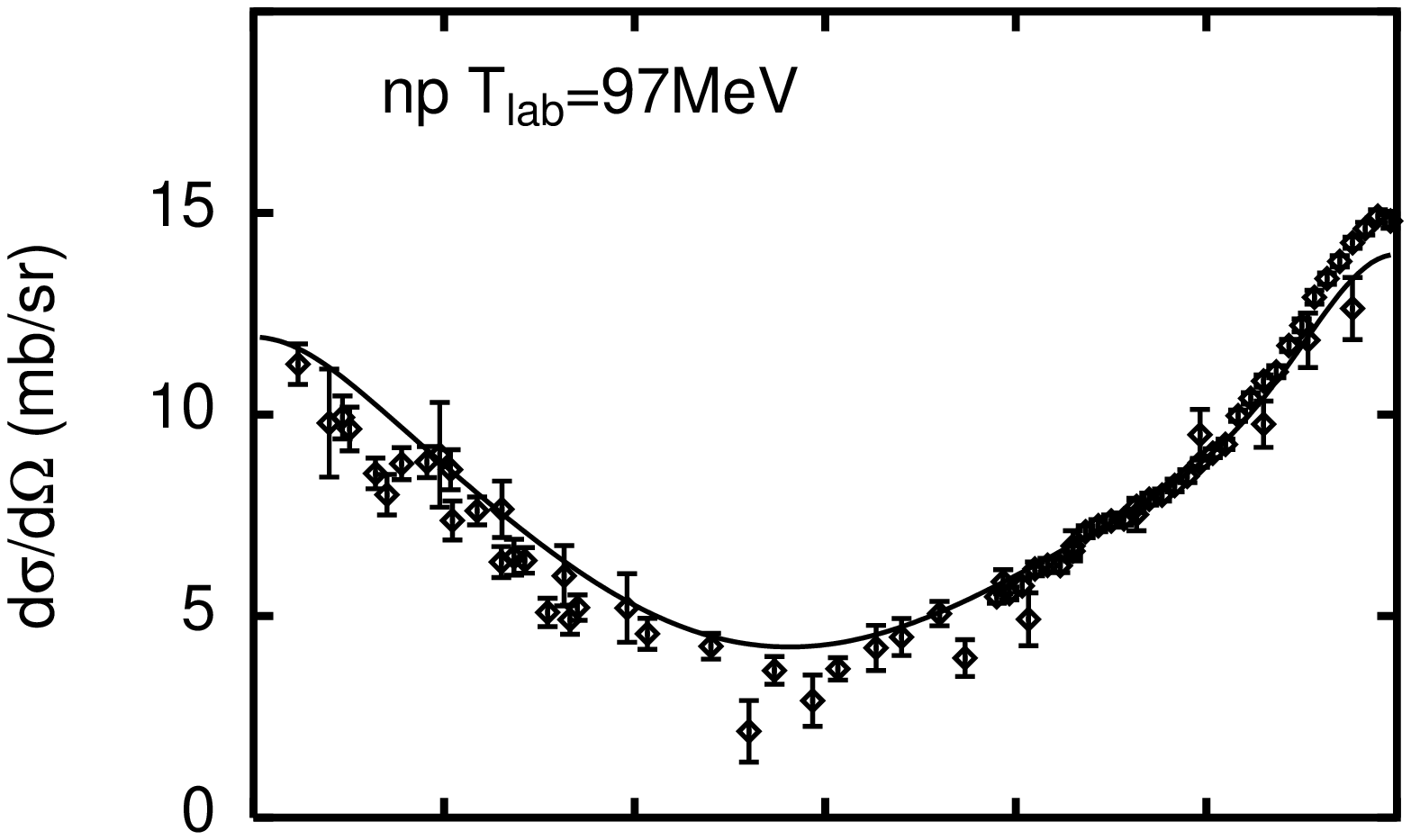}
\end{minipage}~%
\hspace{-29.49mm}
\begin{minipage}{0.47\textwidth}
\epsfxsize=\textwidth
\epsffile{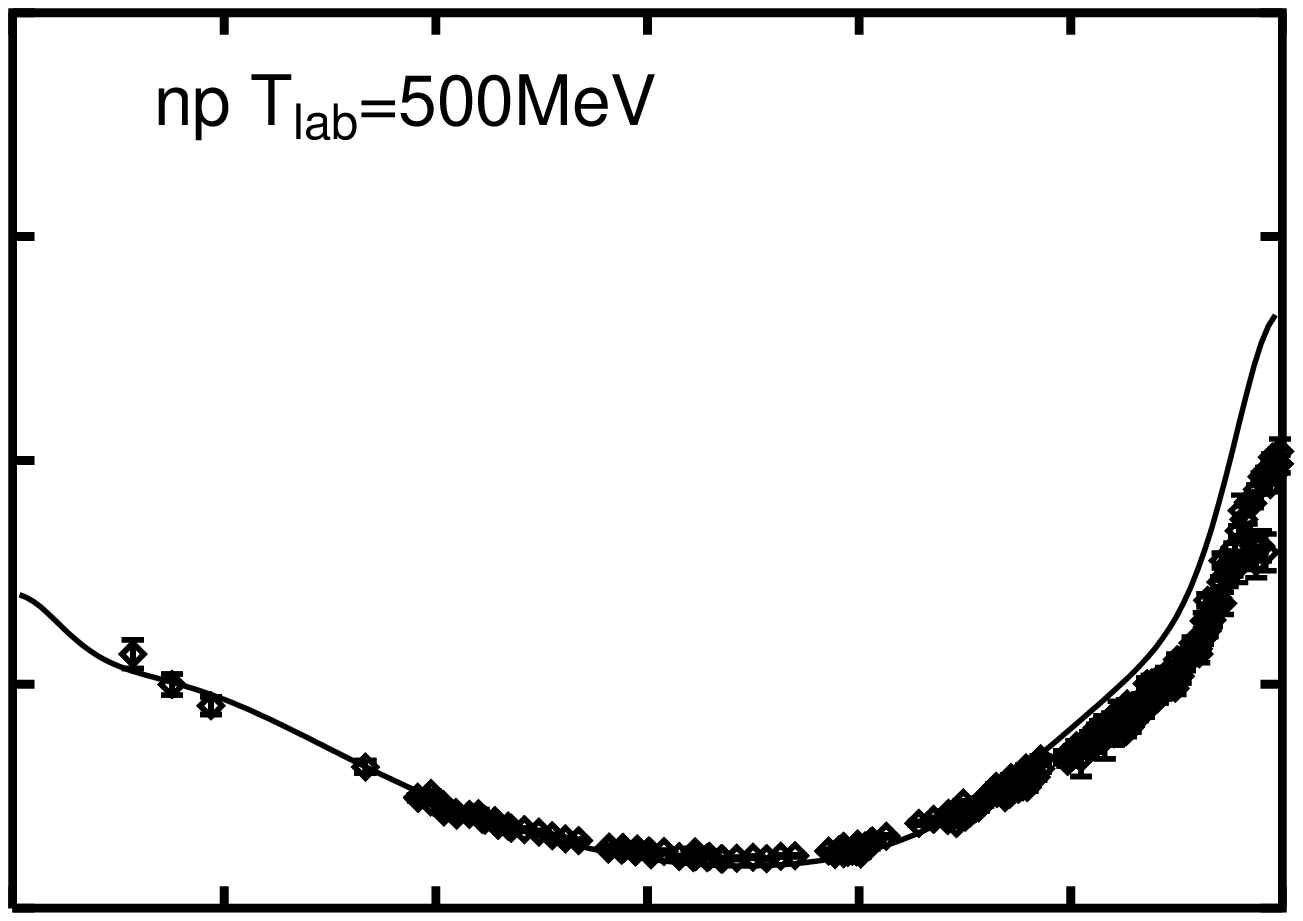}
\end{minipage}
\end{figure}

\vspace{-27.52mm}


\begin{figure}[h]
\begin{minipage}{0.47\textwidth}
\epsfxsize=\textwidth
\epsffile{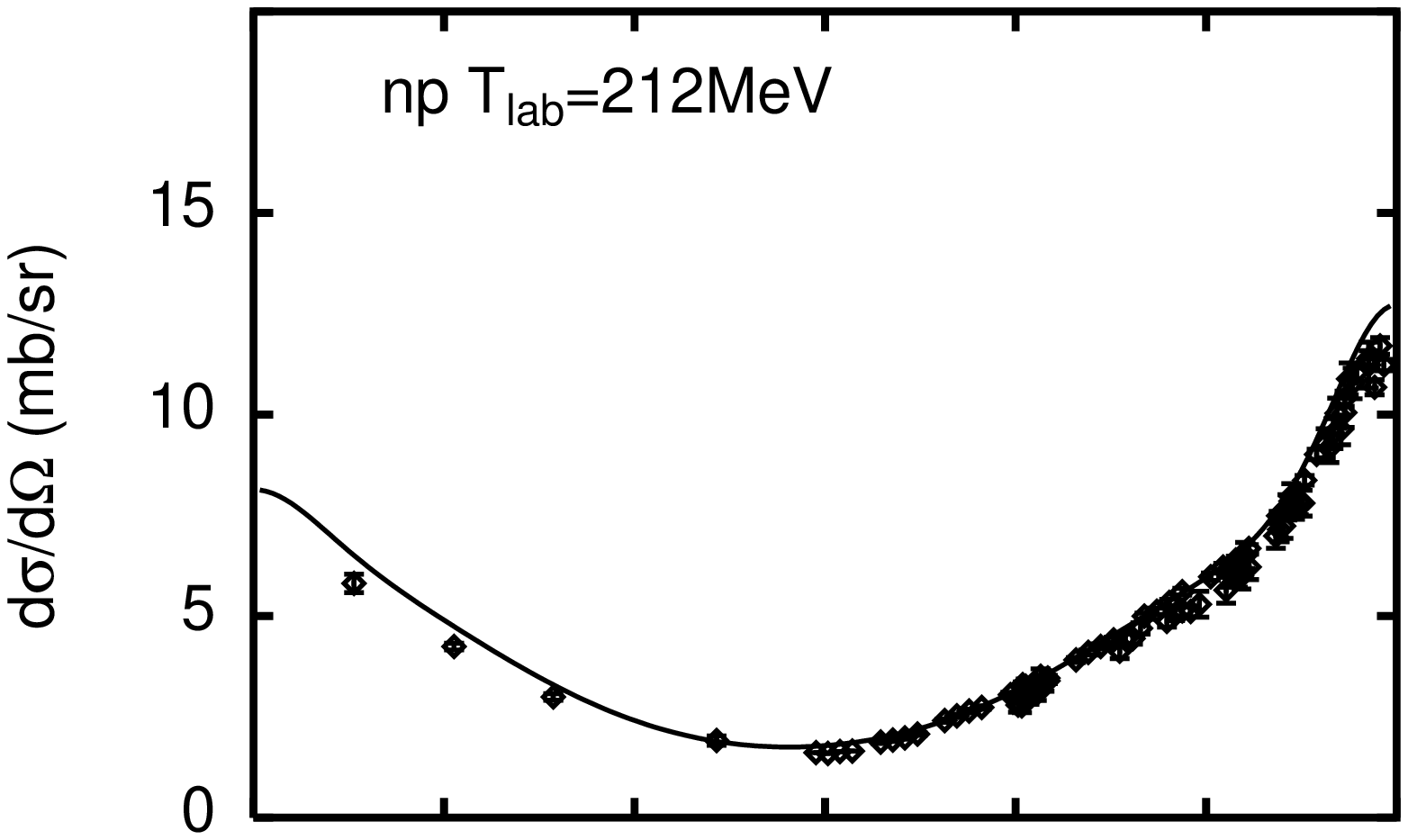}
\end{minipage}~%
\hspace{-29.49mm}
\begin{minipage}{0.47\textwidth}
\epsfxsize=\textwidth
\epsffile{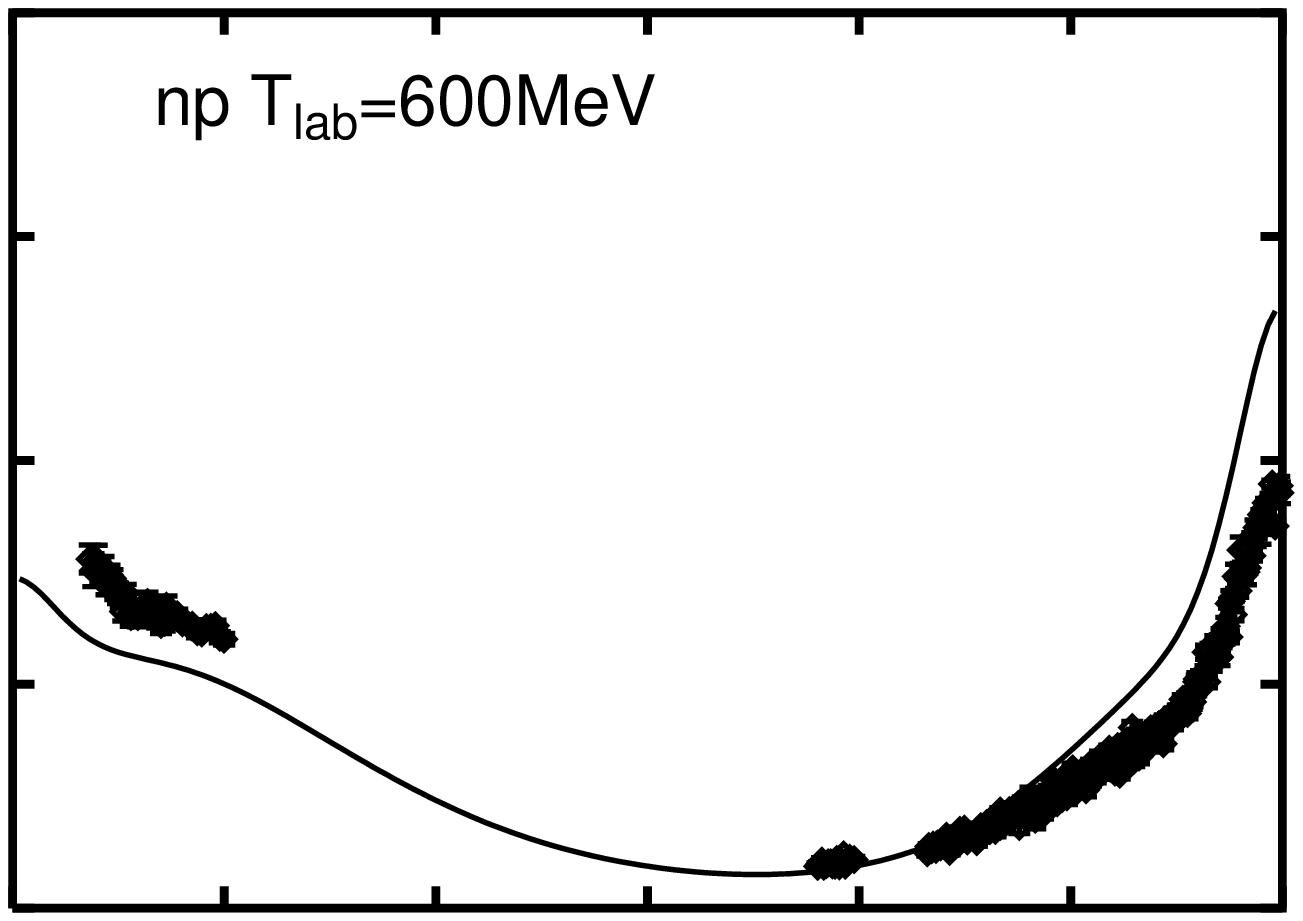}
\end{minipage}
\end{figure}

\vspace{-27.52mm}


\begin{figure}[h]
\begin{minipage}{0.47\textwidth}
\epsfxsize=\textwidth
\epsffile{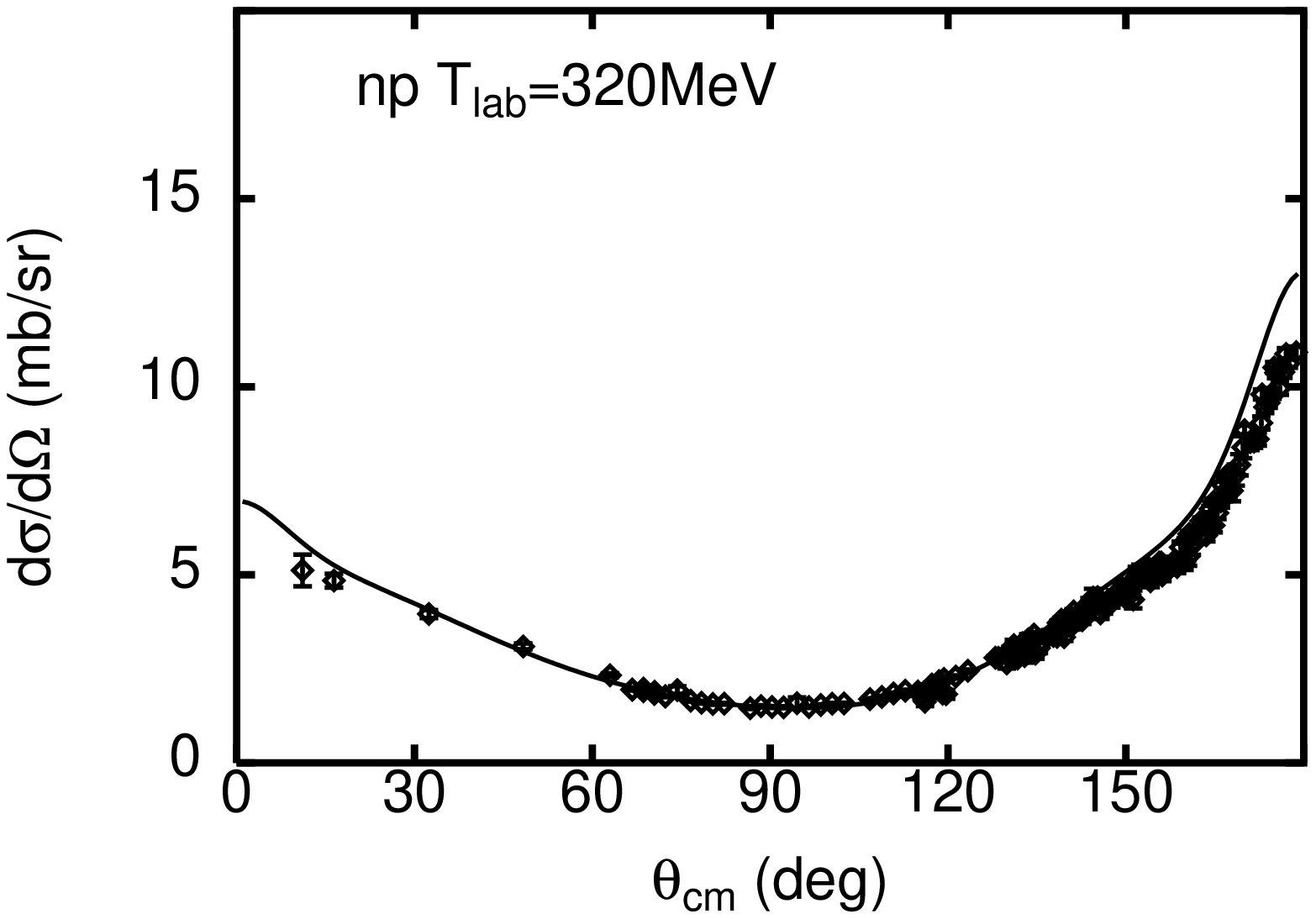}
\end{minipage}~%
\hspace{-29.49mm}
\begin{minipage}{0.47\textwidth}
\epsfxsize=\textwidth
\epsffile{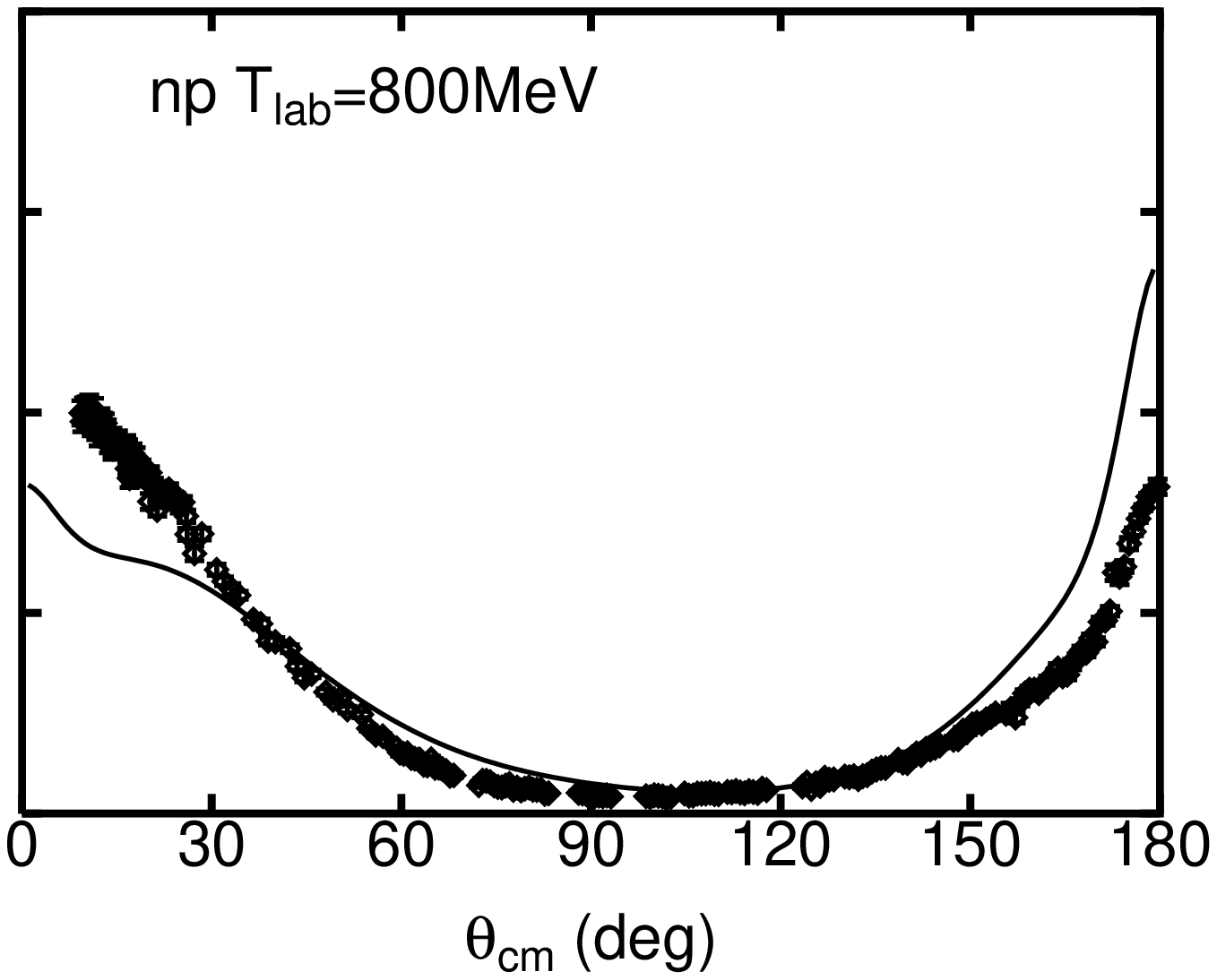}
\end{minipage}
\bigskip
\caption{
Calculated $np$ differential cross sections
compared with the experiment \protect\cite{SAID}.
Calculation is performed using fss2 with the full pion-Coulomb
correction included.
}
\label{npdif}
\end{figure}

\clearpage

\begin{figure}[h]
\begin{minipage}{0.47\textwidth}
\epsfxsize=\textwidth
\epsffile{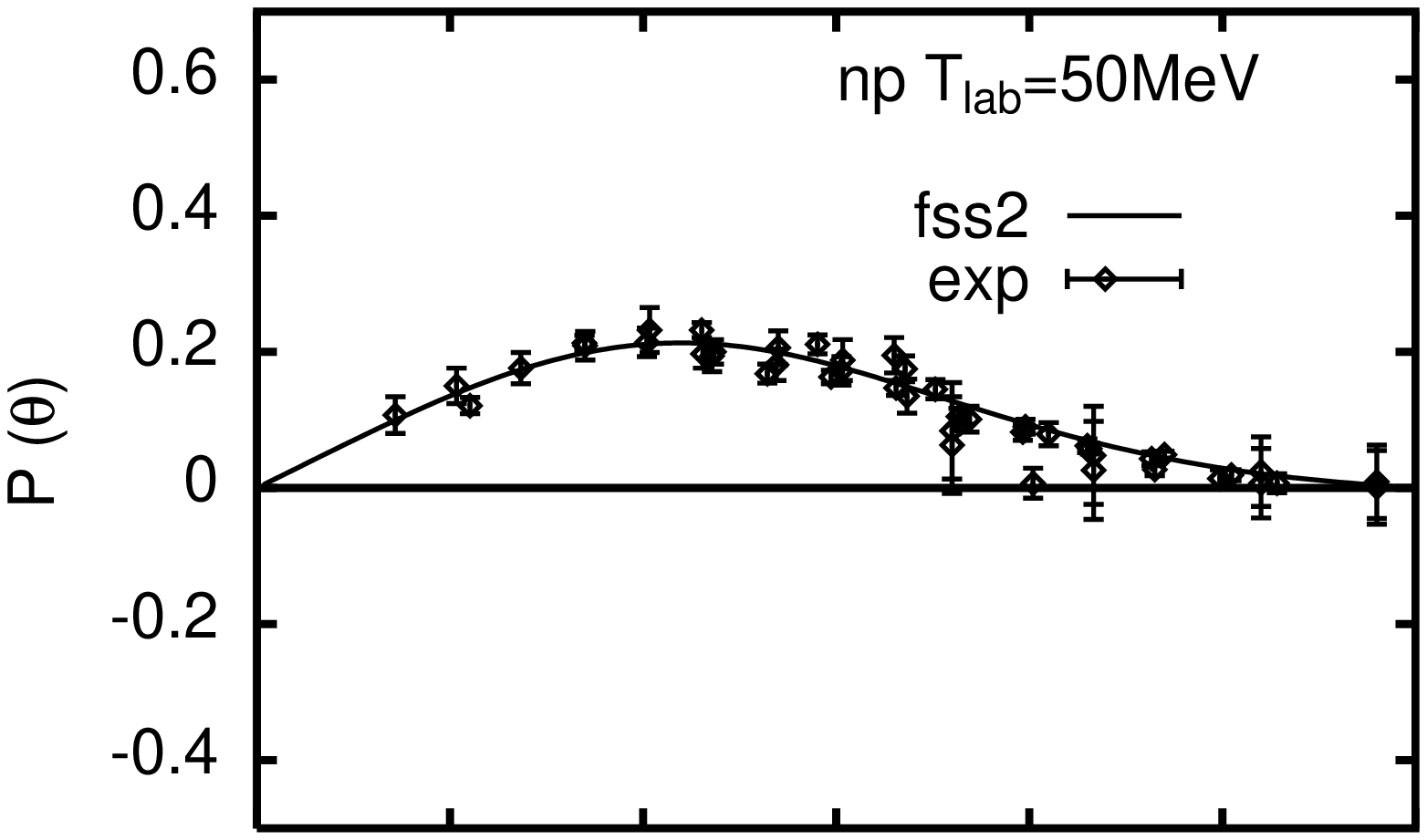}
\end{minipage}~%
\hspace{-29.49mm}
\begin{minipage}{0.47\textwidth}
\epsfxsize=\textwidth
\epsffile{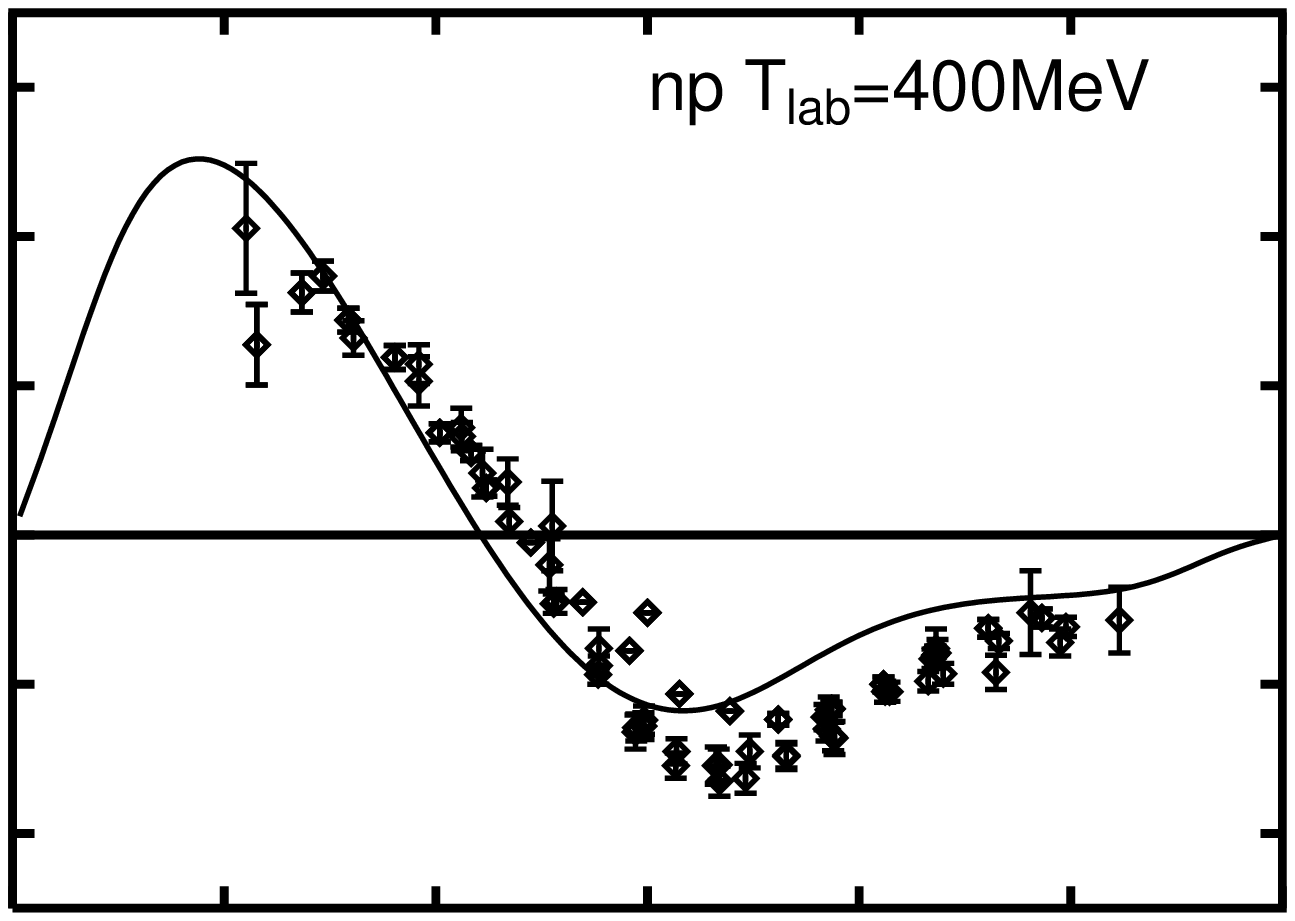}
\end{minipage}
\end{figure}

\vspace{-27.52mm}

%
\begin{figure}[h]
\begin{minipage}{0.47\textwidth}
\epsfxsize=\textwidth
\epsffile{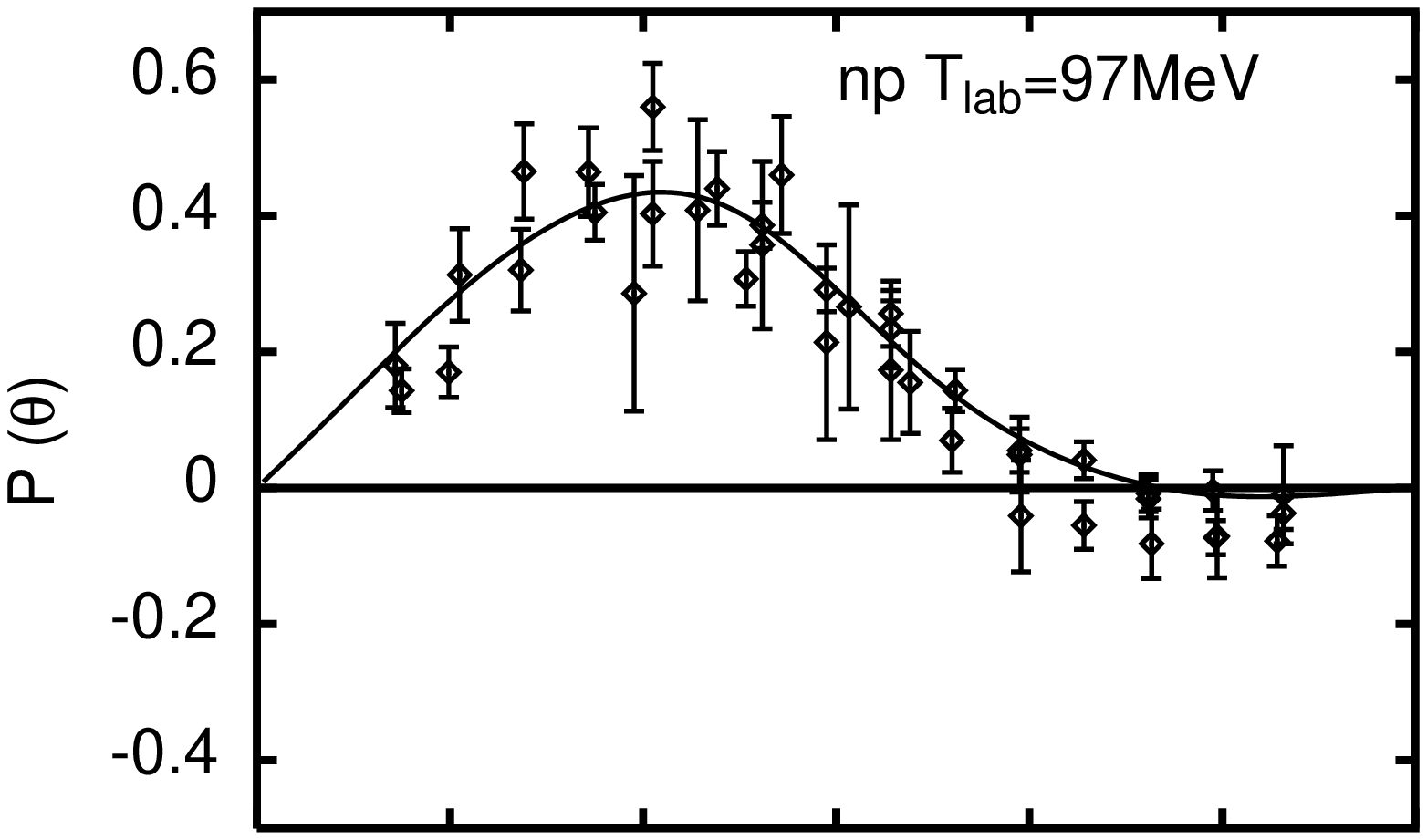}
\end{minipage}~%
\hspace{-29.49mm}
\begin{minipage}{0.47\textwidth}
\epsfxsize=\textwidth
\epsffile{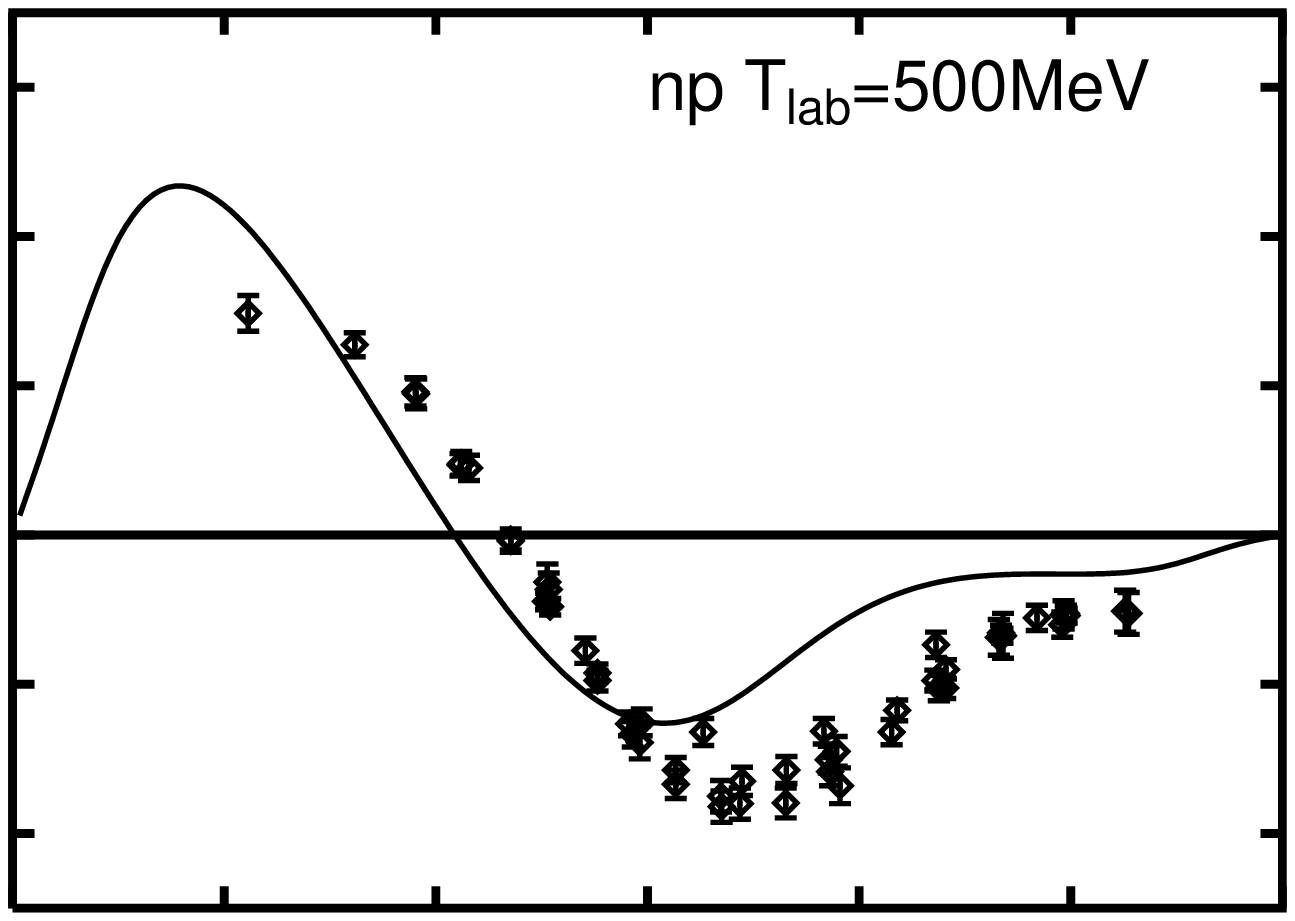}
\end{minipage}
\end{figure}

\vspace{-27.52mm}

\begin{figure}[h]
\begin{minipage}{0.47\textwidth}
\epsfxsize=\textwidth
\epsffile{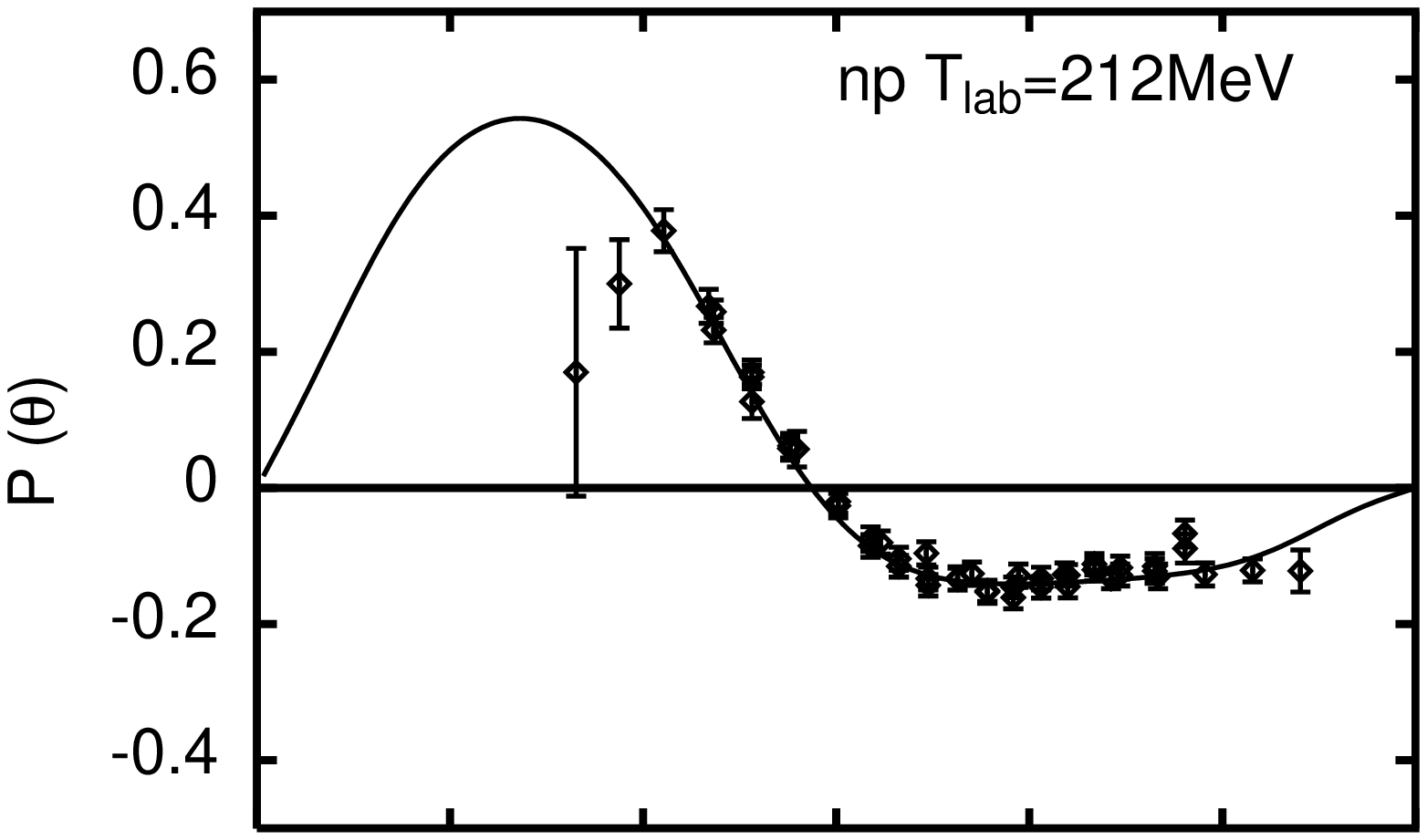}
\end{minipage}~%
\hspace{-29.49mm}
\begin{minipage}{0.47\textwidth}
\epsfxsize=\textwidth
\epsffile{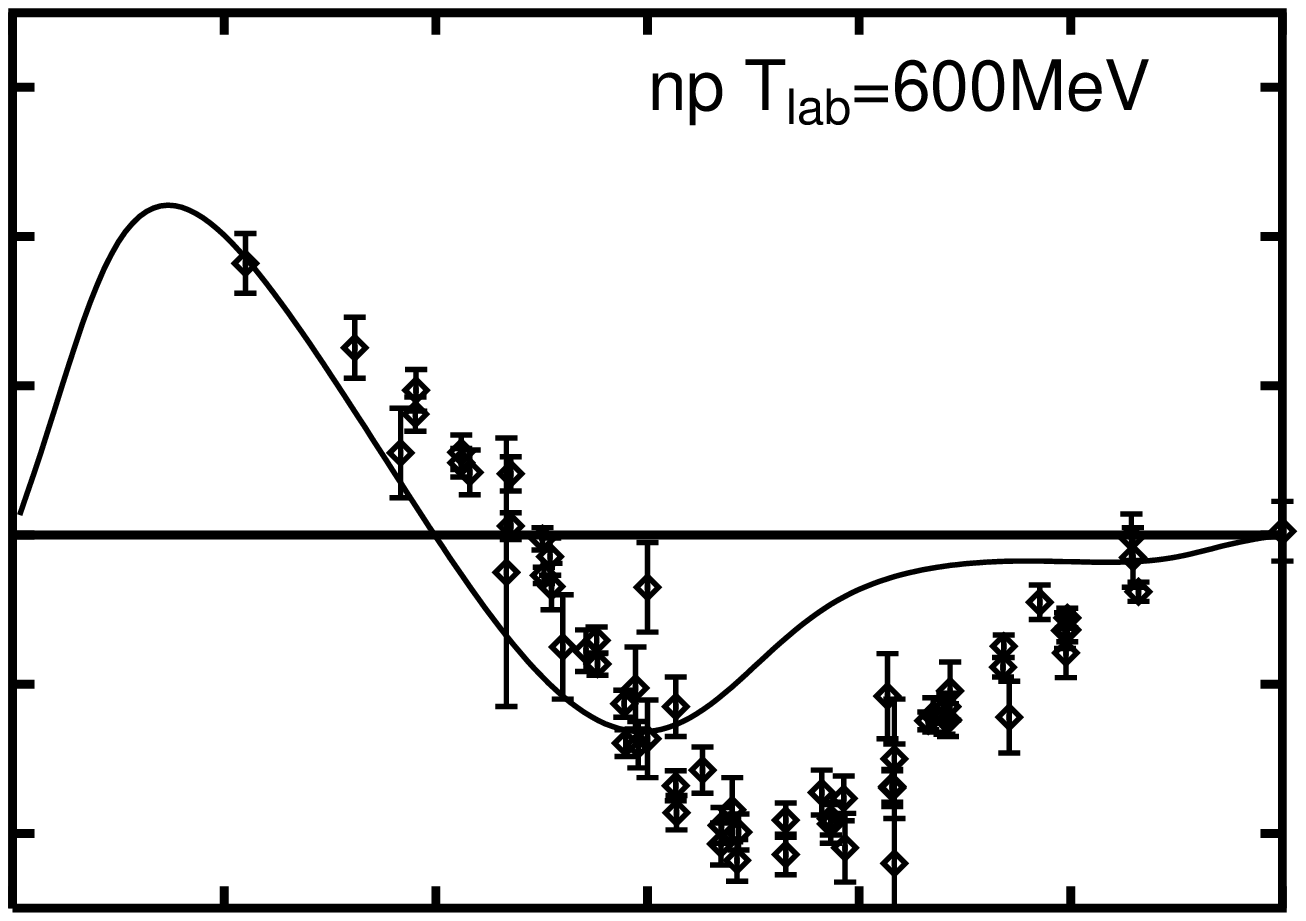}
\end{minipage}
\end{figure}

\vspace{-27.52mm}

\begin{figure}[h]
\begin{minipage}{0.47\textwidth}
\epsfxsize=\textwidth
\epsffile{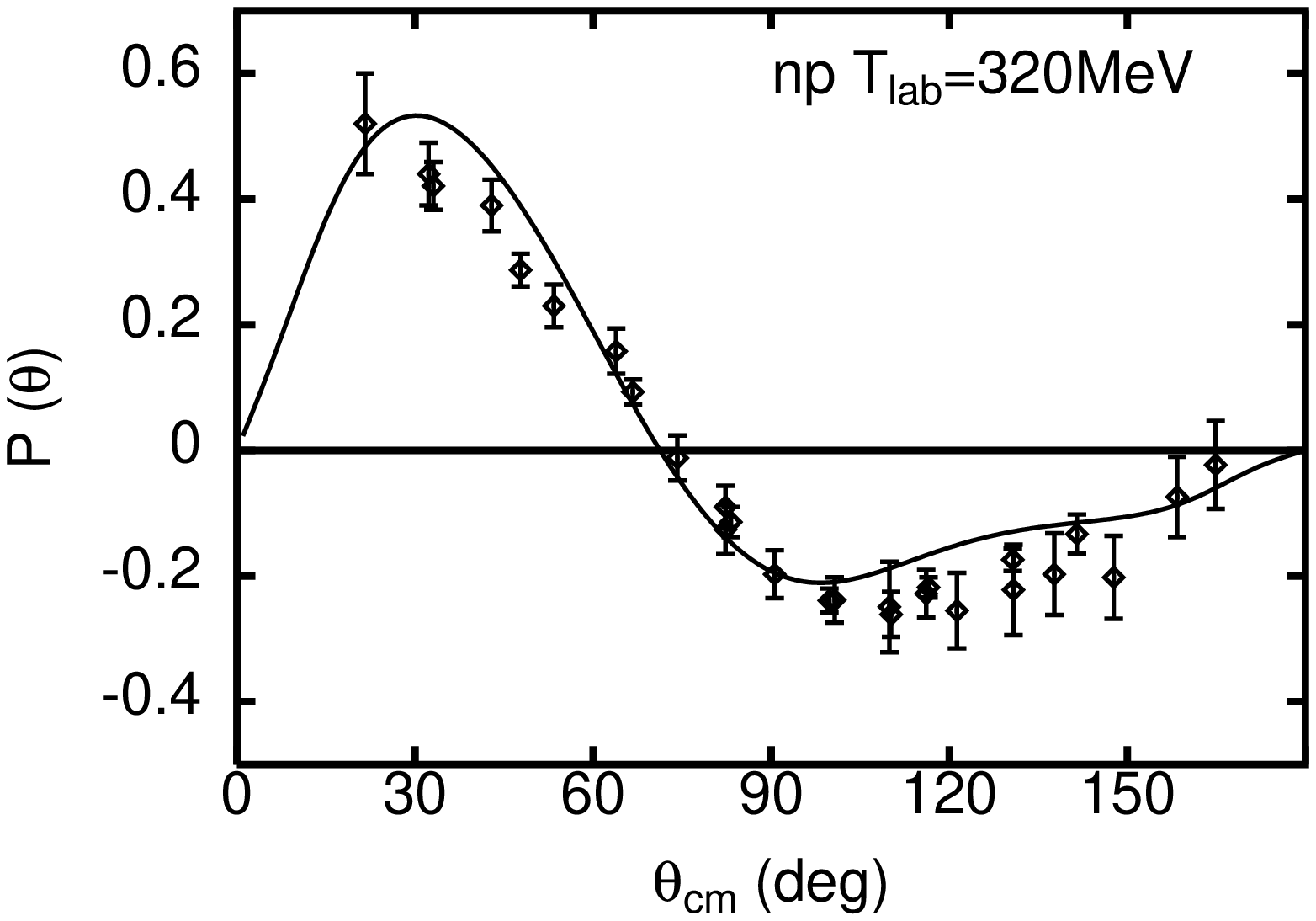}
\end{minipage}~%
\hspace{-29.49mm}
\begin{minipage}{0.47\textwidth}
\epsfxsize=\textwidth
\epsffile{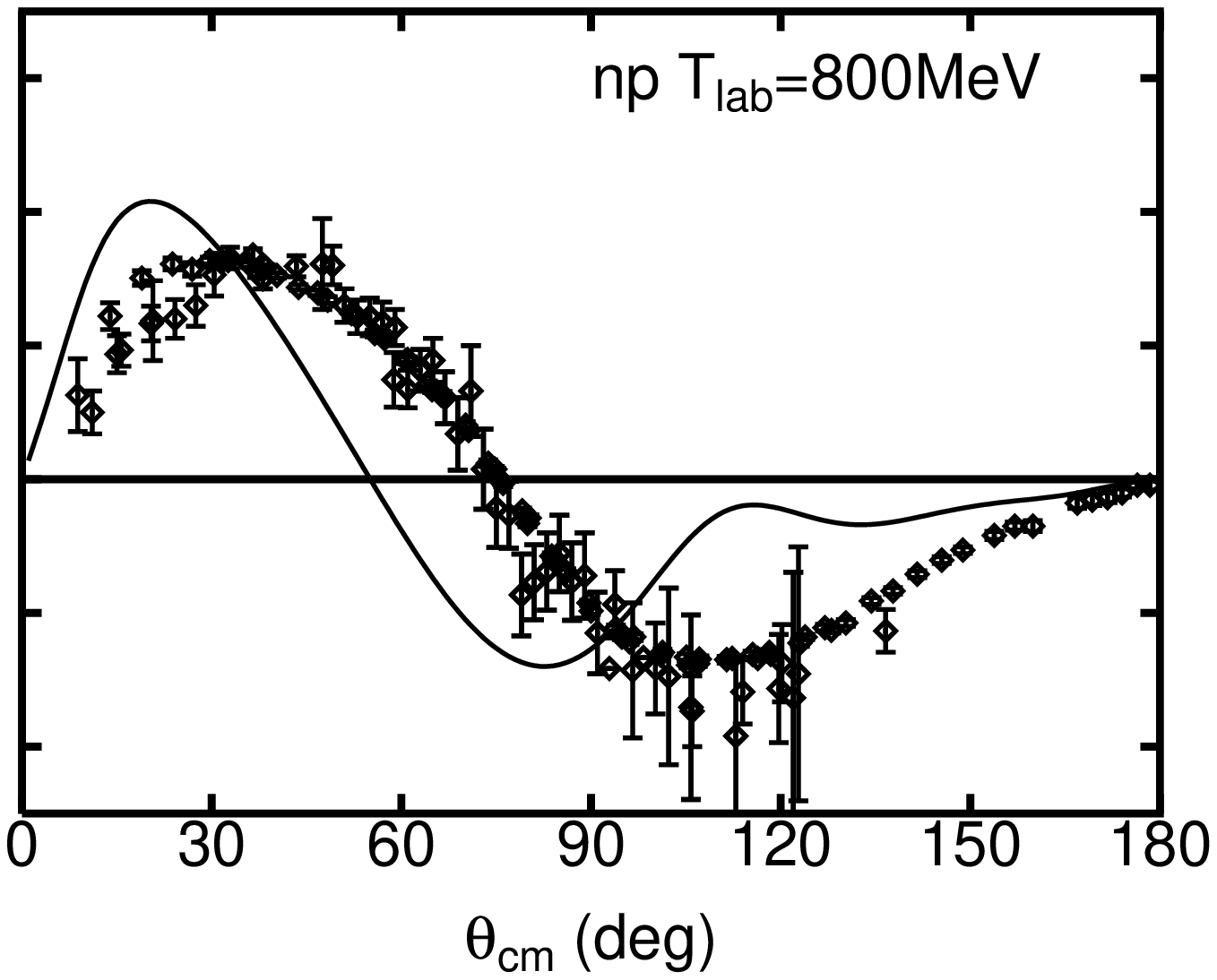}
\end{minipage}
\bigskip
\caption{
The same as Fig.\,\protect\ref{npdif} but for the $np$ polarization.}
\label{nppol}
\end{figure}


\begin{figure}[h]
\begin{minipage}{0.47\textwidth}
\epsfxsize=\textwidth
\epsffile{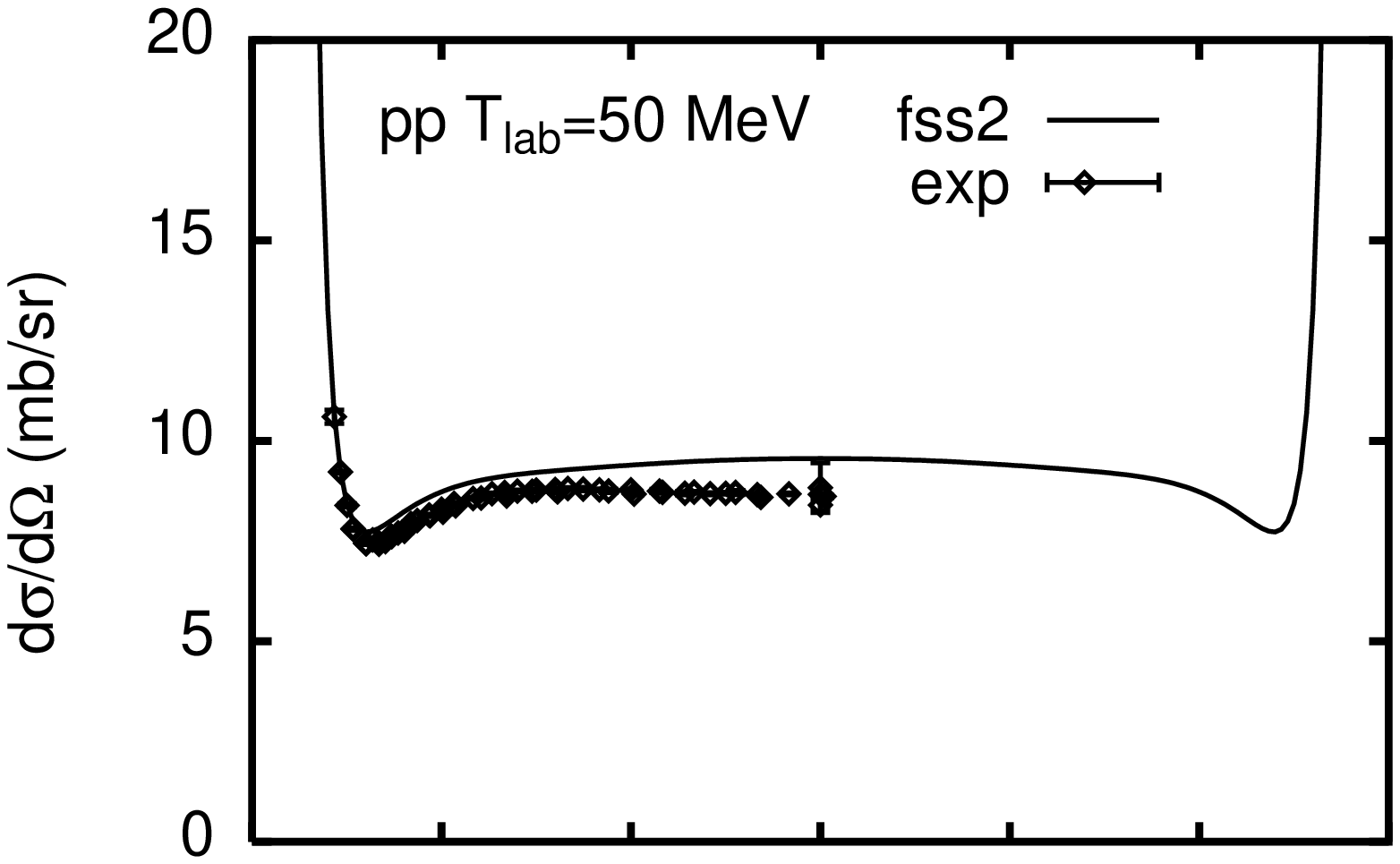}
\end{minipage}~%
\hspace{-29.49mm}
\begin{minipage}{0.47\textwidth}
\epsfxsize=\textwidth
\epsffile{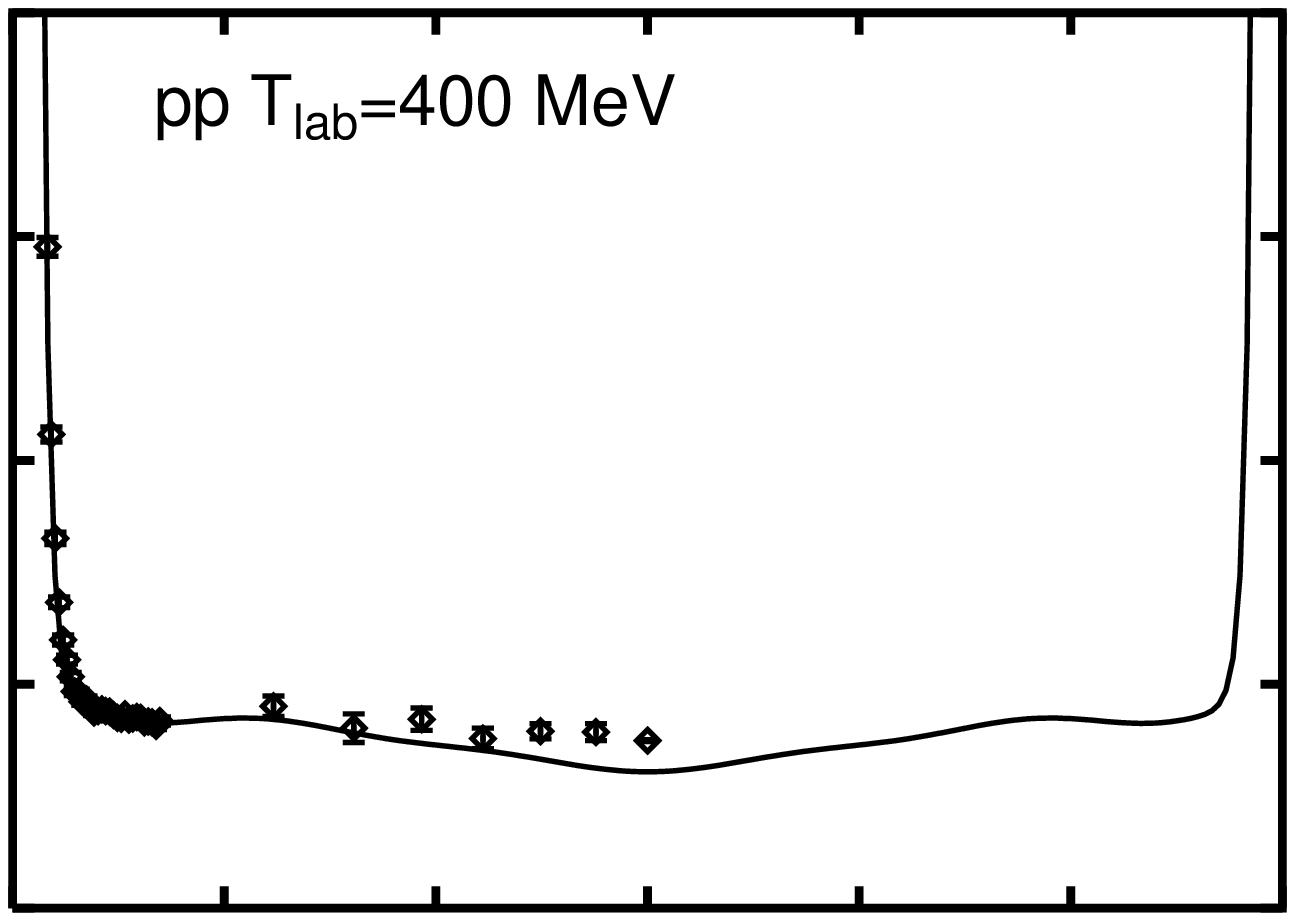}
\end{minipage}
\end{figure}

\vspace{-27.52mm}

\begin{figure}[h]
\begin{minipage}{0.47\textwidth}
\epsfxsize=\textwidth
\epsffile{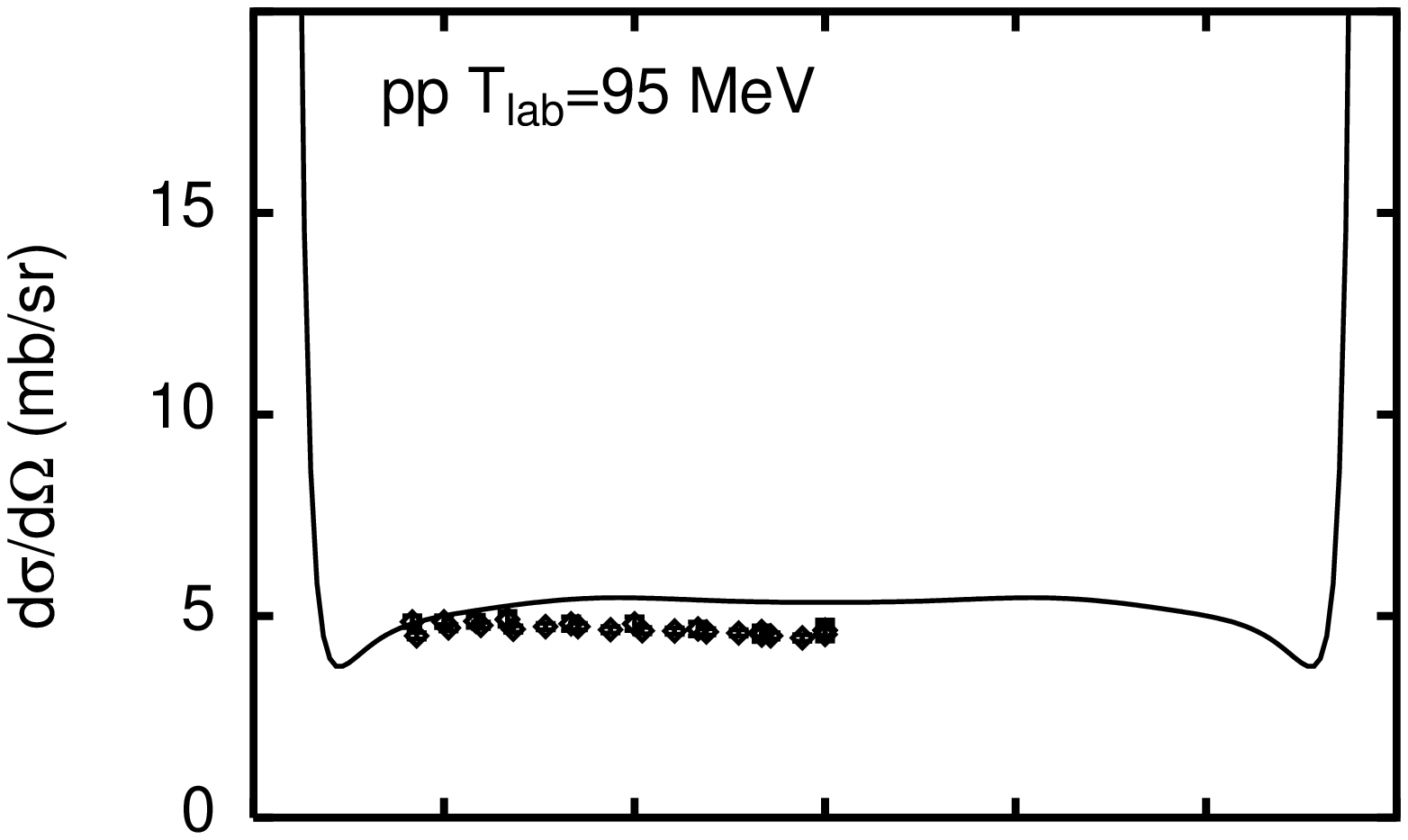}
\end{minipage}~%
\hspace{-29.49mm}
\begin{minipage}{0.47\textwidth}
\epsfxsize=\textwidth
\epsffile{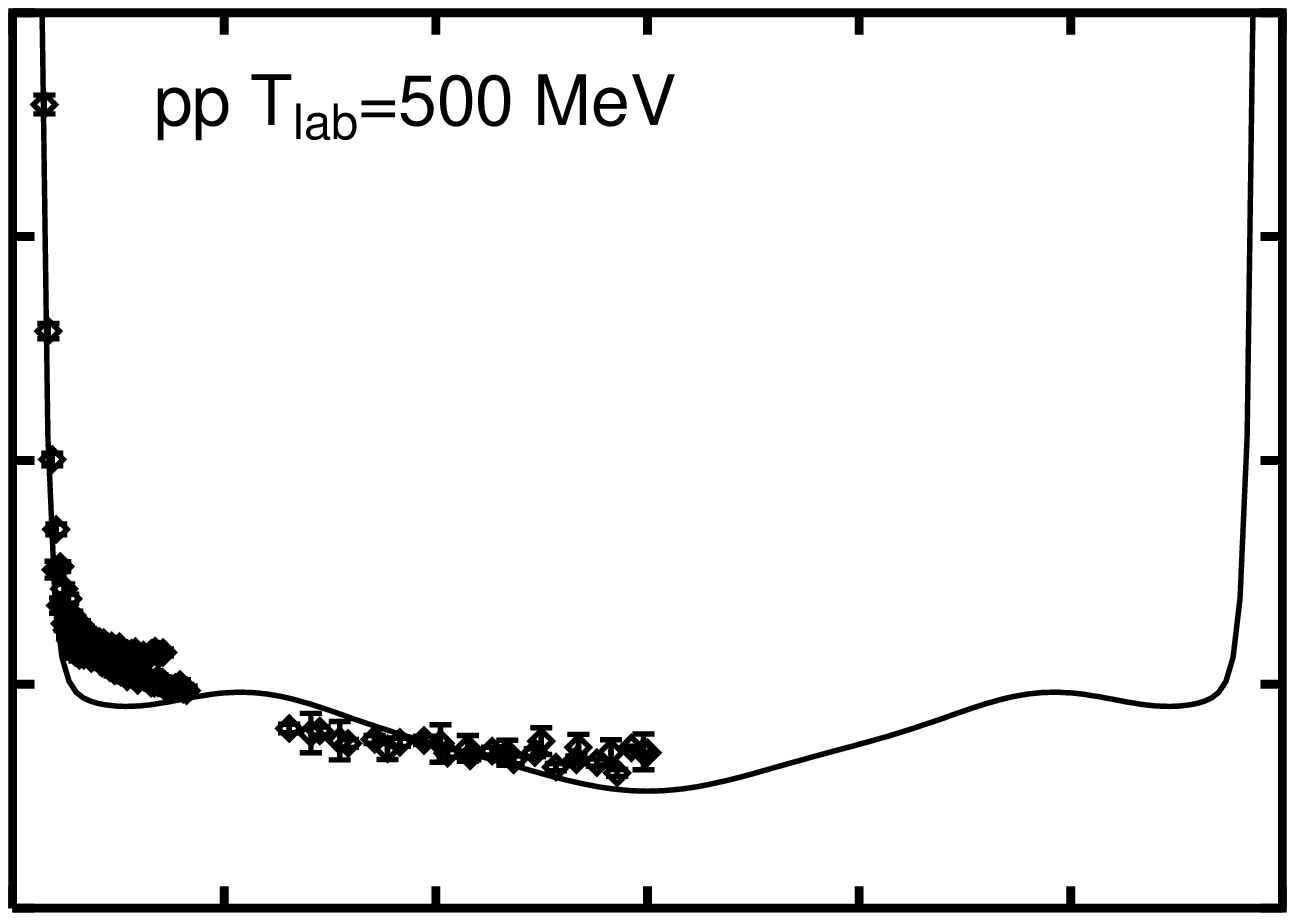}
\end{minipage}
\end{figure}

\vspace{-27.52mm}

\begin{figure}[h]
\begin{minipage}{0.47\textwidth}
\epsfxsize=\textwidth
\epsffile{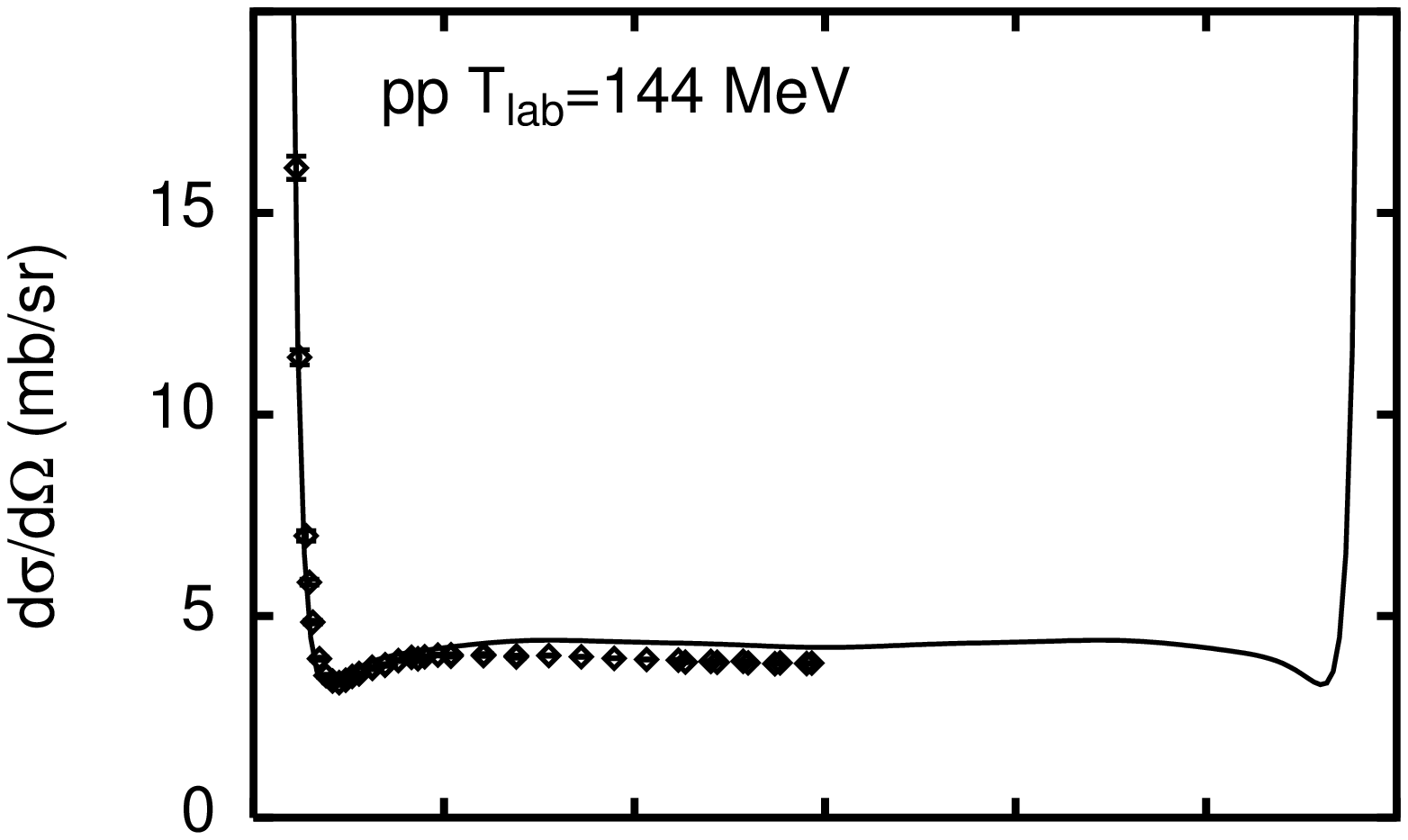}
\end{minipage}~%
\hspace{-29.49mm}
\begin{minipage}{0.47\textwidth}
\epsfxsize=\textwidth
\epsffile{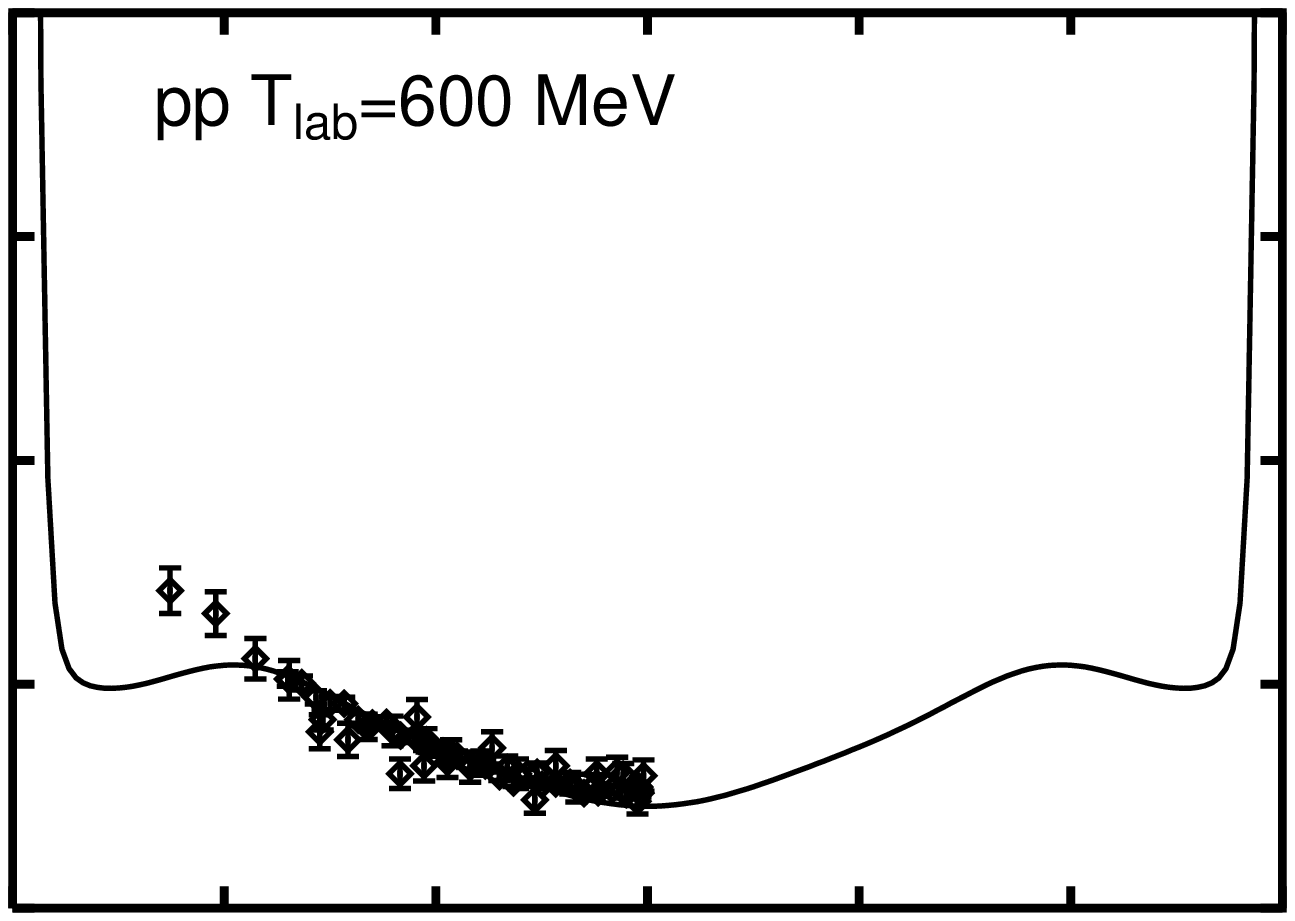}
\end{minipage}
\end{figure}

\vspace{-27.52mm}

\begin{figure}[h]
\begin{minipage}{0.47\textwidth}
\epsfxsize=\textwidth
\epsffile{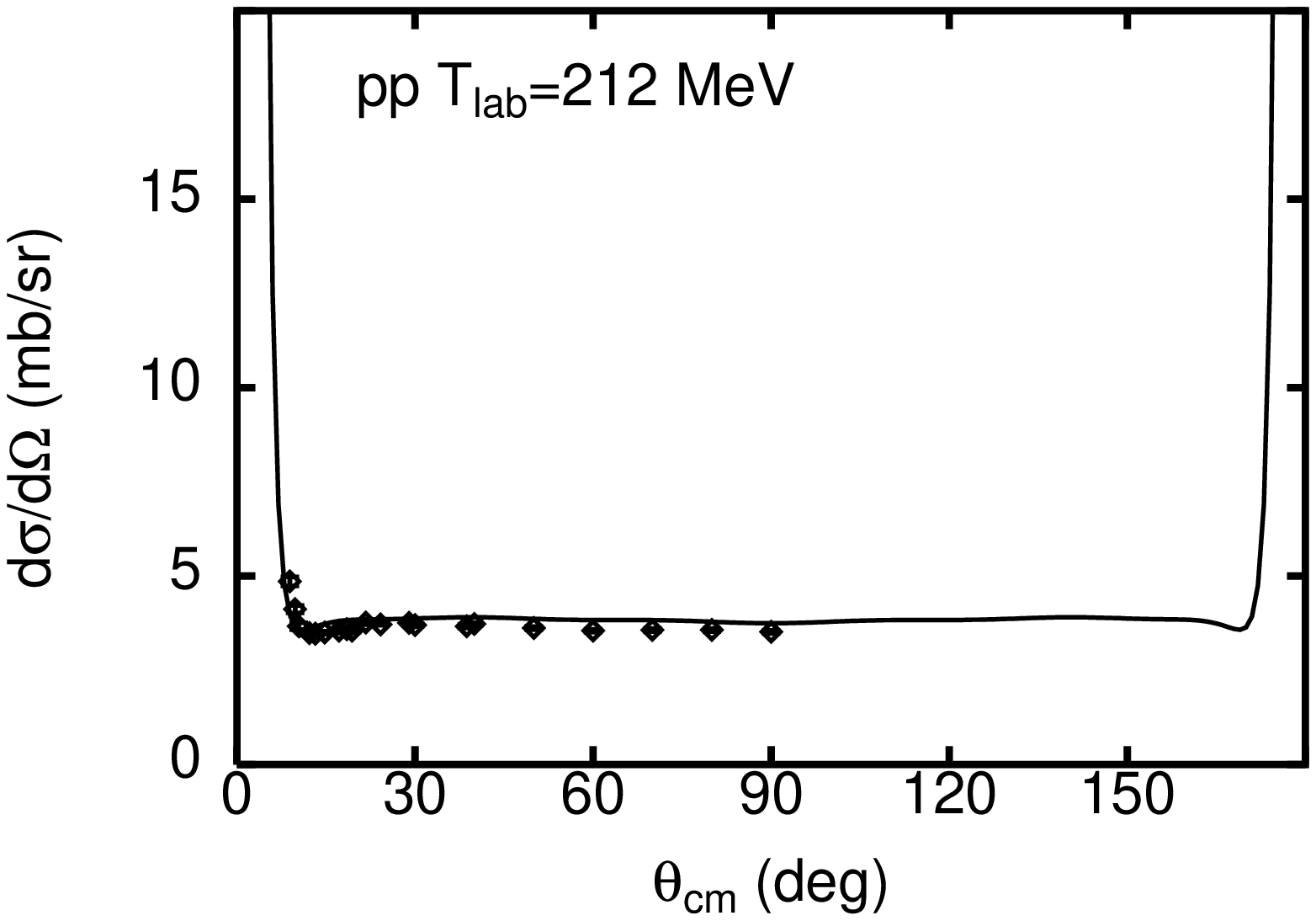}
\end{minipage}~%
\hspace{-29.49mm}
\begin{minipage}{0.47\textwidth}
\epsfxsize=\textwidth
\epsffile{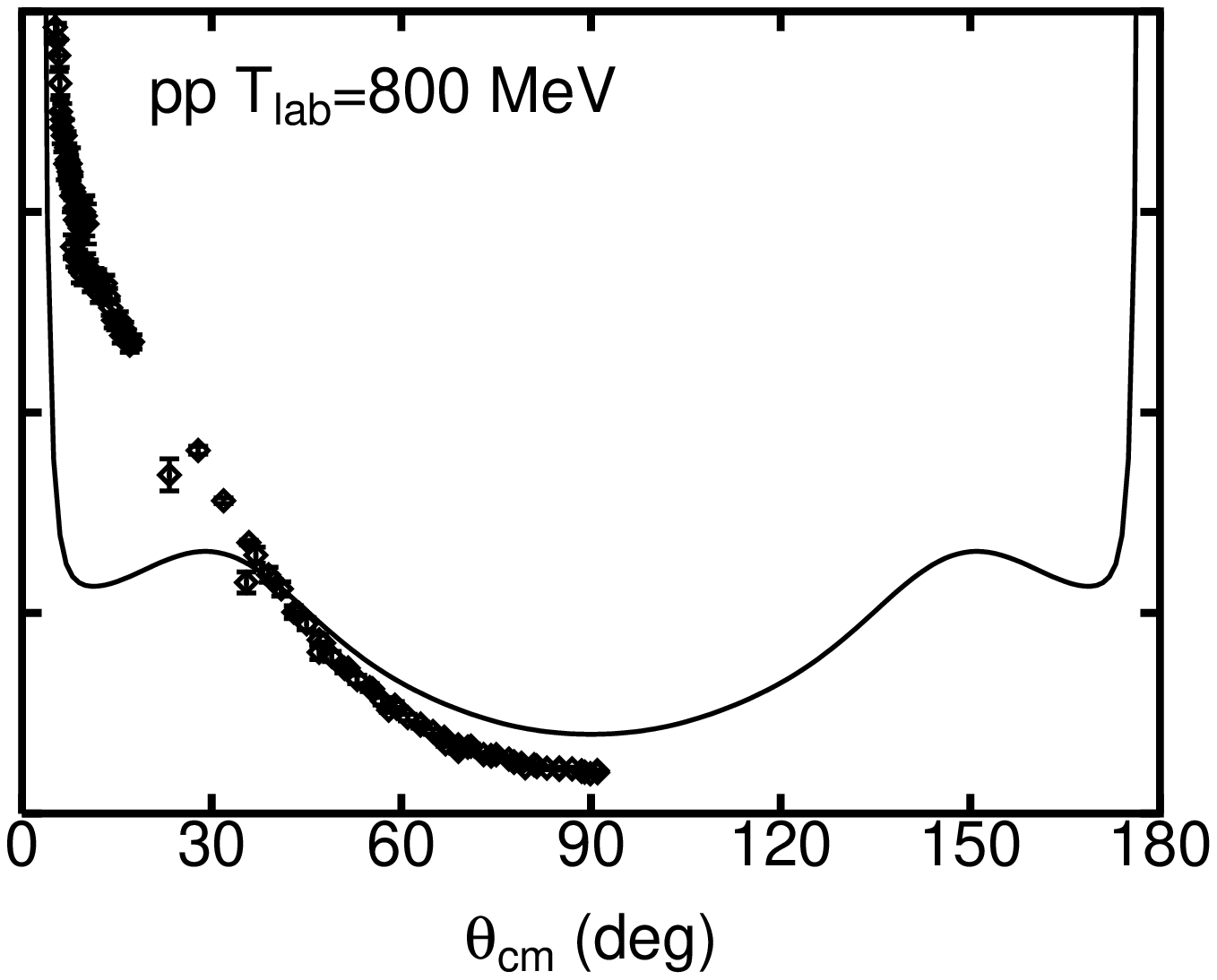}
\end{minipage}
\bigskip
\caption{
The same as Fig.\,\protect\ref{npdif} but for $pp$ differential
cross sections.}
\label{ppdif}
\end{figure}


%
\begin{figure}[h]
\begin{minipage}{0.47\textwidth}
\epsfxsize=\textwidth
\epsffile{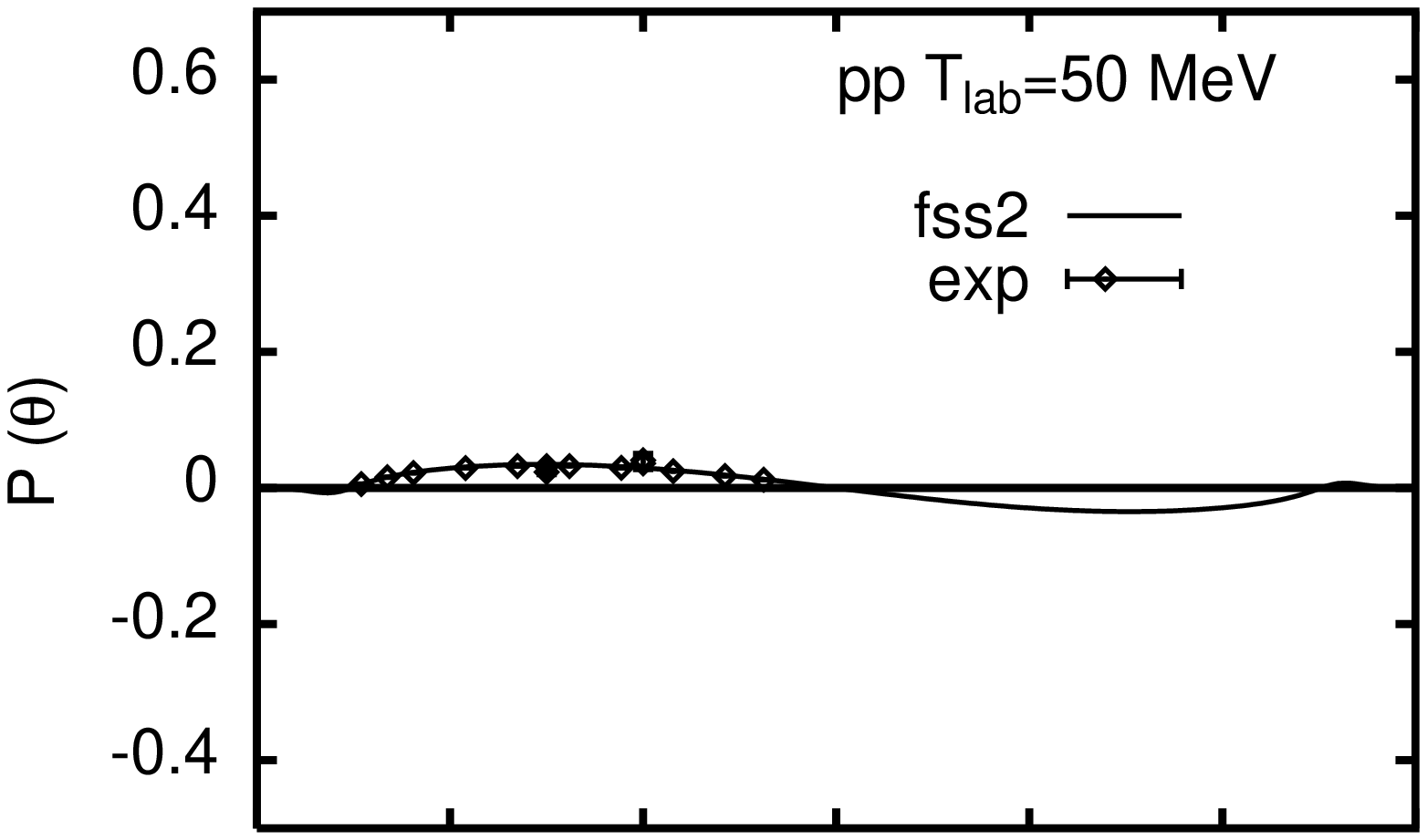}
\end{minipage}~%
\hspace{-29.49mm}
\begin{minipage}{0.47\textwidth}
\epsfxsize=\textwidth
\epsffile{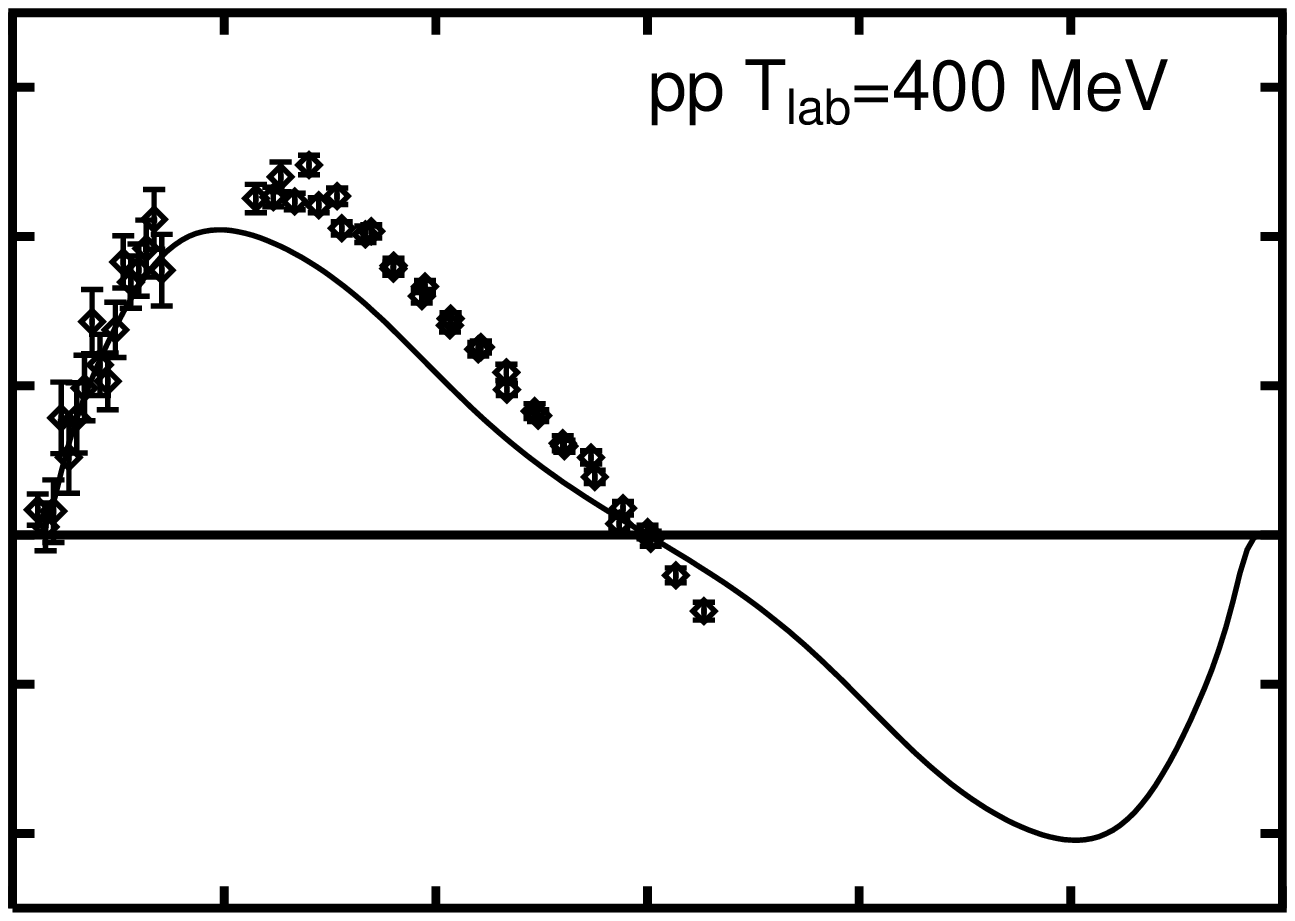}
\end{minipage}
\end{figure}

\vspace{-27.52mm}


\begin{figure}[h]
\begin{minipage}{0.47\textwidth}
\epsfxsize=\textwidth
\epsffile{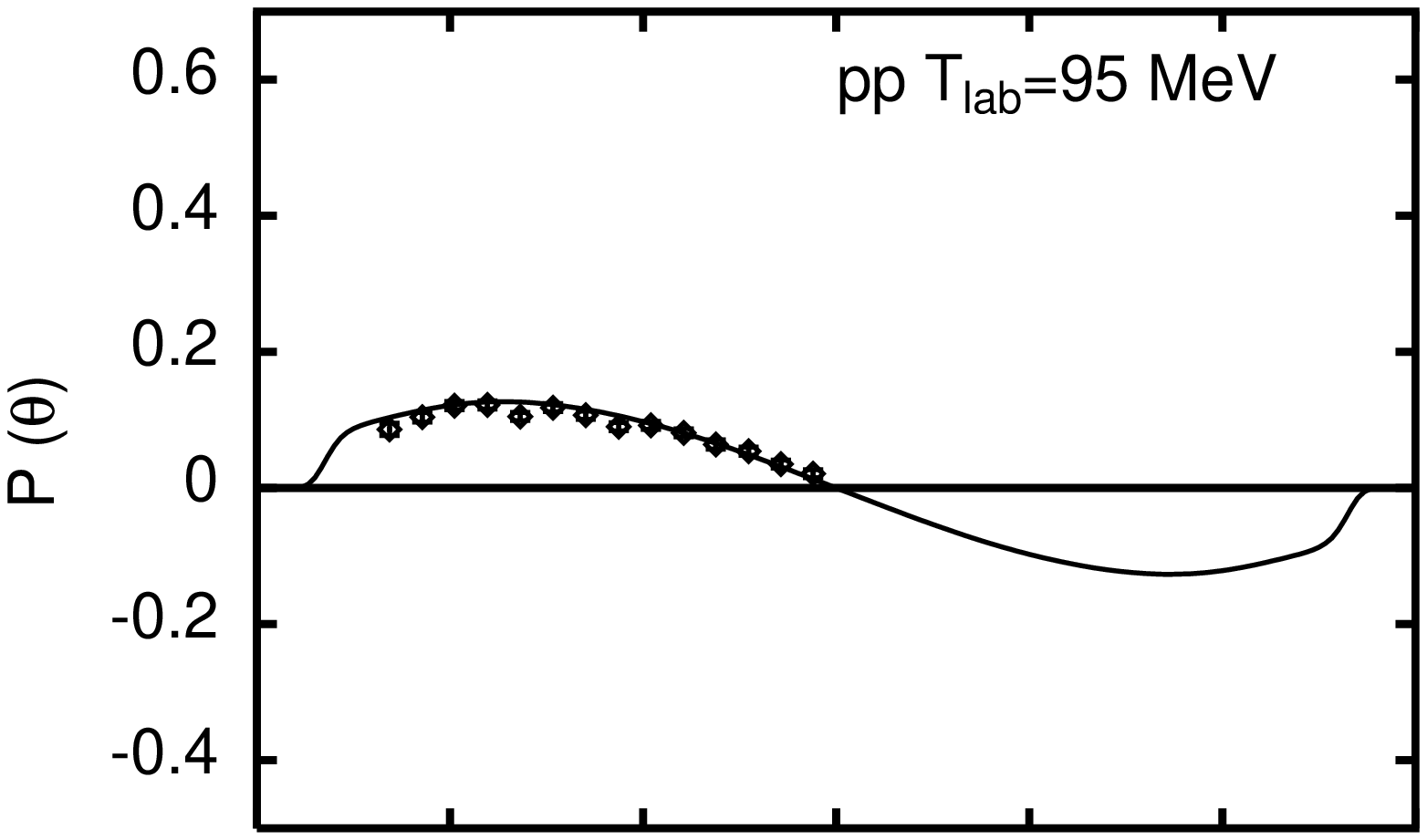}
\end{minipage}~%
\hspace{-29.49mm}
\begin{minipage}{0.47\textwidth}
\epsfxsize=\textwidth
\epsffile{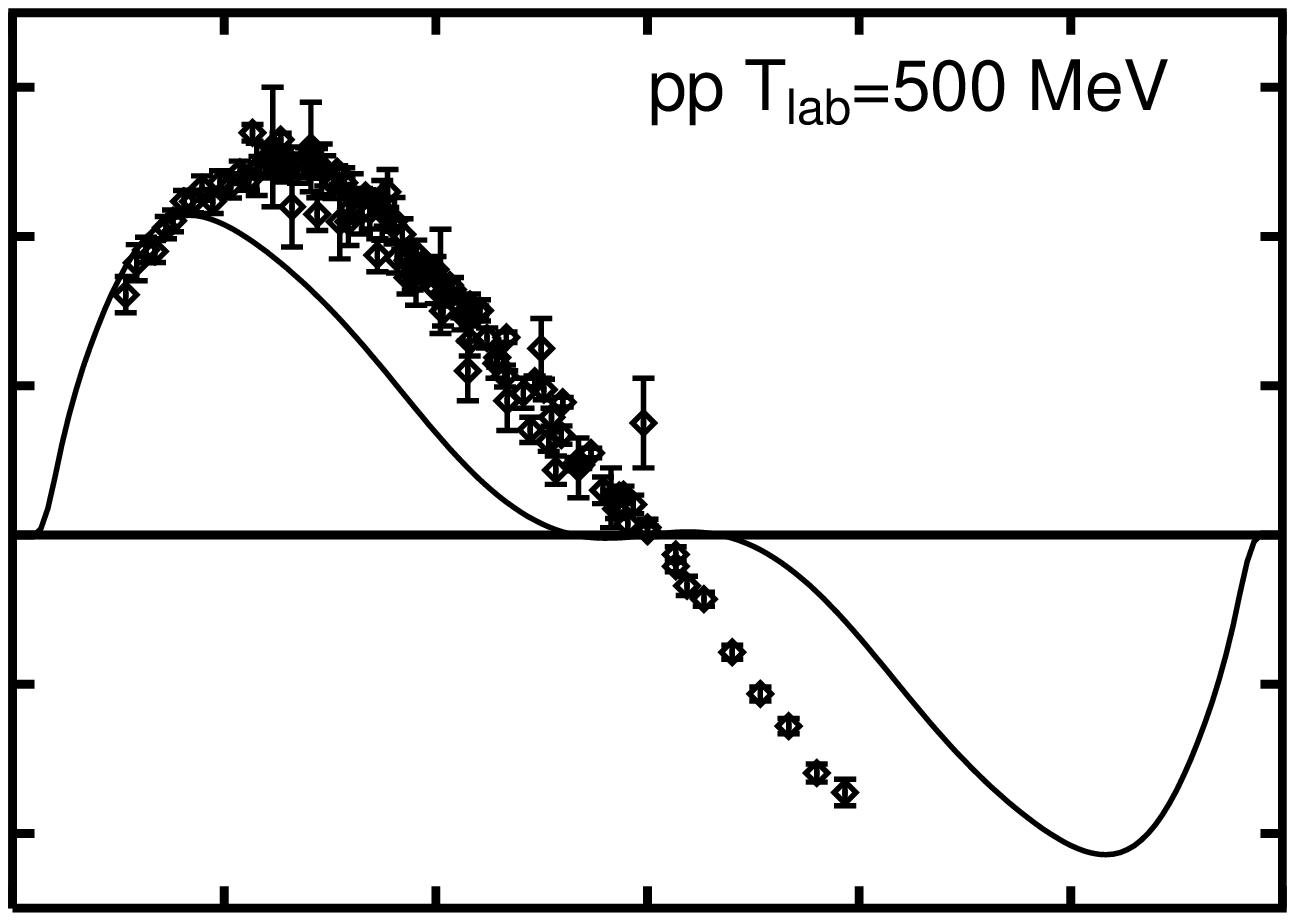}
\end{minipage}
\end{figure}

\vspace{-27.52mm}

\begin{figure}[h]
\begin{minipage}{0.47\textwidth}
\epsfxsize=\textwidth
\epsffile{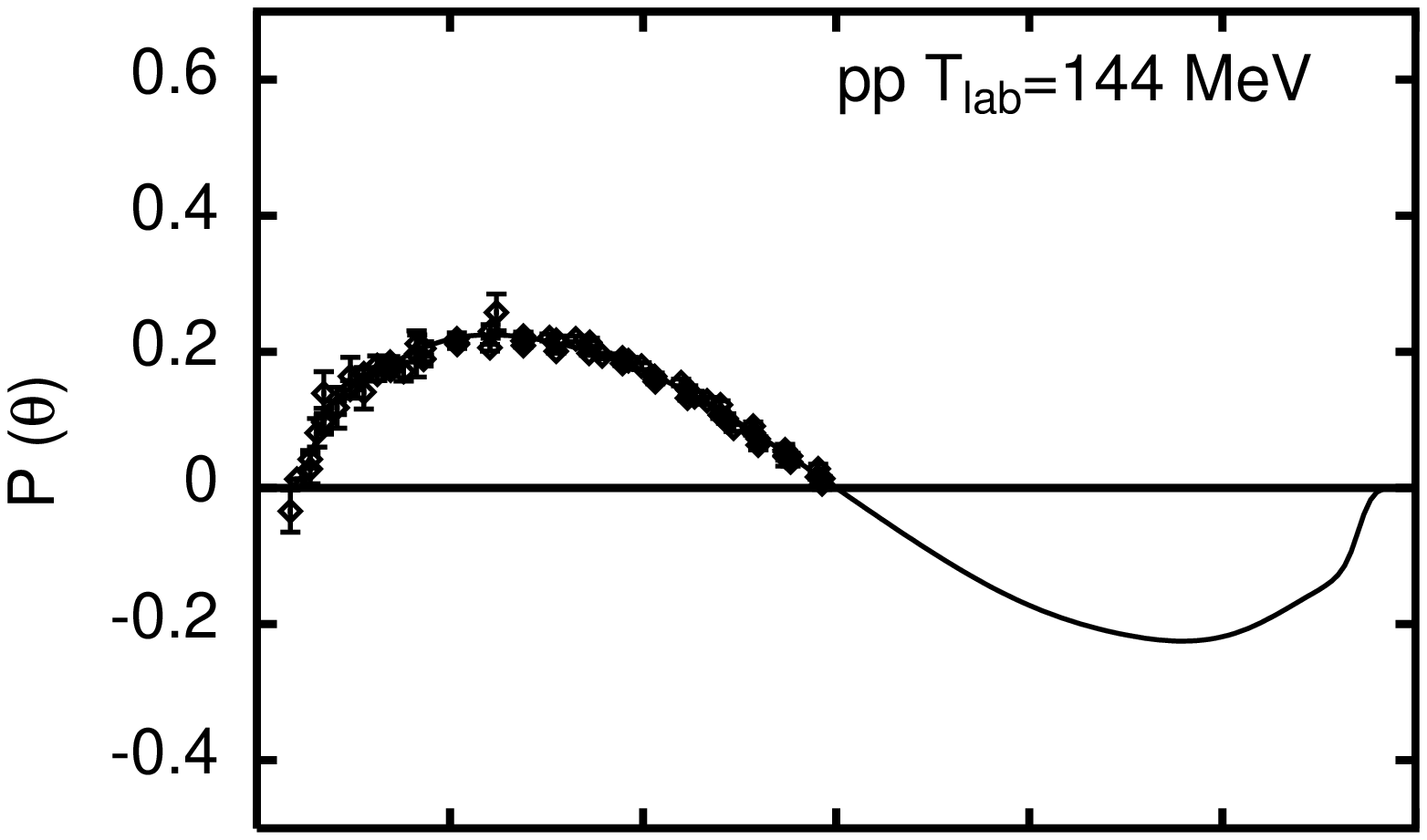}
\end{minipage}~%
\hspace{-29.49mm}
\begin{minipage}{0.47\textwidth}
\epsfxsize=\textwidth
\epsffile{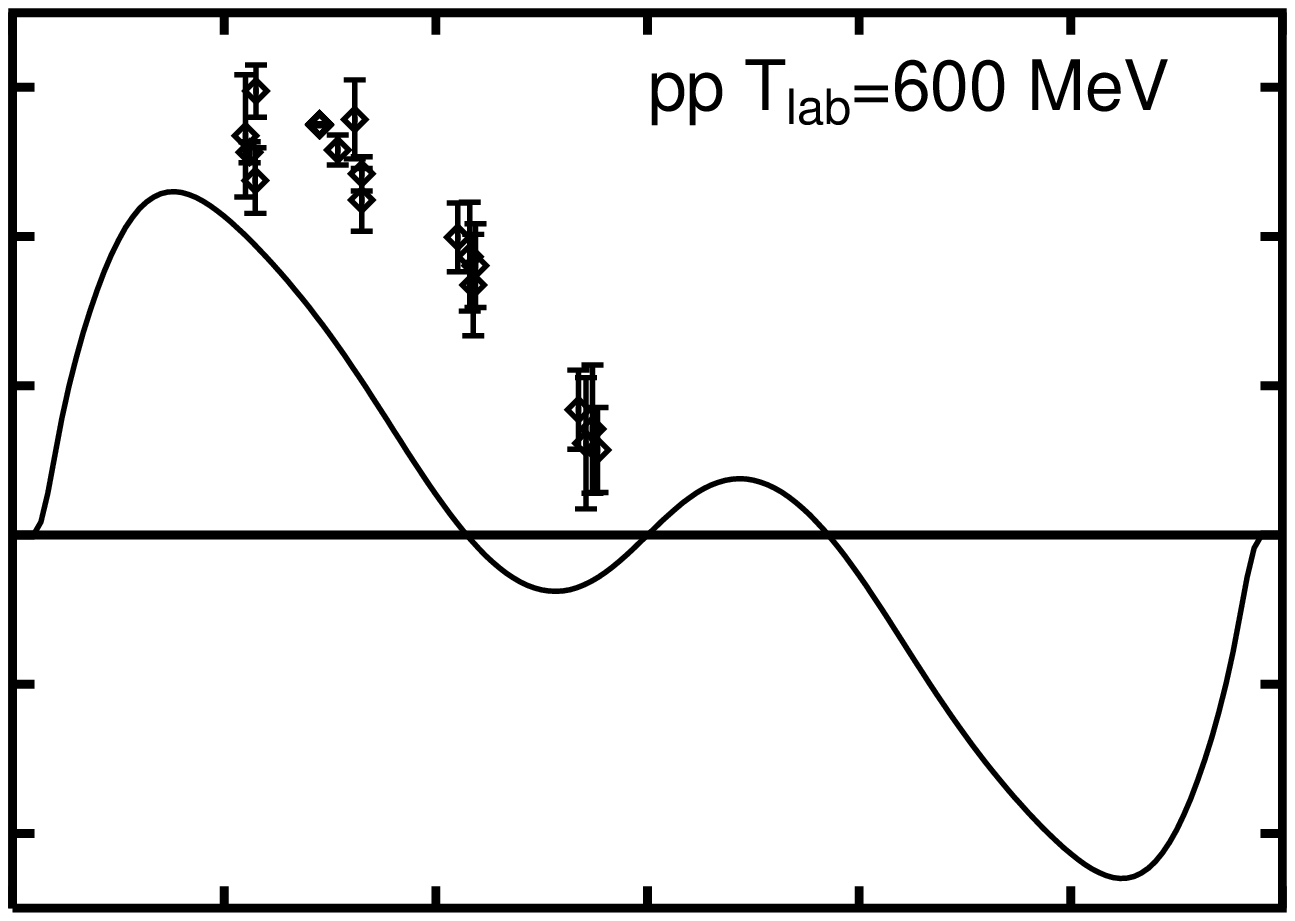}
\end{minipage}
\end{figure}

\vspace{-27.52mm}

\begin{figure}[h]
\begin{minipage}{0.47\textwidth}
\epsfxsize=\textwidth
\epsffile{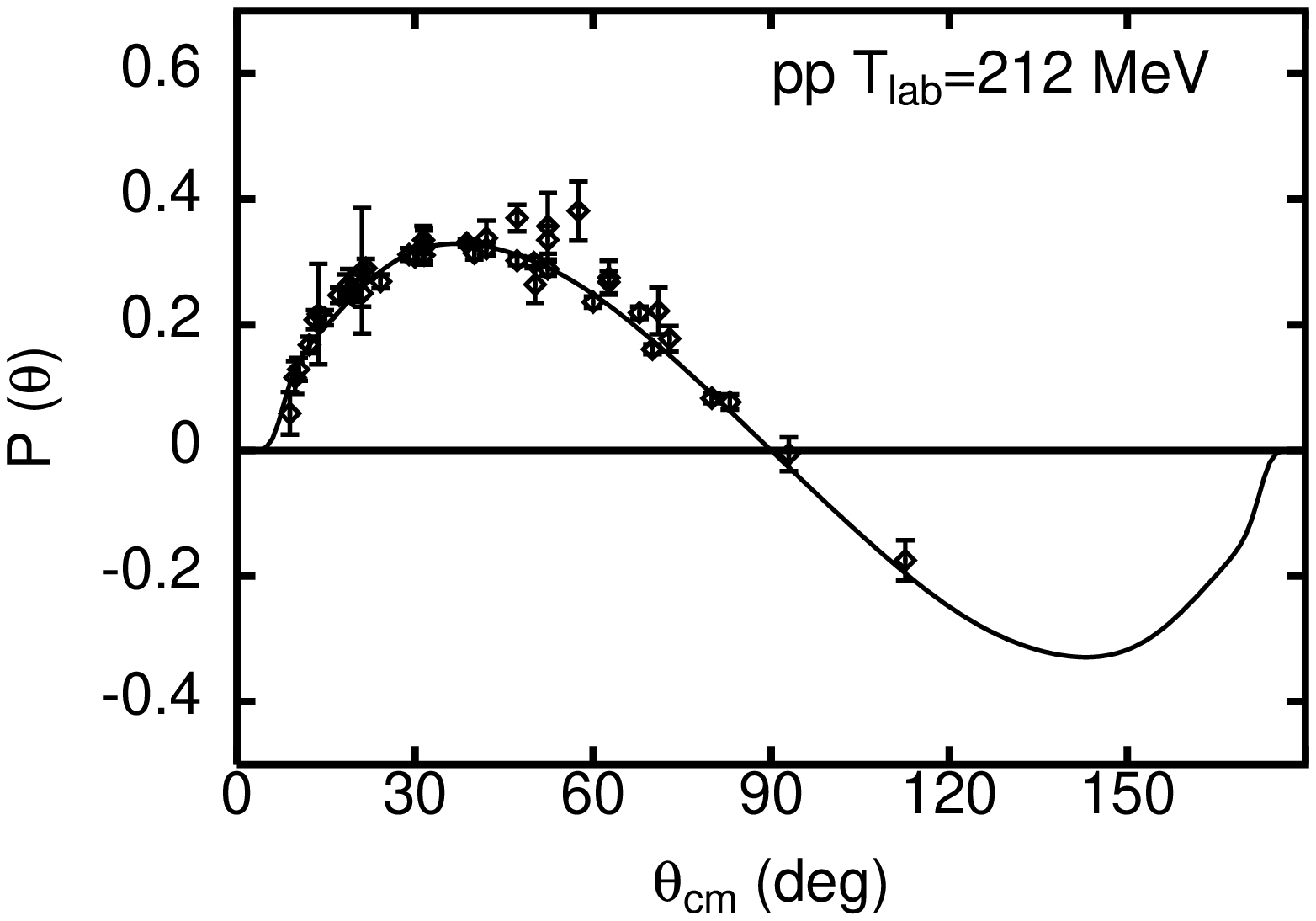}
\end{minipage}~%
\hspace{-29.49mm}
\begin{minipage}{0.47\textwidth}
\epsfxsize=\textwidth
\epsffile{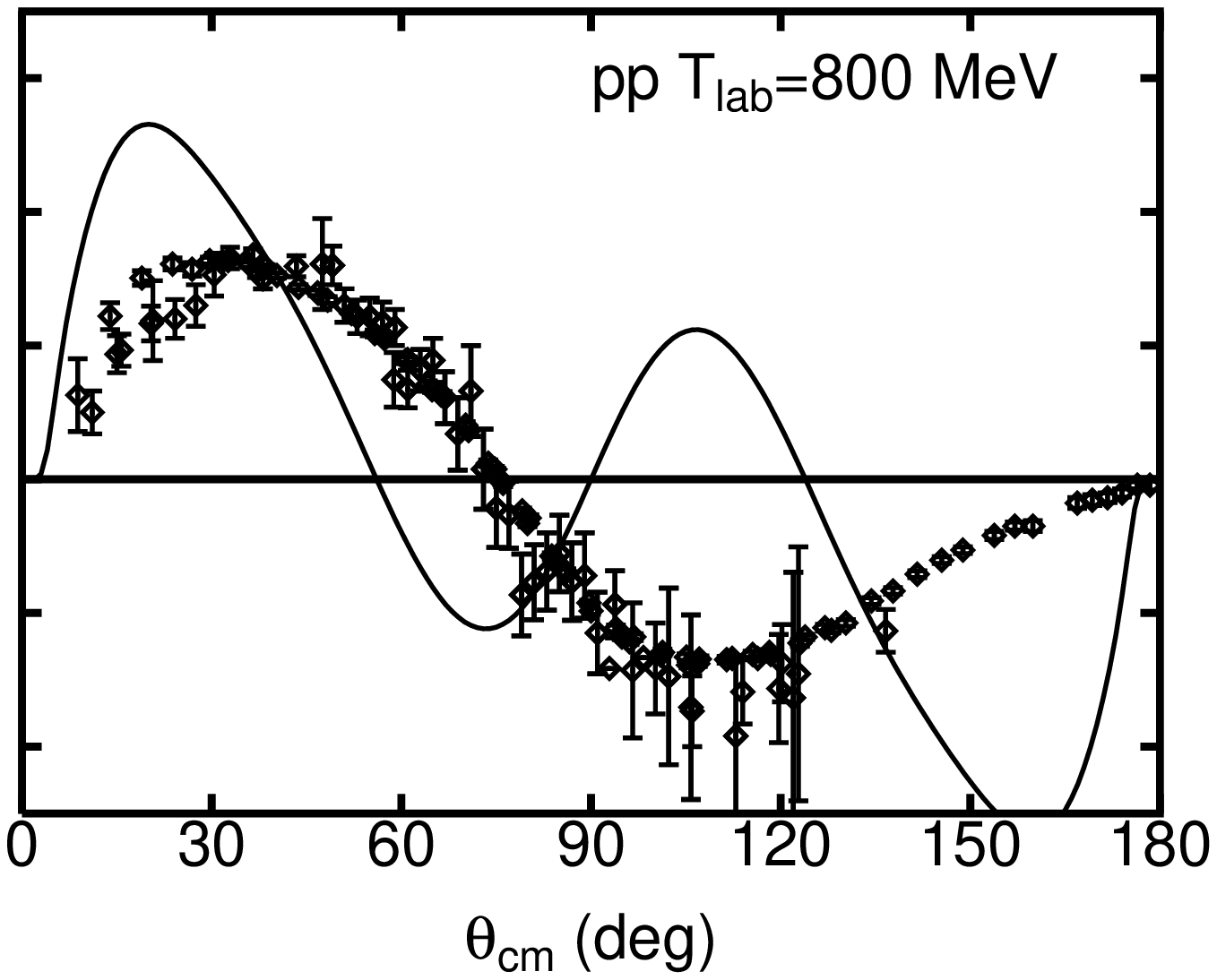}
\end{minipage}
\bigskip
\caption{
The same as Fig.\,\protect\ref{npdif} but
for the $pp$ polarization.}
\label{pppol}
\end{figure}

\clearpage

\begin{figure}[h]
\begin{minipage}{0.47\textwidth}
\epsfxsize=\textwidth
\epsffile{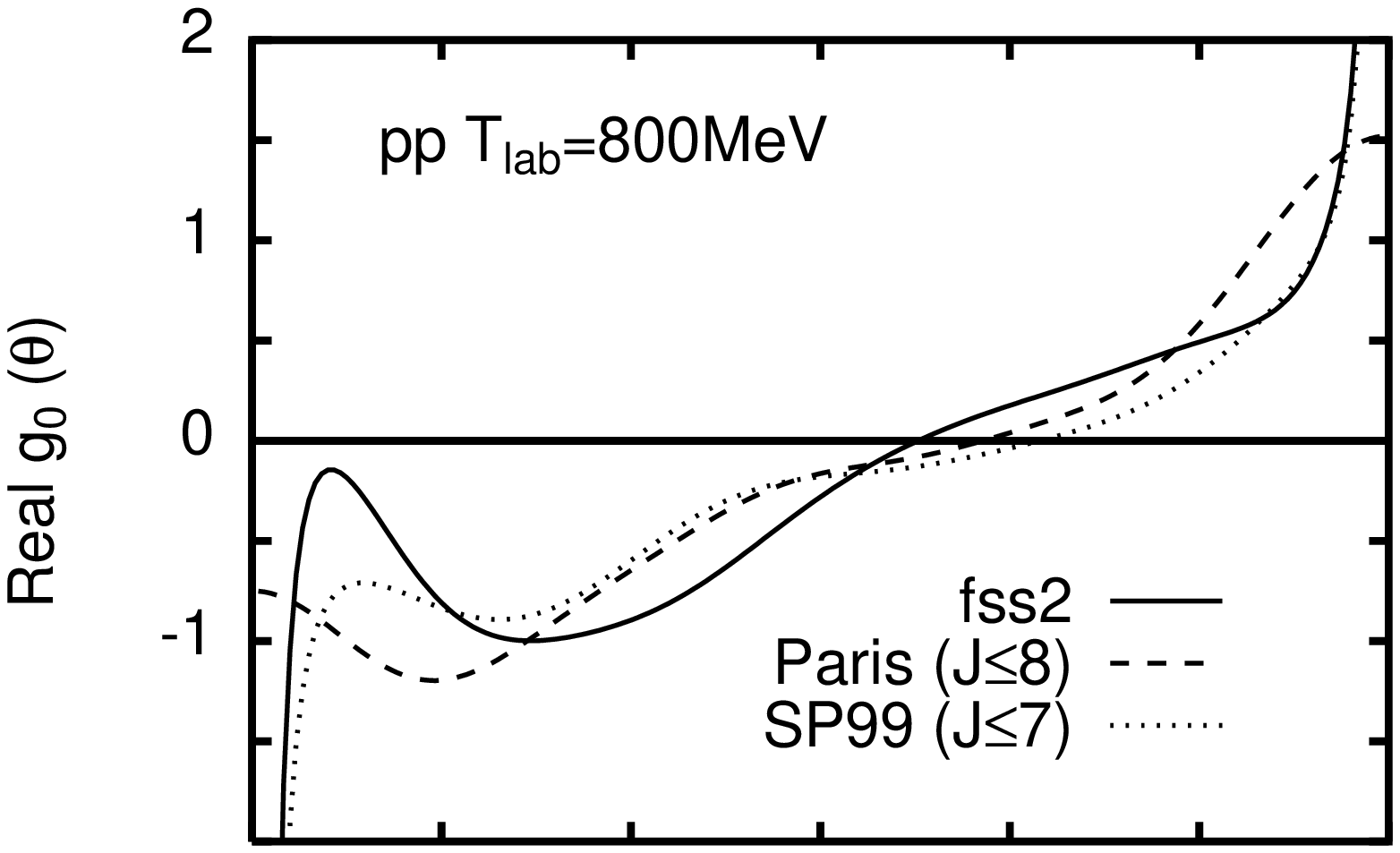}
\end{minipage}~%
\hfill~%
\begin{minipage}{0.47\textwidth}
\epsfxsize=\textwidth
\epsffile{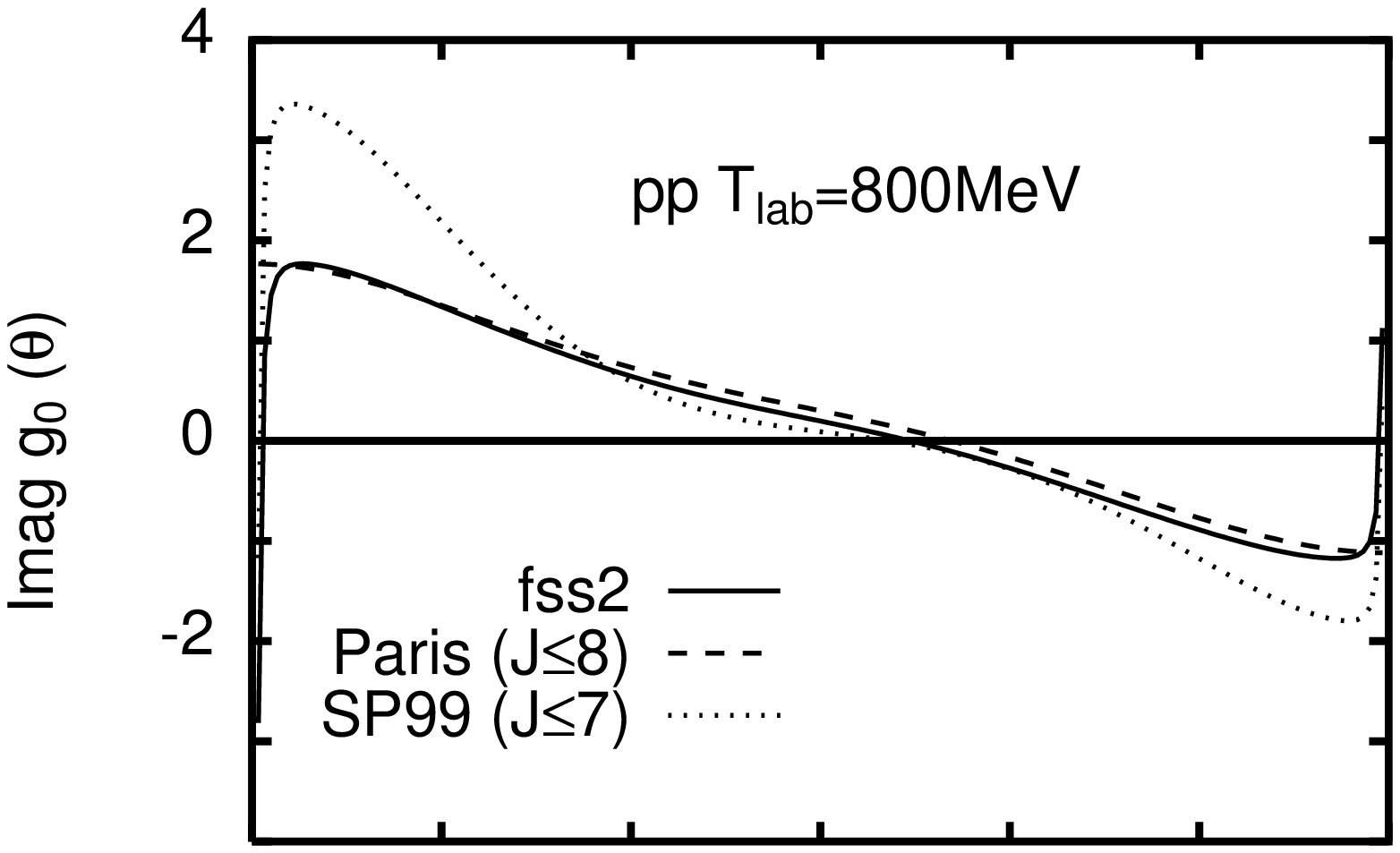}
\end{minipage}
\end{figure}

\vspace{-27.52mm}

\begin{figure}[h]
\begin{minipage}{0.47\textwidth}
\epsfxsize=\textwidth
\epsffile{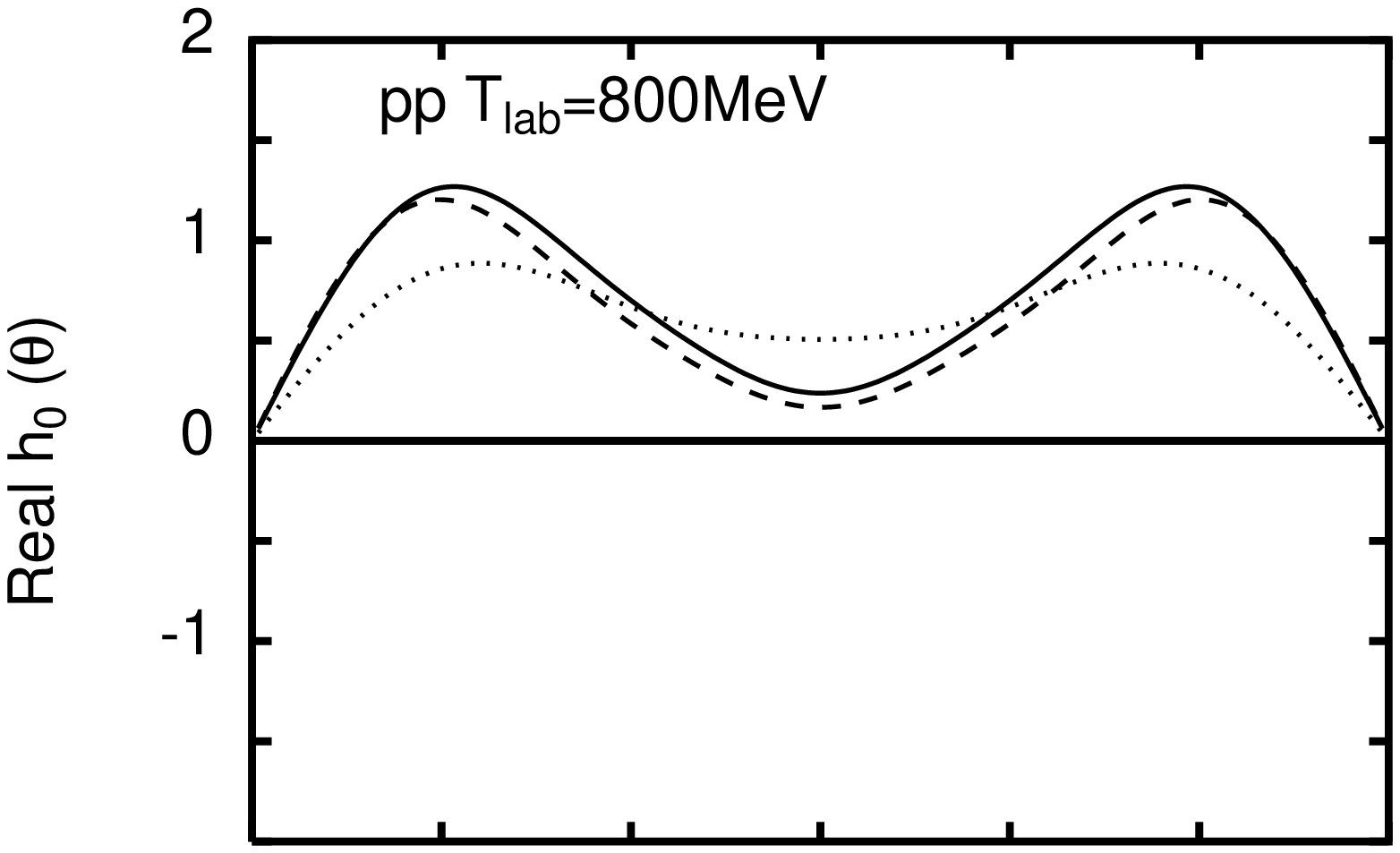}
\end{minipage}~%
\hfill~%
\begin{minipage}{0.47\textwidth}
\epsfxsize=\textwidth
\epsffile{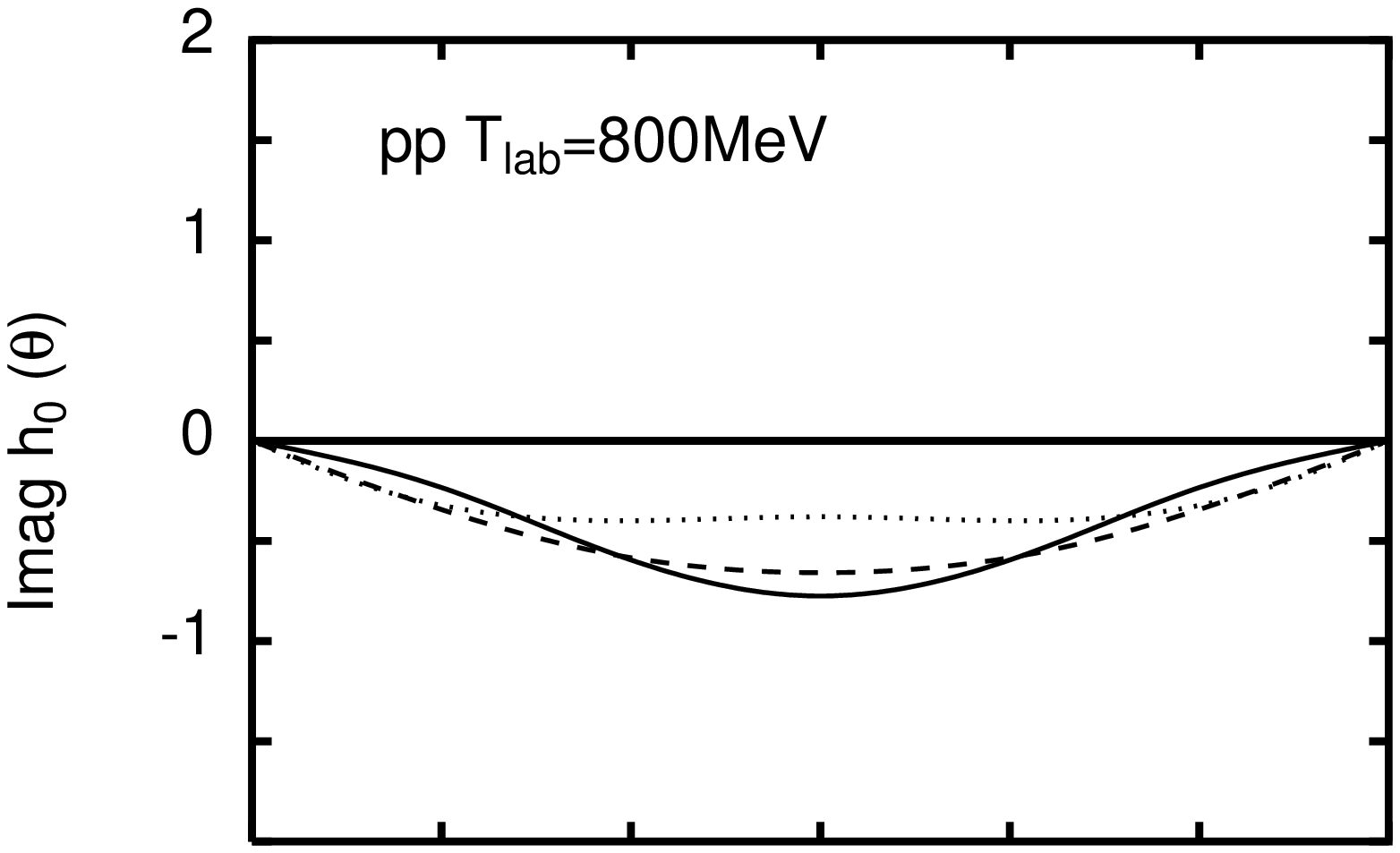}
\end{minipage}
\end{figure}

\vspace{-27.52mm}

\begin{figure}[h]
\begin{minipage}{0.47\textwidth}
\epsfxsize=\textwidth
\epsffile{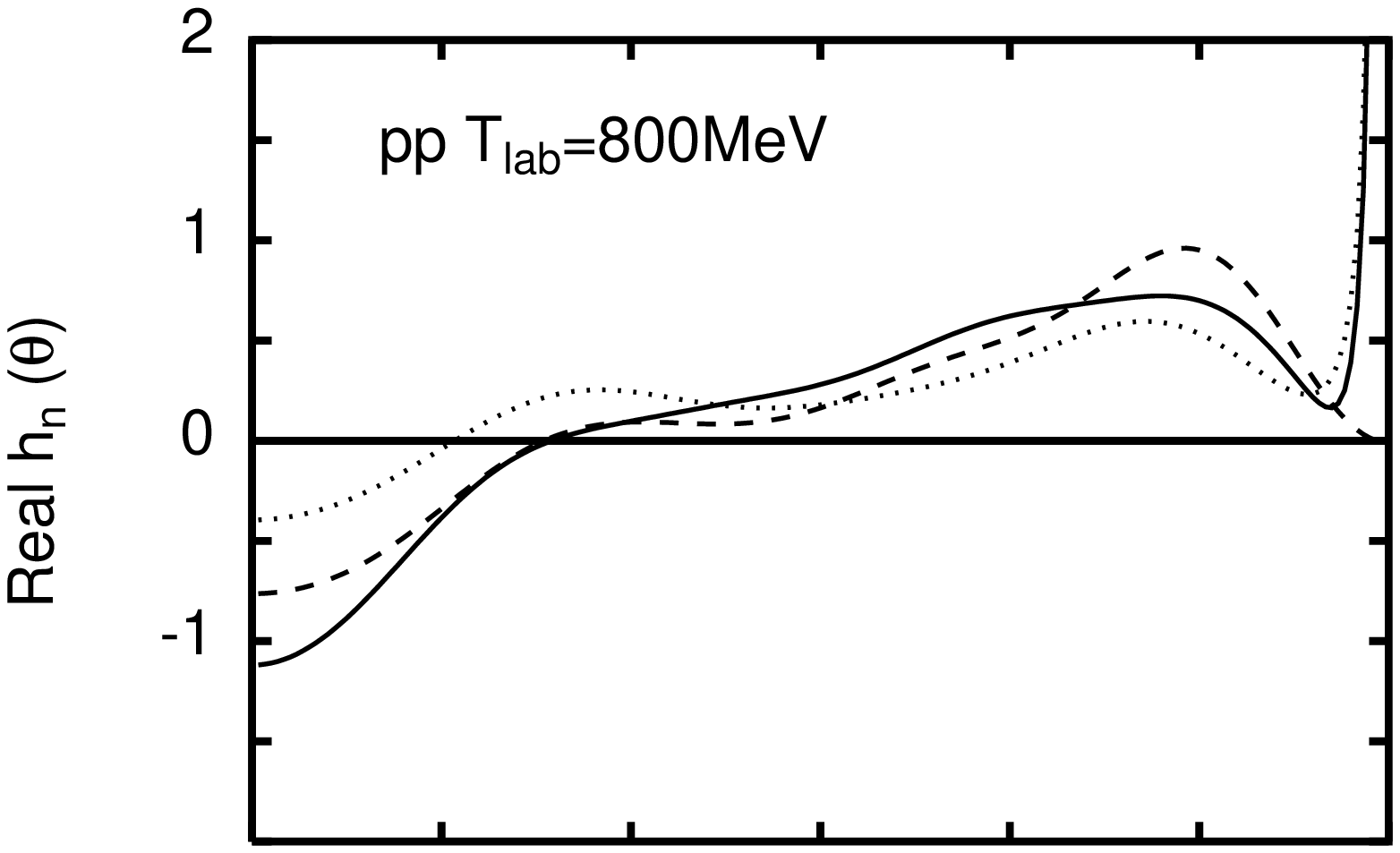}
\end{minipage}~%
\hfill~%
\begin{minipage}{0.47\textwidth}
\epsfxsize=\textwidth
\epsffile{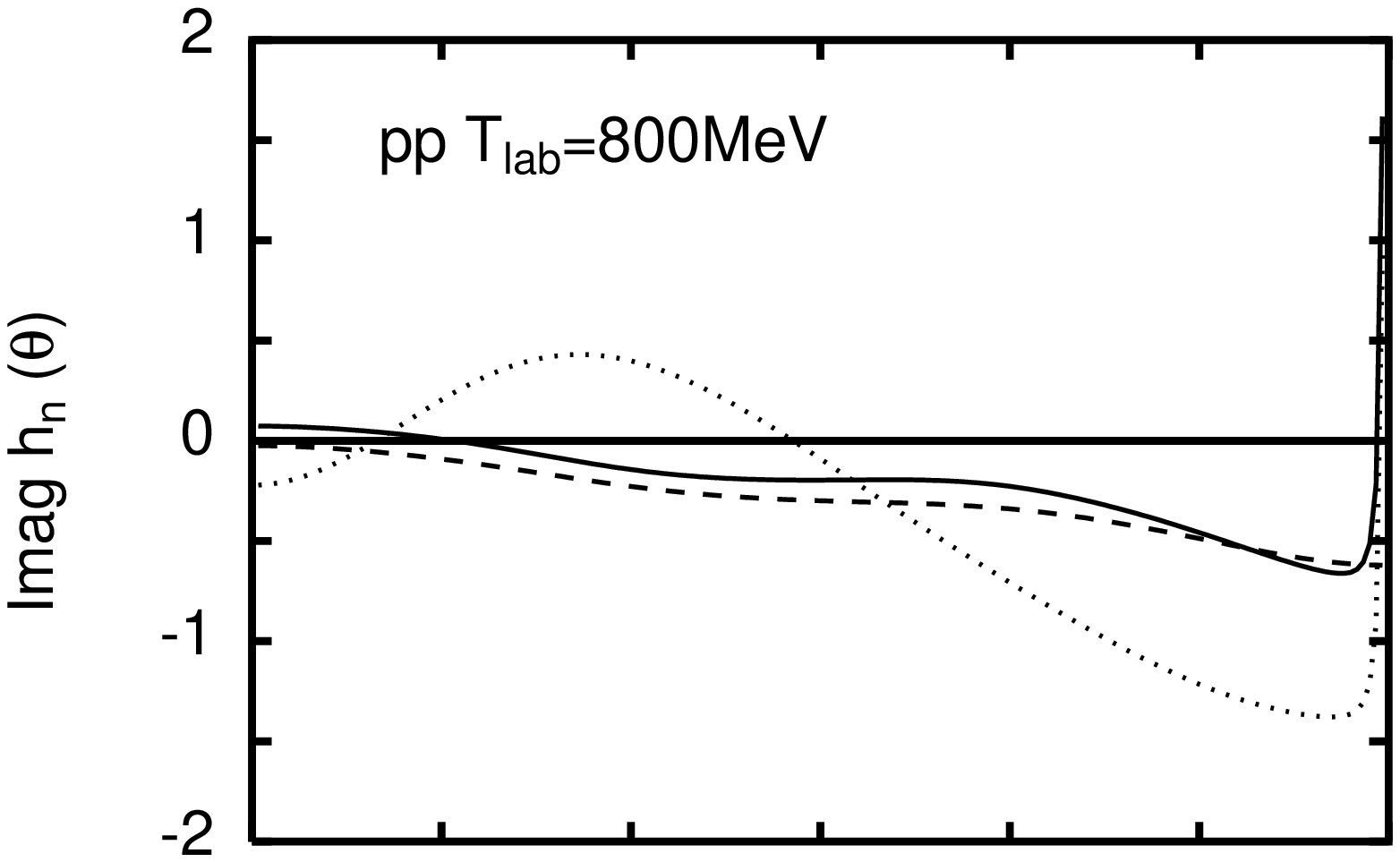}
\end{minipage}
\end{figure}

\vspace{-27.52mm}

\begin{figure}[h]
\begin{minipage}{0.47\textwidth}
\epsfxsize=\textwidth
\epsffile{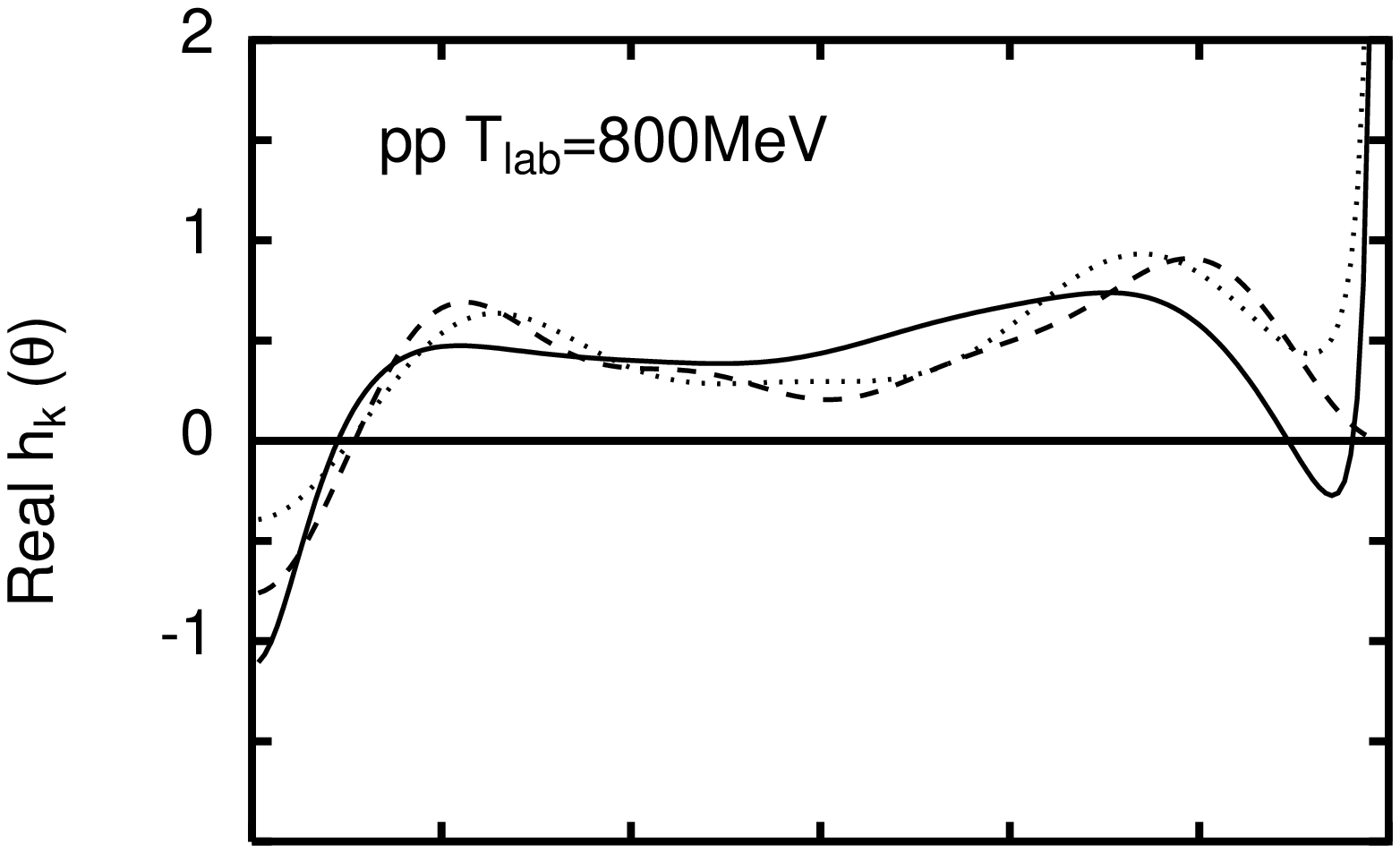}
\end{minipage}~%
\hfill~%
\begin{minipage}{0.47\textwidth}
\epsfxsize=\textwidth
\epsffile{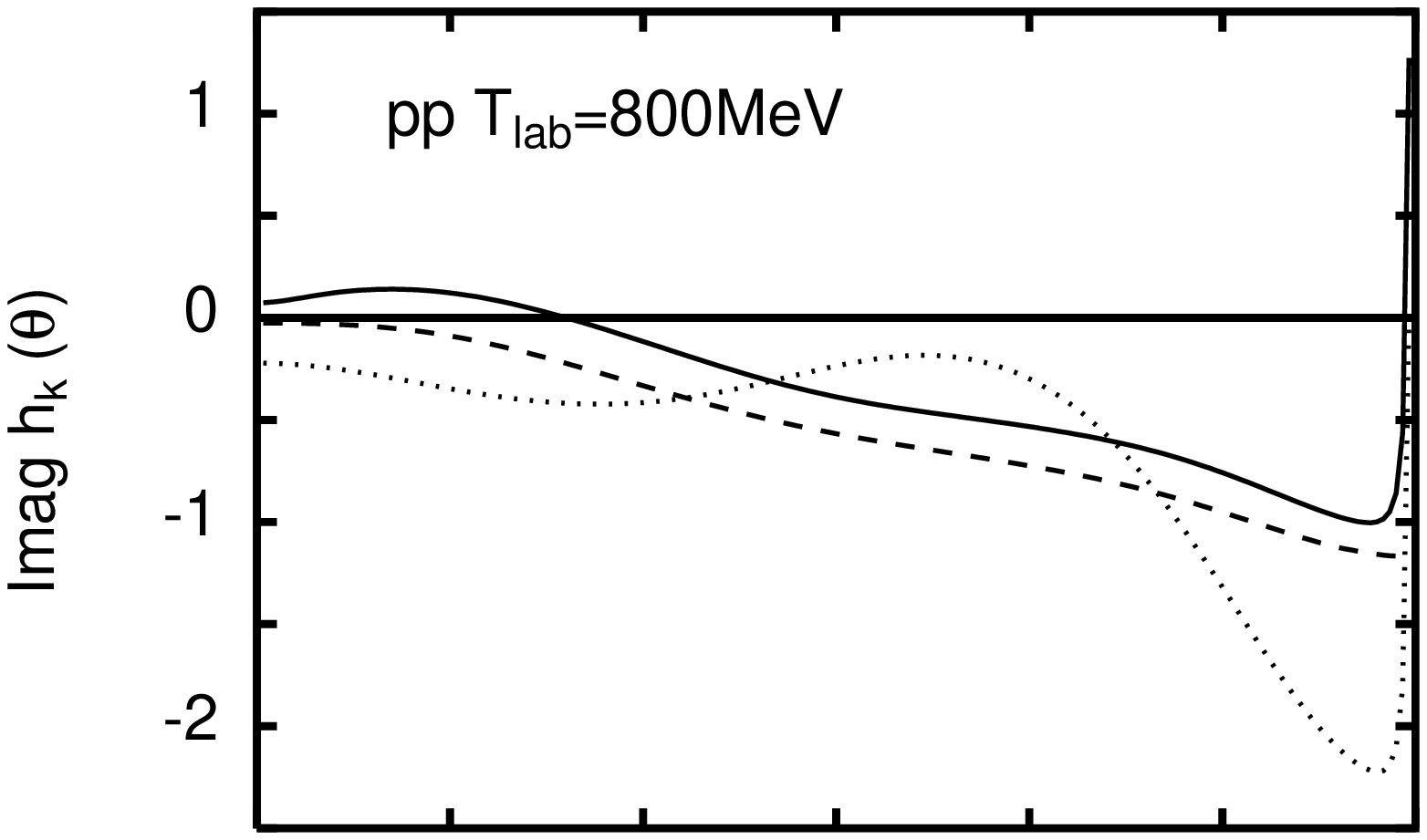}
\end{minipage}
\end{figure}

\vspace{-27.52mm}

\begin{figure}[h]
\begin{minipage}{0.47\textwidth}
\epsfxsize=\textwidth
\epsffile{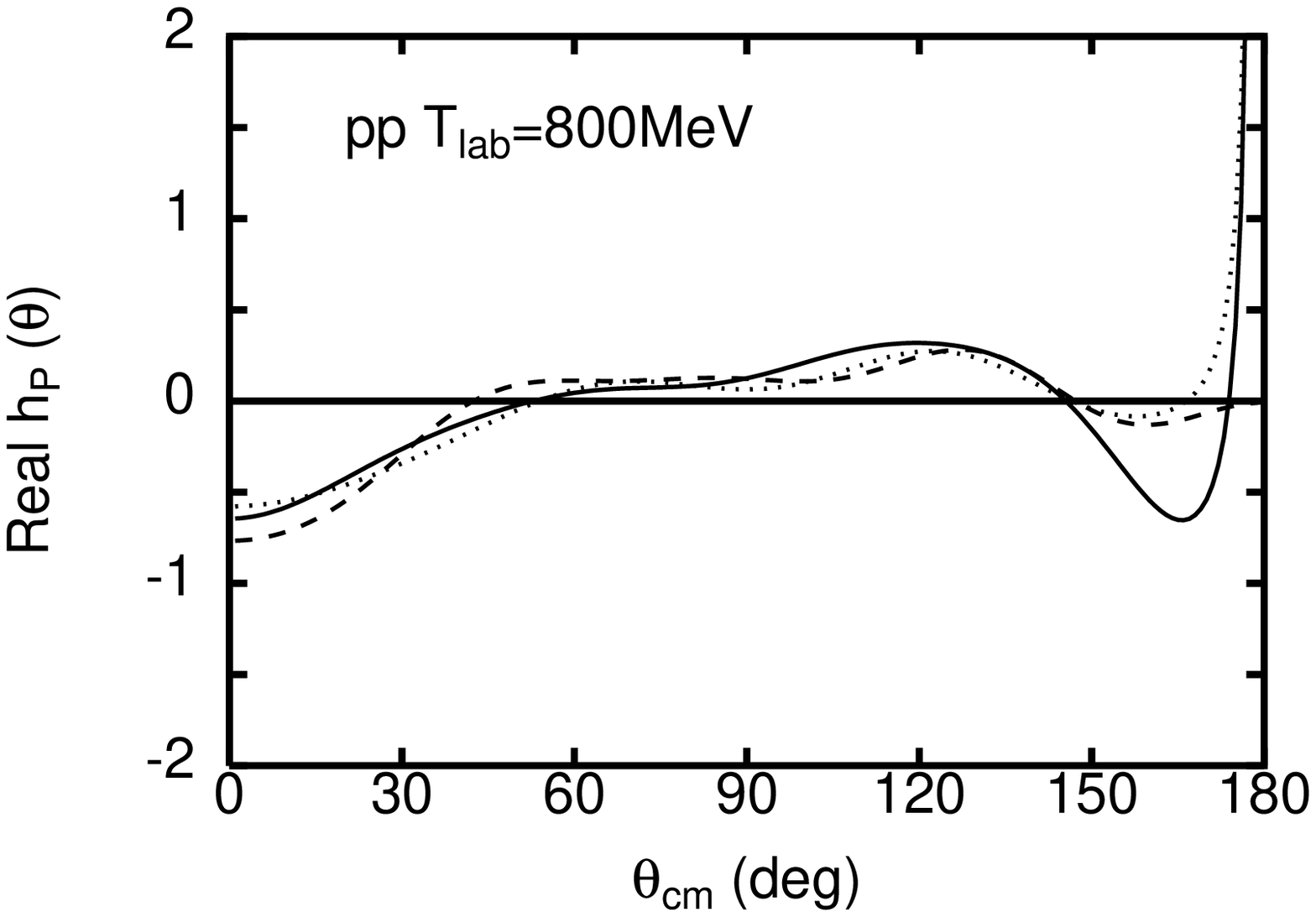}
\end{minipage}~%
\hfill~%
\begin{minipage}{0.47\textwidth}
\epsfxsize=\textwidth
\epsffile{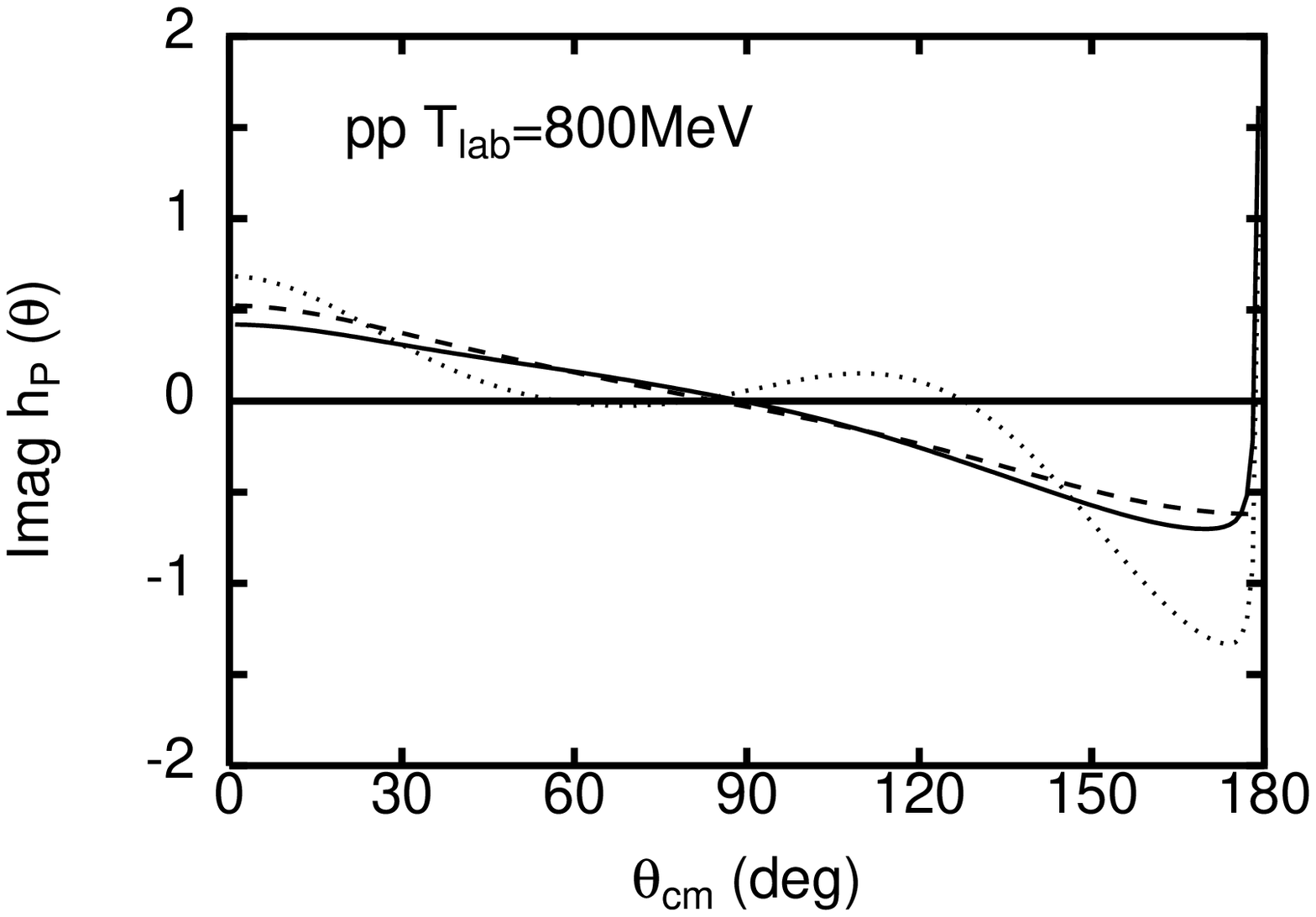}
\end{minipage}

\bigskip
\caption{
The five invariant amplitudes
for the $pp$ scattering at $T_{\rm lab}=$ 800 MeV,
calculated by fss2 (solid curves),
the Paris potential \protect\cite{PARI} (dashed curves)
and the empirical phase shifts SP99 \protect\cite{SAID} (dotted
curves). The Coulomb force is included in fss2 and SP99,
but not in the Paris potential.
}
\label{inv}
\end{figure}

\clearpage

\ \ 
\vspace{-20mm}

\begin{figure}[h]
\begin{minipage}{0.47\textwidth}
\epsfxsize=\textwidth
\epsffile{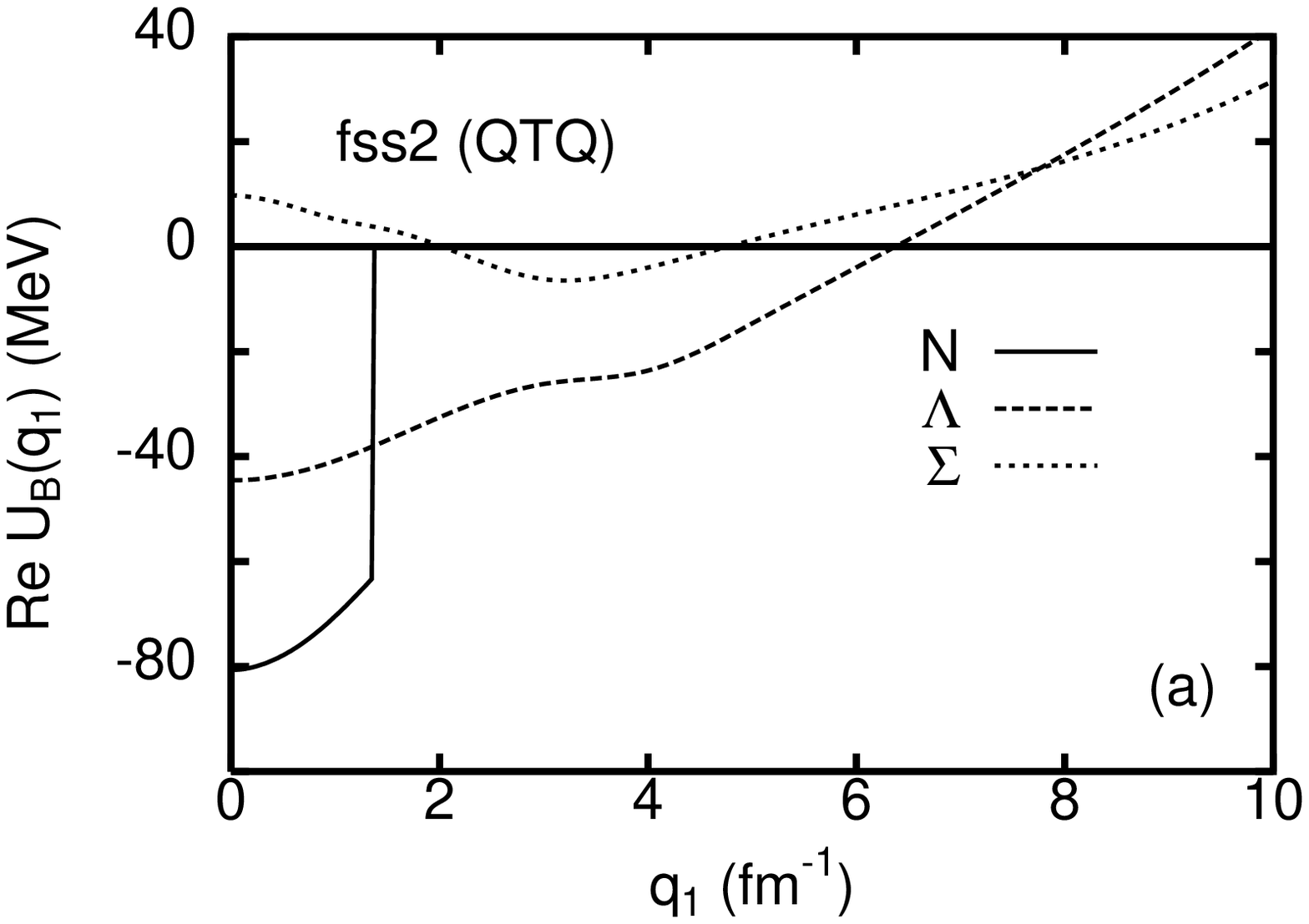}
\end{minipage}~%
\hfill~%
\begin{minipage}{0.47\textwidth}
\epsfxsize=\textwidth
\epsffile{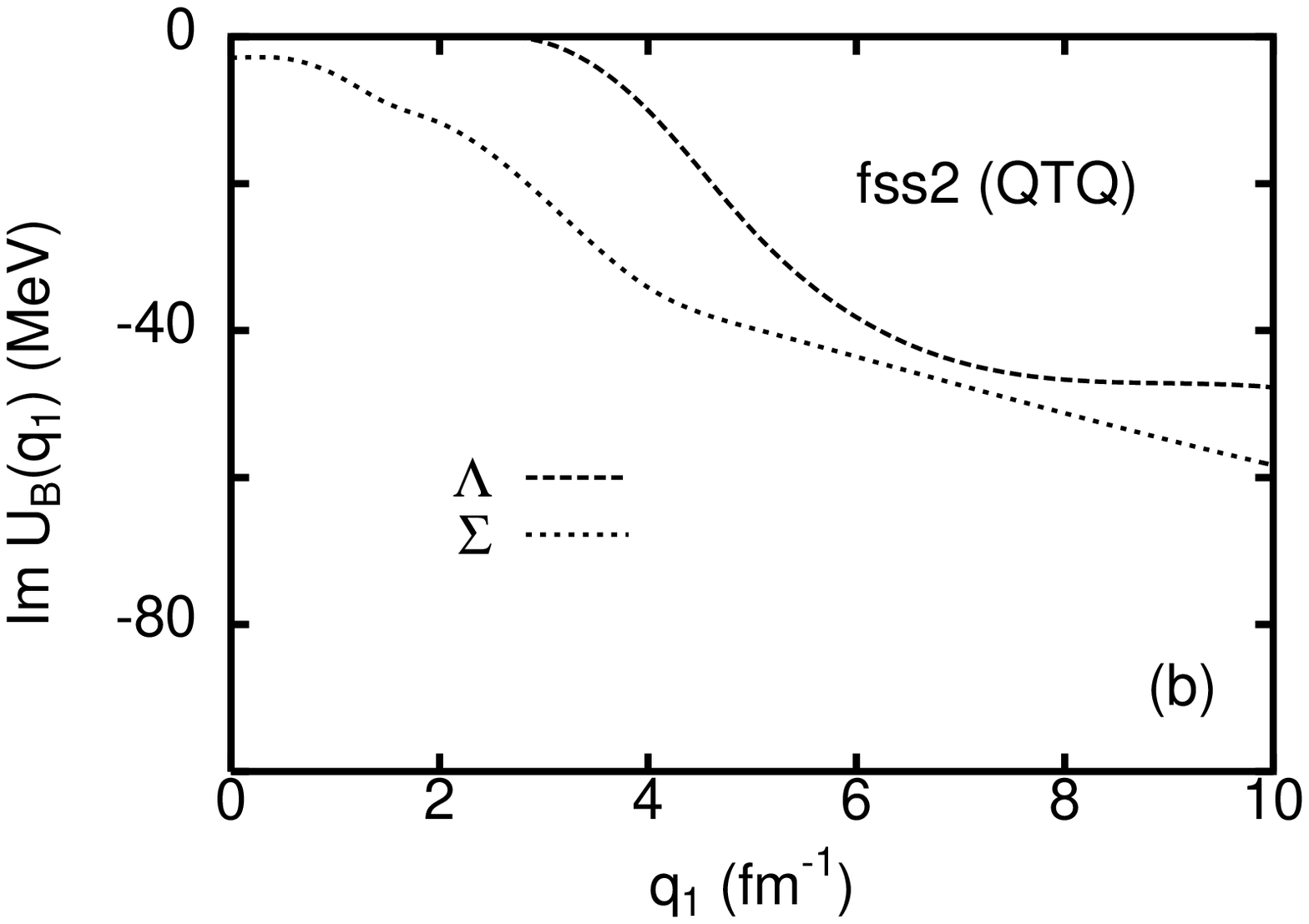}
\end{minipage}
\end{figure}

\vspace{-16mm}

\begin{figure}[h]
\begin{minipage}{0.47\textwidth}
\epsfxsize=\textwidth
\epsffile{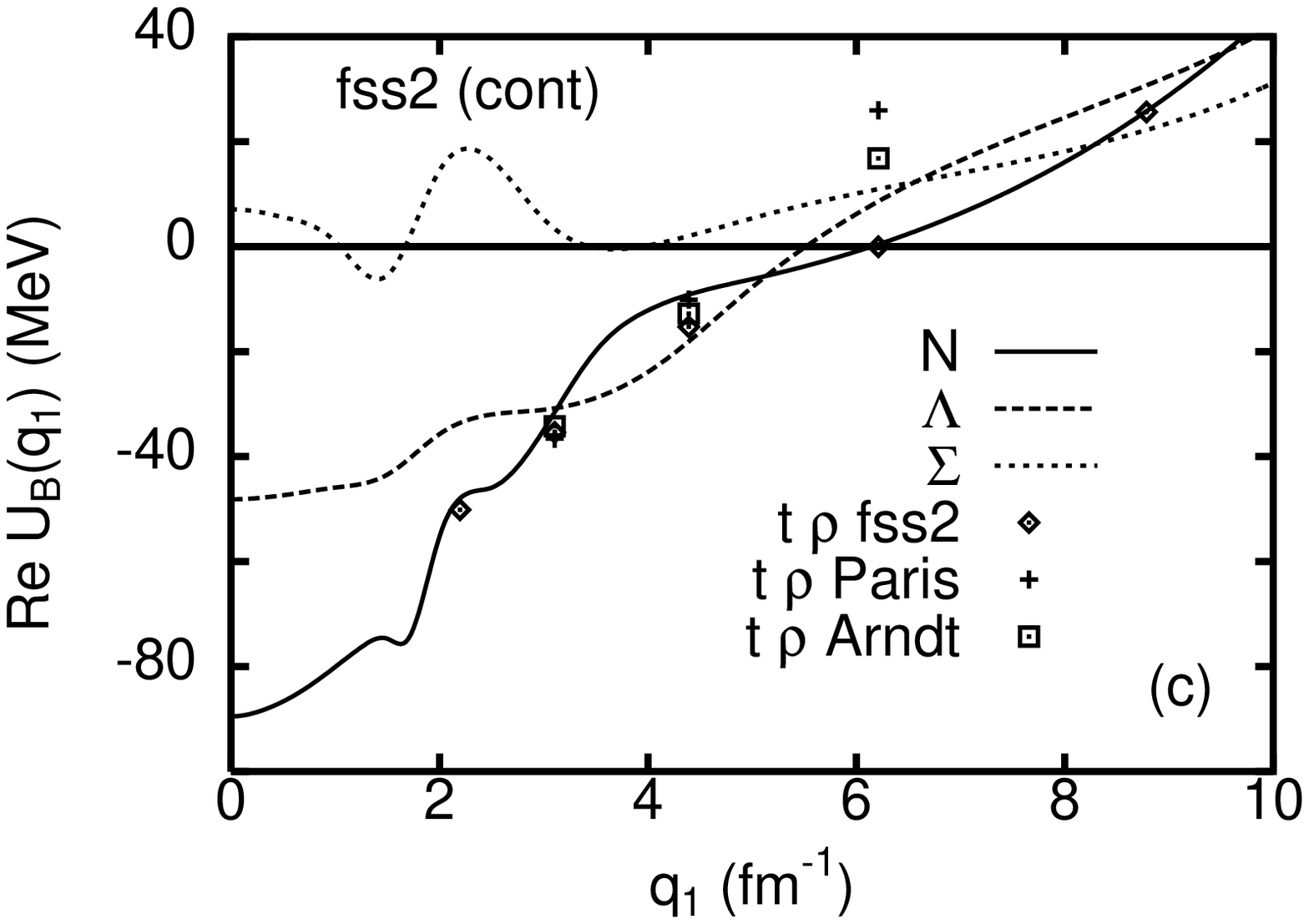}
\end{minipage}~%
\hfill~%
\begin{minipage}{0.47\textwidth}
\epsfxsize=\textwidth
\epsffile{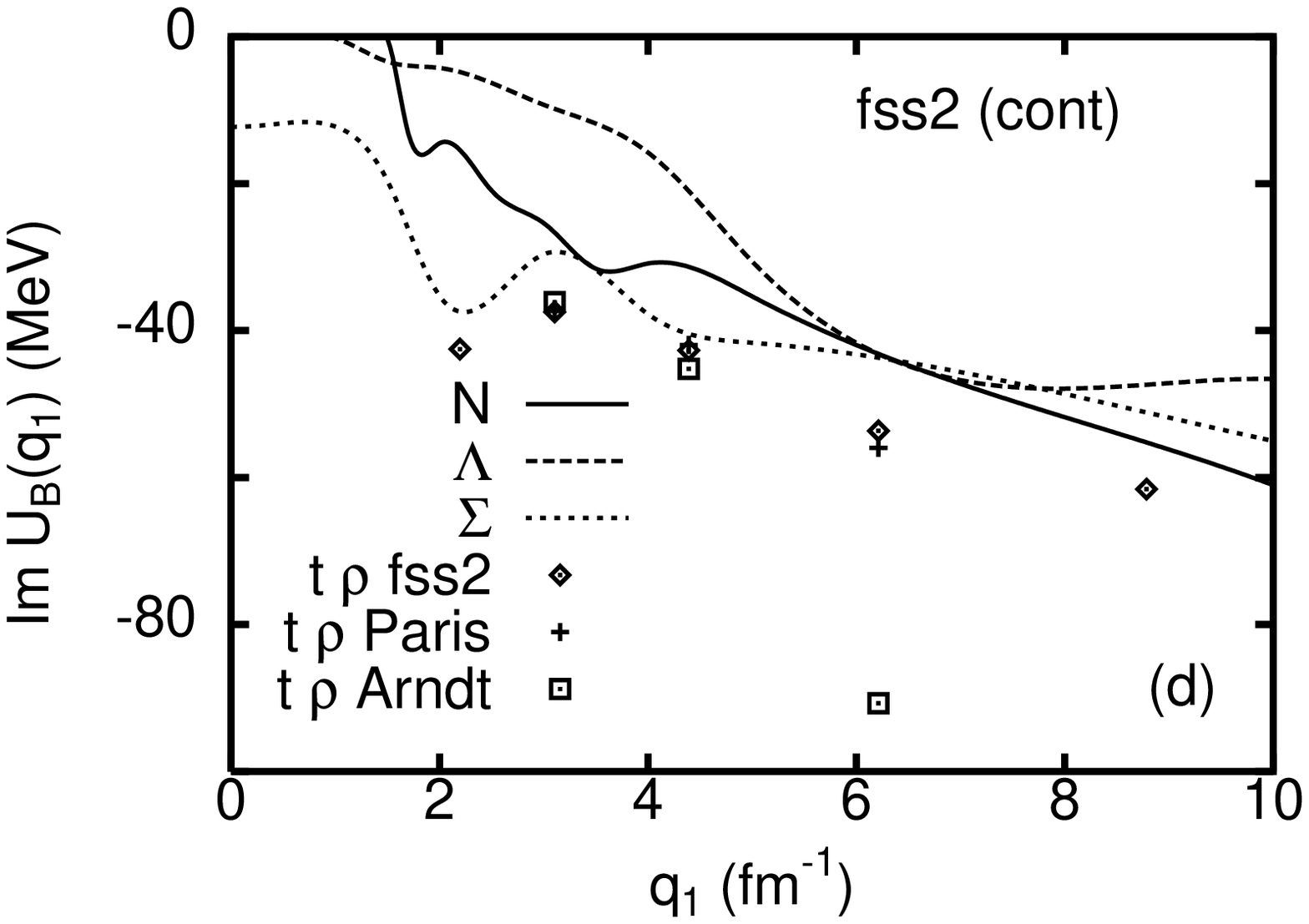}
\end{minipage}
\bigskip
\caption{
(a) The momentum dependence of the s.p. potentials $U_B(q_1)$
predicted by the $G$-matrix calculation of fss2.
The $QTQ$ prescription is used for intermediate spectra.
The real part $\Re e\,U_B(q_1)$ is shown.
(b) The same as (a) but for the imaginary part $\Im m\,U_B(q_1)$.
(c) The same as (a) but in the continuous prescription
for intermediate spectra.
The nucleon s.p. potentials obtained
by the $t^{\rm eff}\rho$ prescription are also shown
with respect to the $T$-matrices of fss2,
the Paris potential \protect\cite{PARI}
and the empirical phase shifts SP99 \protect\cite{SAID}.
The momentum points selected correspond to $T_{\rm lab}=100$,
200, 400, 800, and 1600 MeV for the $NN$ scattering.
The partial waves up to $J\leq 8$ are included in fss2 and
the Paris potential, and $J\leq 7$ in SP99.
(d) The same as (c) but for the imaginary part $\Im m\,U_B(q_1)$.
}
\label{trho}
\end{figure}

\begin{figure}[h]
\epsfxsize=0.47\textwidth
\epsffile{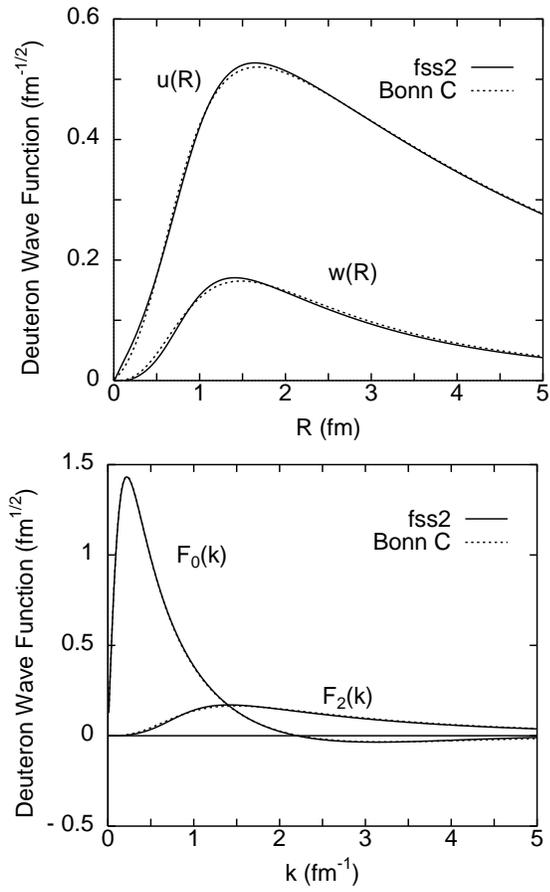}
\bigskip
\caption{
The deuteron wave functions predicted by fss2 (solid curves) and
by Bonn model-C \protect\cite{MA89} in the coordinate (upper) and
momentum (lower) representations.
}

\label{deutf}
\end{figure}

\begin{figure}[h]
\epsfxsize=0.47\textwidth
\epsffile{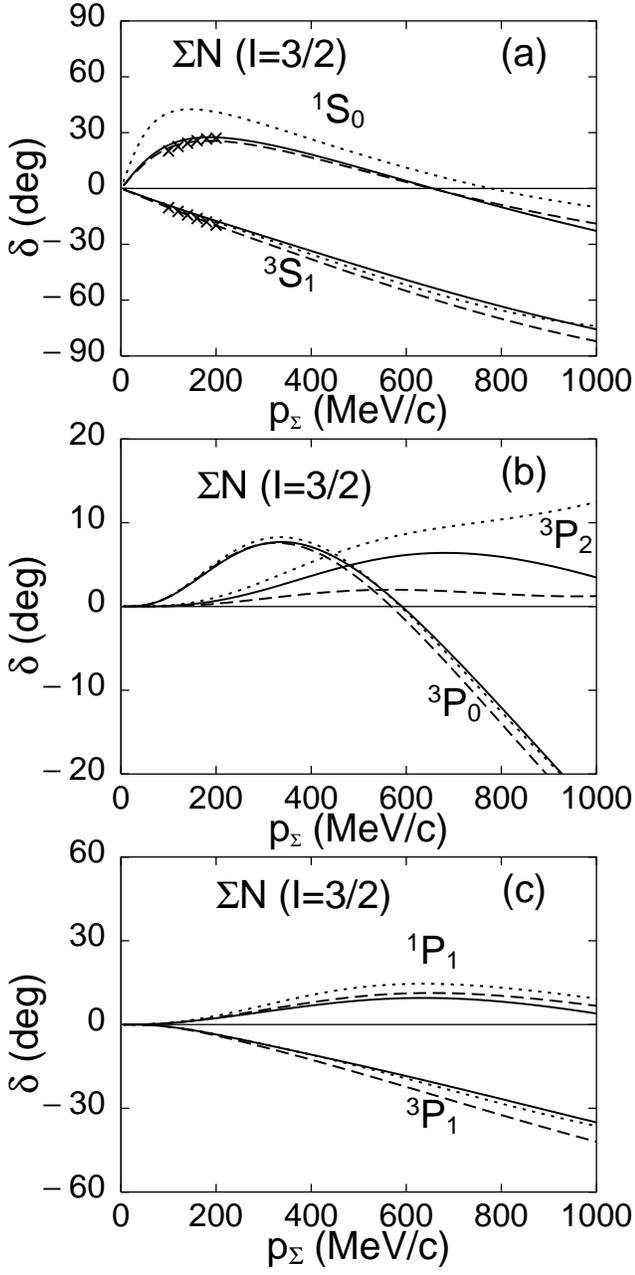}
\bigskip
\caption{The $S$- and $P$-wave phase shifts 
of the $\Sigma N(I=3/2)$ system, predicted by fss2 (solid curves).
(a): $^1S_0$, $^3S_1$, (b): $^3P_0$, $^3P_2$,
and (c): $^1P_1$, $^3P_1$.
Results by FSS (dashed curves) and RGM-H (dotted curves) are
also displayed.
}
\label{phsp}
\end{figure}

\begin{figure}[h]
\epsfxsize=\textwidth
\epsffile{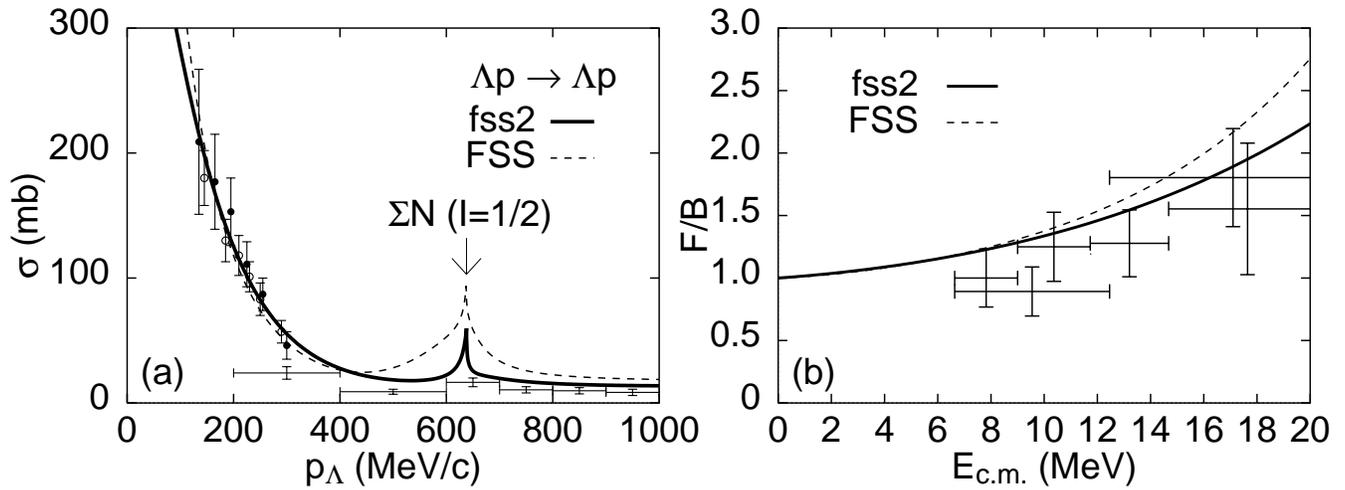}
\bigskip
\caption{(a) The total cross sections and (b) the forward
to backward ratio of the differential cross
sections ($F/B$) for the $\Lambda p$ elastic scattering,
predicted by fss2 in the isospin basis.
The results by FSS are also shown in dashed curves.
The experimental data are cited
from Refs. \protect\cite{alex68,sechi68,kadyk71}. 
}
\label{lato}
\end{figure}

\begin{figure}[h]
\epsfxsize=\textwidth
\epsffile{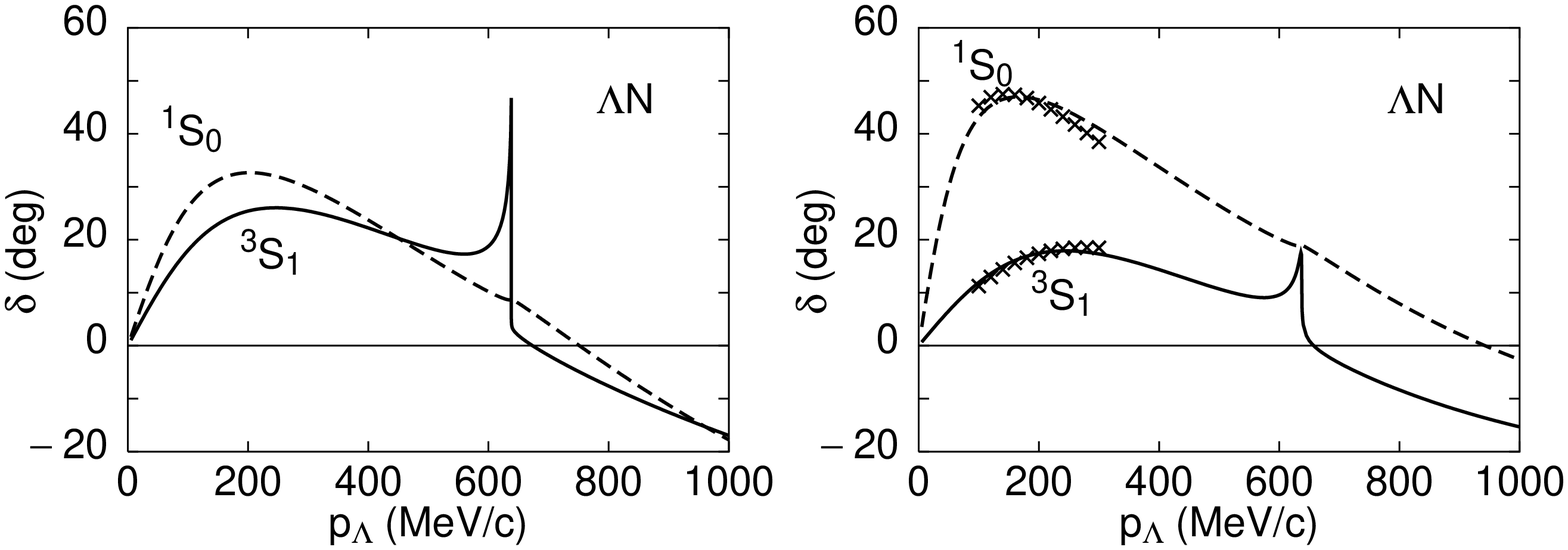}
\bigskip
\caption{The $\Lambda N$ $^3S_1$ and $^1S_0$ phase shifts
calculated by fss2 (left) and by FSS (right) in the isospin basis.}
\label{phlam1}
\end{figure}

\begin{figure}
\epsfxsize=\textwidth
\epsffile{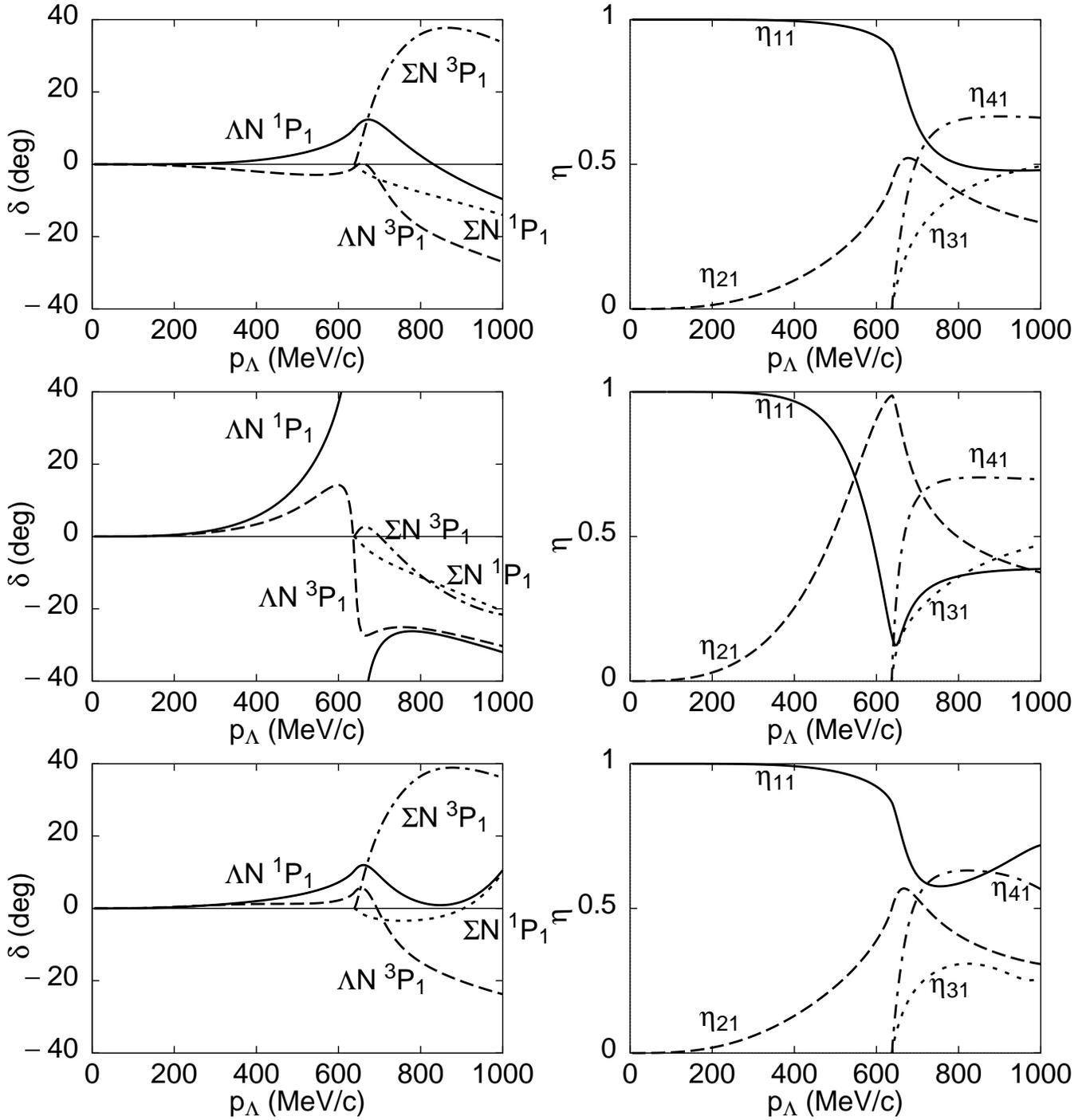}
\bigskip
\caption{The $S$-matrix of the $\Lambda N$ - $\Sigma N(I=1/2)$
$\hbox{}^1P_1$ - $\hbox{}^3P_1$ coupled-channel system,
predicted by our three models fss2 (upper), FSS (middle)
and RGM-H (lower) in the isospin-basis.
The diagonal phase shifts $\delta_{ii}$ (left-hand side),
and the reflection and transmission
coefficients $\eta_{i1}$ (right-hand side) defined
by $S_{ij}=\eta_{ij}e^{2i\delta_{ij}}$ are displayed.
The channels are specified by 1: $\Lambda N$ $\hbox{}^1P_1$,
2: $\Lambda N$ $\hbox{}^3P_1$, 3: $\Sigma N$ $\hbox{}^1P_1$
and 4: $\Sigma N$ $\hbox{}^3P_1$.
}
\label{phlam2}
\end{figure}

\begin{figure}[h]
\epsfxsize=0.47\textwidth
\epsffile{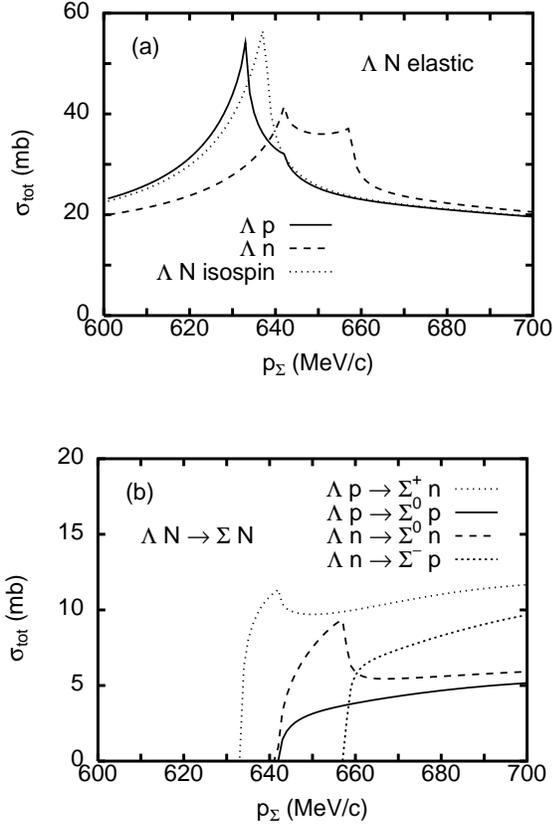}
\bigskip
\caption{(a) The $\Lambda N$ elastic total cross sections
and (b) $\Lambda N \longrightarrow \Sigma N$ reaction total
cross sections in the threshold region, predicted by fss2
with the full pion-Coulomb correction.
In (a), the $\Lambda N$ elastic total cross section
in the isospin basis (without the Coulomb force) is
also shown by the dotted curve.
}
\label{cusp}
\end{figure}

\begin{figure}[h]
\epsfxsize=\textwidth
\epsffile{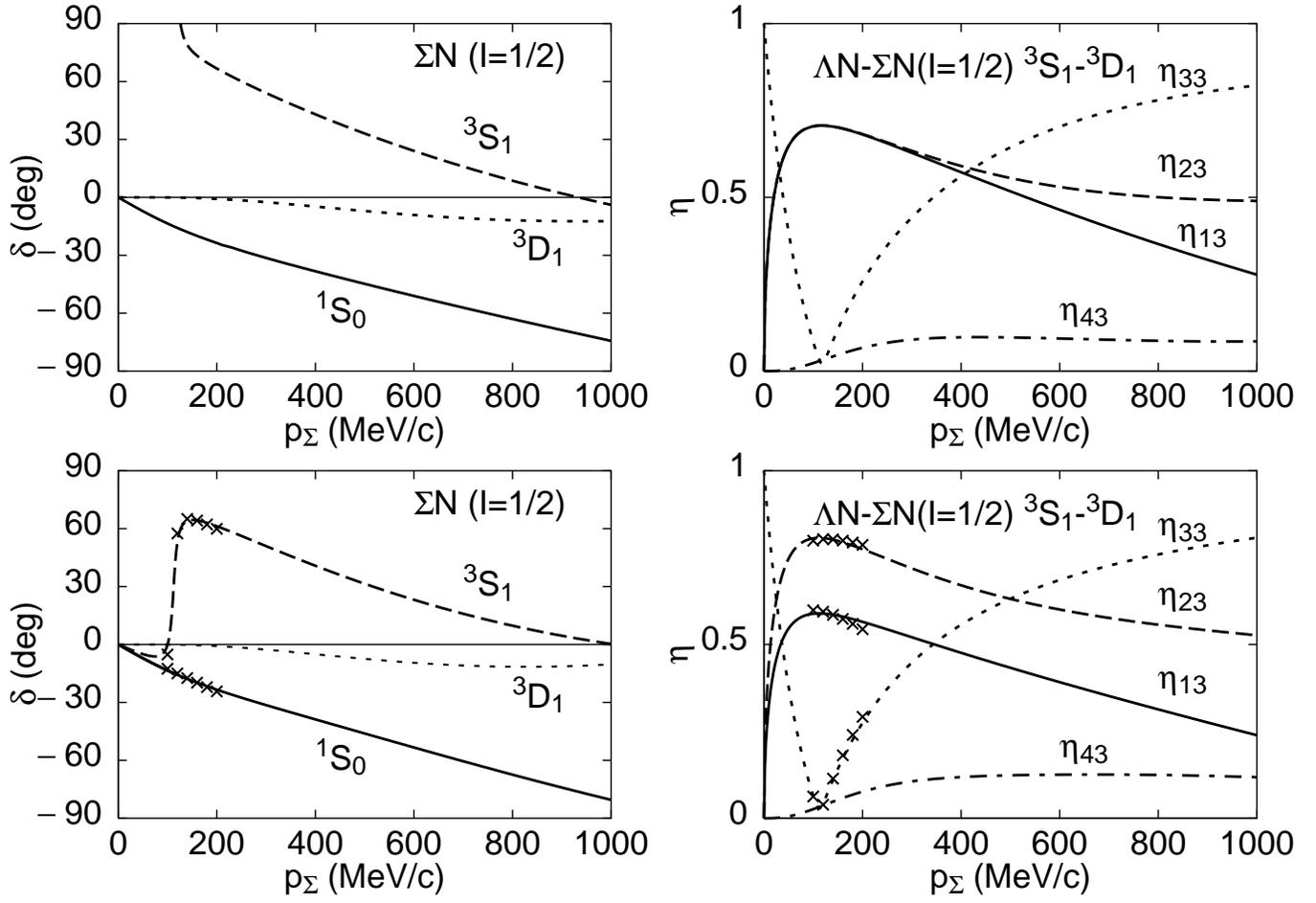}
\bigskip
\caption{The same as Fig.\,\protect{\ref{phlam2}} but
for the $^1S_0$ phase shift and for the $S$-matrix
of the $\Lambda N$ - $\Sigma N (I=1/2)$
$\hbox{}^3S_1$ - $\hbox{}^3D_1$ coupled-channel state.
In the coupled-channel system, the channels are specified
by 1: $\Lambda N$ $\hbox{}^3S_1$, 2: $\Lambda N$ $\hbox{}^3D_1$,
3: $\Sigma N$ $\hbox{}^3S_1$ and 4: $\Sigma N$ $\hbox{}^3D_1$.
The upper figures display the result given by fss2,
while the lower by FSS.
}
\label{phsgn}
\end{figure}


\begin{figure}[h]
\epsfxsize=0.47\textwidth
\epsffile{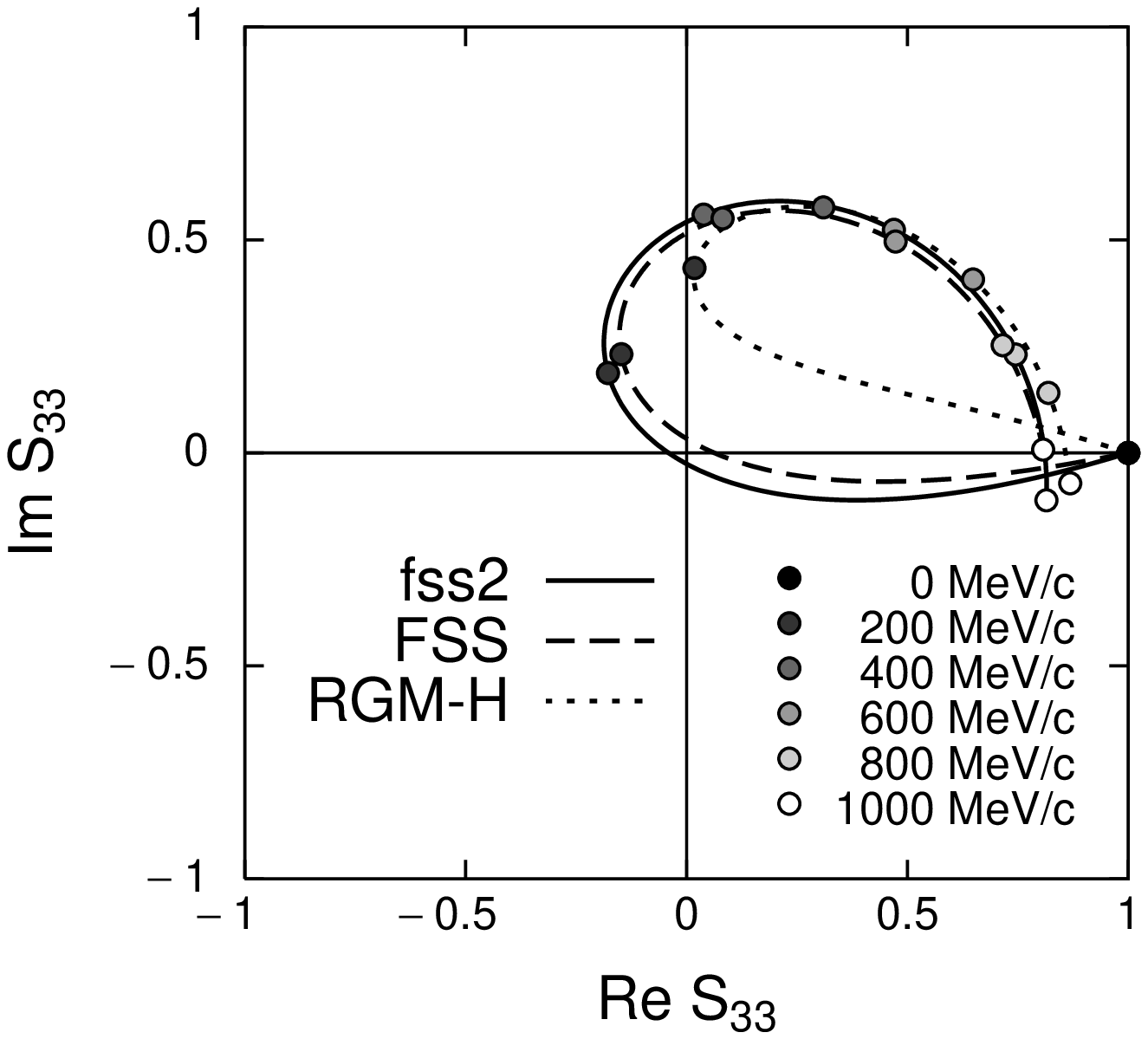}
\bigskip
\caption{Argand diagram of the $S$-matrix element $S_{33}$ for
the $\Lambda N$ - $\Sigma N (I=1/2)$
$\hbox{}^3S_1$ - $\hbox{}^3D_1$ coupled-channel system.
The channel $i=3$ corresponds to the $\Sigma N (I=1/2)$
$\hbox{} ^3S_1$ channel.
The range of the incident momentum is $0 \le
p_{\Sigma}\le 1000$ MeV/$c$.
Results given by fss2 (solid curve), FSS (dashed curve) and
RGM-H (dotted curve) are displayed.
}
\label{sgsmat}
\end{figure}

\begin{figure}[h]
\begin{minipage}{0.47\textwidth}
\epsfxsize=\textwidth
\epsffile{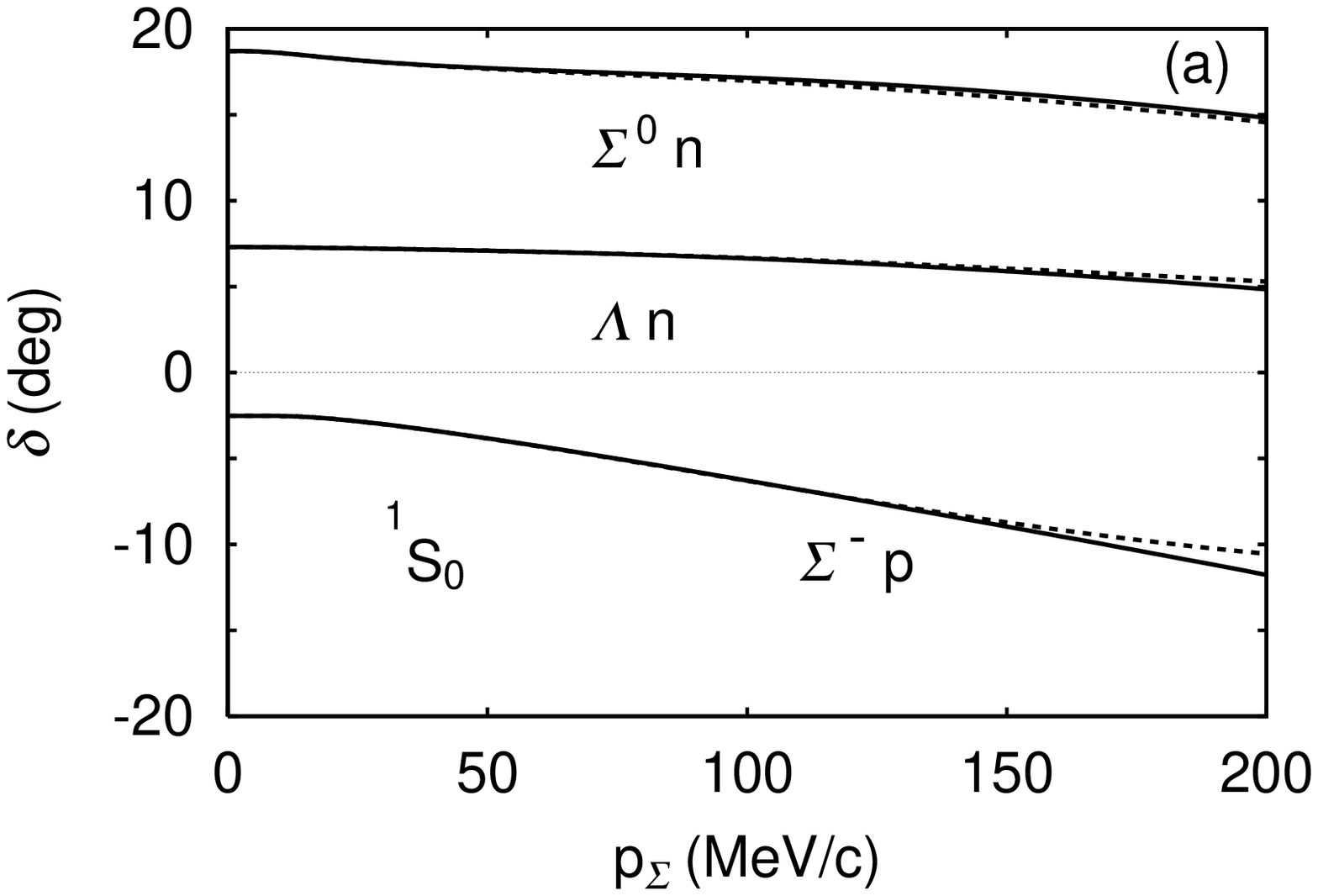}
\end{minipage}~%
\hfill~%
\begin{minipage}{0.47\textwidth}
\epsfxsize=\textwidth
\epsffile{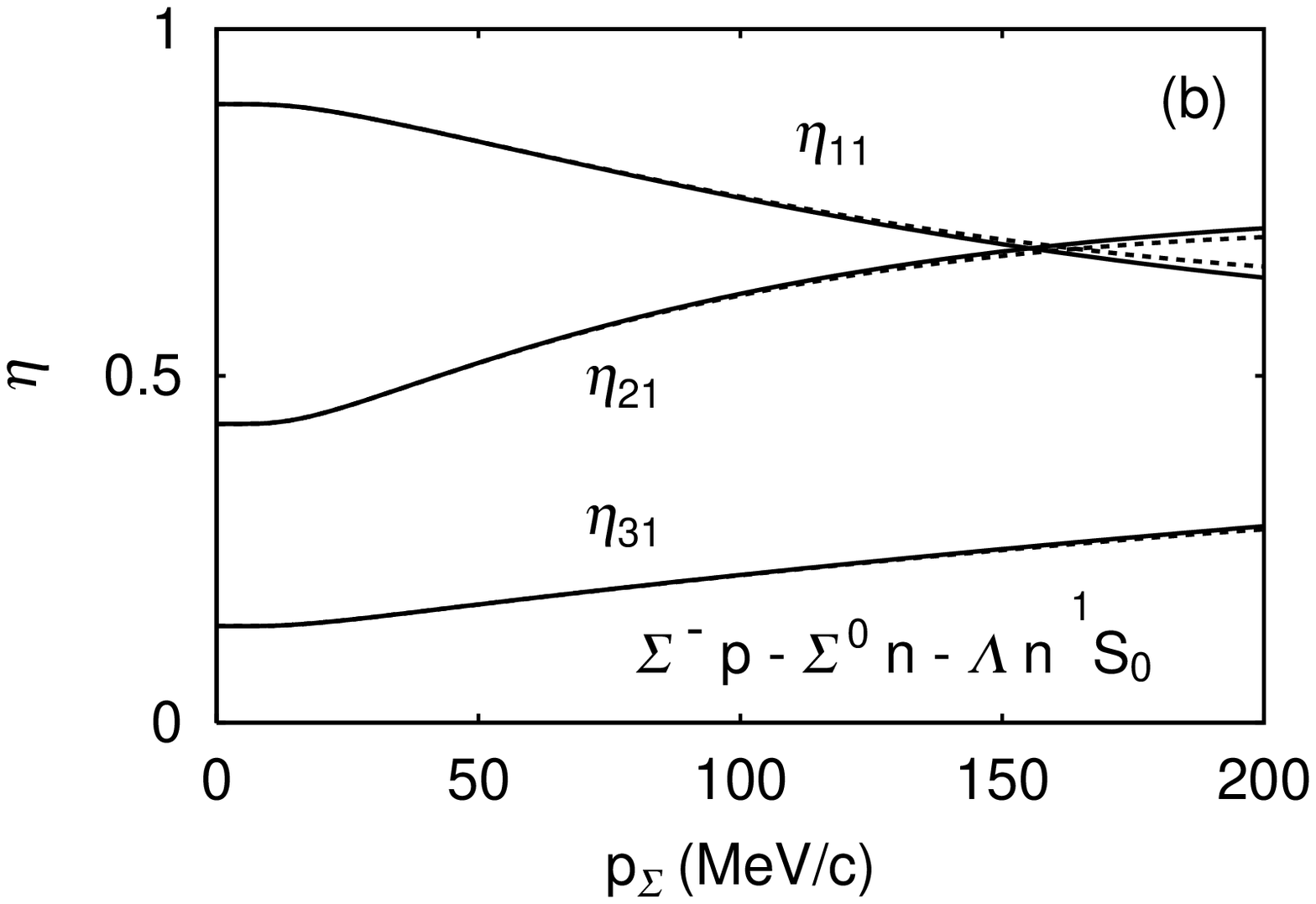}
\end{minipage}
%
\bigskip
\caption{(a) The low-energy $\Sigma^- p$, $\Sigma^0 n$
and $\Lambda n$ phase shifts
for the $\hbox{}^1S_0$ channel, predicted by fss2,
as a function of the incident momentum $p_\Sigma$.
The solid curves include the full pion-Coulomb correction
in the particle basis,
while the dashed curves are obtained
in the effective range approximation
using the parameters in Table \protect\ref{effect2}.
(b) $\Sigma^- p$ reflection and transmission coefficients $\eta_{i1}$
for the $\hbox{}^1S_0$ channel, predicted by fss2.
The final channel $i$ is specified
by 1: $\Sigma^- p$, 2: $\Sigma^0 n$ and 3: $\Lambda n$.
The solid curves denote the full calculation,
while the dashed curves the effective range approximation.
}
\label{part1}
\end{figure}

\begin{figure}[h]
\begin{minipage}{0.47\textwidth}
\epsfxsize=\textwidth
\epsffile{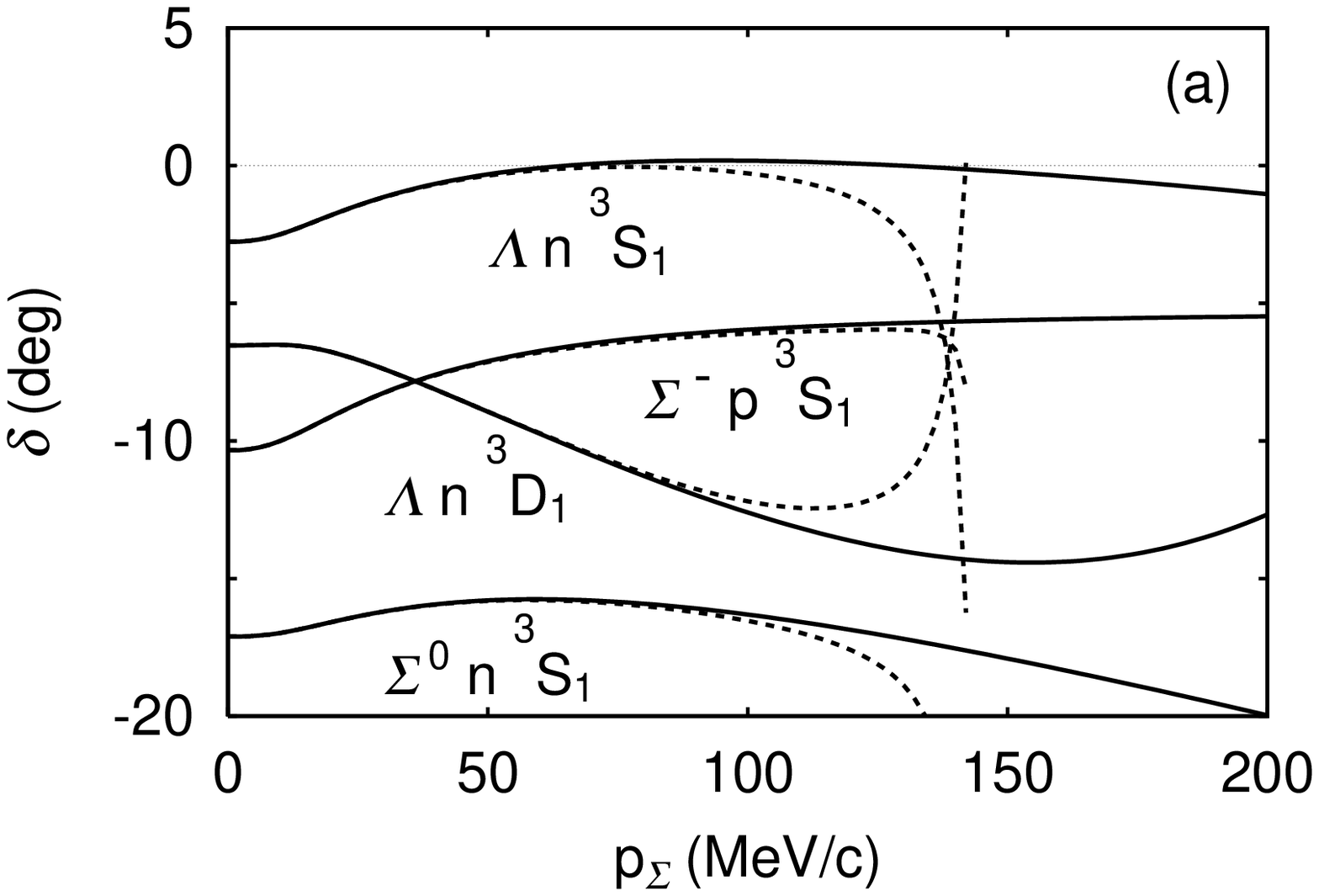}
\end{minipage}~%
\hfill~%
\begin{minipage}{0.47\textwidth}
\epsfxsize=\textwidth
\epsffile{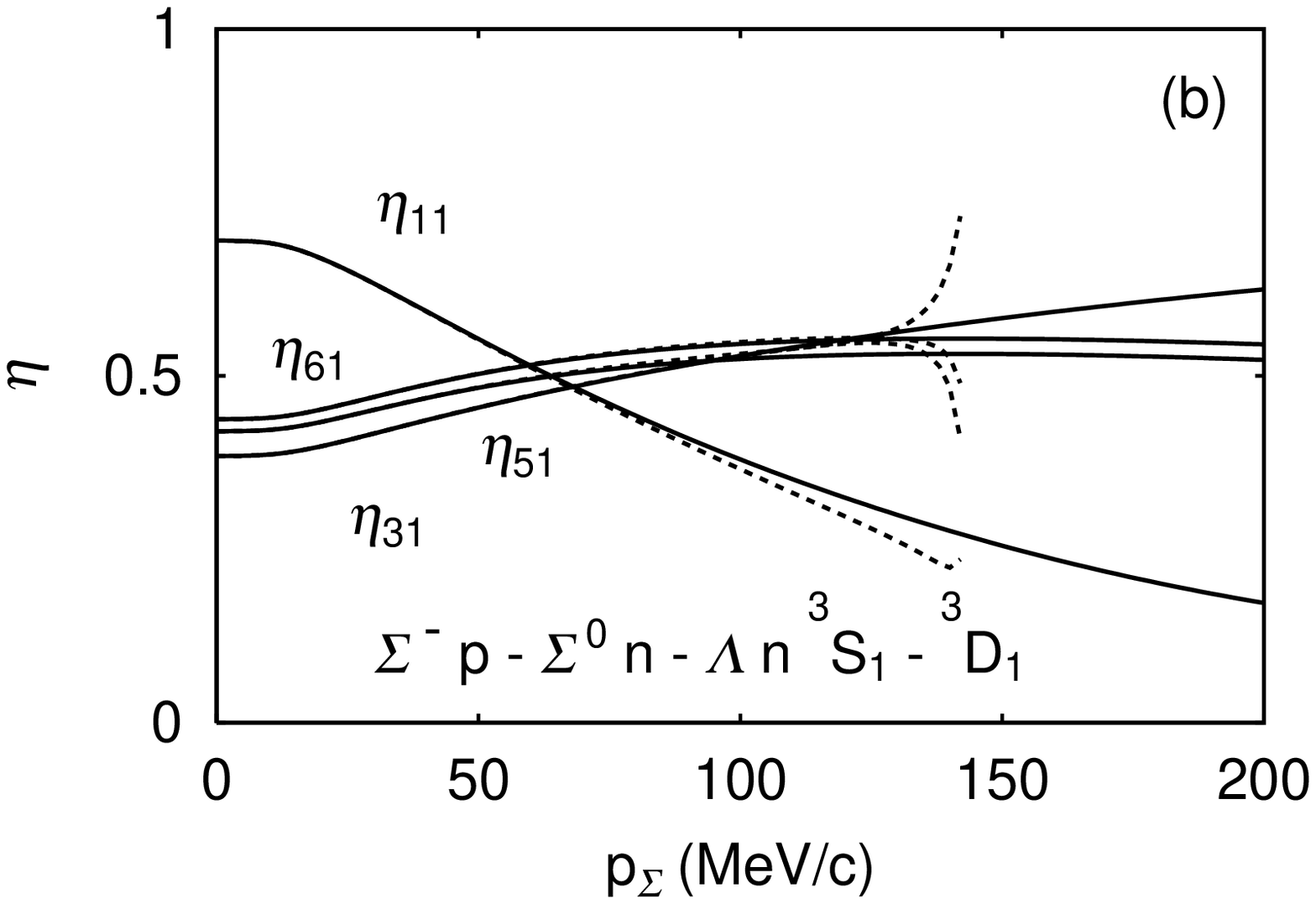}
\end{minipage}
\bigskip
\caption{
(a) $\Sigma^- p$, $\Sigma^0 n$ and $\Lambda n$ phase shifts
for the $\hbox{}^3S_1+\hbox{}^3D_1$ state, predicted by fss2,
as a function of the incident momentum $p_\Sigma$.
$\Sigma^- p$ $\hbox{}^3D_1$ phase shift is not shown,
since it is smaller than $0.1^\circ$.
The full pion-Coulomb correction is included in the solid curves.
The dashed curves are obtained
in the effective range approximation
using the parameters in Table \protect\ref{effect3}.
(b) $\Sigma^- p$ reflection and transmission coefficients $\eta_{i1}$
for the $\hbox{}^3S_1+\hbox{}^3D_1$ channel, predicted by fss2.
The final channel $i$ is specified
by 1: $\Sigma^- p$ $\hbox{}^3S_1$, 2: $\Sigma^- p$ $\hbox{}^3D_1$,
3: $\Sigma^0 n$ $\hbox{}^3S_1$, 4: $\Sigma^0 n$ $\hbox{}^3D_1$,
5: $\Lambda n$ $\hbox{}^3S_1$ and 6: $\Lambda n$ $\hbox{}^3D_1$.
$\eta_{21}$ and $\eta_{41}$ are not shown, since they are very small.
The solid curves include the full pion-Coulomb correction,
while the dashed curves are obtained in the effective range
approximation.
}
\label{part2}
\end{figure}

\begin{figure}[h]
\begin{minipage}{0.47\textwidth}
\epsfxsize=\textwidth
\epsffile{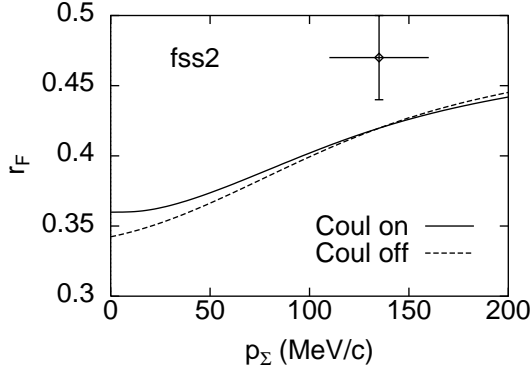}
\end{minipage}
\caption{
$\Sigma^- p$ inelastic capture ratio in flight $r_F$,
predicted by fss2, as a function of the incident
momentum $p_\Sigma$.
Calculation is made in the particle basis
with (solid curve) and without (dashed curve) the
Coulomb force.
}
\label{flight}
\end{figure}


\begin{figure}[h]
\begin{minipage}{0.47\textwidth}
\epsfxsize=\textwidth
\epsffile{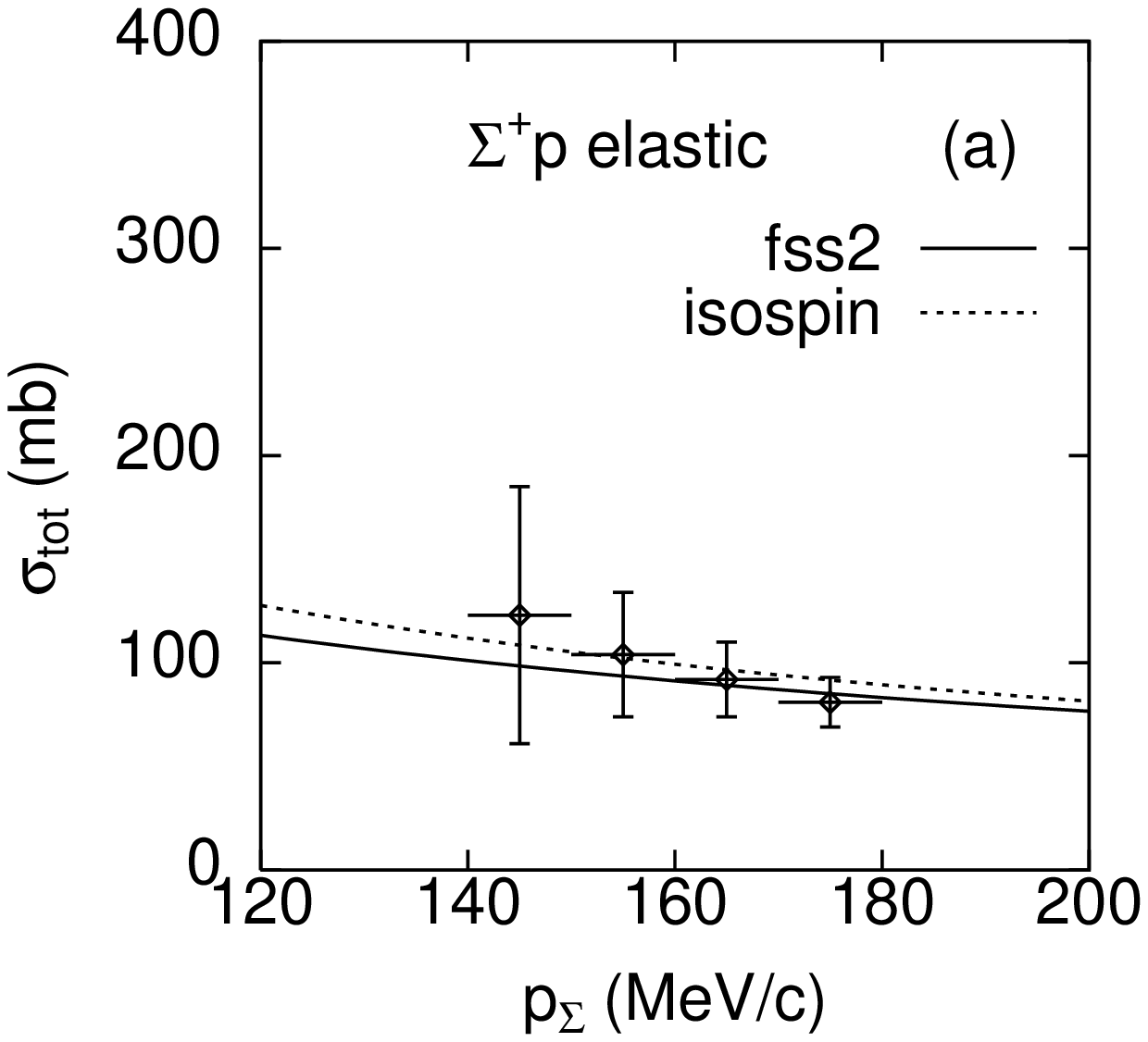}
\end{minipage}~%
\hfill~%
\begin{minipage}{0.47\textwidth}
\epsfxsize=\textwidth
\epsffile{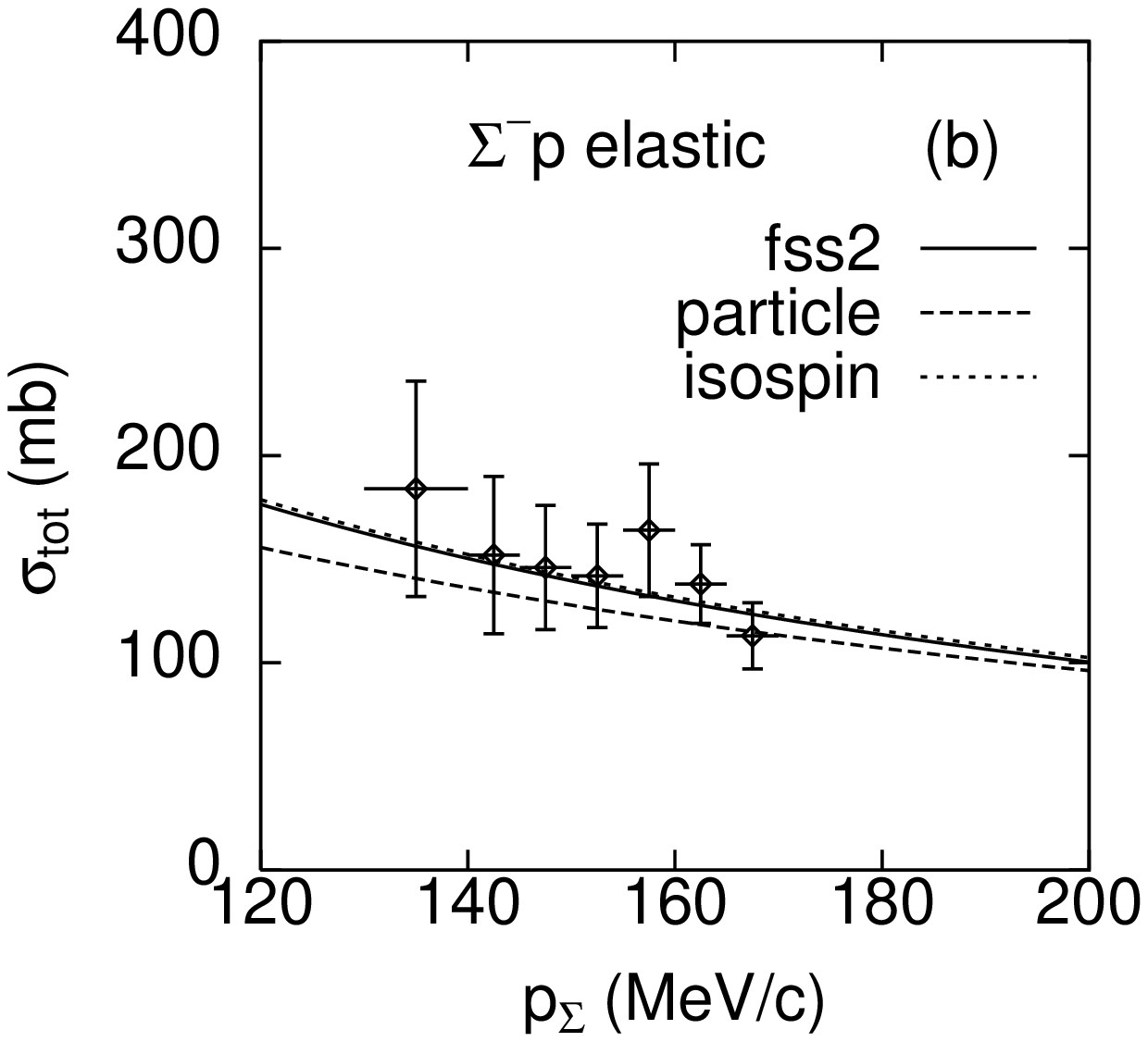}
\end{minipage}
\end{figure}

\vspace{-15mm}

\begin{figure}[h]
\begin{minipage}{0.47\textwidth}
\epsfxsize=\textwidth
\epsffile{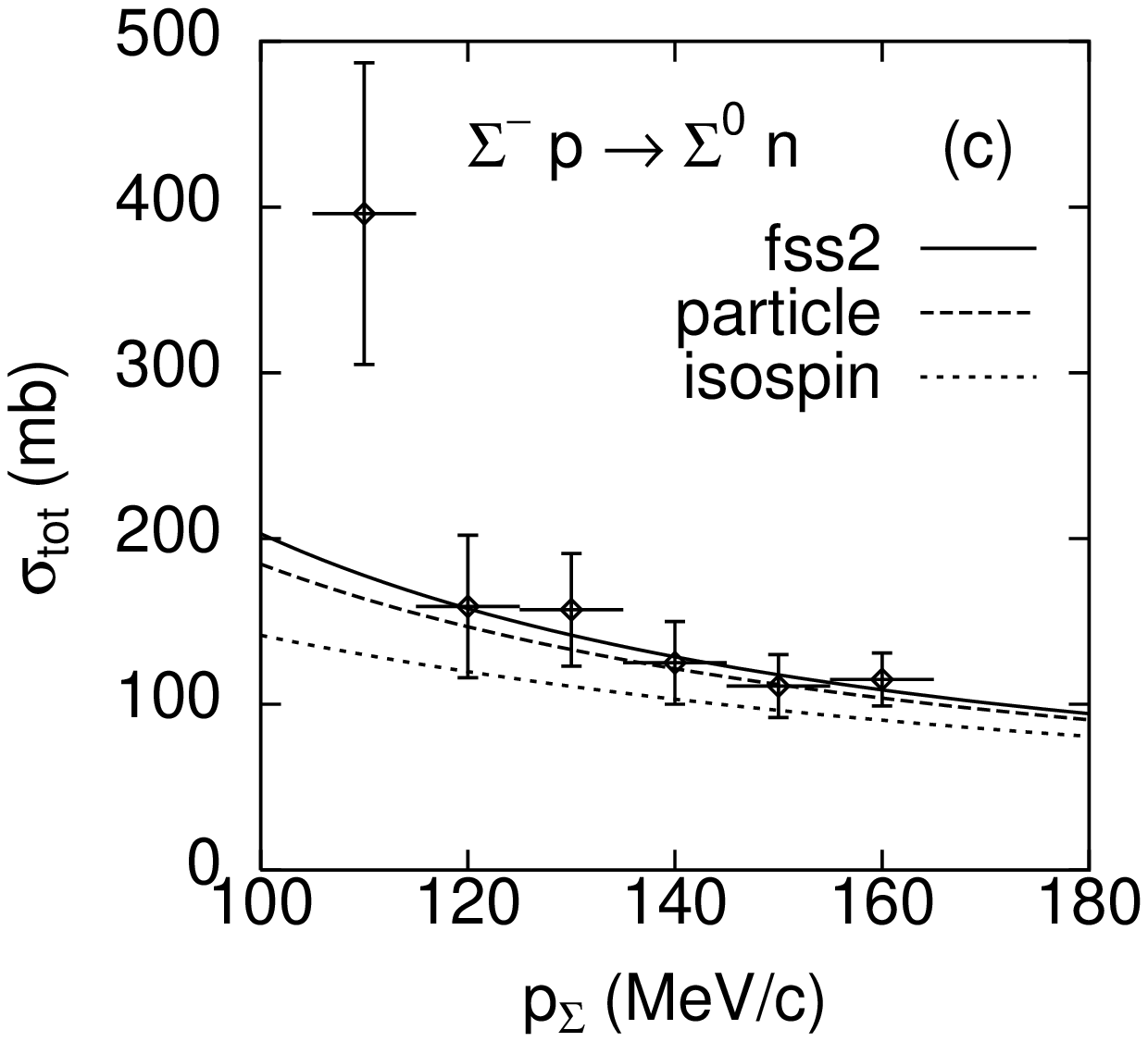}
\end{minipage}~%
\hfill~%
\begin{minipage}{0.47\textwidth}
\epsfxsize=\textwidth
\epsffile{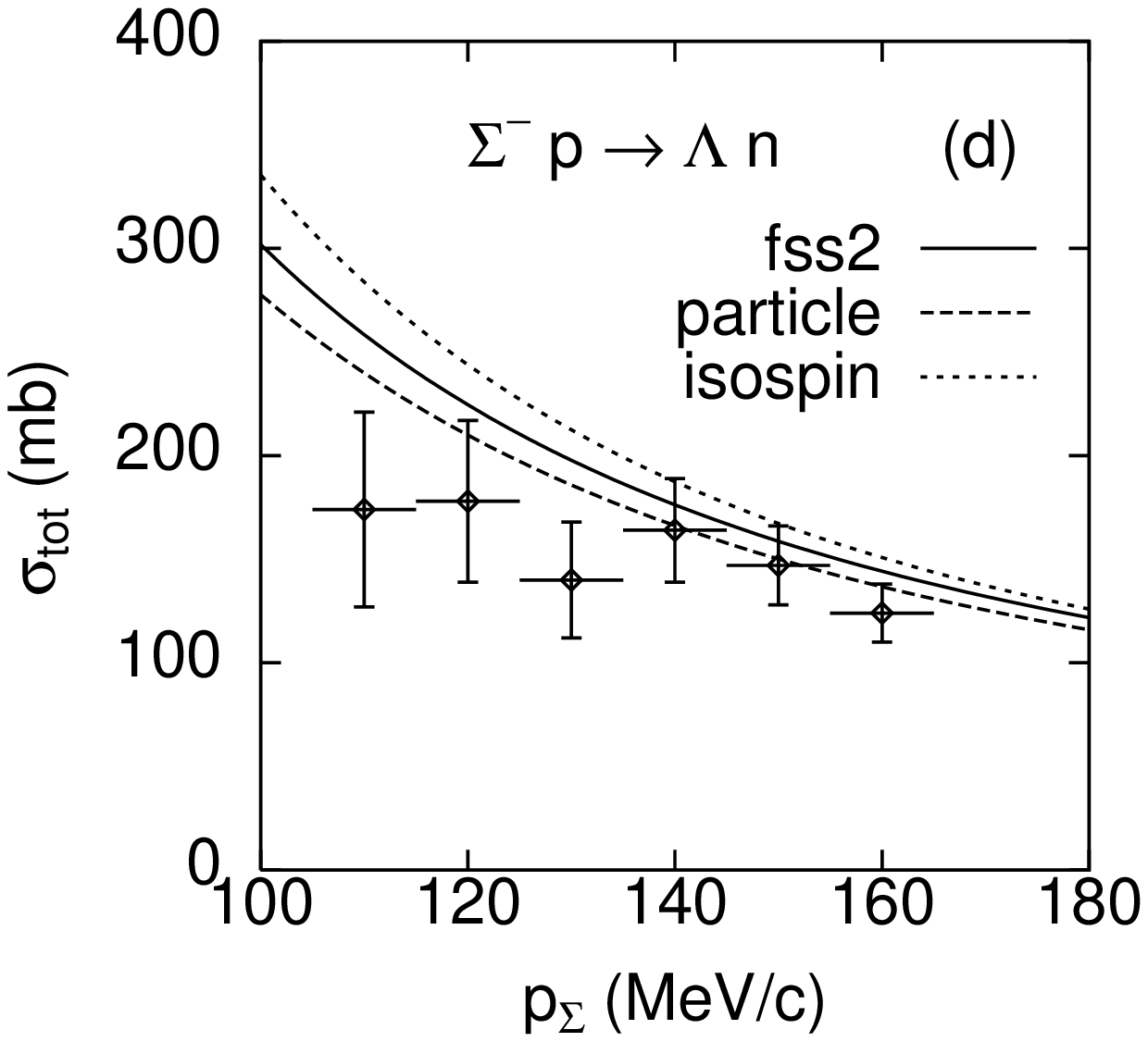}
\end{minipage}
%
\caption{
Calculated low-energy of $\Sigma^+ p$ and $\Sigma^- p$ scattering
total cross sections by fss2 (solid curves),
compared with experimental data:
(a) $\Sigma^+ p$ elastic,
(b) $\Sigma^- p$ elastic, (c) $\Sigma^- p \rightarrow \Sigma^0 n$
charge-exchange, (d) $\Sigma^- p \rightarrow \Lambda n$ reaction
cross sections.
Predictions in the particle basis without the Coulomb
force (dashed curves), and those in the isospin
basis (dotted curves) are also shown.
The experimental data are taken
from \protect\cite{EI71} for (a) and (b),
and from \protect\cite{EN66} for (c) and (d).
}
\label{sgto}
\end{figure}

\clearpage

\begin{figure}[h]
\begin{minipage}{0.47\textwidth}
\epsfxsize=\textwidth
\epsffile{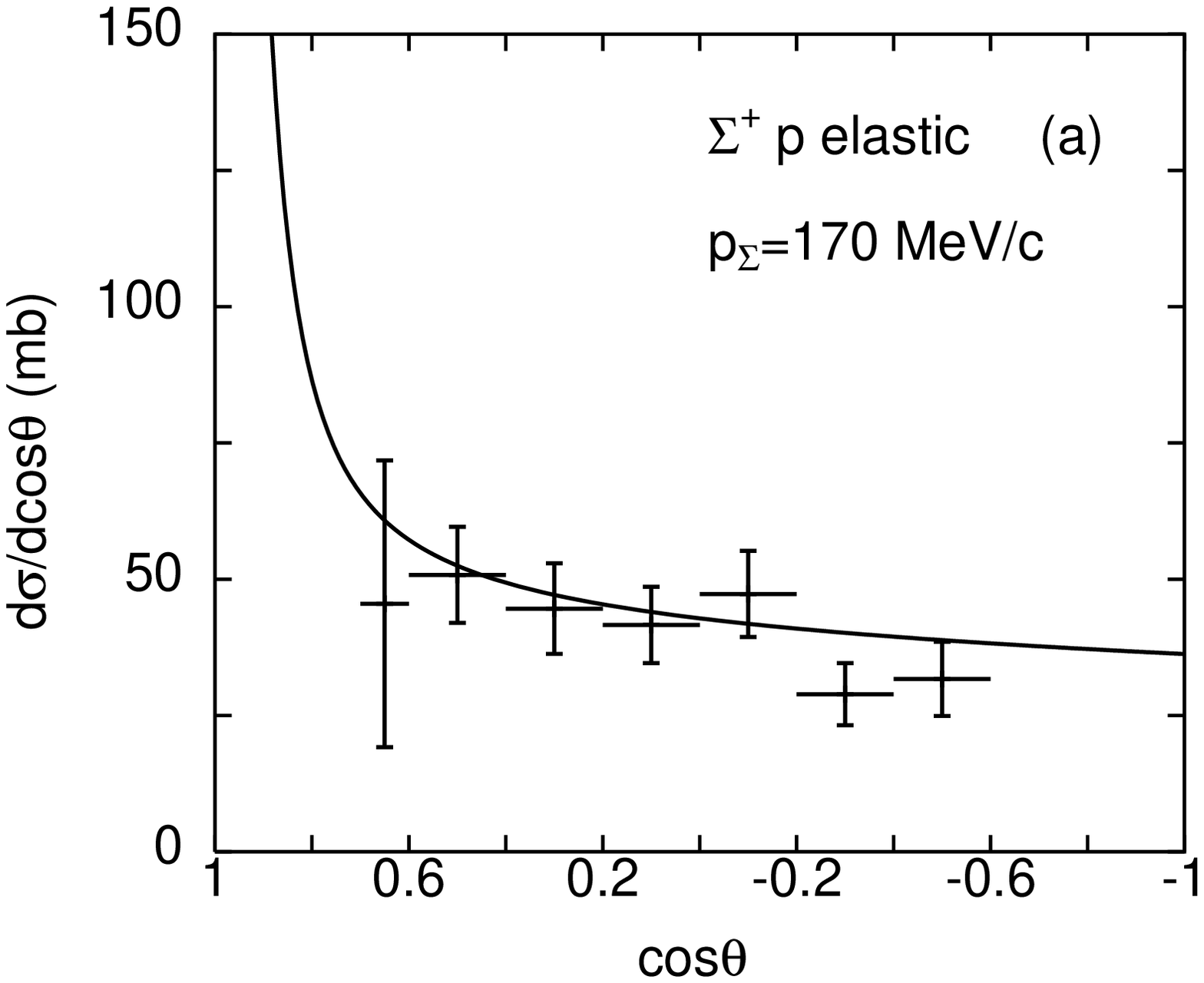}
\end{minipage}~%
\hfill~%
\begin{minipage}{0.47\textwidth}
\epsfxsize=\textwidth
\epsffile{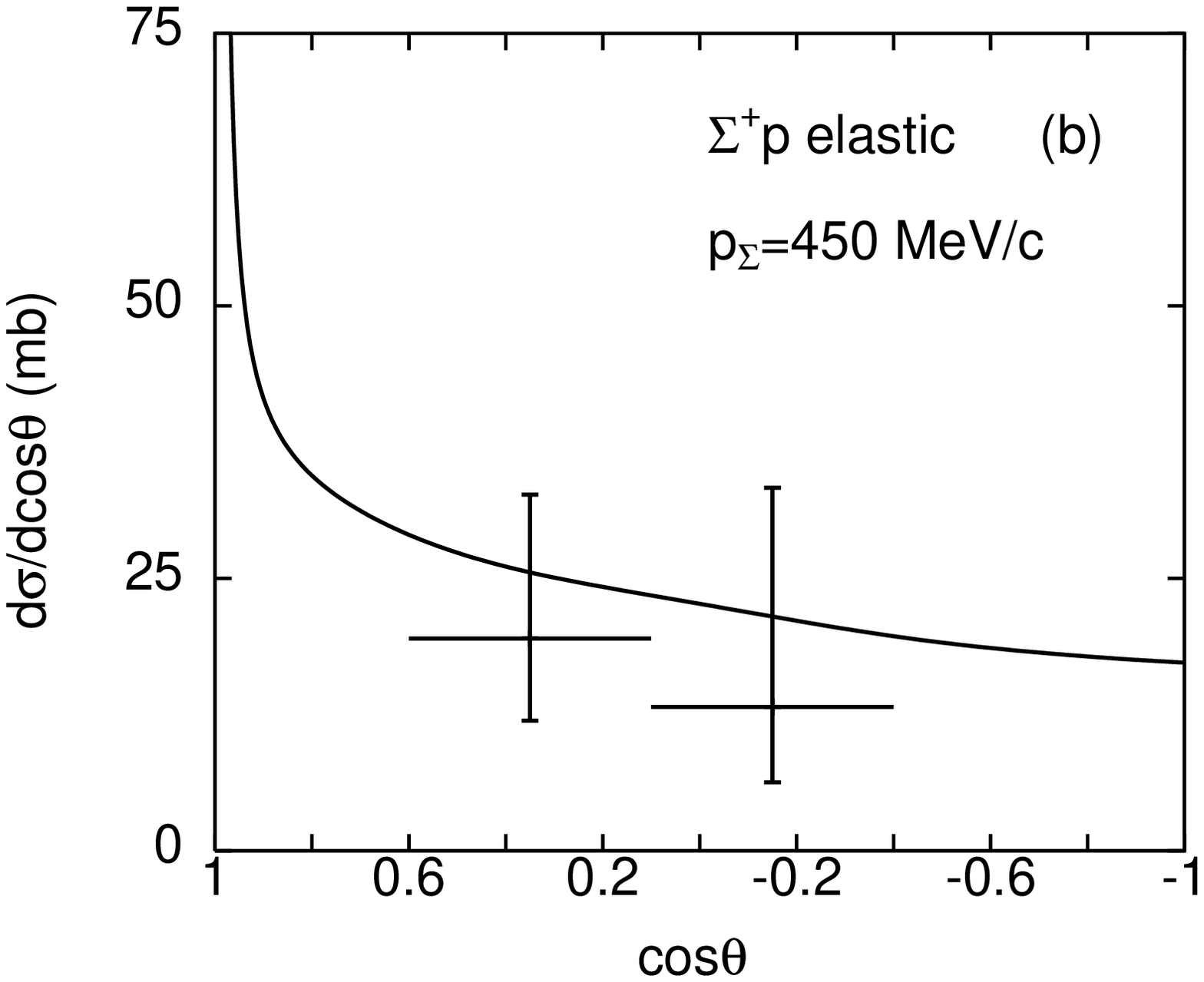}
\end{minipage}
\end{figure}

\vspace{-8mm}

\begin{figure}[h]
\begin{minipage}{0.47\textwidth}
\epsfxsize=\textwidth
\epsffile{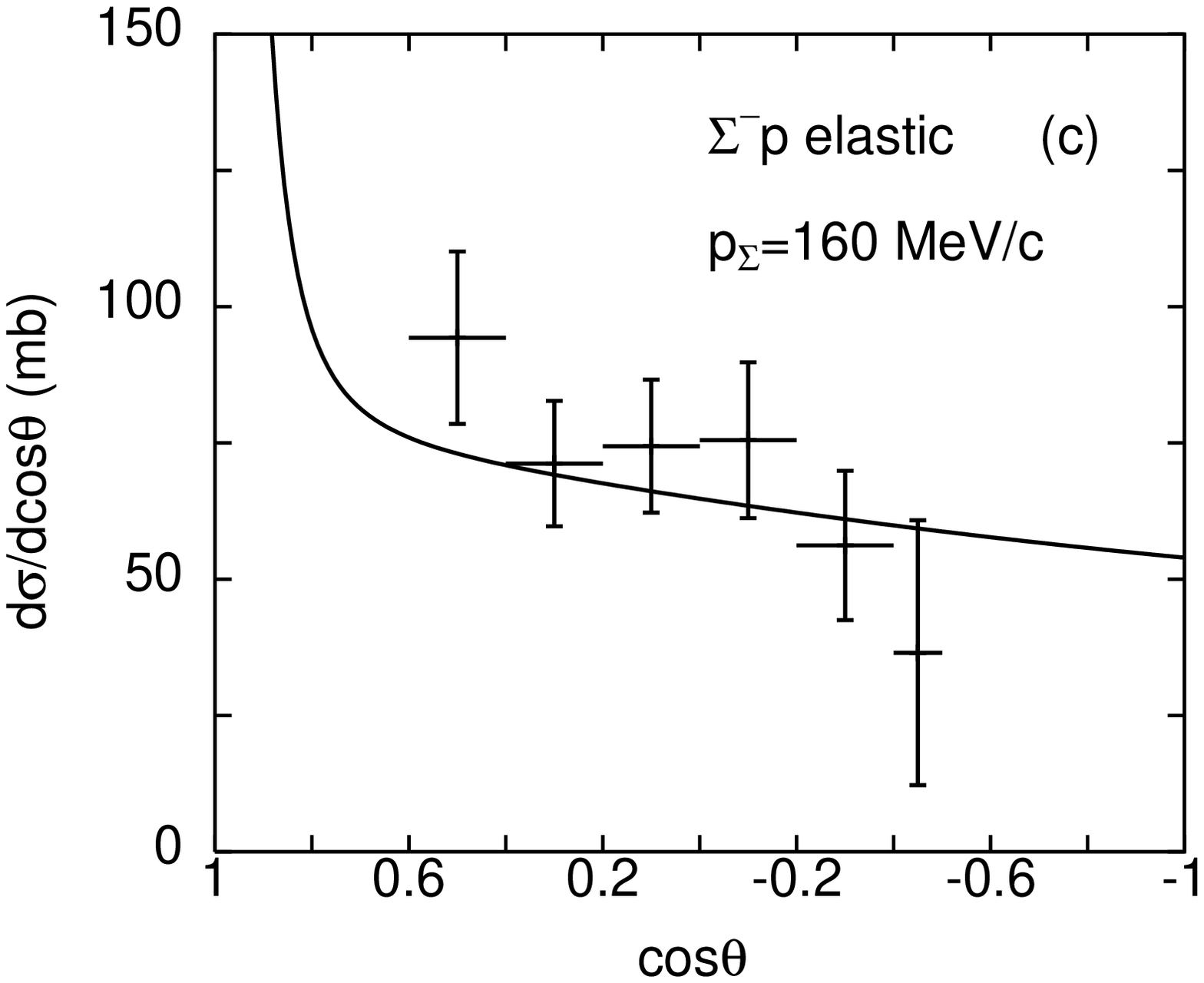}
\end{minipage}~%
\hfill~%
\begin{minipage}{0.47\textwidth}
\epsfxsize=\textwidth
\epsffile{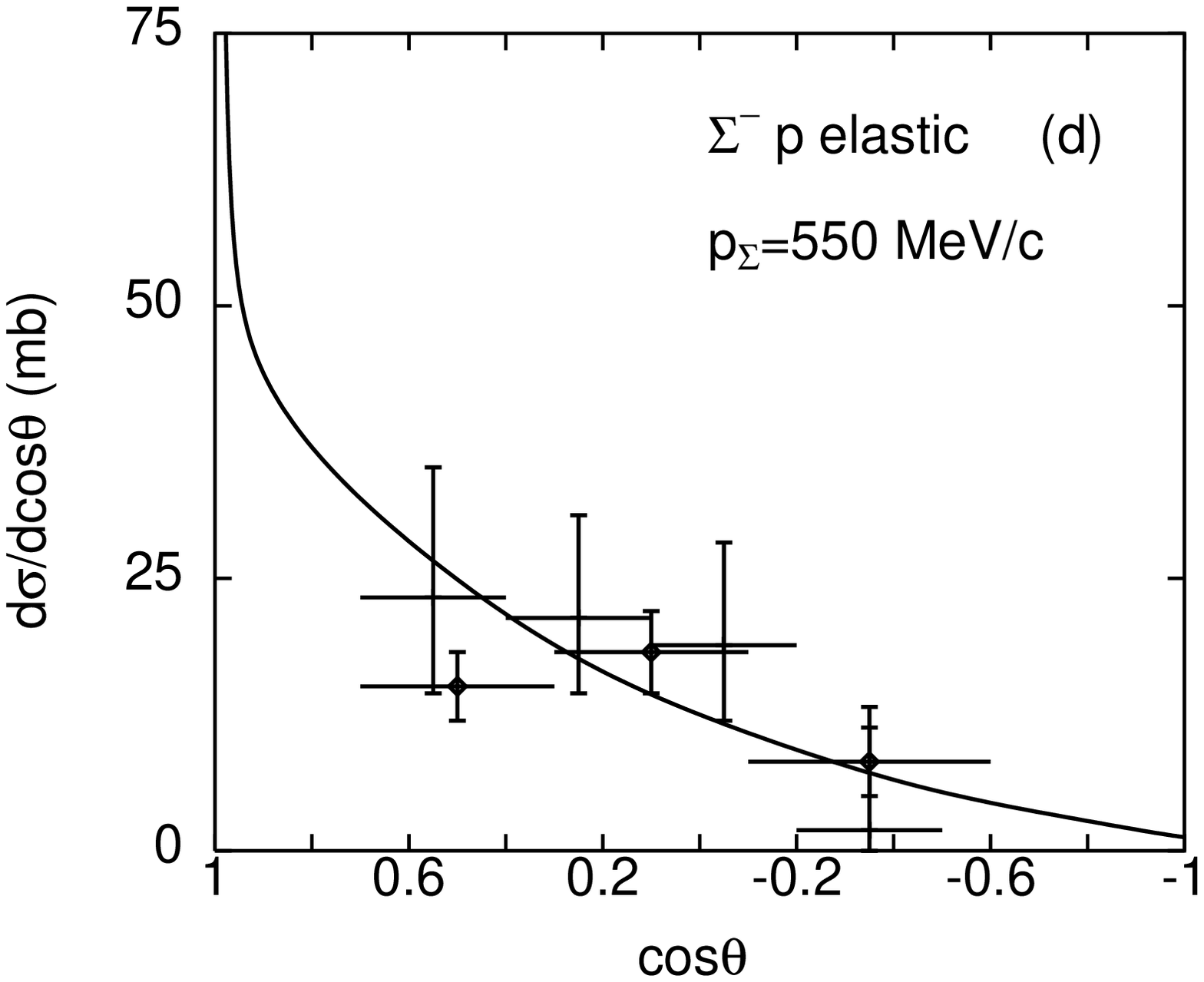}
\end{minipage}
\end{figure}

\vspace{-8mm}
\begin{figure}[h]
\begin{minipage}{0.47\textwidth}
\epsfxsize=\textwidth
\epsffile{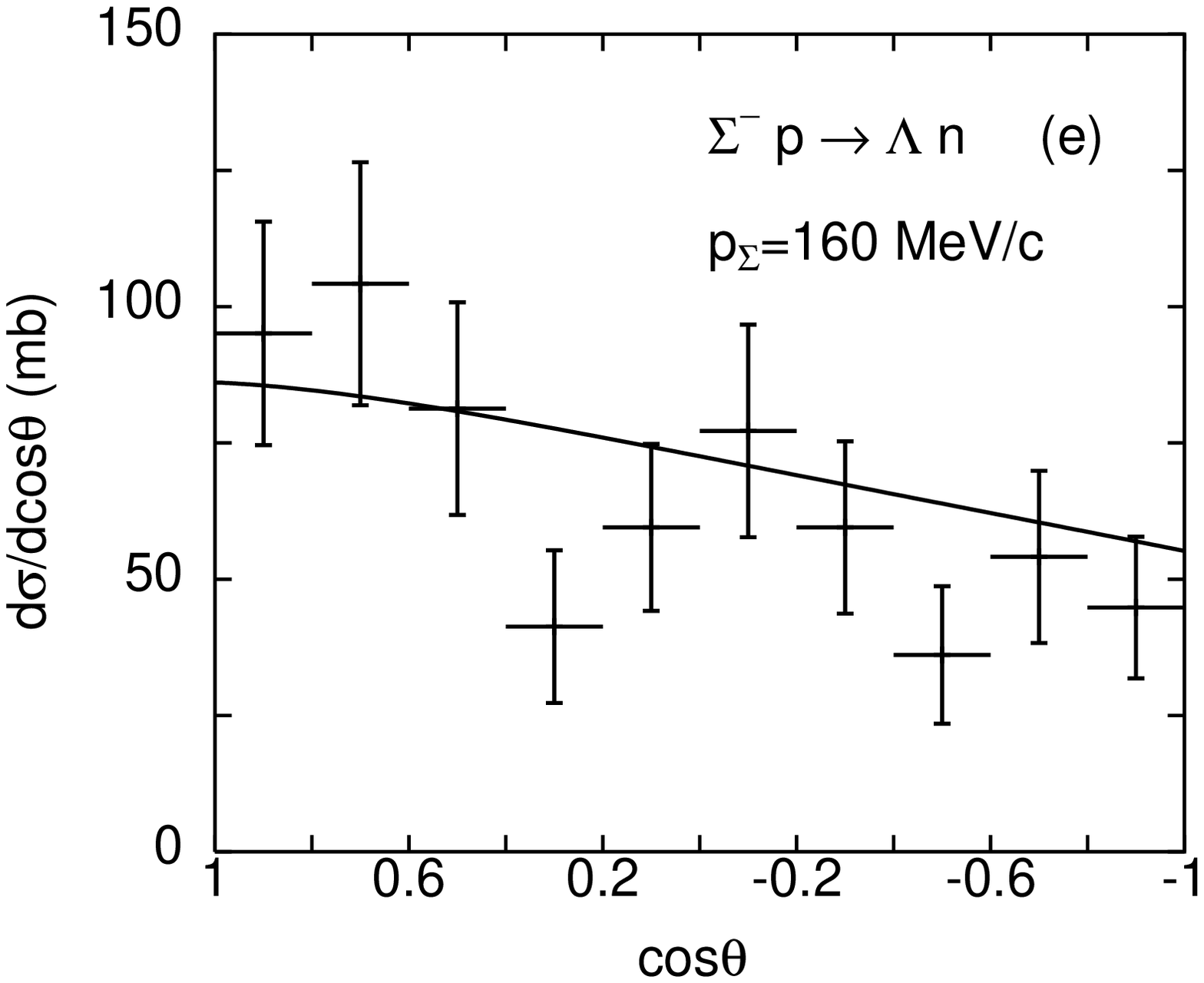}
\end{minipage}~%
\hfill~%
\hspace{10mm}
\begin{minipage}{0.47\textwidth}
\end{minipage}
\bigskip
\caption{
Calculated $\Sigma^+ p$ and $\Sigma^- p$ differential cross sections
by fss2, compared  with the experimental angular distributions:
(a) $\Sigma^+ p$ elastic scattering at $p_\Sigma=170~\hbox{MeV}/c$,
(b) the same as (a) but at $p_\Sigma=450~\hbox{MeV}/c$,
(c) $\Sigma^- p$ elastic scattering at $p_\Sigma=160~\hbox{MeV}/c$,
(d) The same as (c) but at $p_\Sigma=550~\hbox{MeV}/c$,
(e) $\Sigma^- p$ $\rightarrow$ $\Lambda n$ differential cross sections
at $p_\Sigma=160~\hbox{MeV}/c$.
The experimental data are taken
from \protect\cite{EN66} for (a), \protect\cite{E251} for (b),
\protect\cite{EI71} for (c) and (e), and \protect\cite{E289} for (d).
}
\label{yndif}
\end{figure}

\clearpage

\begin{figure}
\begin{figure}[h]
\begin{minipage}{0.47\textwidth}
\epsfxsize=\textwidth
\epsffile{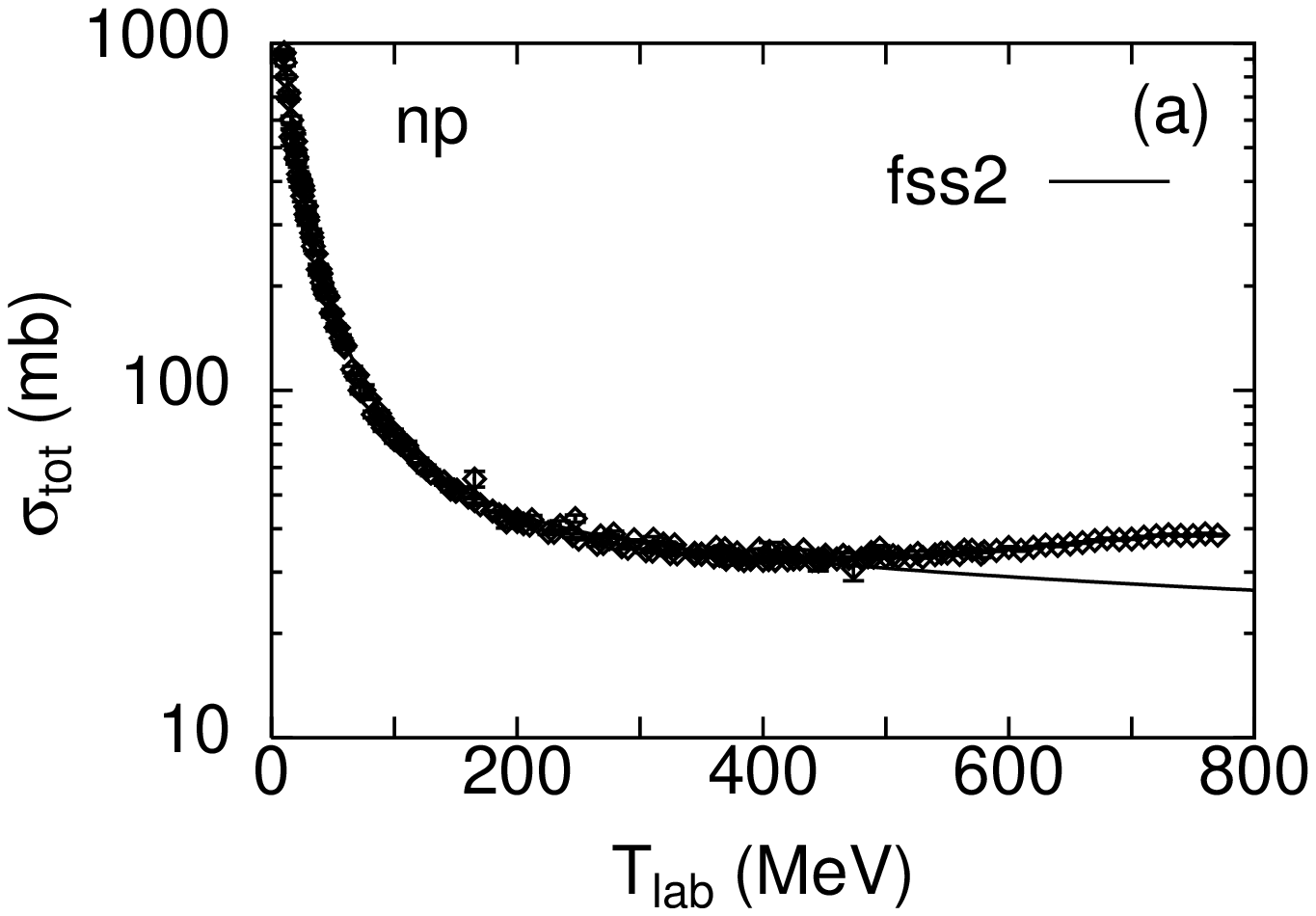}
\end{minipage}~%
\hfill~%
\begin{minipage}{0.47\textwidth}
\epsfxsize=\textwidth
\epsffile{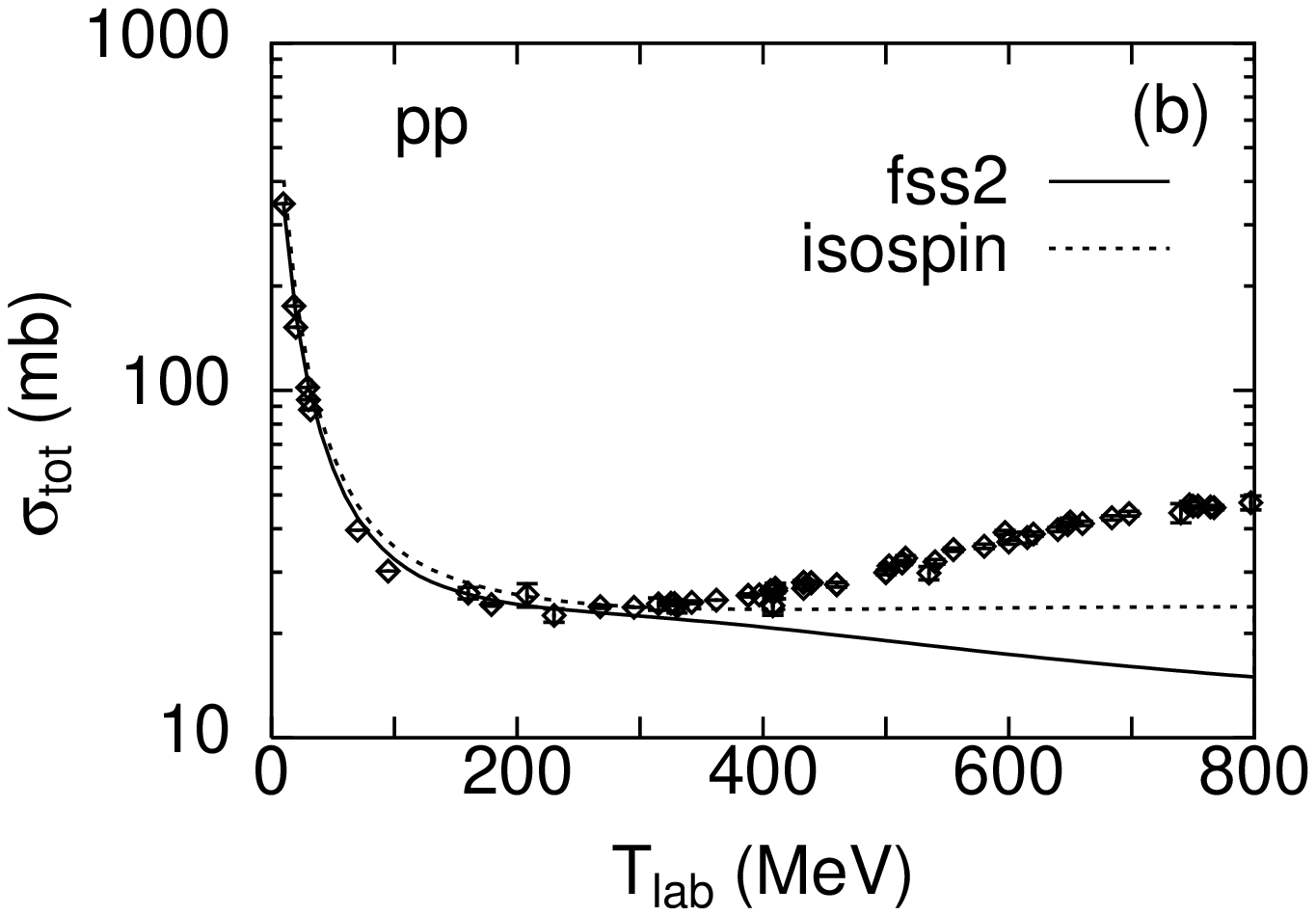}
\end{minipage}
\end{figure}

\vspace{-15mm}

\begin{figure}[h]
\begin{minipage}{0.47\textwidth}
\epsfxsize=\textwidth
\epsffile{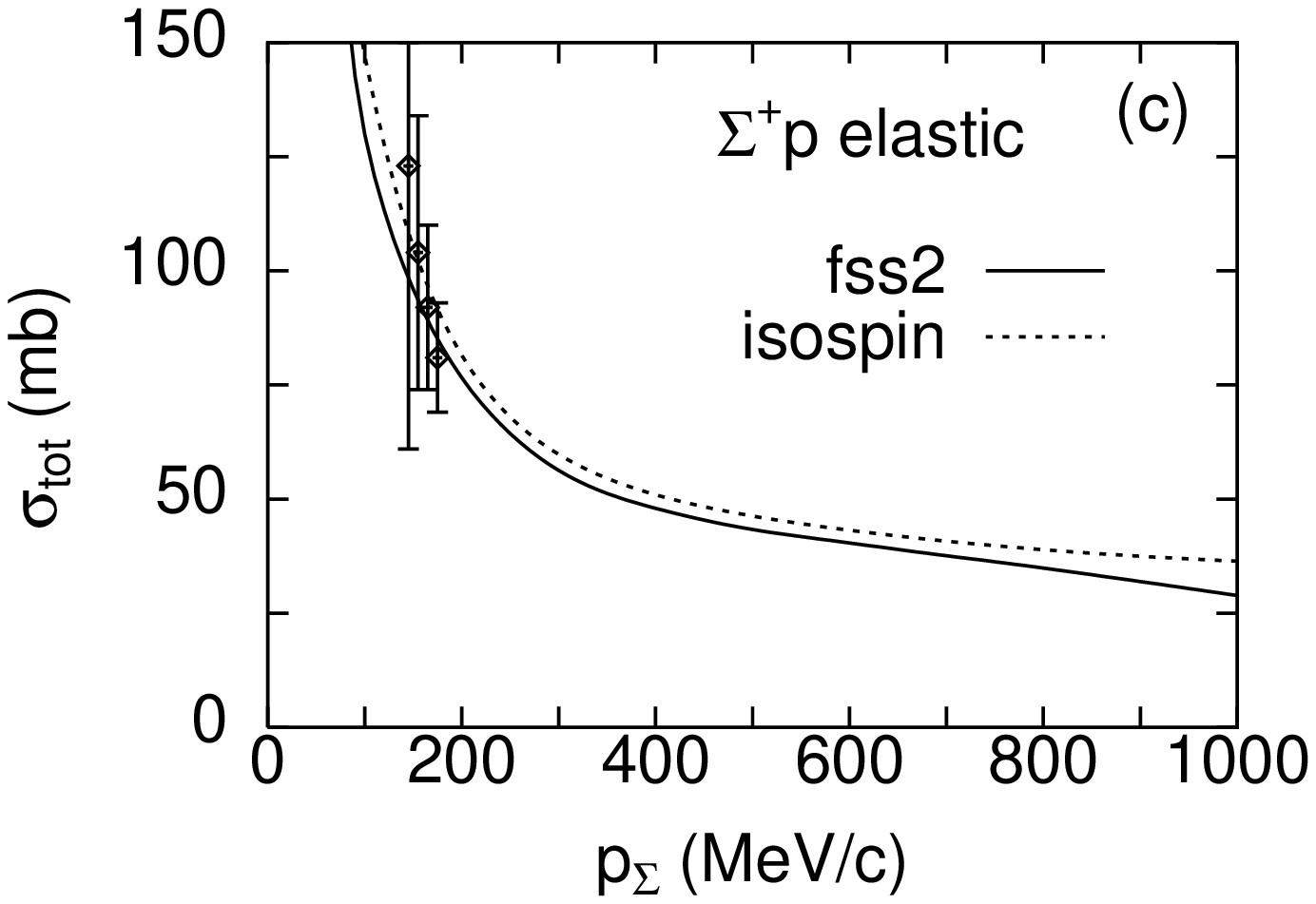}
\end{minipage}~%
\hfill~%
\begin{minipage}{0.47\textwidth}
\epsfxsize=\textwidth
\epsffile{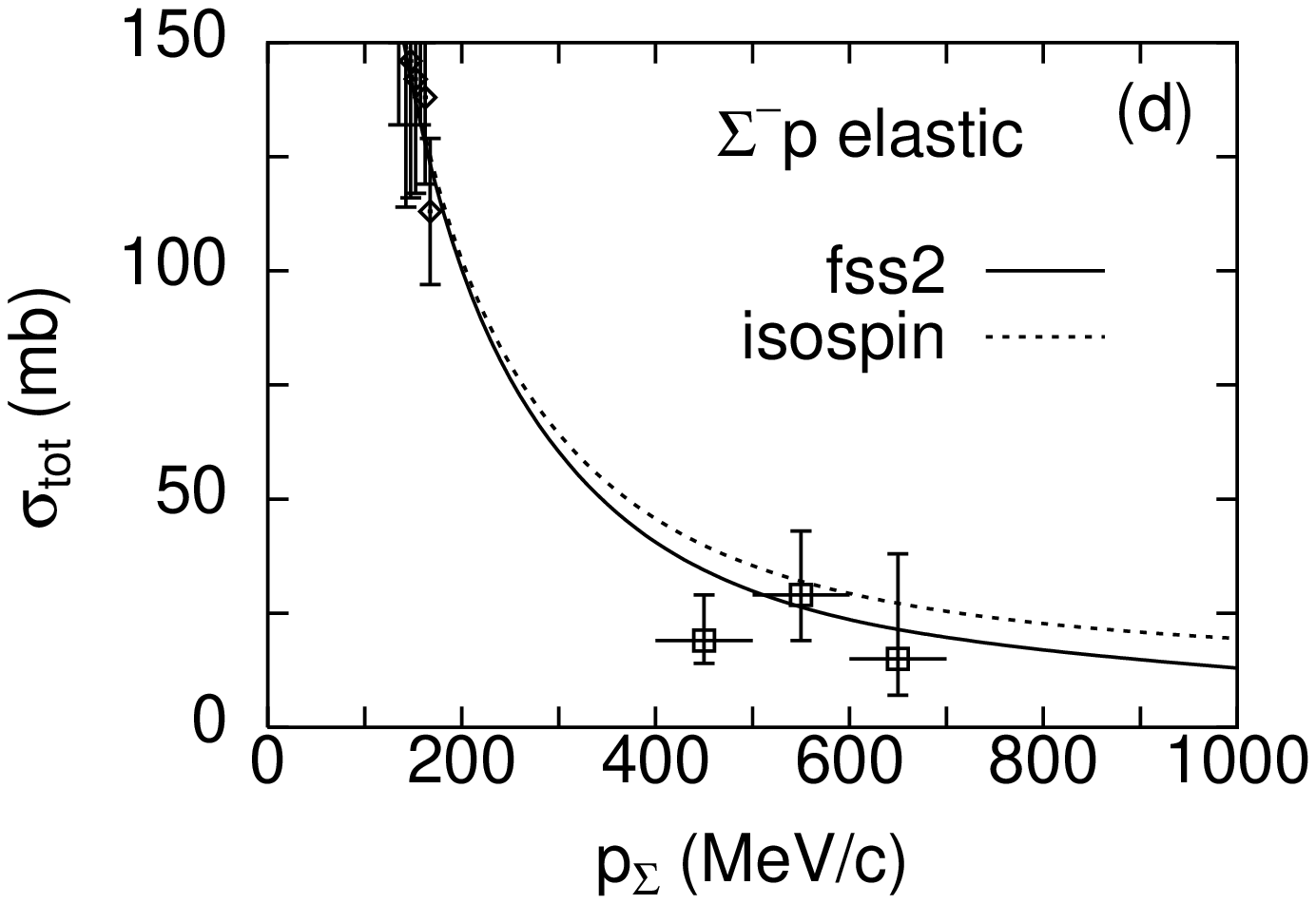}
\end{minipage}
\end{figure}

\vspace{-15mm}
\begin{figure}[h]
\begin{minipage}{0.47\textwidth}
\epsfxsize=\textwidth
\epsffile{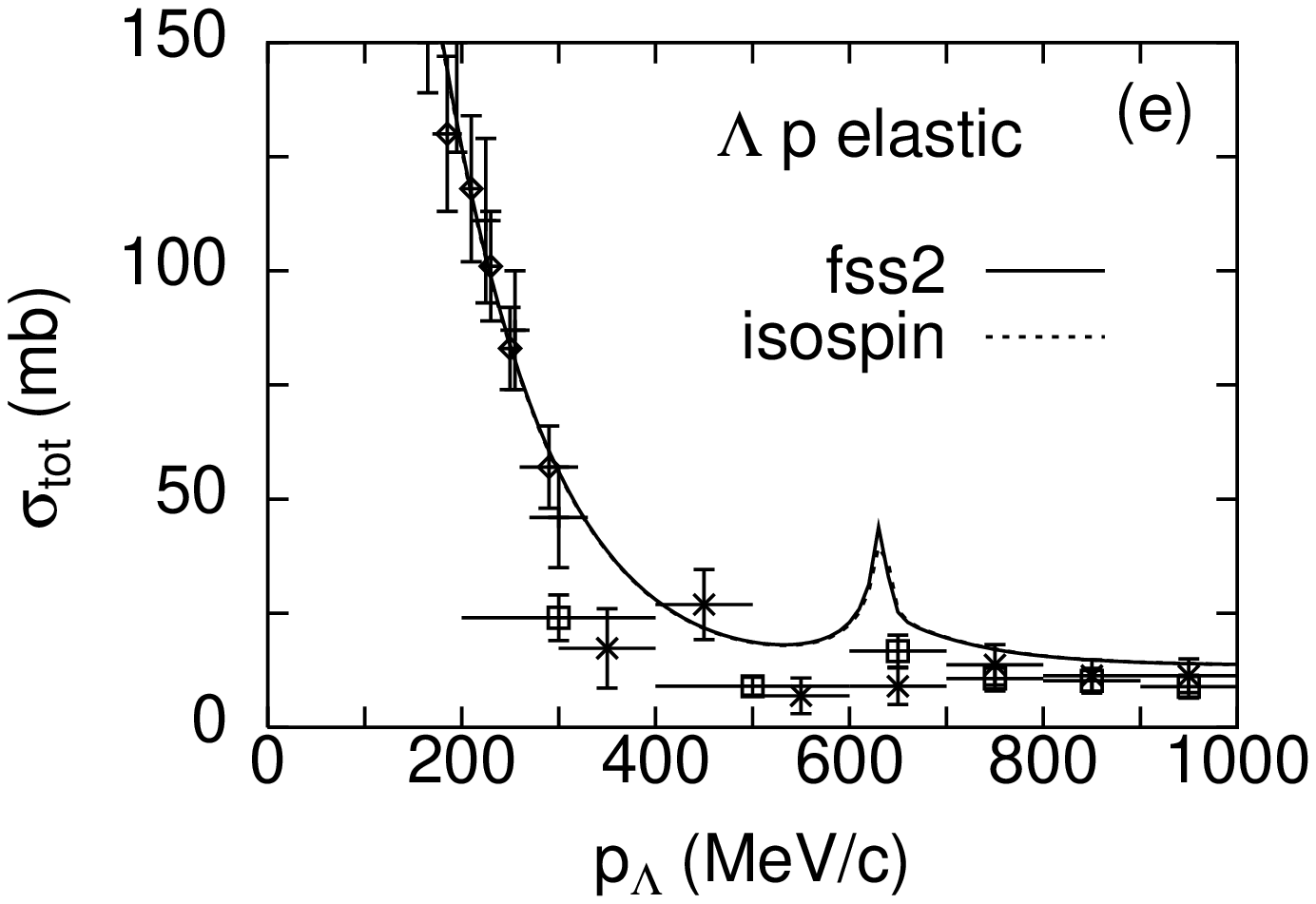}
\end{minipage}~%
\hfill~%
\begin{minipage}{0.47\textwidth}
\epsfxsize=\textwidth
\epsffile{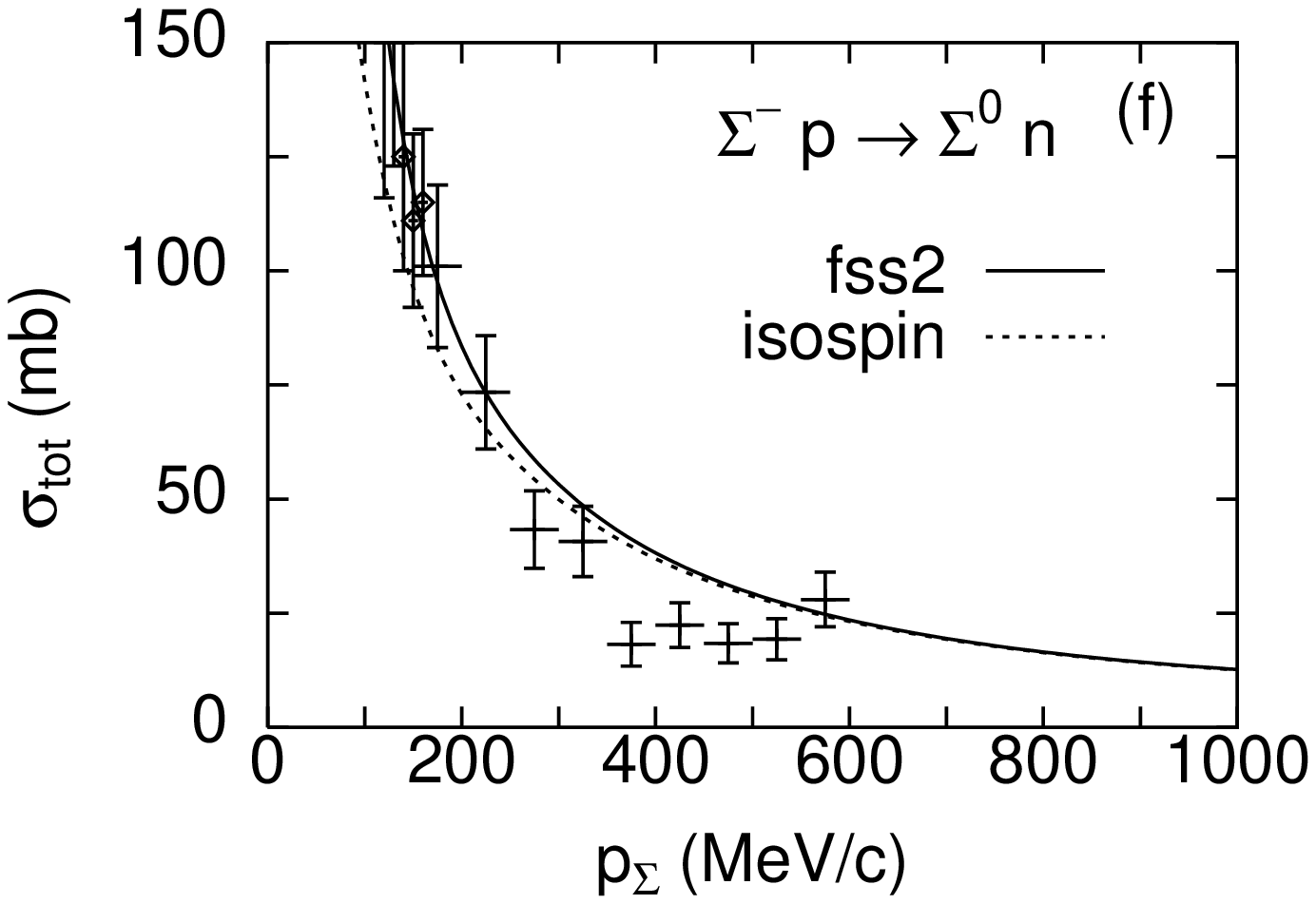}
\end{minipage}
\end{figure}
\vspace{-15mm}
\begin{figure}[h]
\begin{minipage}{0.47\textwidth}
\epsfxsize=\textwidth
\epsffile{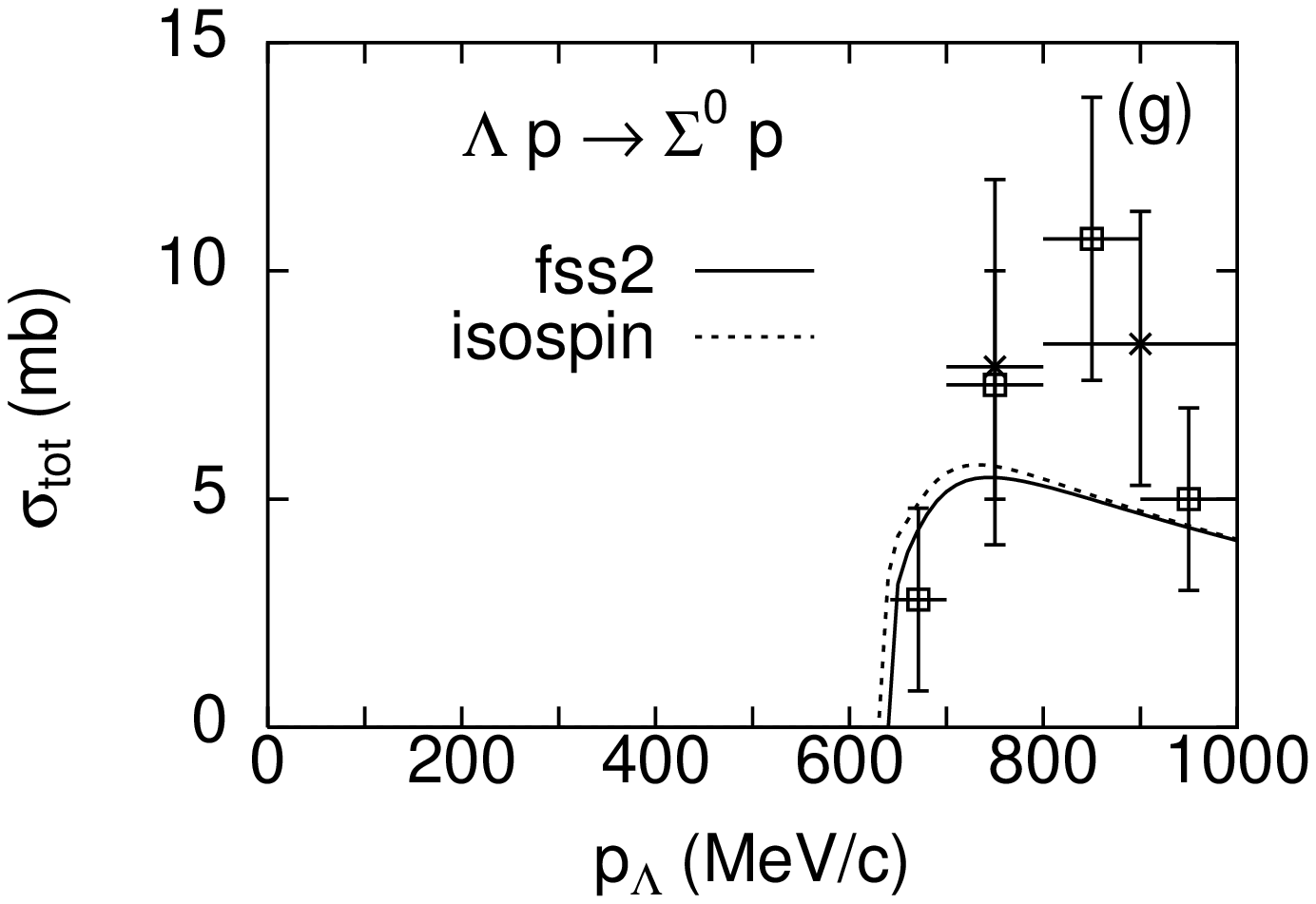}
\end{minipage}~%
\hfill~%
\begin{minipage}{0.47\textwidth}
\epsfxsize=\textwidth
\epsffile{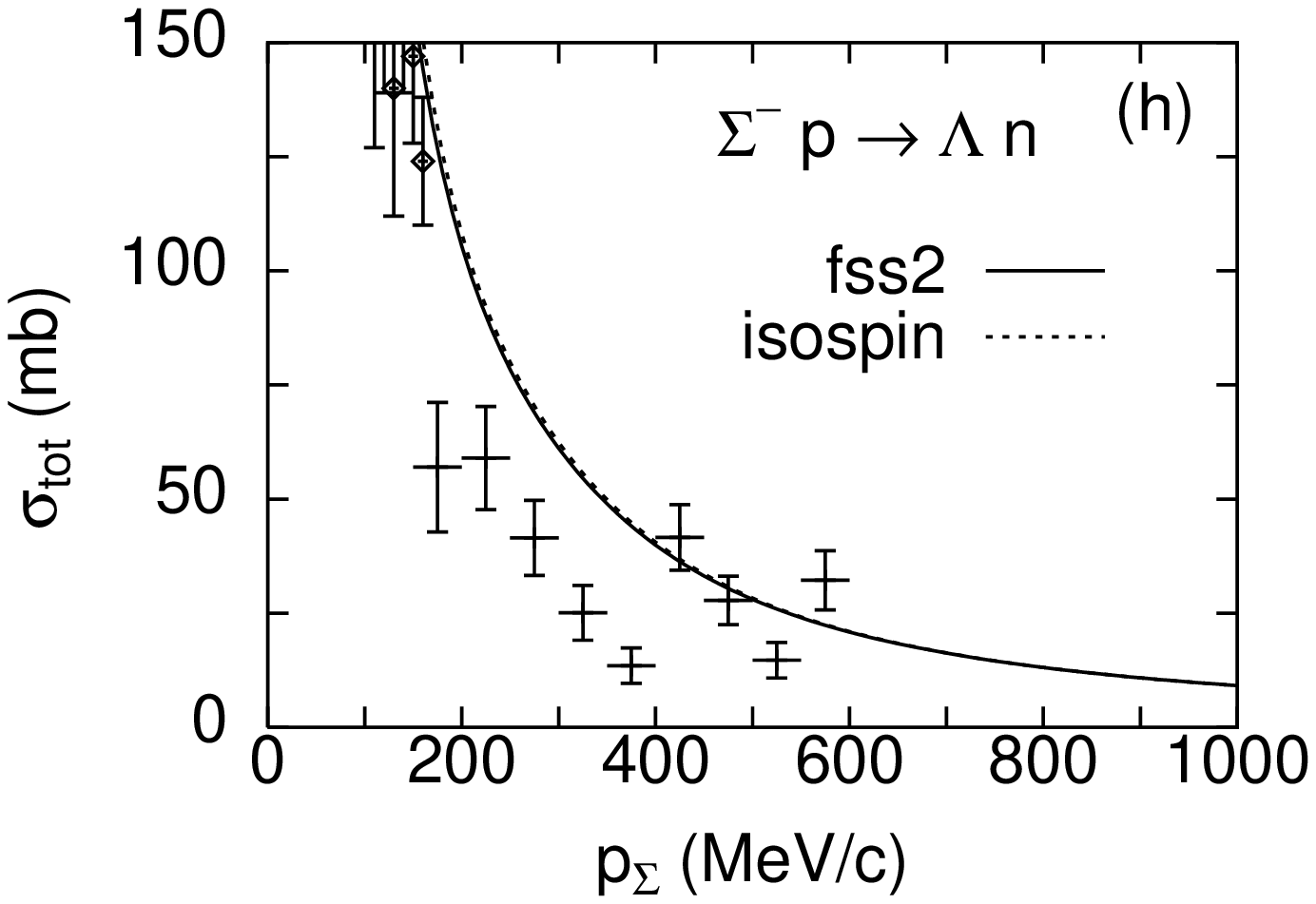}
\end{minipage}
\end{figure}
%
\caption{
Calculated $NN$ and $YN$ total cross sections by fss2,
compared  with the experimental data.
The solid curves denote the full calculation
and the dotted curves the calculation in the isospin basis.
The experimental data are taken
from \protect\cite{SAID} for $NN$,
\protect\cite{alex68}, \protect\cite{sechi68},
\protect\cite{kadyk71} for $\Lambda p$,
\protect\cite{EN66} for $\Sigma^+ p$, and \protect\cite{EI71},
\protect\cite{ST70}, \protect\cite{E289}
for $\Sigma^- p$ scattering.
}
\label{tot}
\end{figure}

\begin{figure}
\epsfxsize=0.47\textwidth
\epsffile{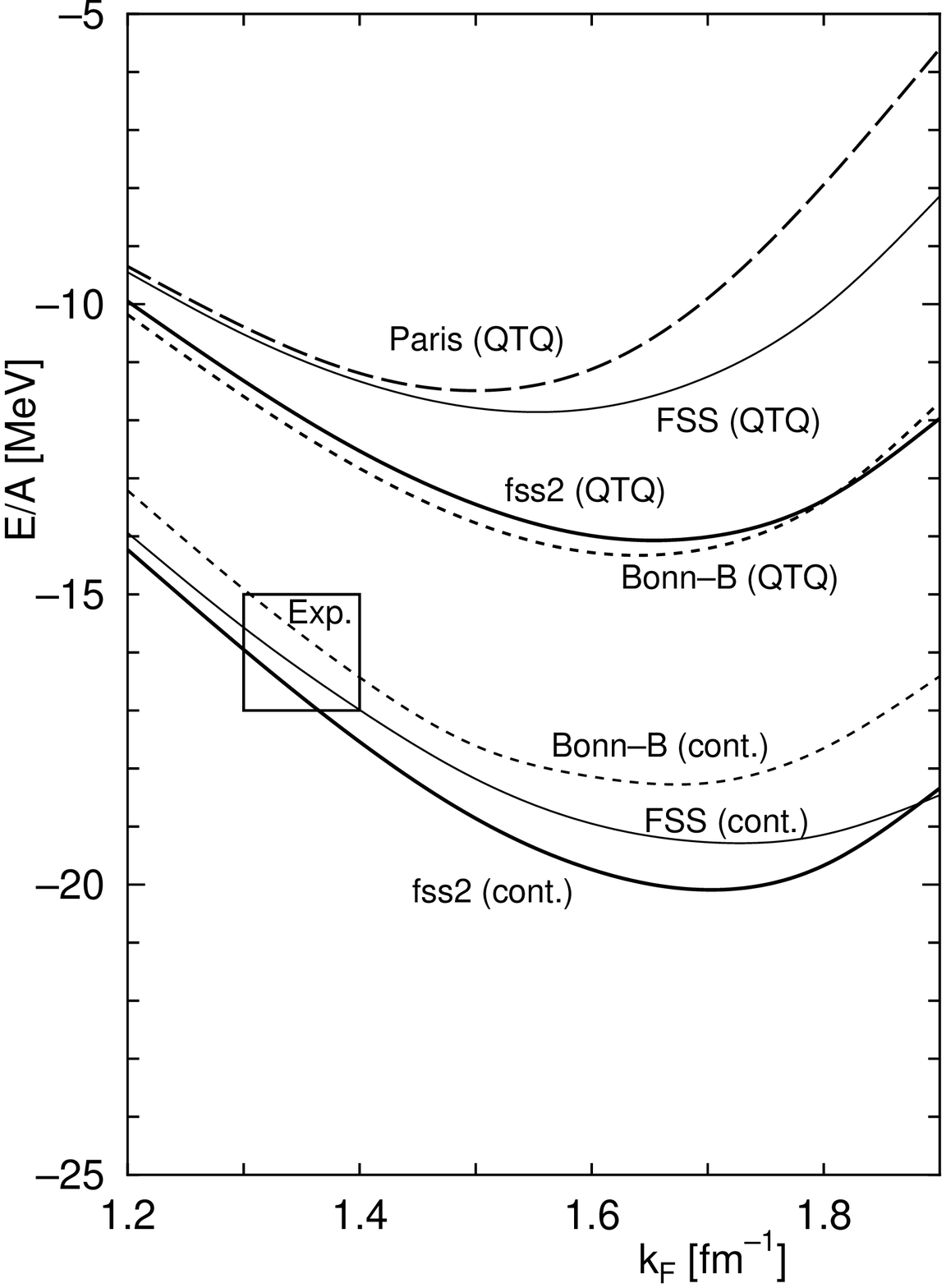}
\vspace{-20mm}
\bigskip
\caption{
Nuclear matter saturation curves obtained for fss2 and FSS,
together with the results of the Paris
potential \protect\cite{PARI} and
the Bonn model-B (Bonn-B) potential \protect\cite{MA89}.
The choice of the intermediate spectra is specified by "QTQ" and "cont.",
respectively. The result for the Bonn-B potential in the continuous
choice is taken from the non-relativistic calculation
in \protect\cite{BM90}.
}
\label{matter}
\end{figure}

\clearpage

\begin{figure}
\noindent
\begin{minipage}{0.4\textwidth}
\epsfxsize=\textwidth
\epsffile{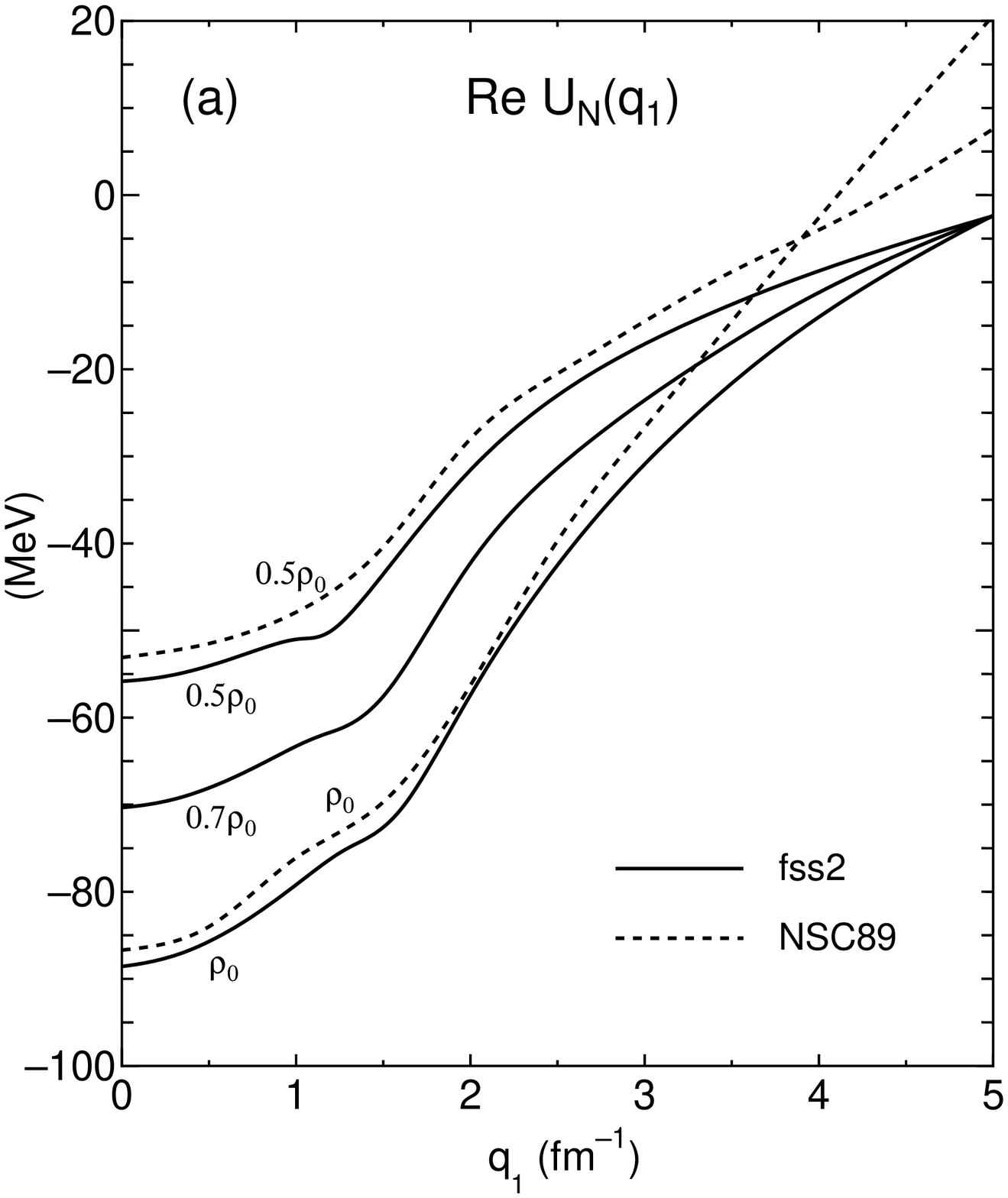}
\end{minipage}~%
\hfill~%
\begin{minipage}{0.4\textwidth}
\epsfxsize=\textwidth
\epsffile{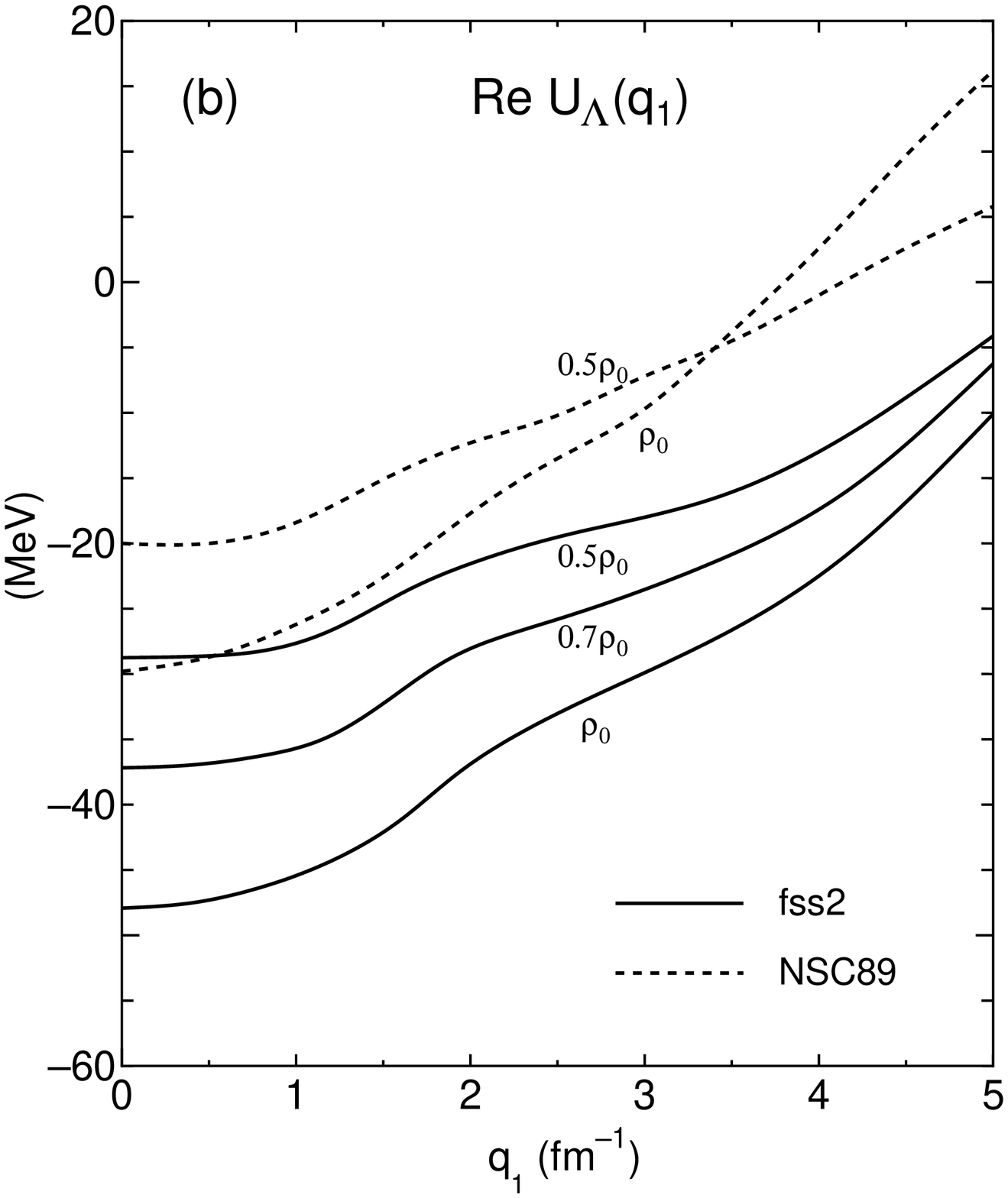}
\end{minipage}
\bigskip
\hfill~%
%
\begin{minipage}{0.4\textwidth}
\epsfxsize=\textwidth
\epsffile{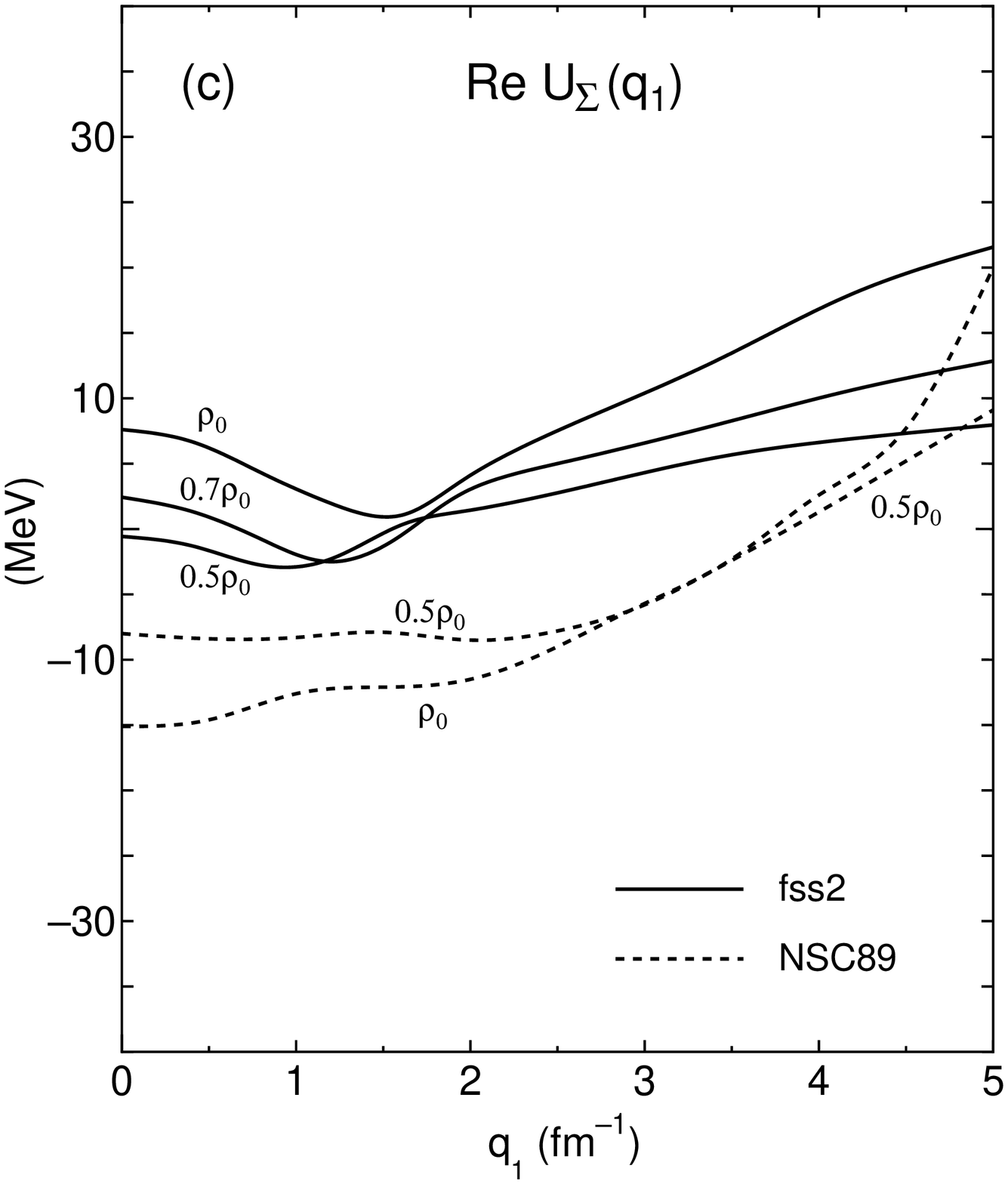}
\end{minipage}~%
\hfill~%
\begin{minipage}{0.4\textwidth}
\epsfxsize=\textwidth
\epsffile{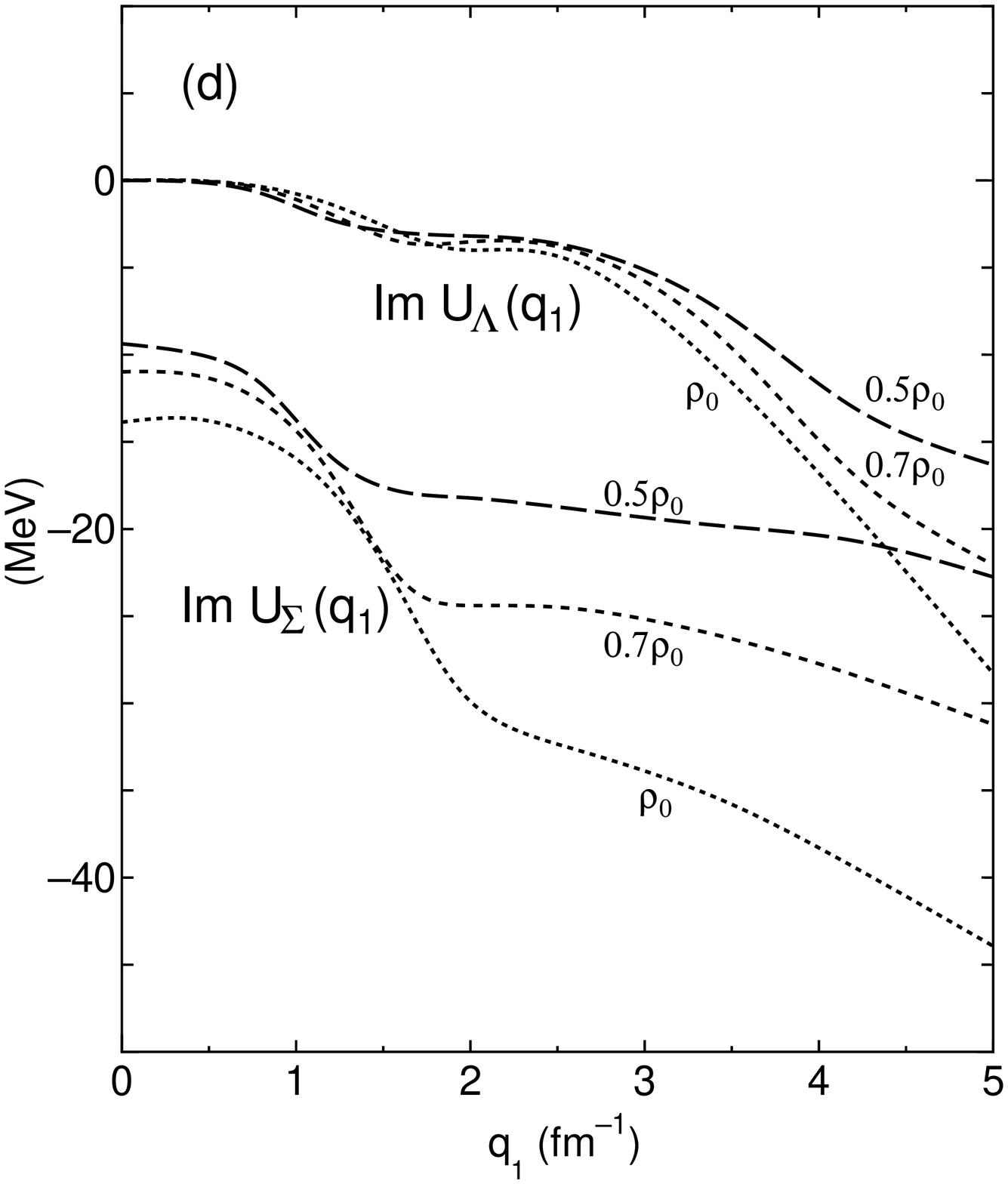}
\end{minipage}
\bigskip
\bigskip
\caption{
(a) The nucleon s.p. potential $U_N (q_1)$ in nuclear
matter in the continuous choice for intermediate spectra.
Predictions by fss2 for three densities $\rho= 0.5\,\rho_0,
~0.7\,\rho_0$ and $\rho_0$ are shown.
Here the normal density $\rho_0$ corresponds
to $k_F =1.35~\hbox{fm}^{-1}$.
The dashed curve is the result achieved by Schulze {\em et al.}
\protect\cite{SCHU} with
the Nijmegen soft-core $NN$ potential NSC89 \protect\cite{NSC89}.
(b) The same as (a) but for the $\Lambda$ s.p. potential
$U_{\Lambda}(q_1)$.
(c) The same as (a) but for the $\Sigma$ s.p. potential
$U_{\Sigma}(q_1)$.
(d) The same as (a) but for the imaginary part
of the $\Lambda$ and $\Sigma$ s.p. potentials $U_B(q_1)$
predicted by fss2.
}
\label{spnuc}
\end{figure}

\end{document}